\documentstyle{mn}

   \title[GEMINI GMOS-IFU spectroscopy of BAL+IR+FeII QSOs: II. IRAS 04505-2958]
         {GEMINI 3D spectroscopy of BAL+IR+Fe II QSOs: \\
         II. IRAS 04505$-$2958 an explosive QSO with hypershells\\
         and a new scenario for galaxy formation and galaxy end
}

\author[L\'{i}pari et al.]
   {S. Lipari$^{1}$,  M. Bergmann$^{2}$, S.F.Sanchez$^{3}$, B. Garcia$^{4}$,
   R. Terlevich$^{5,6}$, E. Mediavilla$^{4}$,    
   \newauthor
Y. Taniguchi$^{7}$,  W. Zheng$^{8}$, B. Punsly$^{9}$, K. Jahnke$^{10}$,
A. Ahumada$^{1,}$$^{ 11}$, D. Merlo$^{1}$\\
$^{1}$ C\'ordoba Observatory and CONICET, Laprida 854, 5000 C\'ordoba, Argentina.\\
$^{2}$ Gemini Observatory, La Serena, Chile.\\
$^{3}$ Calar Alto Observatory, C/Jesus Durban Remon 2-2, E-04004 Almeria, Spain.\\
$^{4}$ Instituto de Astrofisica de Canarias, 38205 La Laguna, Tenerife, Spain.\\
$^{5}$ Institute of Astronomy, Madingley Road, Cambridge CB3 OHA, UK. \\
$^{6}$ Instituto Nacional de Astrofisica Optica y Electronica (INAOE), Puebla, Mexico.\\
$^{7}$ Research C. for Space \& Cosmic Evolution, Ehime Univ., Matsuyama 790-8577, Japan.\\
$^{8}$ Depart. of Physics and Astronomy, John Hopkins Univ., Baltimore, MD 21218, USA. \\
$^{9}$ Centre for Relativistic Astrophysics, Univ. of Rome La Sapienza, Italy and USA.\\
$^{10}$ Max-Planck-Institute fur Astronomie, Konigstuhl 17, D-69117, Hidelberg, Germany. \\
$^{11}$ ESO Post-Doc, Santiago and Paranal, Chile.
}

\date{Received     ;
      in original form }

\pagerange{\pageref{firstpage}--\pageref{lastpage}}
\pubyear{2005}

\begin{document}

\maketitle

\label{firstpage}

\begin{abstract}

From a study of BAL + IR + Fe {\sc ii} QSOs (using deep Gemini GMOS-IFU
spectroscopy) new results are presented, for IRAS 04505$-$2958.
Specifically, we have studied in detail the outflow (OF) process, at two
large galactic scales:
(i) two blobs/shells at radius r $\sim$1.1 and 2.2 kpc; and
(ii) an external hypergiant shell at r $\sim$11 kpc.
In addition, the presence of two very extended hypergiant
shells at r $\sim$60--80 kpc  is discussed.

From this GMOS study the following main results were obtained:
(i) For the external hypergiant shell   
the kinematics GMOS maps of the ionized gas ([O {\sc ii}], [Ne {\sc iii}],
[O {\sc iii}], H$\beta$) show a small scale bipolar OF, with 
similar properties to those observed in the prototype of exploding
external super shell:  NGC 5514.
(ii) Three main knots --of this hyper shell S3-- show the presence of a
young starburst.
(iii) The  two internal shells show  OF components with typical properties
of nuclear shells.
(iv) The two blobs and the hyper shell are aligned at PA $\sim$
131$^{\circ}$ showing bipolar OF shape, at $\sim$10--15 kpc scale.
In addition,
the more external shells (at $\sim$60--80 kpc scale) are aligned at
PA $\sim$ 40$^{\circ}$ with also bipolar OF shape (perpendicular to
the more internal OF).
(v) A strong blue continuum and multiple emission line components were
detected in all the GMOS field.

The new GMOS data show a good agreement with an extreme + explosive OF
scenario for IRAS 04505-2958; in which part of the ISM of the host galaxy was
ejected (in multiple shells).
This extreme OF process could be also associated with
two main processes in the evolution of QSOs:
(i) the formation of companion/satellite galaxies by giant explosions; and
(ii) to define the final mass of the host galaxy, and even if the
explosive nuclear outflow is extremely energetic, this process could
disrupt an important fraction of the host galaxy.
Finally, the generation of UHE cosmic rays and neutrino/dark--matter
--associated with HyNe in explosive BAL + IR + Fe {\sc ii} QSOs-- is
discussed.

\end{abstract}



\section{INTRODUCTION}
\label{intro}

There is increase observational evidence confirming
that galactic outflow (OF), broad absorption line (BAL) processes,
super/hypernova (SN/HyN) explosions and the associated shells play
important roles in galaxy and QSO evolution and formation
(specially at high redshift, in the young universe; see 
Frye, Broadhurst, Benitez 2002;  Steidel et al. 2000; Taniguchi \& Shioya
2000; Dawson et al. 2002; Ajiki et al. 2002; Iwamuro et al. 2002;
Maiolino et al. 2003, 2004a,b; L\'{i}pari et al 2005a;
L\'{i}pari \& Terlevich 2006; Smith et al. 2007, 2008).

At low z, HST images and 3D spectroscopic data of the
interesting class of composite BAL + IR + Fe {\sc ii} QSOs
shows in practically all of these objects  {\it ``giant and
shells with circular--symmetric shape in the external borders
(with their centre at the
position of the nucleus)"}; which are associated  with
strong OF processes and giant explosive events (see for details
L\'{i}pari et al. 2003, 2005a, 2007a,b, 2008, 2009). \\


\subsection{\bf Evolutionary IR Colour-Colour Diagram}
\label{IRdiagram+QSO}

The IR colour-colour diagram is an important tool to detect
and discriminate different types of activity in the nuclear regions of
galaxies. Thus this diagram is also important for the study of
possible links between different phases of galaxy and QSO evolution.
Using this IR colour-colour diagram
[$\alpha$${(60,25)}$ vs. $\alpha$${(100,60)}$],
L\'{i}pari (1994) found that the IR colours  of  $\sim$10 extreme  IR +
Fe {\sc ii} QSOs are distributed between the power law (PL) and the
black-body (BB) regions: i.e., the {\it transition area}.
From a total of $\sim$10 IR transition IR + Fe {\sc ii} QSOs four systems
show {\bf low ionization BALs}. Therefore, we already
suggested that low ionization BALs + IR + Fe {\sc ii} QSOs
could be associated with the {\it young phase of the QSO evolution}.

Using the data base of more than 50 IR mergers and IR QSOs
with OF and galactic winds (GW),
L\'{i}pari et al. (2005a; their Fig. 15) showed the IR
energy distribution
for IR mergers and IR QSO with OF.
This diagram shows:

\begin{enumerate}

\item
All the IR mergers with low velocity OF and starburst are located very
close to the BB area.

\item
The standard QSOs and radio QSOs are located around the PL
region.

\item
All the BAL + IR + Fe {\sc ii} QSOs are located in the transition region,
in  a clear sequence: from
Mrk 231 (close to the BB area) $\to$  IRAS 07598+6508
$\to$ IRAS04505--2958 $\to$ IRAS 21219-1757
$\to$ IRAS/PG 17072+5153 and IRAS 14026+4341 (close to the PL area) $\to$
standard QSOs. \\

\end{enumerate}

\noindent
{\bf The IR Diagram and the BAL system in IRAS 04505-2958:}

Using this IR colour-colour diagram, L\'{i}pari et al.\ (2005a; their Fig.\ 15),
found the BAL system in IRAS~04505$-$2958. For the BAL detection we used
the fact that IRAS 04505-2958 is located exactly in the sequence of
BAL + IR + Fe {\sc ii} QSOs:
between the BAL QSOs IRAS 07598+6508 and IRAS 21219-1757/IRAS 17072+5153.

The spectra of IRAS 04505-2958  show: (i)  clearly 
BAL system, and (ii) strong Fe {\sc ii} emission. Moreover,
several authors already showed that the dominant IR source (IRAS 04505-2958)
is likely associated with the QSO (see Section 11, for a detailed
discussion about this point).

The BAL system in IRAS 04505-2958 is relatively narrow and very similar
to those detected in Mrk 231.
The standard definition of BAL QSOs (Weymann et al. 1991) is based in
the measurement of  the equivalent width of the C {\sc iv}$\lambda$1550
resonance absorption line system (called balcity index: BI).
Hall et al. (2002) proposed a less restrictive index to include a wider
range of  line widths.

More recently, Hamann \& Sabra (2003) strongly advocated the use of
simple quantitative indices (for BAL systems), and they
proposed the following definition:
{\bf BAL QSOs have continuous absorptions $>$ 2000 km s$^{-1}$}.
The UV HST-FOS spectra of the QSO-core of
IRAS 04505-2958 (see Fig. 8c) clearly shows that the C {\sc IV}$\lambda$1550
absorption fits this criteria. Since the absorption start at
$\lambda$1986 \AA,  and the continuous absorptions reach at least
$\lambda$1973 \AA: thus the range of continuous absorptions is
2515 km s$^{-1}$.


\subsection{BAL + IR + Fe {\sc ii} QSOs and Hypernova}
\label{BAL+IR+FEII-QSO}

Some of the observational results obtained for \emph{nearby BAL QSOs},
such as extreme IR and Fe\,{\sc ii} emission,
strong blue asymmetry/OF in H$\alpha$, radio quietness, and
very weak [O\,{\sc iii}]$\lambda$5007 emission
(Low et al. 1989; Boroson \& Meyers 1992; L\'{\i}pari et al. 1993, 1994,
2003, 2005a; Turnshek et al. 1997),
 can be explained in the framework of the starburst + AGN + OF scenario.
In our study of Mrk\,231 and  IRAS\,0759+6559
(the nearest extreme BAL + IR + GW + Fe\,{\sc ii}  QSOs), we
detected typical characteristics of  young QSOs with extreme
nuclear starburst.
In particular, for  Mrk 231 we found evidence that the BALs
systems are associated with the composite nature of the nuclear regions:
i.e., OF generated by explosive SN events and  the radio-jet
(L\'{i}pari et al. 2005a, 2009; Punsly \& Lipari 2005).

For BAL + IR + Fe {\sc ii} QSOs we suggested that these QSOs could be
young, and composite QSOs at the {\bf end phase of an extreme starburst}.
At the final stage of an ``extreme starburst", i.e. type {\sc ii}
SN/HyN phase ([8--60]\ $\times10^{6}$\,yr from the initial burst;
Terlevich et al.\ 1992, 1993) powerful galactic winds, super/hypergiant
galactic shells, BAL systems, extreme Fe\,{\sc ii} emission, large
amount of dust, and strong IR emission can appear
(L\'{i}pari \& Terlevich 2006; L\'{i}pari et al. 2003).

The first starburst phase (0--3 $\times$ 10$^6$ yr: which is dominated by
hot main sequence stars with H{\sc ii} regions)  is associated with
the presence of large amount of dust and extreme IR emission (Terlevich et
al. 1993; Franco 2009, private communication).

\noindent
{\bf Hypernovae in IR QSOs:}

Theoretical works suggested that type {\sc ii} SN/HyN generate the blowout
phase of the supergiant shells and bubbles (Norman \& Ikeuchi 1989).
However,
in dusty nuclear regions of IR  QSOs and mergers + shells
(with $A_V\sim$\,10-1000\,mag; see Genzel et al. 1998), the
presence of type {\sc ii} SNe/HyNe could be detected only for
the nearest IR merger and QSO: Arp 220 and NGC 7469.
Which were detected using the largest
--very long baseline-- radio interferometry (VLBI) array
(see  Lonsdale et al. 2006; Parra et al. 2007; Colina et al. 2001).

A very interesting point about the radio-SN/HyN found in Arp 220 and
NGC 7469 is that almost all these HyNe are of the type {\sc ii}n
(i.e., their progenitors are massive stars, which explode
in  a dense circumstellar medium generated by their stellar wind).
These unusual highly luminous core-collapse radio-SNe/HyNe are 
implying a different stellar initial mass function (with a large number of
massive stars) in the nuclei of IR QSOs and mergers.


\subsection{Shells  in BAL + IR + Fe {\sc ii} QSOs}
\label{shells+OF+HyN}

Using Gemini, La Palma-WHT and HST observations we are studying 
shells associated to outflowing ``shocked" material.
Which have properties very different  to ring and arcs 
associated with tidal tails and loops in galactic collisions.

3D high resolution spectroscopic data give  clear evidences of strong
OF processes; mainly from the study of multiple emission line
components, kinematics maps and  emission line ratios
plus colour maps with structures associated with shocks 
(L\'{\i}pari et al. 2004a,b,d, 2005a, 2006, 2007a,b, 2008, 2009)

The presence of multiple concentric expanding supergiant shells
in young and composite BAL + IR + Fe {\sc ii} QSOs (specially shells
with their centre at the position of the nucleus and with highly symmetric
circular external--borders) could be
associated  with giant  symmetric explosive events (L\'{i}pari et al
2003, 2005a, 2007a,b). Moreover, only an explosive scenario 
could explain the exponential shape of the  variability curve
observed in the BAL  system-{\sc iii} of Mrk 231
(L\'{i}pari et al. 2005a, 2009).

Furthermore, we found --for IR QSO and Mergers--
in the shells of Mrk 231 and NGC 5514 plus
in the nuclei of PG 1535+547, IRAS 01003-2238 and IRAS 22419+6049:
spectral features of massive WR stars and OF.
These WR stars are  progenitors of core--collapse super and
hypernova.  Theoretical works  suggested that SNe/HyNe from
massive progenitors are probably the only objects that could generate the
rupture phase of the bubbles, in the QSOs nuclei and in the main 
knots of the shells (Norman \& Ikeuchi 1999; Tomisaka \& Ikeuchi 1988).

Thus, ``circumnuclear and external shells and arcs"
could be associated  with:
(i) the final phase of the galactic--wind, i.e., the blowout 
of the galactic bubbles (Tomisaka \& Ikeuchi 1988; Norman
\& Ikeuchi 1989; Suchkov et al. 1994); and
(ii) galaxy collisions: i.e., tidal tails, rings, loops, etc.
For distant QSOs  (and even for some nearby QSOs/galaxies)
it is difficult to discriminate between these two types of structures.
However,
it is well known that the velocity fields of mergers and galaxies in 
interaction show emission line components with difference of velocities
$\Delta$V $<$ 500-600 km s$^{-1}$;
and in extreme OF the multiple components show diferences of velocities
$\Delta$V $>$ 700 km s$^{-1}$.
Theoretical results obtained for galactic wind --associated with
strong starbursts-- show multiple OF components with even $\Delta$V $>$
2000 km s$^{-1}$ (Suchkov et al. 1994).
This is one of the more clear difference between these two types of
shells and arcs.


\subsection{IRAS 04505-2958}
\label{intro3}

The mid and far IR emission of IRAS 04505-2958 was associated with
a luminous quasar, at z $=$ 0.286, with L$_{FIR} =$ 3.55 $\times$
10$^{12}$ L$_{\odot}$ and M$_V =$ --25.8
(de Grijp et al. 1987, 1992; Low et al. 1988, 1989; Hutching \& Neff 1988;
L\'{i}pari et al. 2003, 2005a, 2007a,b, 2009; L\'{i}pari \& Terlevich 2006;
Kim et al. 2007; Zhou  et al. 2007; Letawe, Magain \& Courbin 2008).

The first optical images and spectroscopy of this IR source (obtained
by Hutching \& Neff 1988 and Low et al. 1989) showed a bright QSO, a close
foreground G star (at 2$''$ to the NW, from the QSO) plus a possible
tidal tail to the SE (also at $\sim$ 2$''$, from the QSO). 
HST WFPC2  images by Boyce et al. (1996) show that the possible SE
``tail" is a complex structure. They suggested that this 
structure could be associated with a galaxy with ring shape,
which is interacting with the QSO host galaxy.
Some authors suggested that the possible ring  galaxy could be
the main source of ultraluminous IRAS emission, instead of the QSO
(Canalizo \& Stockton 2001; Magain et al. 2005; Merrit et al.
2006; and others).

From a detailed study of
IR QSOs and mergers with strong OF, BALs and Fe\,{\sc ii} emission
(using HST morphological and 3D spectroscopic data)
L\'{i}pari et al. (2003, 2005a, 2007a,b, 2009)
suggested that the SE tail/ring 
structure --in IRAS\,04505-2958-- could
be a very large scale  shell, with an extension of $\sim$20--30 kpc.
Which could be associated  to an extreme nuclear OF process.
In the present paper, a strong starburst was detected in this shell.
Thus, probably the observed IRAS emission could be
associated with the QSO plus the shell.
IRAS 04505-2958
was already included in our published data base of BAL + IR + Fe {\sc ii}
QSOs (L\'{i}pari et al. 2005a; see also L\'{i}pari et al. 2003, 2007b, 2009;
L\'{i}pari \& Terlevich 2006).

From a study of host galaxies in a sample of 17 QSOs,
Magain et al. (2005) found --from this sample of QSOs--
that only in the case of the QSO HE 0450-2958
the host galaxy was not detected. They suggested that the
host galaxy of this QSO could be dark or absent
(i.e., a naked QSO).
Several authors analysed the theoretical scenarios
for a naked QSO in IRAS 04505-2958 (Haehnlet, Davies \& Rees 2005;
Merrit et al. 2006; Hoffman \& Loeb 2006). In addition,
Merrit et al. (2006) derived the mass of the super massive black
hole (SMBH) of the QSO, considering that this QSO is a  high luminosity
version of narrow line Seyfert 1 AGNs. They obtained a low
value for the mass of the SMBH, thus they suggested that the host
galaxy of this QSO could be less massive and less luminous than the
previously assumed values.

Thus,  from different points of view IRAS 04505-2958 is one of the
more interesting QSO.
Throughout the paper, a Hubble constant of H$_{0}$ = 75 km~s$^{-1}$
Mpc$^{-1}$ will be assumed.
For IRAS 04505-2958 we adopted the distance of $\sim$1144 Mpc. 
This distance was obtained from the redshift of the  narrow emission
lines (see Section 5): with a mean value of redshift
z = 0.2860 and cz = 85800 $\pm$10 km~s$^{-1}$
(the angular scale is 0.1$'' \approx$550 pc).

\clearpage

\section{THE PROGRAMME AND EXPLOSIVE MODEL OF BAL + IR + FE II QSO }
\label{progra+model}                               

In order to study and discuss the GMOS results obtained for
IRAS 04505-2958 it is important to
summarize --previously-- some observational and theoretical results
obtained in the programme  of BAL + IR + Fe {\sc ii}
QSOs, and in the study of explosive  models for QSOs.

In our observational study of BAL + IR + Fe {\sc ii} QSOs and IR
mergers/QSOs with OF we have combined  high resolution HST images
and 3D spectroscopic data (using Gemini+GMOS,  La Palma William
Herschel Telescope+Integral and Calar Alto+PMIS) for: \\

\noindent
{\it (I) Nearby IR QSOs and mergers with OF + shells}. For NGC 5514, Arp 220,
NGC 2623,  NGC 3256, and others.\\

\noindent
{\it (II) Nearby BAL + IR + Fe {\sc ii} QSOs}. For Mrk 231, IRAS 04505$-$2958,
IRAS 17072+5153, IRAS 07598+6508, IRAS 14026+4341, IRAS 21219-1757, etc. \\

\noindent
{\it (III) BAL + IR + Fe {\sc ii} QSOs at medium and high redshift }.
For SDSS 030000.58+004828.0, SDSS 143821.40+094623.2 (both at z $>$ 0.5), and
Sub-mm low ionization  BAL SDSS-QSOs (at z $>$ 2.0).

The general goal of these observational programmes is to study the kinematics,
physical conditions and morphology of the gas and the stars in 
BAL + IR + Fe {\sc ii} QSOs. In Paper I, we have explained the 
particular goals of these programme, which  can be summarized as
follows: \\

\noindent
{\it (i) To study the physic of composite OF and BAL processes:}
Those associated with
(a) supergiant explosive events, likely
generated by Hyper Novae and multiple SNe; and
(b) bipolar OF probably generated by sub-relativistic jets.

\noindent
{\it (ii) To investigate the role of hypergiant explosive events
in the formation and end of galaxies:}
Specially, to study the effect of super/hyper explosive events in the
formation of companion and satellite galaxies and also in the host
galaxies of BAL QSOs.

\noindent
{\it (iii) To analyze the role of HyN in the generation of ultra high
energy CR and Neutrinos:}
We have a special interest to study the Astrophysical consequeces of
one the main component of the explosive model for QSOs: the HyN
explosions and their role in the posible generation of CR and Neutrino.
In addition, the observational data of explosive BAL + IR + Fe {\sc ii}
QSOs could help to test the different  theoretical models for the
generation of CR and Neutrinos.


\subsection{\bf Explosive Model for QSOs: the interaction Black Hole  \&
Starburst}
\label{evolmodel}

An evolutionary, explosive and composite model was proposed for
QSOs and the  formation and evolution of galaxies. This scenario involves:
a SMBH, a nuclear starburst (SB), 
an extreme OF and an accretion region
(L\'{i}pari \& Terlevich 2006; L\'{i}pari 1994; L\'{i}pari et al. 1994,
2003, 2005a, 2007a,b, 2008, 2009).
This model is based --in part-- in the main evolutionary sequences of
IR mergers, IR QSOs, elliptical galaxies, etc derived from  the
study of the IR colour-colour diagram.

In this scenario a bi-parametric   {\bf evolutionary} model for AGNs
was proposed.
Intrinsic parameters like the BAL,  Fe\,{\sc ii}/BLR intensity,
NLR size and luminosity, IR emission and radio luminosity, all evolve
with a time scale of less than $10^8$ yr.
Young AGNs are obscured BAL and strong Fe\,{\sc ii} + IR emitters
with relatively narrow line BLR and a compact and faint NLR.

In this model, IR mergers fuel and generate extreme star formation
processes and  AGNs, resulting in strong dust/IR emission and large
number of SN and HyN events (likely located in starburst rings or toroids).
The more energetic of these explosives events, will generate super/hypergiant
expanding shells, bubbles and extreme OF processes.

\noindent
{\bf Giant Explosions Associated with the Interaction of
SMBH + Starburst:} \\

In general, we have suggested that in {\it composite QSOs and AGNs}
the interaction --in the nuclear regions-- of four main processes:
the star formation process, the SMBH/AGN, OF and the accretion process
(of the ISM gas) could generate special condition --in the accretion
regions-- for the formation of
very massive stars and the associated giant explosive events, i.e. hypernovae.

More specifically,  a theoretical study of this composite scenario was
performed by Collin \& Zahn (1999). They developed a model for the outer
--gravitationally unstable regions-- of accretion disks around SMBHs of
10$^6$--10$^{10}$ M$_{\odot}$ and primeval abundance. They studied 
the evolution of the star formation in a gaseous marginally disk
showing that unstable fragments collapse rapidly to compact objects (mainly
protostars). Which then accrete at high rates. In less than 10$^6$ yrs
they acquire a mass of a few tens of M$_{\odot}$ (according to the process
suggested by Artymowicz et al. 1993). These massive stars explode as SNe/HyNe.
The shells of SNe/HyNe break out of the disk producing very strong OF.
The disk is able to support a large number of massive stars and SN/HyN.
In addition, the giant SN generate neutron stars, which can undergo
other high rate accretion process: leading to other very powerful HyN
explosions.

Then, in a second step, they assume that the region of the periphery of
the disk provide  a quasi stationary mass inflow during the lifetime of
quasars (i.e. $\sim$10$^8$ yrs). The whole mass transport is ensured
by the SN/HyN, inducing a transfer of angular momentum to the exterior. \\

\noindent
{\bf Low and Extreme Velocity OF:} \\

L\'{\i}pari et al.\ 2005a, 2004a,b, 2003) present a data base with the main
properties of more than 50 IR mergers and IR QSOs with OFs/galactic winds.
Ussing this data base two interesting results were found: \\

(i) ``Low velocity OF"  (LVOF, $V_{\rm LVOF} <$
700\,km\,s$^{-1}$, L\'{\i}pari et al.\ 2003, 2004a,b, 2005a) were found only
in IR mergers with  Starburst and LINER properties.
This result is consistent with those obtained by Lutz, Veilleux \& Genzel
(1999) and Veilleux, Kim, \& Sanders (1999): they detected that the main
source of ionization in Luminous IR Galaxies and Ultra Luminous IR Glaxies
--derived from {\itshape ISO}, optical and near IR polarimetry observations--
is LINERs associated  with starbursts and shocks (originated in
galactic winds). \\

(ii) ``Extreme velocity OF"  (EVOF, $V_{\rm EVOF} >$ 700
km s$^{-1}$) were found only in IR QSOs/AGNs with composite nuclear source:
AGN/QSO + Starburst.
Thus, we suggested that the interaction between  QSOs/AGNs and  Starburst
 could generate extreme velocity OF (associated with giang explosive/HyN
 events).  \\

\noindent
{\bf Strong Fe\,{\sc ii} emission:} \\

In the explosive and composite model for QSOs/AGNs,
the observed properties of QSOs with Fe\,{\sc ii} emission
can be understood as an evolutionary sequence, in which the observed differences
between strong and weak Fe\,{\sc ii} emitters and the observed correlations
with Fe\,{\sc ii}  are related --at least in part-- to
evolutionary changes in the SN, compact SNR activity and the development of
the OF + NLR.

In the composite model,
the BLR could be  produced  --in part-- in compact SNR (cSNR)
and the observed emission lines are the product of reprocessing of shock
radiation by two high density thin shells and the ejecta (Terlevich et al.
1992). In this model the abundances of the ionized
gas emitting the BLR lines are the abundances of the envelope of the star
and the SN ejecta and not the abundances of the ISM.

In addition, L\'{i}pari (1994) and
Lawrence et al. (1997) already proposed that the nuclear OF is a main
process, that could explain some of the Fe {\sc ii} correlations and
properties, observed in AGNs and QSOs.
More recently,
using our database of IR QSOs with galactic winds and OF we
found a correlation between the Fe\,{\sc ii}$\lambda4570$/H$\beta$ vs.
velocity of OF (see Fig. 30 of L\'{i}pari et al.
2004d). We suggested that a probable explanation for the link between
the extreme Fe\,{\sc ii} emission and the extreme velocity OF is that
 both are associated to the interaction of the star formation processes
 and the AGN, that generate extreme explosive and HyN events. \\

Thus, from our programme of observational and theoretical studies of
the evolution of IR QSOs and galaxies, we suggested the following sequences
and evolutionary--links: 

\noindent
{\it IR merger + starburst + GW  $\to$
IR + BAL + Fe\,{\sc ii} + shells QSOs 
(at the end phase of a starburst: with SN/HyN)  $\to$
standard QSOs and ellipticals          $\to$
galaxy remnants}.


\subsection{\bf Explosive Model for QSOs:
Narrow Line AGNs.}
\label{evolmodel}

Narrow line Seyfert 1  are AGNs characterized by optical spectra with:
narrow H-Balmer lines (500 $<$ FWHM $<$ 1500 km s$^{-1}$),
strong or extreme optical Fe {\sc ii} emission
(Fe {\sc ii}$\lambda$4570/H$\beta$ $>$ 1), and
weak [O {\sc iii}]$\lambda$5007 ([O {\sc iii}]$\lambda$5007/H$\beta$
$<$ 3).  See for more details 
Osterbrock \& Pogge (1985);  Halpern \& Oke (1987); Goodrich (1989).

L\'{i}pari (1994) found that the prototypes of narrow line Seyfert 1
AGNs (I Zw 1, Mrk 507 and Mrk 957/5C 03.100) show --in the
IR colour colour diagram-- composite and transition properties.
Moreover,
he found that all the NLS1 of the sample are located in a second sequence
of {\bf transition AGNs/NLS1s}.
This sequence is similar (parallel) to that detected for transition luminous
BAL + IR + Fe {\sc ii} QSOs, but with lower values of $\alpha$(60,25) and
starting in the starburst region (the sequence of transition QSOs start
in the area of ultra luminous IR galaxies).
Mrk 507 and Mrk 957/5C 03.100 are located in the sequence of transition-NLS1
and inside the starburst area (of this IR diagram).

Thus,  several authors suggested:
(i) a link between NLS1 and IR emission + BAL + starburst systems;  and
(ii) that composite and transition NLS1 could be  young systems with a
very high rate of accretion in their super massive BH
(L\'{i}pari 1994; Lawrence et al. 1997; Brandt \& Gallagher 2000; Mathur 2000a,b;
Boller et al. 1996, 1993; Boroson 2002; L\'{i}pari \& Terlevich 2006;
Komossa 2008; and others).

From the theoretical point of view,
L\'{i}pari \& Terlevich (2006) already studied the main steps for the
formation and evolution
of the NLR in a strong galactic wind OF-process, which is associated with
a luminous nuclear starburst + AGN.
In this evolutionary explosive model
the observed line ratios, FWHM and size of the NLR evolve on time scales
comparable to the time scale for the wind development. This time scale
will depend on the rate of energy input, size of the SN region and
on the details of the gas distribution.

Thus, from our observational and theoretical results (of IR QSOs and mergers
with OF) we suggested that young QSOs/AGNs --with relatively narrow lines--
will evolve:

\noindent
{\it from Narrow Line young BAL + IR + Fe {\sc ii} QSOs/AGNs  $\to$ Broad
Line standard QSOs/AGNs}.


\subsection{Theory of Galactic Winds and Shells (generated by
SNe/HyNe)}\label{mod2}

Galactic winds and outflows have been detected  in starburst and
Seyfert/AGN galaxies (see Heckman et al. 1987, 1990, 2000; Veilleux et al.
2002).
IR QSOs and mergers often show strong and extreme nuclear starbursts, with very
powerful galactic winds/OF (Heckman et al. 1987, 1990, 2000;
L\'{\i}pari et al. 1994, 1997, 2000, 2003, 2004a,b,c,d, 2005a, 2008, 2009).

The understanding of galactic winds associated with
starbursts and explosive events was improved  by the use of
theoretical and numerical models (see Strickland \& Stevens 2000; Suchkov
et al. 1994, 1996; Mac Low, McCray \& Norman 1989; Tomisaka \& Ikeuchi 1988).
Theory suggests four main phases for GWs associated with starbursts
(Heckman et al. 1990; Lehnert \& Heckman 1995, 1996):

\begin{itemize}

\item
{\it Phase I}:
A GW results when the kinetic energy of the ejecta supplied by
multiple  supernovae and winds from massive stars is high enough to excavate
a cavity in the centre of a starburst.
At this point the kinetic energy is converted into thermal energy.

\item
{\it Phase II}: As the bubble expand and sweeps up the ambient gas, it will
enter the `radiative phase' (Castor, McCray \& Weaver 1975). The
bubble will  then collapse --due to radiative cooling-- into a
{\it `thin shell'}.

\item
{\it Phase III}: After the shell was formed, its evolution is strongly
dependent on the input physics. If the cooling rate in the interior is high,
then the expanding bubble could stop the expanding process
(Tomisaka \& Ikeuchi 1988).

\item
{\it Phase IV}: If other probable dynamical and thermal conditions are
considered (e.g. Suchkov et al. 1994; MacLow et al. 1989), the shell can
{\it `break up'}. After this break up the host interior become a freely
expanding wind, and the bubble then {\it `blows out'}.
In the blow out phase the optical emission comes from obstacles,
such as clouds and  shell fragments, which are immersed and shock--heated
by the OF.

\end{itemize}

Tenorio-Tagle et al. (1999, 2003a,b, 2005, 2006);
Silich et al. (2004, 2005, 2008); Tenorio-Tagle \& Bodenheimer (1988);
Heiles (1987, 1992)
give further details and references of theoretical and observational studies
of giant shells, bubbles and rings, associated  with multiple explosion
of type {\sc ii} SN/HyN.

\noindent
{\bf Role of SN/HyN in the generation of Super Shells:} \\

Theoretical studies suggested that  mainly type {\sc ii} SN/HyN generate
the blowout phase of the supergiant bubbles (Norman \& Ikeuchi 1989;
Suchkov et al. 1994; Strickland \& Stevens 2000).
The presence of  Wolf Rayet features in  IR QSOs (in their nuclei and
shells) is  indicative of a large number of massive stars,
which are one of the main progenitor o type {\sc ii} SN.
Several groups detected these WR features in the nuclei and shells of
IR QSOs and IR mergers
(Armus et al. 1988; Conti 1991; L\'{i}pari et al. 1992, 2003, 2004d, 2005a,
2009).
We note that the highest value of WR emission known in a
WR galaxy or QSOs was
detected in a ultraluminous IR QSO with extreme OF:  IRAS 01003-2238
(see Armus et al. 1988; L\'{i}pari et al. 2003).

In the last decade,
interesting observational and theoretical results were found in the field
of giant SN/HyN. Specifically,
HyN were detected associated with gamma ray bursts (GRB), radio-HyN in
nearby IR mergers and QSOs + shells (in Arp 220, NGC 7469), HyN associated
with extreme massive stars like Eta Carinae (SN/HyN 2006gy, SN/HyN 2006tf),
etc.
From the theoretical point of view, several groups developed interesting
models that even explain  --for HyN-- their observed luminous light curves,  
broad emission line spectra  and the strong radio plus gamma ray
emissions (see for references Nomoto et al. 2008, 2007a,b,c, 2006).
We already explained that HyN is one of the main components in the
explosive and composite model for BAL + IR + Fe {\sc ii} + shells QSOs.
This theme --the role of HyN, specially in explosive BAL QSOs--
will be analysed and discussed in details in Section 14.


\subsection{Previous Explosive Models:}

The presence of {\it extreme explosions, OF and galactic-winds}
--associated mostly to extreme star
formation processes-- must be considered in the develope of
theoretical models  for galaxy and QSO formation and evolution.
More specifically, 3 main theoretical explosive models were already
proposed:

\begin{enumerate}

\item
Ikeuchi (1981) suggested that QSOs were formed and they exploded mainly at
the cosmological redshift Z $>$ 4. The shock waves propagate through
the gaseous medium generated cooled shells (at the shock
fronts). Which are split into galaxies of mass of 10$^{10-11}$ M$_{\odot}$.

\item
Ostriker \& Cowie (1981) have proposed a galaxy formation picture in which
(after redshift 100) small seed perturbation are supposed to collapse, giving
rise to a explosive release of energy from the deaths of the first generation
of stars (Population {\sc iii}).
This energy drives a blast wave into the surrounding gas.
Thereby sweeping up a shell of shocked material, which eventually cools.
These cool shells are  split into galaxies

\item
Berman \& Suchkov (1991) proposed a hot/explosive model for galaxy
formation. They suggested that the period of major star formation
of protogalaxies (or even giant galaxies) is preceded by an evolutionary
phase of a strong galactic wind. Which is driven by the initial burst
of star formation that enriches the protogalaxy with metals.
Thus this event revert from a process of contraction to expansion.
Specifically, the result of this process is the ejection of enriched
material from the outer part of the protogalaxy, while the inner part,
after a delay of few Gyr, finally contract and cools down to form the
galactic major stellar component. \\

\end{enumerate}

More recently,
Kawakatu et al. (2003) studied the proto-QSO evolution and super massive
black hole growth, using hydrodynamic models with OF.
They found that a ultra luminous IR galaxy phase (in which the host
galaxy is the dominant source of luminosity, i.e. IR galaxies and mergers with
starbursts) precedes {\bf a galactic wind epoch}: i.e.,
{\it young and composite IR +  OF/GW mergers and QSOs}.
This would be a transition state to the
AGN--dominated phase (i.e. to the standard QSO phase).

Thus,
this last theoretical evolutionary path is almost identical to the
observational sequence found --in our programme-- for BAL + IR +
Fe {\sc ii} QSOs (using the IR colour-colour diagram): see sub-Sections
 1.1 and 2.1.


\subsection{Young Low Ionization BAL + IR + Fe {\sc ii}  QSOs}
\label{bal1.2}

Low et al. (1989) and Boroson \& Meyers (1992) found that IR QSOs
contain a 27\% low-ionization BAL QSO fraction compared with 1.4\%
for  optically selected high-redshift QSOs sample (Weymann et
al. 1991).
The high fraction of IR QSOs and mergers showing properties of low
ionization BAL systems  could be
explained by the large fraction of extreme OF with multiple giant shells,
detected in these IR systems. Probably these shells were originated in
the starburst phase of type {\sc ii} SN/HyN.

In the last decades,
two main interpretation for the occurrence of BALs have been proposed:
 the orientation and evolution hypothesis.
Observational evidence supporting the orientation hypothesis come from
spectral comparison of BAL and non-BAL QSOs (Weyman et al. 1991) and
polarization studies (Hines \& Will 1995; Goodrich \& Miller 1995).
Evidence in favour of the evolution hypothesis comes largely
from the high number of low ionization BALs detection in IR + Fe {\sc ii}
QSOs and mergers.
Further support for the evolution hypothesis has been provided for
radio observations of BAL QSOs, which are inconsistent with
{\it only orientation schemes} (Becker et al. 2000, 1997).

Recently,  from a study of a very large sample of 37644 Sloan Digital
Sky Survey (SDSS)  QSOs, from the third Data Release (DR3) and
for all redshift (in the range: 0 $<$ z $<$ 5): White et al. (2006) found
that the radio properties of the rare class of low ionization BALs QSOs are
different to the group of non-BAL QSOs + high ionization BAL QSOs, at all
redshift. They suggested that this result could
be explained  in the frame work of an evolutionary scenario
for BAL QSOs, in close agreement with the model proposed by
L\'{i}pari \& Terlevich (2006).

\noindent
{\bf Low Ionization BAL QSOs at Very High Redshift:} \\

Maiolino et al. (2004a,b, 2003)  presented near-IR spectra
of eight of the more distant QSO (at 4.9 $<$ z $<$ 6.4). Half of these
QSOs are characterised by strong UV BAL systems
(at C {\sc iv}, Mg {\sc ii}, Si {\sc iv}, Al {\sc iii} lines): i.e. mainly
low ionization BAL QSOs.
Although the sample is small, the large fraction of BAL QSOs suggests that
the accretion of gas, the amount of dust and the presence of OF processes
are larger (in these objects) than in standard QSO at z $<$ 4.0.
They also suggested that the very high amount of dust was generated by
early explosions of giant SNe (Maiolino et al. 2004b).

Dietrich et al. (2002) and Barth et al.
(2003) discussed that in order to obtain a good fit of the UV
emission lines Mg {\sc ii} + Fe {\sc ii} in very high redshift QSOs,
they need to include a strong blushifted component (they explain that
this component was used without a physical explanation).
In particular, Barth et al. (2003, their Fig. 2) show
this strong blue OF component in the  Mg {\sc ii} line, for
the Fe {\sc ii}-QSO SDSS J114816.64+525150.3. This object is
one of the younger known QSO, with a redshift z =  6.4.
A similar OF component was found (by us), in the line Mg {\sc ii},
in the extreme Fe {\sc ii} + IR  QSO: PHL 1092.

L\'{i}pari et al. (2005a)
already suggested that this type of blue component observed
in the Mg {\sc ii} emission line --in very high redshift QSOs and with
very high OF velocities-- is associated with extreme OF processes.
Thus, the results obtained from QSOs  at very high redshift
are very similar to our
results for  BAL + IR  + Fe {\sc ii} QSOs at low redshift.
Moreover, we have already  point out the inportance of 
the detection of a high fraction of QSOs with BAL system in young IR +
GW/OF + Fe {\sc ii} + BAL QSOs at
low redshift and  in  QSOs at very high redshift (at z $\sim$ 6.0;
Maiolino et al. 2003, 2004a,b).

\clearpage


\section{OBSERVATIONS}
\label{obser}


\subsection{Gemini GMOS-IFU observations} \label{gmosobservations1}

The three-dimensional (3D) deep optical spectroscopy of  the QSO and the
3 more internal shells of IRAS 04505-2958  were obtained
during four photometric nights in  October 2005, December 2005, and February
2007, at the  8.1 m telescope in Gemini South Observatory.  
The telescope was used with the  Gemini Multi Object Spectrograph (GMOS)
in the integral field unit mode (IFU; Allington-Smith
et al. 2002).
The spectra cover all the optical wavelength range:
from 3400 \AA to 9500 \AA.
The observations were made in photometric conditions with
seeing in the ranges $\sim$0.4--0.5$''$ (in the observing runs of 2005
December 25 and 26; and 2007 February 14) and $\sim$0.9$''$ (in 2005
October 7, and part of 2007 February 14).
For detail of each observation runs, see Table 1.

The data were obtained with the IFU in one slit mode (blue), which provide
a field of 3.5$'' \times$ 5.0$''$ ($\sim$ 20 $\times$ 30 kpc)
for the science data.
With this observing configuration, the GMOS IFU is comprised of 750 fibres;
each spans a 0\farcs2 hexagonal region of the sky.
Five hundred fibres make up the 3.5$'' \times$ 5.0$''$  science field of view;
and 250 fibres make up a smaller, dedicated sky field, which is fixed at 1$'$
of distance of the science position (Allington-Smith et al. 2002).

We used the following gratings in GMOS: R831, B600 and R400, which have 
$\sim$40, $\sim$120 and $\sim$200 km s$^{-1}$ of spectral resolutions,
respectively.
The GMOS Y-axis was aligned at the position angle PA $=$ 131$^{\circ}$;
which is the direction of the  hypergiant shell (from the QSO-core).

Very deep 3D spectra were obtained for the observations of the R400 and B600
gratings, for this bright QSO. The typical
exposure time --for R400 and B600-- were of 1 hour (see Table 1).
These very deep observations were performed 
in order to study:
(i) multiple components in the OF process, and
(ii) the stellar population in the  knots of the expanding shells.


\subsection{Reduction and Analysis of the Gemini GMOS--IFU data}
\label{reductions2}

The following software package were used to reduce and to analyse the
GMOS data:
{\sc R3D + EURO3D}\footnote{{\sc R3D} is the imaging
analysis software facility developed at Calar Alto Observatory (Sanchez \&
Cardiel 2005; Sanchez 2006; Sanchez et al. 2006a,b).
{\sc EURO3D} visualization tool is a software
package for integral field spectroscopy, developed by EURO3D Research
Training Network (Sanchez 2004)};
{\sc IRAF}\footnote{{\sc IRAF} is the imaging analysis software 
developed by NOAO};
{\sc GEMINI}\footnote{{\sc GEMINI} is the reduction
and analysis software facility developed by Gemini Observatory}; and
{\sc STSDAS}\footnote{{\sc STSDAS} is the reduction
and analysis software facility developed by STScI}.

The 3D GMOS spectroscopic observations were reduced using  a modified
version of R3D software package (Sanchez \& Cardiel 2005; Sanchez 2006).
This reduction process was performed following the standard procedure:
(1) the data were bias subtracted;
(2) the location of the spectra were traced using continuum lamp
exposures obtained before each target exposure;
(3) the fiber-to-fiber  response at each wavelength was determined from a
continuum lamp exposure;
(4) wavelength calibration was performed using arc lamp spectra and the
telluric emission line in the science data;
(5) the sky background spectrum was estimated before subtraction by
averaging spectra of object free areas;
(6) the calibration flux was done using the observation of standard stars;
and a total of $\sim$11000 spectrum --of IRAS 04505$-$2958 and sky-- were
reduced and calibrated, using this technique
(see for more details L\'{i}pari et al. 2009, hereafter Paper {\sc i}).

To generate two-dimensional maps of any spectral feature (intensity,
velocity, width, etc.) the IDA  software tool were used
(Garc\'{\i}a-Lorenzo, Acosta-Pulido, \& Megias-Fernandez 2002). The IDA
interpolation is performed using the IDL standard routine TRIGRID, which
uses a method of bivariate interpolation and smooth surface fitting for
irregularly distributed data points (Akima 1978).
Maps generated in this way are presented in the next Sections.

The emission line components were measured and
decomposed using Gaussian profiles by means of a non-linear
least-squares algorithm described in Bevington (1969). In
particular, we used the software {\sc SPECFIT}\footnote{{\sc SPECFIT} was
developed and is kindly provided by Gerard A. Kriss.}, and SPLOT from the
{\sc STSDAS} and {\sc IRAF} packages, respectively.
An example of SPECFIT deblending, using three components for each  emission
lines
(H$\beta$, [O {\sc iii}]$\lambda$4959 and [O {\sc iii}]$\lambda$5007)
in IRAS\,01003$-$2238, was shown in figure 2 of L\'{\i}pari et al.\
(2003). We note that in each GMOS spectrum the presence of OF components and
multiple emission line systems were confirmed by detecting these systems
 in at least two or three different
emission lines (at H$\alpha$, H$\beta$, H$\gamma$, H$\delta$,
[N {\sc ii}]$\lambda$6583, [N {\sc ii}]\ $\lambda$6548,
[S {\sc ii}]$\lambda\lambda$6717/6731,
[O {\sc iii}]\ $\lambda$5007, [O {\sc ii}]\ $\lambda$3727, etc).

For the study of the kinematics, the {\sc ADHOC}\footnote{{\sc ADHOC} is
a 2D/3D kinematics analysis software developed by Dr. J. Boulesteix at
Marseille Observatory.}
software package was also used.  For the analysis of the errors/$\sigma$ in
the kinematics,  the prescriptions suggested by Keel (1996) were used.

The main parameters of the spectra (i.e., the fluxes, equivalent widths,
S/N, errors/$\sigma$, etc) were measured  and their errors analysed
using different software tasks described previously: i.e.
R3D, Euro-3D, Gemini-GMOS, IRAF, IDA, INTEGRAL, STSDAS, SPECFIT,
GALFIT-3D, etc. In general,
we follow for the analysis of the errors/$\sigma$, S/N, etc the mathematical
algorithms described in detail by Bevington (1969) and Roederer (1963).

\subsection{GMOS-IFU PSF}

\label{psf}

In Paper {\sc i}, a detailed analysis of the PSF (in the GMOS-IFU data) was
already performed, for the GMOS observations of Mrk 231.
For the GMOS-IFU data of IRAS 04505-2958, we have performed a similar study
of the PSF; specially for the spectra obtained with high spatial
and spectral resolution.

In particular,
the PSF was carefully obtained for the core of the QSO
IRAS 04505-2958, using the H$\alpha$ and H$\beta$ broad line emission.
This PSF was derived using 
the GMOS-IFU B600 and R831 data, and for the observation obtained with
the best seeing of our Gemini GMOS data (of 0.4$''$ FWHM).

Using the obtained PSF 
the contributions of the nuclear core-PSF at spatial offset
of 0.2$''$, 0.4$''$, 0.6$''$ and 0.8$''$ (from the core) were measured.
We found that these contributions --at 0.2$''$, 0.4$''$, 0.6$''$ and 0.8$''$--
are:  52, 11.5, 3.7 and 1.0 per cent (respectively), of the peak/core.

Therefore, from these results it is important to remark two main points:

\begin{itemize}

\item
An empirical limit for the extension of the wing of the PSF is
r $\sim$1.0$''$.
This limit  is almost the same that we found --in Paper {\sc i}-- from the
study of the PSF for the GMOS data of the BAL + IR + Fe {\sc ii} QSO
Mrk 231 (with similar seeing of $\sim$ 0.4$''$ and B600 spectra).

\item
The results for the extension of the PSF suggest that
the contribution of the PSF at offset of 0.2$''$ is important:
52 per cent.
Thus at this offset (0.2$''$) we need to consider the contribution of the
PSF-core (if it is required). For offset of 0.4$''$ the contribution
of the PSF-core is low: only 11.5 per cent.

\end{itemize}


\subsection{HST--WFPC2 and ACS broad band images and HST/FOS
spectroscopy (archive data)}

\label{hstobservations}

Optical {\itshape HST\/} Wide Field Planetary Camera 2 (WFPC2) archival images 
of IRAS 04505$-$2958 were analysed, which include broad-band images positioned on the
Planetary Camera (PC) chip with scale of 0\farcs046\,pixel$^{-1}$,
using the filter
F702W (6895 \AA, $\Delta\lambda$ 1389 \AA, $\sim$R Cousin filter).

Optical {\itshape HST\/} Advanced Camera for Surveys (ACS) archival observations
 were analysed, obtained with the High Resolution Channel (HRC).
They include images with the filter
F606W (5907 \AA, $\Delta\lambda$ 2342 \AA, $\sim$V Cousin filter).
The scale is 0\farcs027\,pixel$^{-1}$.

{\itshape HST\/} FOS aperture spectroscopy of this QSO was obtained,
from the HST archive (at ESO Garching).
The spectra were taken with the
G190H ($\lambda\lambda$1575--2320 \AA) gratings and the blue detector.
The G190H observation was made with the effective aperture of
4\farcs3 $\times$ 1\farcs4; and the spectra have a resolution of $\sim$4
\AA, FWHM.
A summary of the {\itshape HST\/} observations is presented Table 1.

\clearpage


\section{Morphology of the hyper+super shells and the QSO}
\label{res.3-morphol.map-SHELL}

In this paper, we will study --using GMOS-IFU data--  the
multiple shells system detected in the QSO IRAS 04505-2958
(L\'{i}pari et al. 2003, 2005a, 2007a,b, 2009).
Specially, we will analyze  the properties and the nature of the extended
object found close to the QSO.
This extended and complex  structure (which is located to the south-east,
at a radius --from the QSO-- of r$_{min.} \sim$ 1.5$''$ and r$_{max.}
\sim$2.5$''$) could be associated to a 
{\it hypergiant shell} (referred as S3), centered at the position
of the QSO. We proposed
that S3 was probably generated by  nuclear explosive/HyN events,
similar to those detected in Mrk 231, IRAS 17002+5153, and
IRAS 07598+6508 (L\'{i}pari et al. 1994, 2005a, 2008, 2009; L\'{i}pari 1994).

Using GMOS data,
the properties and the nature of two inner/nuclear shells
(at r $\sim$ 0.2$''$ and 0.4$''$, $\sim$1.1 and 2.2 kpc from the QSO)
will be also studied. In addition,  
the presence of a very extended hypergiant shells at r $\sim$  15$''$
($\sim$80 kpc)  and a possible shell at r $\sim$  10-12$''$
($\sim$55--66 kpc) will be discussed. These two external shell
were already reported by Hutching \& Neff (1988), from CFHT data.

%

\subsection{The four main super+hypergiant shells:  HST, GMOS,
and CFHT images}
\label{results-3.1}


Figs. 1a, b, c and d present high resolution HST WFPC2 and ACS  
broad-band images and contours obtained in the optical wavelengths through
the filters WFPC2-F702W ($\sim$R) and ACS-F606W ($\sim$V).
These HST images show:
(i) the QSO,
(ii) the main supergiant galactic
shell S3, which is located at a radius r of $\sim$11 kpc, from the
QSO (and showing several bright  knots), and
(iii) a field star.

In addition, the panels of Figs. 1 show --in orange colour--
the observed GMOS field (covering an area of
$\sim$3\farcs5$\times$5\farcs0, $\sim$20 kpc $\times$ 30 kpc).
The GMOS frame was centered close to the middle position between the QSO 
and the  extended shell S3 (at r $\sim$ 11 kpc), and at
the position angle PA $\sim$ 131$^{\circ}$.
These HST images (without any smoothing or filtering process) show that
the QSO contours have a structure different to the HST-PSF (of the field
G star). The presence of two  nuclear shells could explain the 
structure of the QSO contours.


The deep HST WFPC2-F702W image of this QSO (Fig. 2) 
shows  the very extended shell S3.
Fig. 2 depicts that the {\it external--border} of  S3 is
symmetric,  with circular shape, and with the centre at the position of
the QSO. This plot was performed using  a scale of fluxes starting from
very low values of flux (thus, this figure shows  almost the complete
emission associated with this shell).
We remark that  S3 shows very extended emission, at scale of $\sim$ 15-20
kpc around the QSO. This is an interesting point, specially  in order to
explain the GMOS emission line maps (Fig. 5). These maps
show H$\beta$, [O {\sc iii}]$\lambda$5007, [O {\sc ii}]$\lambda$3727,
[Ne {\sc iii}]$\lambda$3869 emissions in almost all the observed GMOS
field.


Two interesting results were already found --in the literature--
in relation with the proposed hypergiant shell scenario for IRAS 04505-2958.
In particular:

\begin{itemize}

\item
Hutching and Neff (1988, their Fig. 1) and  Fig. 3  show the presence of
an arc at r $\sim$15$''$ (80 kpc, from the north to the north-east),
in their CFHT R image.
In addition, they show a faint possible arc at r $\sim$10-12$''$
(55-66 kpc, to the south-west) with also several knots. This possible
arc is located in the south-west direction: i.e., in the opposite direction
(from the QSO), to the more extended arc.

Fig. 3 shows that the positions of these two faint external
arcs are consistent with a bipolar OF.
This external bipolar OF (at PA = 40$^{\circ}$) is almost perpendicular to
the direction of the internal bipolar OF (at PA = 131$^{\circ}$, for the
shell S3).

\item
From a study of host galaxies in QSOs (by decoupling and subtracting the
QSO/PSF images from HST ACS data), Magain et al. (2005) detected
close to IRAS 04505-2958 a partial blob at r $\sim$ 0.3$''$,
to the north-west, without other clear evidence of the host galaxy.
The spectra of this blob show the surprising result of the absence of
continuum emission.

Using the HST and GMOS data (see Fig. 4, Section 5 and Table 5)
we found that this partial blob  is --from the morphological point of view--
very similar to the multiple nuclear shells detected in Mrk 231.
In addition, several areas in this blob were analysed using GMOS
spectra, these data show double peaks in the main components of the
emission lines (plus multiple weak blue/OF peaks), which are probably
associated with two shells at r $\sim$ 0.2$''$ and 0.4$''$
($\sim$1.1 and 2.2 kpc; see Fig. 4 and Section 5).
We call these nuclear shells S1 and S2, respectively.

\end{itemize}


Thus,  
these works suggest at least the presence of four (or five) super/hypergiant
shells. Namely:

\begin{enumerate}

\item
{\bf Blobs or shells S1 and S2:} \\

These two partial shells are located at radius r $\sim$ 0.2 and 0.4$''$
($\sim$1.1 and 2.2 kpc), from the QSO-core.
Which are more intense and clear in the north-west
region. A knot and a filament were also detected at the radius of the
shell S1, in the north-east and north directions, respectively.
Probably, we are observing only partially these two shells by the effect of
a bipolar OF process. \\

\item
{\bf Shell S3:} \\

An extended and bright hypergiant shell --with symmetric and circular
external--border--  at r $\sim$  2.0$''$
(11 kpc, from the QSO) was already detected by L\'{i}pari et al. (2003,
2005a, 2007a,b, 2009) and L\'{i}pari \& Terlevich (2006).
The total extension of S3 is at least 
$\sim$30 kpc. This shell is observed in the south-east region, from the QSO.

We have already proposed that this hypergiant shell was
generated by explosive and composite hyperwinds (L\'{i}pari et al. 2003,
2005a, 2007a,b, 2009; L\'{i}pari \& Terlevich 2006).
This explosive process with giant shells is similar to those
observed in  similar BAL + IR + Fe {\sc ii} QSOs, like:
Mrk 231 (L\'{i}pari et al. 2005a, 2003, 1994); IRAS 17002+5153 (L\'{i}pari et al.
2003, 2008; L\'{i}pari 1994); IRAS 07598+6508 (L\'{i}pari 1994; L\'{i}pari et al.
2003, 2008); etc.

The deep HST-WFPC2 R broad band image of IRAS 04505-2958 (Fig. 2) shows
--for S3-- a clear external--border with  circular shape and with the
centre at the position of the QSO. \\

\item 
{\bf Shell  S4:}\\

The presence of an external supergiant
shells at r $\sim$ 15$''$ ($\sim$80 kpc, with the centre at the position
of the QSO)
was already proposed by Hutching and Neff (1988).
They suggested that this arc has red colour.

A wide field CFH R broad band image of IRAS 04505-2958 (Fig. 3, adapted
from Hutching \& Neff 1988) shows that the shell S4 has also a knotty
structure. This arc is extended from the north to the NE, ending at
a brighter knot.
Hutching \& Neff (1988) explained that this very extended arc
S4 is faint but clear in the R CFHT-image. 
In their B CFHT-images the knots were observed, but the arc
was not detected. They proposed
that the arc and their knots are associated with region of strong
reddening. \\

\item
{\bf Shell S5 candidates:} \\

Fig. 3 also shows
the presence of an extended weak structure with also arc shape
at r $\sim$  10-12$''$  ($\sim$55-66 kpc, from the QSO),
to the south-west.
We call this extended structure as a shell candidate (S5).
An interesting point is that S5 is located in the opposite
direction of the shell S4 (from the QSO).

\end{enumerate}

Table 2 and Figs. 1b, and 4  present the location  of the strong
knots observed with GMOS inside of  the 3 more
internal super and hypergiant shells (S1, S2, and S3).
In the next sub-Sections,
the  physical and kinematical properties of these  knots
--and several selected external regions--
  will be analysed. 

Using the borders of these shells,
we show in Figs. 4 and 3 the probable limits for the internal and external
bipolar OFs. In particular,
for the internal OF (at 10--15 kpc scale) we used the borders of the
shells S1+S2 and S3 (Fig. 4),
and for the external OF (at 60--80 kpc scale) the borders of S4 and S5
(Fig. 3). For these internal and external
bipolar OFs (at PA = 131$^{\circ}$ and 040$^{\circ}$, respectively; which
are perpendicular) the following total opening angles  of
$\sim$55$^{\circ}$ and $\sim$95$^{\circ}$ were measured, respectively.

For the BAL + IR + Fe {\sc ii} QSO Mrk 231, we already found an
interesting result at radio wavelengths in relation with the extreme
OF or hyperwind  + multiple symmetric shells scenario:
a very extended radio emission  (of $\sim$ 50 kpc) was
detected, which is aligned with the position angle of the bipolar OF.
For IRAS 04505-2958, Feain et al. (2007) found in their 6208 MHz ATCA
radio data an extended radio emission with bipolar structure.
This radio emission shows  three peaks or lobes of radio emission, with
a main peak at the position of the QSO, and two symmetric peaks (with
the centre at the position of the QSO). One of these symmetric peaks is
located close to S3 and the other in the opposite direction.
This extended radio structure  --of $\sim$20 kpc--
is aligned at almost the same position angle of the
internal bipolar OF or hyperwind (at PA = 131$^{\circ}$).
Furthermore, Feain et al. (2007) found that the radio emission obeys the
far-IR to radio continuum correlation, implying that the radio emission is
energetically dominated by star formation activity. In particular, they
detected that at least 70 per cent of the radio emission is associated with
the star formation process; and the contribution from the QSO --to the
radio emission-- is less that 30 per cent. Therefore, the radio data of
Feain et al. (2007) suggest the presence of some star formation around the
QSO and/or the  host galaxy.
However, more recently these authors (Papadopoulos et al. 2008) suggested
that the star formation process is  probably located in S3, since
they found that the CO J $=$ 1-0  emission is located mainly
in this shell S3.

In addition, using ESO-VLT+VIMOS data, Letawe et al. (2008) found similar
extended [O {\sc ii}]$\lambda$3727, [O {\sc iii}]$\lambda$5007 and
H$\alpha$+[N {\sc ii}] emission aligned at almost the same position angle
of the internal bipolar OF  (at PA = 131$^{\circ}$).
Thus, these  HST + CFHT + Radio + ESO--VLT morphological
results show a good agreement with the hyperwind scenario (with
multiple hyper shells).

%
\subsection{Interesting external regions}
\label{results-3.2}

Four interesting external regions in the GMOS field of IRAS 04505-2958
were also analysed.
Two external regions (R1 and R4) are associated with two emission knots
detected close to the QSO and  S3, respectively.
The other two external regions (R2 and R3) are located
at the external border of  S2 and S3, respectively.
In the region R2 we found that the emission lines are very weak,
thus we measured two areas -very close- in this region (R2a, and R2b).
The detailed positions of all the regions are given in Table 2, 
and in Figs. 1b and 4.

These external regions (specially R2 and R3) were selected
in order to study the extended  OF process (i.e. multiple OF
systems and the emission lines ratios).


%

\subsection{High resolution  GMOS  maps and images}
\label{results-3.3}


Figures 5a, b, c, and d show
the [O {\sc iii}]$\lambda$5007, [O {\sc ii}]$\lambda$3727
[Ne {\sc iii}]$\lambda$3869 and H$\beta$
emission line images, obtained from the GMOS data.
These figures show strong emission lines  from the QSO,
the circumnuclear regions and also weak emissions from S3.


From these GMOS images or maps, we remark the following interesting
features,

\begin{enumerate}

\item
The  [O {\sc iii}]$\lambda$5007 and [O {\sc ii}]$\lambda$3727
emission line maps show  strong emission from the
QSO and  in two filament aligned in the direction of
the external regions R1 and R4.
In addition, several weak emissions were observed in the area
of the hypergiant shell S3. \\

\item
The  [Ne {\sc iii}]$\lambda$3869  emission line map depicts
emission with similar structure of the
[O {\sc iii}]$\lambda$5007 and [O {\sc ii}]$\lambda$3727
maps. \\

\item
The  H$\beta$
emission map shows clear emission from the QSO and several
weak knots in the circumnuclear regions.

\end{enumerate}

These areas with clear emission in the maps will be analysed together
with the emission lines ratios and kinematics maps.

\clearpage


\section{Deep GMOS-IFU spectra of the hyper+super shells
and the QSO}
\label{res4-Spectra.shell.OF}


\subsection{Multiple Outflow components in the QSO
and the shells S1 and S2 }
\label{res-4.2}

Using the 3D GMOS spectra --of IRAS 04505$-$2958-- obtained with
moderate (B600, R400) and high (R831) spectral resolutions,  a detailed
study of multiple emission line components was performed; 
in order to analyse to OF process in the QSO-core and in the shells.
In particular, we have studied the stronger emission lines
H$\alpha$, H$\beta$, H$\gamma$, H$\delta$, [O {\sc iii}]$\lambda$5007,
[O {\sc ii}]$\lambda$3727, [N {\sc ii}]$\lambda$6583, etc; and
the strong absorption lines H$\beta$, H$\gamma$, H$\delta$,
etc.

In the QSO-core and  the circumnuclear shells S1 and S2
(of IRAS 04505$-$2958), we found multiple and strong emission
lines systems.  In particular,
the panels of Fig. 6  show the presence of these multiple OF systems
(in the emission lines).

From this study --of multiple components for the QSO-core and
circumnuclear shells S1 and S2-- the following main and OF components
were found:

\begin{enumerate}

\item
{\it  Main Component (MC-EMI)}:

In the QSO-core and the circumnuclear shells S1 and S2
--of IRAS 04505$-$2958--  a
strong emission line component (MC-EMI) was detected;
plus several OF.
The main ELC was measured and deblended using the software
SPLOT (see Section 3).

In the QSO-core, for the MC-EMI a redhift Z $=$ 0.28600
(85800 $\pm$15 km s$^{-1}$) was measured (using  the narrow
[O {\sc iii}]$\lambda$5007, [O {\sc ii}]$\lambda$3727 emission lines).\\

For the study of the H-Balmer lines --in the QSO-core--  the main
component was  decomposed in a intermediate and broad sub-components.
A detailed analysis and discussion of these sub-components will be
presented in Section 10.

\item
{\it Blue Outflow Components (OF-EB)}:

We found several blue OF components, in the
ionized gas. Specifically,
in the strong emission lines (like H$\alpha$, H$\beta$, H$\gamma$,
H$\delta$, H$\epsilon$, [O {\sc iii}]$\lambda$5007,
[O {\sc ii}]$\lambda$3727, [Ne {\sc iii}]$\lambda$3869, etc),
we found a number of 3-5 blue OF components.
The range of velocities measured for these OF systems
[$\Delta$V = $\Delta$V(OF) - $\Delta$V(MC)] is: from
$\sim$300  to 3000 km s$^{-1}$ (in the blue OF systems).

From these OF systems, three strong blue  OF components
were detected and analysed.
These OF components  are described in  Table 3.

\item
{\it Red Outflow Component (OF-ER1)}:

In the QSO-core and the circumnuclear shells and regions
we have found a  weak but clear red OF component and we
have  measured OF-ER1 a redhifts
Z $=$ 0.291500 (87450 $\pm$20 km s$^{-1}$),
$\Delta$V $=$ V(OF-ER1) - V(MC-EMI) $=$ +1650 $\pm$35 km s$^{-1}$.

\end{enumerate}

The very high values of  velocities found in the multiple emission
line components ($\Delta$V $>$ 500 km s$^{-1}$) --in the QSO
IRAS 04505-2958, S1 and S2--
could be associated mainly/only with an extreme and probably
explosive OF process (see for details Sections 1 and 2).


\subsection{Multiple OF components in the hyper shell S3 }
\label{res-4.3}

{\bf Emission Lines:} \\

In the shell S3  --of  IRAS 04505$-$2958--
also strong and multiple emission lines systems were detected.
In particular,
the panels of Fig. 7  show the presence of these multiple OF systems
(in several emission lines) in the knots of the shell S3.

From the study of multiple components --for the shell S3-- the
following main and OF components were detected:

\begin{enumerate}

\item
{\it  Main Component (MC-S3-EMI)}:

In the shell S3, for the MC-S3-EMI a redhift at Z $=$ 0.2865
(85950 $\pm$20 km s$^{-1}$) was obtained. \\

\item
{\it Blue Outflow Components (OF-S3-EB)}:

Again we found several blue OF components in the hypergiant shell S3.
In the strong emission lines, like H$\alpha$, H$\beta$, H$\gamma$,
H$\delta$, [O {\sc iii}]$\lambda$5007, a number of 3-4 blue OF
components were found.
The range of velocities measured for these OF systems is: from
$\sim$300 to 1500 km s$^{-1}$ (in the blue  OF systems).

Three of these blue and red OF emission lines systems were observed more
clearly and strong. These blue OF systems are described in  Table 3. \\

\item
{\it Red Outflow Component (OF-S3-ER1)}:

In the main knots of S3  
we have found a strong blue OF component and we
have  measured OF-S3-ER1 a redhifts
Z $=$ 0.295500 (87150 $\pm$30 km s$^{-1}$),
$\Delta$V $=$ V(OF-S3-ER1) - V(MC-S3-EMI) $=$ +1200 $\pm$45 km s$^{-1}$.

\end{enumerate}

\noindent
{\bf Absorption Lines:} \\

{\it  The Main Component (MC-S3-ABS): }

In the region of the shell S3 --of IRAS 04505$-$2958--
the presence of an interesting
stellar absorption line system was already noted by
Canalizo \& Stockton (2001), Merrit et al. (2006) and others.

From the strong H$\beta$, H$\gamma$, H$\delta$, H$\epsilon$, H$_{8}$,
and H$_{9}$, absorption lines, we have measured for MC-S3-ABS
a redshift similar --within the errors-- to the MC-S3-EMI: i.e.
Z $\sim$ 0.2865 (85950 $\pm$20 km s$^{-1}$).

In conclusion, with the spectral resolution of this study we can identify,
in S3 at least 6 different emission line  systems.
Again --for the shell S3-- high values of  velocities were found in the
multiple emission line components ($\Delta$V $>$ 500 km s$^{-1}$),
which could be associated with OF processes.
In addition, we found that
the main component of the emission and absorption lines are at the
same redshift.


\subsection{GMOS spectra in the QSO-core and the shells S1 and S2}
\label{results-4.3spectra}

{\bf QSO-core:} \\

Fig. 8 shows the spectra of the QSO-core.
Tables 4 and 9 include the values of the fluxes, FWHM and the emission line
ratios for the QSO-core (for a pixel of 0.2$''$).

Table 4 shows that the H-Balmer lines were fitted using
different components (intermediate, broad, narrow, and OF components).
This detailed fit was performed in order to study the interesting
optical spectrum of the QSO-core (of IRAS 04505-2958), which shows
Narrow Line Seyfert 1 AGN features.
The mean value of the FWHM of the H-Balmer
Emission lines (H$\alpha$, H$\beta$, H$\gamma$, etc) in the QSO-core is
$\sim$1050 $\pm$25 km s$^{-1}$; for a fit of one main component plus
several OF. In Section 10 the other fits of the H-Balmer lines --using
more components-- will be analysed (specially their physical nature).

The value of the FWHM
of the  narrow emission lines ([O {\sc iii}]$\lambda$5007) is
$\sim$630 $\pm$20 km s$^{-1}$ (with the peak blueshifted by -100
$\pm$25 km s$^{-1}$, from the peak of the Balmer H lines).
In addition,  
the FWHM of the narrow line [O {\sc ii}]$\lambda$3727) is
$\sim$480 $\pm$25 km s$^{-1}$. \\

\noindent
{\bf Shells S1 and S2:} \\

Since, the supergiant shells S1 and S2 are located very close to
the QSO-core (at $\sim$0.2 and at 0.4$''$) we have measured very
carefully the emission line systems of these two shells (using one
of the best code to deblend emission lines: SPECFIT; see
for details Section 3).
The results of the  study of the emission lines in S1 and S2 are included
in Tables 5 and 9.
The emission line ratios of Table 9  show --for S1 and S2-- values clearly
consistent with ionization associated with the QSO plus shocks 
(see for details Fig. 13a, b in the next Section 6).


\subsection{GMOS spectra in the hyper shell S3}
\label{results-4.4spectra}

It is important to study in detail the main knots detected in the
multiple hypergiant shell S3 with high resolution 3D spectroscopic
data: since they are  the best and brightest tracers of the expanding
super bubbles (see Paper {\sc i}, for details and references of our
previous studies --using 3D-Spectroscopy-- of the main knots in the
expanding shells of the BAL + IR + Fe {\sc ii} QSO Mrk 231 and in the
IR merger NGC 5514).

Figures 9 and 10 show  the individual 3D GMOS spectra of the
main knots of the shell S3 for different wavelength ranges.
Tables 6, 7 and 9 depict the values of the fluxes and FWHM of the
emission lines, the Equivalent Width (EqW) of the absorption stellar
system, and the emission line ratios, respectively (for the main knots
of the shell S3).

In order to study the GMOS spectra of the main knots of the hypergiant
shell we used the following technique (described in more details in
Paper {\sc i}): (i) first the main
knots of the shell were selected, from the high
spatial resolution HST WFPC2 and ACS images; (ii) using the HST offset
positions --from the QSO-core-- $\Delta\alpha$ and $\Delta\delta$ (and
the corresponding offset position in the GMOS rotate-field: $\Delta X$
and $\Delta Y$), of all the main knots, then
we selected the closest GMOS individual spectrum.
Thus, the offset --in Table 2-- were derived from the
nearest GMOS spectra of the corresponding knot peaks.
In addition, we have verified also
that the nearest spectrum --corresponding to each knot-- shows the strongest
value of continuum and line emission (for all the area of each knot).

In addition,
it is important to note that only in the very deep GMOS
3D data (with 1800 $\times$ 2 seconds of total exposure time, see Table 1)
the spectra depict high quality, even with S/N $>$ 3 in the weak
OF components of the shells. Which is
required in order to study some weak knots and region of the
hyper+supergiant shell S3: K1, K3 and R3.

From this detailed study of the main knots of the hypergiant shell S3 of
IRAS 04505$-$2958 (see Tables 6, 7 and 9), we remark the following
 main results: \\

\begin{enumerate}

\item
\noindent
{\bf Absorption Lines:}\\

We detected two type of absorption spectra in the main knots of the
shell S3 (see Table 7 and Fig. 9).

\begin{itemize}

\item
The spectra of the Knots S3-K1, K2, and K3 (and also the close region R4)
show in the blue wavelength range weak --or even absent--  absorption
H-Balmer lines. Specifically,
the H$\gamma$ and H$\beta$ absorption are absent in these 3 knots
(see Table 7).

\item
The spectra of the Knots S3-K4 and K5 show
strong absorption H-Balmer lines: from H$\beta$, H$\gamma$, to H$_{11}$
(see Table 7).

\end{itemize}

These GMOS-IFU results will be analysed in detail in Section 9
(using theoretical and observational templates of stellar populations). \\

\item
\noindent
{\bf Emission Lines:}  \\

\begin{itemize}

\item
The emission spectra of the Knots S3-K1, K2, and K3 (and also the close
region R4)
contain very strong OF emission line components (see Table 6).
These components (the OF and MC systems) show
LINER properties associated with shocks plus H {\sc ii} regions.

These are typical OF features associated with shocks of low
and high
velocities in a dense medium (similar to those observed in
the OF of SNR and Herbing--Haro objects; Heckman et al. 1990;
Binette, Dopita, Tuohy 1985; Canto 1984; Shull \& McKee 1979).

\item
The spectra of the Knots S3-K4 and K5 show
 OF and MC systems with only
LINER properties associated with shocks.

\item
In the main knots of S3 we found OF components with high values
of velocities, of  $\Delta$V $\sim$ --[400-1500] km s$^{-1}$.

\end{itemize}

\end{enumerate}

A detailed study of the emission line ratios and kinematics of  these 
knots of S3 will be present in the next Sections.


\subsection{GMOS spectra in some interesting Regions}
\label{results-4.5spectra}

Figs. 11 and 12 show the spectra of the selected external regions (see
Table 2 and Section 4, for details about the location of each region).
Tables 8 and 9 include the values of the fluxes, FWHM and emission lines
ratios of these regions.

From these Figures and Tables of the selected external regions  of
IRAS 04505$-$2958, we remark the following results:

\begin{itemize}

\item
The spectra of the regions R1 and R4  show clearly a blue component in
the continuum emission, at the wavelength ranges:
[O {\sc ii}]$\lambda$3727--H$\gamma$,
and H$\beta$ + [O {\sc iii}]$\lambda$5007.

\item
The spectra of the regions R2a and R2b --at the external border of the shell
S2-- show that this blue component is weak (in the continuum emission,
at the [O {\sc ii}]$\lambda$3727--H$\gamma$,
and H$\beta$ + [O {\sc iii}]$\lambda$5007 wavelength ranges).

\item
The spectra of the region R3  depict  even a clear drop 
in the blue continuum emission
(specially, at the [O {\sc ii}]$\lambda$3727--H$\gamma$ wavelength range).

\end{itemize}

Letawe et al. (2008) studied -using ESO-VLT+FORS2 Multi Slit MXU
2D-spectroscopy- three external regions, with a slit width of
1$''$. They call these regions R1, R2, and R3 (in order to avoid
problem of notation we will used the following notation for these
areas: L-R1, L-R2, and L-R3).
Their regions L-R1 and L-R2 are located relatively close to our external
areas R3 and R2, respectively
They found that these two regions show only emission lines, which is
a similar result to that found in this paper for these areas. In addition,
for the region L-R2, Letawe et al. (2008) found that the ionization is
probably associated with the AGN, and we found that the ionization is
mainly generated by shocks. The difference could be explained by the fact
that  their slit width is 1$''$, and thus they are probably including a
contribution from --or close-- to the blobs (in these blobs our GMOS
spectra --with 0.2$''$ of spatial resolution-- show ionization by the AGN).

\clearpage


\section{The ionization structure of  the hyper+super shells}

\label{res5-elr-SHELLS}

Using the emission lines ratios (ELR) obtained from the 3D GMOS data
(which cover the QSO and 3 super/hypergiant shells) we have studied
in detail the ionization and the physical conditions in IRAS 04505-2958;
specially in order to compare these  results with those obtained previously
for similar BAL + IR + Fe {\sc ii} QSOs and mergers, with 
strong OF process.

This study was performed in two steps:
first the individual GMOS spectra of the main knot of the shells and the
external regions were analysed in detail using the
log [S {\sc ii}]/H$\alpha$ vs. log [O {\sc i}]/H$\alpha$ and
log [S {\sc ii}]/H$\alpha$ vs. log [O {\sc iii}]$\lambda$5007/H$\beta$
ELR-diagrams (of physical conditions).
Then the GMOS-IFU ELR maps were studied.


\subsection{The emission line ratios diagram for the hyper+super shells}
\label{res5-emlr1}

For the study of the physical conditions and the OF process in the shells
and in several selected external regions (inside of the GMOS field,
of IRAS 04505-2958), 
the log [S {\sc ii}]$\lambda$6717+31/H$\alpha$ vs.
log [O {\sc i}]$\lambda$6300/H$\alpha$, and
log [S {\sc ii}]$\lambda$6717+31/H$\alpha$ vs. 
log [O {\sc iii}]$\lambda$5007/H$\beta$ 
ELR-diagram were used. The first diagram
is an important tools for the analysis of OF processes and
associated shocks (see Heckman et al. 1987, 1990; Dopita 1995).

The panels of Figure 13a, b show these two diagrams
for the 3 observed  hyper+supergiant shells and the selected
external regions.
In Fig. 13a,b the values of emission lines ratios (for the main knots of
these shells and selected external regions) were obtained from Table 9.
It is interesting to remark the following main points,

\begin{enumerate}

\item
Almost all the knots and areas of the 3 observed hyper/supergiant shells
(S1, S2, S3)
are located in 
the log [S {\sc ii}]$\lambda$6717+31/H$\alpha$ vs.
log [O {\sc i}]$\lambda$6300/H$\alpha$ diagram in the
area of SNR + HH (i.e., the shocks
area), or in the transition/composite region between
SNR+HH and H {\sc ii} regions.
Thus in these areas the OF process play a main role. \\

\item
Some knots of the shells S1, S2, and S3, the ELRs 
show a position inside the SNR+HH (pure shock) area of
this diagram.  In particular,
the following knots and areas are located in the shock region:
S1-A1, S2-A1, S3-K4, S3-K5; and the external regions R1, R2a, R2b and R3.
This fact is consistent with the presence
of strong [S {\sc ii}] and [O {\sc i}] emission; and thus
it is also consistent with shocks process of low velocities
(Heckman et al. 1990; Dopita \& Southerland 1995). \\

\item
In addition, for the shells S1, and S2 the
log [S {\sc ii}]$\lambda$6717+31/H$\alpha$ vs. 
log [O {\sc iii}]$\lambda$5007/H$\beta$ 
ELRs diagram show that the QSO/AGN is also a
source of ionization (in these circumnuclear shells), together
with shocks.

\item
The knots of the shell S3: S3-K1, S3-K2 and S3-K3 are the only knots located
in the composite or transition areas between shocks and H {\sc ii} regions.
This result is in good agreement with the detection of starburst population
in these knots (see Section 9). \\

\item
The regions R3 and R2a,b  are located
close to the external border of the hypergiant shell S3 and the
super shell S2, respectively.
Thus, their ELR are consistent with shock associated
with the OF process in these shells. Furthermore,
the ELR maps (see the next sub-section) also show structures associated
with very large scale shocks and outflows. \\

\end{enumerate}


\subsection{Mapping with GMOS the ionization structure}
\label{res5-emlr2}

Figs. 14a, b and c show the 3D maps
(of $\sim$3\farcs5$\times$5\farcs0, $\sim$20 kpc $\times$ 28 kpc,
 with a spatial sampling of 0\farcs1)
of the emission line ratios
[S {\sc ii}]${\lambda 6717 + 31}$/H$\alpha$,
[N {\sc ii}]$\lambda$6583/H$\alpha$ and
[O {\sc iii}]$\lambda$5007/H$\beta$.
These maps were constructed using the techniques described in Section 3
and for the main component of the emission lines.

Figs. 14a, b, c  show interesting features. We note the following:

\begin{enumerate}

\item
Coincident with almost the border of the more extended super and hyper
shells S2 and S3, 
both maps show arcs and knots with high values ($>$ 0.8) in the
[S {\sc ii}]${\lambda 6717 + 31}$/H$\alpha$ and [N {\sc ii}]/H$\alpha$
emission line ratios (ELR).

These arcs could be associated  with shock processes at the
border of the super and hypergiant shells S2 and S3.
L\'{i}pari et al. (2004a,d, 2005a) already discussed 
that the [S {\sc ii}]/H$\alpha$ map is one of the best tracer
of shocks processes. \\

\item
The GMOS  [N {\sc ii}]/H$\alpha$ map
shows  several knots in the arcs (which show high values
of  emission line ratio).  \\

\item
The [O {\sc iii}]\,$\lambda 5007$/H$\beta$ map
depicts several areas of high values of the ELR, associated
with the circumnuclear regions and the more internal shells (S1 and S2). \\

\end{enumerate}

Thus, in almost all the border of the shells  of
IRAS 04505-2958 the
GMOS-IFU  [S {\sc ii}]/H$\alpha$ and [N {\sc ii}]/H$\alpha$  maps show
high values, which are consistent with
an ionization process produced mainly by  shock-heating in
the outflowing gas of the expanding supergiant shells
(L\'{i}pari et al. 2004a,d, 2005a;  Dopita \& Sutherland 1995;
Dopita 1995, 1994; Heckman 1980, 1996; Heckman et al.\ 1987, 1990).

Similar results --ELR associated with large scale shocks-- were obtained in
the 3D spectroscopic studies of the OF nebula and supergiant shells/bubbles
of NGC 2623, NGC 5514, Mrk 231 (L\'{\i}pari et al. 2004a,d, 2005a, 2006) and
NGC 3079 (Veilleux et al. 1994).

\clearpage


\section{GMOS Map of the Blue continuum}

\label{res6-bluecont.map}

In Paper {\sc i} an  interesting GMOS-IFU result was found 
for the BAL + IR + Fe {\sc ii} QSO Mrk 231 using an optical colour
map: {\it only in the galactic wind area the colour map shows a strong
blue continuum component}.
We have performed a similar study for IRAS 04505-2958.

For the study of the colour map of this BAL QSO,
it is important to note that
the spectra of the QSO-core --of IRAS 04505$-$2958-- show a 
strong blue component in the continuum (see Fig. 8).
Thus, an important point  is to analyse the possible contribution of
the PSF QSO-core blue continuum, to the circumnuclear regions.
About this point, we have already explained in Section 3 and
specially in Paper {\sc i}, that the contribution of the QSO-core PSF
 is important only in the nearest
spectra at 0.2$''$ (with a contribution of 50 per cent of the PSF
peak). But at 0.4$''$ from the QSO-core this contribution is only of
11 per cent.

An important point regarding the quality of the GMOS colour maps is
the following: for Mrk 231 this plot (Paper {\sc i}) shows blue colour
in the south nuclear region (from the QSO), which is exactly  the area
that we previously detected an extreme galactic wind, with shells.
This fact could not be associated with any coincidence and/or a
contribution from the QSO-core, since  the optical
spectra of the core of Mrk 231 show a strong red colour in the continuum
(even with a very strong fall --of continuum flux-- at the blue
wavelength range).

In addition, the GMOS colour map and the individual GMOS spectra
of Mrk 231 clearly show/confirm that the contribution from the QSO-core
continuum (PSF) is very weak at 0.4$''$ offset.
Since this colour map and the individual GMOS spectra
depicts very different colours (i.e., continuum shape) at offsets of:
0.2--0.4$''$-south and 0.2--0.4$''$-north, from the QSO-core.
Even, the shape of the continuum at the QSO-core is very different
to those observed at 0.2-04$''$-south and at 0.2--0.4$''$-north.
More specifically,
the continuum is very blue at 04$''$-south,  almost flat at 0.2$''$-south,
red at the QSO core and very red at 0.2-04$''$-north
(see in Paper {\sc i}:  Fig. 6).

Following the technique described in Paper {\sc i}:
first, a basic qualitative study of GMOS spectra was performed. Which
was based in a direct and simple inspection of the continuum shape, at each
spectrum.
The panels of Figure 15 show the sequence of individual spectra
(for the H$\beta$ + [O {\sc iii}] + Fe {\sc ii} wavelength range)
along the position angle  PA $=$ 131$^{\circ}$, and with step of 0.2$''$.
From this qualitative study of the GMOS spectra,
interesting results were found (which are evident in Fig. 15):
in almost all the regions of the GMOS field, of the QSO IRAS 04505-2958,
the spectra show  a strong blue component, in the continuum.
This result (strong blue continuum, in the H$\beta$ + [O {\sc iii}] +
Fe {\sc ii} wavelength range) was verified also at H$\alpha$ and
[O {\sc ii}]--H$\gamma$ wavelength ranges.

A detailed quantitative study of the continuum was performed,
using for this purpose a colour index defined --by us--  as
the difference of fluxes at the border of the wavelength range
of each GMOS CCD (using the B600 grating; see Table 1 for details
of the GMOS observation,  and Allington-Smith et al. 2002 for
details of the GMOS instrument). In particular, we used the
following colour index (for the H$\beta$ region):

\begin{itemize}

\item
For the Visual-Red wavelength range:

\centerline{[Flux($\lambda$6600) -- Flux($\lambda$5750)] $\times$ 10$^{16}$.}

\end{itemize}

Fig. 16a shows the map of this continuum colour index, for the GMOS
field.
This colour map  shows:
(i) in almost all the field (specially around the QSO and in the shells) a
strong blue continuum component;
(ii) only in the external regions of the shell S3 the blue
continuum component is relatively weak. These two interesting
results are more clear and evident in Fig 16b, which shows the
superposition of the GMOS colour map and the HST-WFPC2 R contours.
Moreover, this Fig. 16b depicts that the strong blue continuum
is likely elongated at the same direction of the OF process
(at PA $=$ 131$^{\circ}$). At this direction
we previously suggested  that an extreme bipolar hyperwind
generated the hyper shells S3.
Moreover, at scale of $\sim$30 kpc the extended radio emission
and the narrow emission lines are also aligned at this position angle.

Thus, an interesting theme is to study the possible nature of the
strong blue continuum detected in this paper for IRAS 04505-2958, and
previously in Mrk 231.
In both cases the extended blue continuum components are aligned --and
probably associated-- with the explosive hyperwind/OF processes. This theme
deserve  a specific and detailed study.

\clearpage


\section{GMOS-IFU kinematics of the QSO and the
hyper/super shells}

\label{res7-kinemaSHELLS}

The study of the kinematics of IRAS 04505-2958, in particular
the hyper shell S3, is an important test for the
hyperwind scenario. Specifically, Merrit et al. (2006)
proposed --if the extended object is a ring or interacting
galaxy (as they suggested)-- that the velocity field of this
object will presents clear evidences of circular motion,
since they found an evolved stellar population
of $\sim$ 10$^{8}$ years, in this extended object.

In order to study the kinematics of the ionized gas, in 
the GMOS field of IRAS 04505$-$2958, we have measured the velocities
from the centroids of the strongest narrow emission lines:   
[O {\sc iii}]$\lambda$5007, [O {\sc ii}]$\lambda$3727,
[Ne {\sc iii}]$\lambda$3869;
plus the emission line H$\beta$ (measuring the peak of the
main component; for detail of the H$\beta$ components see the sections
4 and 9).
The fitting of Gaussians and Lorenzians was performed
using the software SPLOT and SPECFIT (see for details Section 3).

First, the kinematics of the {\it main components} of the emission lines
were analysed; and then the presence of multiple OF components
required a more detailed study.
Figs. 17a, b, c and d show  --for the main component, of the ionized
gas-- the [O {\sc iii}]$\lambda$5007, [O {\sc ii}]$\lambda$3727,
[Ne {\sc iii}]$\lambda$3869, and H$\beta$, velocity field
maps.
The GMOS-IFU field includes the QSO and the shells, in
$\sim$3\farcs5$\times$5\farcs0 ($\sim$20 kpc $\times$ 30 kpc),
 with high spatial resolution (sampling of 0\farcs1).
These maps were constructed using the techniques described in Section 3.
In each map the velocity of the QSO-core was used as the reference
velocity.

The isovelocity colour maps (Figs. 17a, b, c, d and e)
show the following characteristics:

\begin{enumerate}

\item
In the region of the hypergiant shell S3, all the velocity
maps show  very complex structures, which are not consistent
with pure circular motion or an interacting or ring galaxy.
Even these GMOS kinematics maps of the shell S3 are different to
those observed  for IR mergers with OF: like NGC 3256,
NGC 2623, etc.

Only the velocity field (VF) map of the ionized gas in the
{\it external super--giant bubble of NGC 5514} (L\'{i}pari et al.
2004d) shows some similarities --in the structures--
to those observed in the hypergiant shell S3.
More specifically,  S3 shows in the
[O {\sc iii}]$\lambda$5007 GMOS VF two lobes of redshifted
velocities, with bi-cone shape. These features are very similar
to those observed in the VF of the external shell, of NGC 5514. \\

\item
H$\beta$, [O {\sc ii}]$\lambda$3727 and [Ne {\sc iii}]$\lambda$3869
VFs maps show also similar structures to the previous maps. However,
the [O {\sc iii}]$\lambda$5007 VF map depicts more clear
substructures.

\item
Fig. 17e was constructed specially  to detect the
{\bf centre of the kinematics  ``bi-cone structure"}, which is clearly
located in the region of the very bright knots --of the shell S3--
S3-K4 and S3-K5.

However, very recently using HST-NICMOS and
ESO-VLT/ISSAC near and mid-IR data Letawe et al. (2009) found a point
source in the central region of the extended object (S3), which is located
very close (at $\sim$0.2--0.3$''$) of the knot S3-K5
They associated
with an AGN, or  --less probably-- with
a compact and unusual extremelly brigth starburst.

Thus,  
the {\bf centre of the kinematics  ``bi-cone structure"}, could be
also 
a compact and unusual extremelly brigth starburst plus/or a AGN.

\end{enumerate}

In addition, Fig. 18 shows the kinematics profile of
H$\beta$ and [O {\sc iii}]$\lambda$5007 through
the QSO-core and at the position angle PA $=$ 131$^{\circ}$.
This plot shows a smooth variation of velocities,
from the QSO-core to the shell S3. Thus, an interesting
possibility is that
a similar physical process is connecting the QSO-core
and S3: i.e., an extreme galactic wind.
Previously,
Merrit el al. (2006) suggested that this continuity --found also in
the 1D spectroscopic ESO-VLT data-- could be explained by the fact
that the extended ring object was not observed  at the position of
the main kinematics axis.

Therefore, from this GMOS kinematics study there are some main points
which are important to remark in order to discuss  the nature of S3:

\begin{enumerate}

\item
In this paper, all the velocity field maps clearly show that the motion in
the extended object is very complex and probably associated with an extrem
OF process.
In addition, the GMOS spectra show in the main knots of S3 multiple emission
line components which could be mainly associated with OF.

\item
The VFs clearly show that the kinematics in S3 might not
be associated with circular motion, or even to the motion of 
the observed VFs of interacting galaxies.

\end{enumerate}

Thus, in this and previous Sections, the GMOS-IFU kinematics, ELR,
colour maps and the morphology results show a good agreement with
the hyperwind/OF scenario.

\clearpage


\section{Stellar Population in the Shell S3: detection of a
young starburst}

\label{res8-stelpopu-SHELLS}

Using  long slit spectra  Canalizo \& Stockton
(2001) and Merrit et al. (2006) already analysed, in the integrated
spectra the shell S3, the presence of an stellar absorption system in
the Balmer H-lines (H$\beta$, H$\gamma$, H$\delta$, H$\epsilon$, H$_{8}$,
H$_{9}$, H$_{10}$ and H$_{11}$). They detected
a A-Type stellar spectra associated with a post-starburst of intermediate
age of $\sim$10$^8$ yr. In addition,
Merrit et al. (2006) proposed that
{\it there is --together with the post-starburst population-- a
residual (in the spectra of S3), which could be
associated with a possible ongoing star formation  process}.

In this paper,
using high resolution deep GMOS-IFU spectra, this post-starburst
and a possible new starburst system will be analysed,
for each main knots of the hyper shell S3.


\subsection{Fitting the Stellar Population using Theoretical Models }

The GMOS spectra of the shell S3 
were analysed using synthetic spectra of H Balmer and He {\sc i}
absorption lines, for starbursts and post-starburst galaxies.
These synthetic and theoretical spectra were developed by Gonzalez Delgado,
Leitherer \& Heckman (1999).

The values of equivalent width of the Balmer H-Lines H$\beta$, H$\gamma$
and H$\delta$ were measured using  the wavelength
windows suggested by Gonzalez Delgado et al. (1999).
Which allowed to compare the measured values with those derived from their
synthetic spectra of H-Balmer and He {\sc i} absorption lines.
The errors ($\sigma$) in the EqW of H$\delta$ are  less than 1.0 \AA.

Table 7 shows the following results,
from the study of the absorption GMOS-spectra of the main knots of the
shell S3:

\begin{enumerate}

\item
{\it Knots S3-K1, S3-K2 and S3-K3:} \\

For H$\delta$ a range of equivalent width (EqW) of 3.5 -- 6.5 \AA, and
FWHM of 460 -- 470 km s$^{-1}$ were measured. \\

\item
{\it Knots S3-K4 and S3-K5:} \\

For H$\delta$  a range of equivalent width (EqW) of 10.0 -- 11.5 \AA, and
FWHM of 570 -- 590 km s$^{-1}$ were observed.  \\

\end{enumerate}

Thus, this study shows a new interesting result: two different ranges
of EqW were detected for the main knots of the hyper shell S3.
These ranges are different if we consider the errors in the
EqW (which are less than 1.0\AA). Furthermore, the knots of
each of these two ranges are located also in two different
areas of the shell.

We also compared the observed EqW of H$\delta$ (of the main knots of
the hyper shell S3) with the grid of EqW of the models
(developed by Gonzalez Delgado et al. 1999).
The used synthetic model corresponds to a cluster with: instantaneous
burst, solar metallicity and  Salpeter IMF, between
M$_{low} =$ 1 M$_{\odot}$ and M$_{up} =$ 80 M$_{\odot}$.
From this study, the following ranges of age were found:

\begin{itemize}

\item
{\it Knots S3-K1, S3-K2 and S3-K3:} ages of 3.5 -- 10.1 Myr; \\

\item
{\it Knots S3-K4, and S3-K5:} ages of 80 -- 140 Myr.   \\

\end{itemize}

Thus, in the knots S3-K1, S3-K2, and S3-K3 the analysis of the GMOS
spectra, using theoretical stellar population models, we found that
the range of ages corresponds to young stars (in a young
starburst).


\subsection{Fitting the Stellar Population using Stellar Cluster
Templates }

The blue H-Balmer absorption spectra (H$\beta$, H$\gamma$, H$\delta$, etc)
were also analysed using a second method:
observational templates spectra of stellar populations
(provided by Piatti et al. 2002; Bica 1988).
These digital templates integrated spectra were obtained for different
ages, from a library of 47 open stellar clusters .
Covering the optical ranges of $\lambda$
3500--7000 \AA, and $\lambda$5800--9200 \AA;  with spectral resolutions of
14 and 17 \AA, respectively.

From this fit of the GMOS spectra of the shell S3 --using the open-cluster
templates-- the following main results were found:

\begin{enumerate}

\item
{\it Knot S3-K4, S3-K5:} \\

For these knots, with strong H-Balmer lines in absorption, we found
the best fit of the GMOS spectra  using a template of 100-150 Myr age:
i.e., the template call Yf (Piatti et al. 2002). Fig. 19a  shows the result
of this fit.\\

\item
{\it Knot S3-K1, S3-K2, S3-K3:}\\

For these knots with an unusual type of H-Balmer absorption spectra (without
H$\beta$ and H$\gamma$ absorptions) we did not obtain a good fitting
using open-cluster templates. However, we have obtained an
excellent fit using a template from a Library of spectra of
Stars (for details see the next sub-section).

\end{enumerate}


\subsection{Fitting the Stellar Population using Stellar
Template }

The blue H-Balmer absorption spectra
 were also analysed using a third method:
observational templates of stats (presented and provided by
Silva \& Cornell 1992).
This is a digital optical stellar library, covering $\lambda$3510--8930\AA,
with a resolution of 11 \AA; for 72 different stellar types.

From this study the following main result was found:\\

\noindent
{\it Knots S3-K1, S3-K2, and S3-K3:}
A good fit of the spectra of these knots
(with  weak H-Balmer lines absorptions, which started at H$\delta$) was
found using the template corresponding to stellar of type
{\bf B1-I}: i.e., super-giant stars of spectral type B1.
Fig. 19b depicts the results of this fitting process.

Thus, this result (using templates from a stellar library) and also those
obtained in sub-section 9.1 suggest that in the knots S3-K1, S3-K2, and S3-K3
the dominant population corresponds to massive blue stars (probably
associated with a young starburst).
This result shows a good agreement
with the study of the emission line ratio since the knots
S3-K1, S3-K2 and S3-K3 all show ELR consistent with composite
properties of shocks plus H {\sc ii} regions.

In addition,
the presence of a young starburst detected in S3 (using the GMOS
data) is in good agreement with the detection of CO J $=$ 1-0 line emission,
in this area. Which implies
a mass M(H$_2$) $\sim$ 2 $\times$ 10$^{10}$ M$_{\odot}$ and high star
formation rate (Papadopoulos et al. 2008). This mass of H$_2$ is at least
a $\sim$ 30 per cent of the dynamical mass in the CO-luminous region.
Recently, Letawe et al. (2009) reported strong reddening in the central
area of S3.

For the knots S3-K4 and S3-K5 (with A-type stellar population),
it is important to remark that an interesting result was found --in
paper {\sc i}-- for the absorption lines of Mrk 231:
when the position of the strong H$\beta$, H$\delta$, H$\gamma$
absorptions were plotted, these strong absorptions
are located close to the external border of the 
 supergiant shells.
Thus, these strong absorptions show  ``arc--shape" distribution
in Mrk 231.

A simple explanation for this
result could be that the OF process --in these shells-- is cleaning the
dust.
Thus, this OF + cleaning process   allow  --probably-- to see 
clearly the absorptions of the A-type stellar population in Mrk 231
 (specially, close to the external borders of the expanding shells).

An interesting point about the  A-type stellar populations
(detected in mergers at low, medium and high redshift; see
Poggianti et al. 1999) is that different works proposed
that the star formation process was truncated in these
mergers (see Balogh et al. 1997).
However, it is not clear the process that could truncate the star
formation.
A interesting explanation for this result is that the OF process
(detected in  a high percent --$\sim$ 75 per cent-- of IR mergers;
L\'{i}pari et al. 2004a)
could be the origin of the truncate star formation.
Since the galactic wind in the last phase --blow--out + free
wind-- could change strongly the kinematics and physical properties
of the different components of the ISM (and even to expel an
important fraction of the ISM). Thus, the star formation --generate
in the ISM-- will change also strongly.

Finally, we note that --in this section-- the study of the
stellar population in the main knots of S3 shows a excellent
agreement between the different theoretical and observational
metods used.

\clearpage


\section{The Emission and Absorption Lines in IRAS 04505-2958}

\label{res9-NLR+BLR}

Using high spatial and spectral resolution GMOS data, we
studied in detail the properties of the  emission
and absorption lines (for IRAS 04505-2958).
This QSO shows spectral features of  narrow
line Seyfert 1 AGN or QSOs.
Several authors already suggested
a link between NLS1 and BAL systems + IR emission (see L\'{i}pari 1994;
Lawrence et al. 1997; Brandt \& Gallagher 2000;  Boller et al. 1993;
Mathur 2000a,b; Boroson 2002; Kawakatu et al. 2007; Popovich et al. 2009). 

Moreover, there is an interesting discussion about the derived mass of
the SMBH in IRAS 04505-2958, using the properties of the emission lines.
In particular, from a study of the profile of
H$_{\beta}$ emission line, Merrit et al. (2006)  derived a 
value of the mass of the SMBH of IRAS 04505-2958, namely
[2--11] $\times$ 10$^7$ M$_{\odot}$.
This value was obtained considering that IRAS 04505-2958 shows
similar features and properties than NLS1 QSOs.
Since this SMBH mass is smaller than that
obtained by Magain et al. (2005) of 8 $\times$ 10$^8$ M$_{\odot}$
(using the magnitude of the QSO, M$_V$ $=$ --25.8),  
Merrit et al. (2006) proposed
that the host galaxy --of this IR QSO--  could be less massive and
less bright than the values previously assumed.

In Section 5, the results of
detailed fits of  the GMOS emission line spectra
were presented (in Tables 4, 5, and 6) for: the QSO-core, the shells
and several external regions.
About the NLR in the QSO-core, it is important to remark that:
using only one component for the fit of H$\beta$, we found that the
final Gaussian/Lorentzian solutions did not fit well the spectra.
Only, including 
a broad, a intermediate and OF emission components, the fit obtained was
correct: Fig. 23a shows this fact very clearly  for H$\alpha$
(since for this line the different components are strong and this
line was observed with the best GMOS spectral resolution R831).
Fig. 23b depicts the fit of H$\beta$ using a broad and an intermediate
 components (the GMOS spectra H$\beta$ was observed with medium spectral
 resolution, B600; in addition in this line the OF components are weak).
Thus we have some differences with the results of  Merrit et al.
(2006) which were derived  using only a single component
for the emitting region of IRAS 04505-2958.


\subsection{The Narrow Line Emission in the QSO--core and the extended
regions}

\label{res9.1-NLR}


In Section 5 several interesting GMOS-IFU results were obtained in relation
with the narrow line (NL) emission; specially for the QSO-core, the
circumnuclear and external regions of IRAS 04505-2958. In particular, \\

\noindent
{\bf At the QSO-core:}

\begin{enumerate}

\item
{\bf H-Balmer and Fe {\sc ii}:} these strong  emission lines
were decomposed, using:

\begin{itemize}

\item
{\it H-Balmer with 1 component + OF:}
a relatively narrow component with a FWHM at H$\beta$
and H$\alpha$ of 1065 $\pm$25 km s$^{-1}$, plus several OF components
(which fit the blueshifted asymmetry);\\

\item
{\it H-Balmer with 3 components + OF:}
a broad, an intermediate, and a narrow plus OF
components, with the following  FWHM at H$\beta$ and H$\alpha$,\\

\noindent
FWHM-H$\beta$-$_{BROAD}$   of  [2050 $\pm$30] km s$^{-1}$ and \\
FWHM-H$\beta$-$_{INTERM.}$ of  [ 780 $\pm$30] km s$^{-1}$.\\

\noindent
FWHM-H$\alpha$-$_{BROAD}$   of  [2150 $\pm$30] km s$^{-1}$ and \\
FWHM-H$\alpha$-$_{INTERM.}$ of  [ 800 $\pm$30] km s$^{-1}$.\\

It is interesting to remark  that
all the H-Balmer Broad components show a blueshift of $\sim$ 500 km s$^{-1}$,
in relation to the corresponding H-Balmer intermediate components. \\

\item
{\it Fe {\sc ii} with 1 component:}
for this emission line we found a
FWHM in the region of Fe {\sc ii}-$\lambda$4570 of [800 $\pm$35] km s$^{-1}$.\\

We found that 
the Fe {\sc ii} emission lines are at the same redshift of the H-Balmer
intermediate components (and also these lines show the same FWHM).
In addition,
Table 4 shows that IRAS 04505-2958 could be considered as a strong
Fe {\sc ii} emitter, since the ratio
Fe {\sc ii}$\lambda$4570$_{INTERM.}/$H$\beta$-$_{INTERM.}$ is larger than 1. \\

\end{itemize}

\item
{\bf [S {\sc ii}]$\lambda$6717-6731 and [N {\sc ii}]$\lambda$6583:} the
[S {\sc ii}]$\lambda$6717-6731 lines --at the QSO-core--
are very weak (almost absent).
In addition, 
the  [N {\sc ii}]$\lambda$6583 line is absent.

The GMOS-IFU spectra of almost all
the BAL + IR + Fe {\sc ii} QSOs (Mrk 231, IRAS 04505-2958,
IRAS 17002+5153, IRAS 07598+6508, etc) show --in the QSO-cores--
very weak NLR or  absent,  at [S {\sc ii}]$\lambda$6717-6731
and [N {\sc ii}]$\lambda$6583. We already associated this fact,
with the QSO OF process: which expel the NLR. \\

\end{enumerate}

\noindent
{\bf At the QSO-core, circumnuclear and external reg.:}

\begin{enumerate}

\item
{\bf [O {\sc iii}]$\lambda$5007:}
this strong emission line has a FWHM of
$\sim$630 $\pm$20 km s$^{-1}$
(with the peak blueshifted by -100 $\pm$25 km s$^{-1}$, from
the peak of the H-Balmer lines). This line shows also OF components.
L\'{i}pari (1994, his Fig. 4) showed that there is an anticorrelation
between the  strength and presence of this line [O {\sc iii}]$\lambda$5007
and the Fe {\sc ii} emission.

We remark that the surveys of the ionized gas using narrow band
 images show frecuently that the [O {\sc iii}]$\lambda$5007
 emission show different extension and location that those found for
 low ionization emission lines.

\item
{\bf [O {\sc ii}]$\lambda$3727:}
the strong [O {\sc ii}] line depicts a FWHM of 
$\sim$480 $\pm$25 km s$^{-1}$, and again, this line depicts OF components.

Figs. 20, 21 and 22 show a very interesting point about the strong
and extended [O {\sc iii}]$\lambda$5007 and [O {\sc ii}]$\lambda$3727
emissions.
In these plots is clear that the emission associated with these
narrow lines are very extended, which were detected  in almost
all  the GMOS-IFU field of $\sim$20 $\times$ 30 kpc.
Letawe et al. (2008) reported a similar result, using ESO VLT+VIMOS  spectra.
These emissions show the highest values of flux in the region of: the
shell S1 and S2, and  to the left border of the GMOS field. 

\end{enumerate}

Thus,
an interesting point about the GMOS results of the NLR observed
in IRAS 04505-2958 is the fact that we are observing at least
3 different NL emission systems:

(a) one strong NL system associated with the H-Balmer lines,
in the QSO-core, with an intermediate FWHM of 800 km s$^{-1}$;

(b) a very weak NL emission associated with the lines
[S {\sc ii}]$\lambda$6717+31 and [N {\sc ii}]$\lambda$6583 
(also in the QSO-core); 

(c) an extended and strong NL system associated with the
[O {\sc iii}]$\lambda$5007 and [O {\sc ii}]$\lambda$3727 emission
(in almost all the GMOS field of 20 $\times$ 30 kpc).\\

In addition,
Fig. 24 shows the FWHM-[O {\sc iii}]$\lambda$5007 map of IRAS
04505-2958. Large values of the FWHM of [O {\sc iii}]$\lambda$5007
were detected, in different regions of the shell S3. This
result could be explained by the presence of weak OF components
in [O {\sc iii}] (which can not be deblended --from the main emission
line component-- using the spectral resolution of GMOS-B600).

A similar  GMOS-IFU result was obtained
for the NLR associated with the weak [S {\sc ii}]$\lambda$6717+31
and [N {\sc ii}]$\lambda$6583 emission in Mrk 231, IRAS 17002+5153 and
IRAS 07598+6508 (Paper {\sc i}; L\'{i}pari et al. 2008).
Furthermore,
we found for the QSO-core of Mrk 231 GMOS spectral evidence of two
weak [S {\sc ii}] narrow emission line systems clearly associated with
OF of low velocity.
A simple explanation --already proposed-- for these very weak nuclear NLR
observed in BAL + IR + Fe {\sc ii} QSOs is that
the extreme OF expel  the NLR.
Even we already suggested that part of the {\it broad line emission region in
Mrk 231} is likely generated by an extreme OF process (L\'{i}pari et al.
2009, 2005a).

Finally, an interesting point is to study if the spectra of extreme OF
associated with giant-SNe/HyN could generate the spectra of NLS1, similar
to IRAS 04505-2958 or to the prototype of this class I Zw 1 (which is 
in addition an extreme Fe {\sc ii} emitter; Lipari 1994).
Fig. 25 shows the spectra of the
type {\sc ii}n SN 1998E (Ruiz \& Suntzeff 2009, private communication),
obtained at CTIO in 1998 January 31, with the 4 mt telescope,
together the spectra of SN 1998E, IRAS 04505-2958 and I Zw 1.
This plot  depicts very similar features in the spectra of SN 1998E
and IRAS 04505-2958. Moreover, this plot shows that
the spectra of SN 1998E and I Zw 1 are almost identical, and both with
extreme Fe {\sc ii} emission!.

Moreover,
L\'{i}pari et al. (2005a, their Fig. 14) showed the superposition of the
spectrum of the {\it radio HyN type {\sc ii}--L 1979c} (observed in 1979
June 26.18; Branch et  al. 1981) and Mrk 231.
Only using colours it is possible to distinguish each spectrum, since they
are almost identical.
Thus, likely
a more constant OF --rather than a single SN-- could explain the very
unusual spectra of: NLS1 in general (even with extreme Fe {\sc ii} emission,
like {\sc i} Zw 1), IRAS 04505-2958 (a NLS1 with strong Fe {\sc ii})
and  Mrk 231 (a QSO with broad H-Balmer lines and extreme Fe {\sc ii}).

Therefore, from all these results we are suggesting that at least part
of the narrow, intermediate and broad  line emissions  are associated
with explosive OF processes: i.e. in giant-SN/HyN + galactic winds + shells.


\subsection{The  Intermediate Emission Line Component at H-Balmer and
Fe {\sc ii} }
\label{res9.2-BLR}

In Table 4, new GMOS results regarding the H-Balmer and Fe {\sc ii}
emissions were presented for the QSO-core, of IRAS 04505-2958.
Specifically, we found that the widths of the H-Balmer H$\beta$ intermediate
components and  the Fe {\sc ii} emission show exactly the same
value of  FWHM of [800 $\pm$ 30] km s$^{-1}$. In addition,
we already noted that the H-Balmer Broad components
are blueshifted (of $\sim$ --500 km s$^{-1}$) in relation to the H-Balmer
intermediate components and the Fe {\sc ii} emission.

A very similar result was found by Popovich et al. (2009),
from a 3D spectroscopic study of the Narrow Line Seyfert 1 AGN  Mrk 493.
Specifically, they found that the widths of the
H-Balmer intermediate emission and  the width of the Fe {\sc ii} show
the same value of FWHM-H$\beta$ $=$ [790 $\pm$ 80] km s$^{-1}$.
Thus, they associated the same origin for the Fe {\sc ii} and
the intermediate H$\beta$ component. In addition,
Popovich et al. (2009)  found that the NLR of Mrk 493
is  ionized by H {\sc ii} regions (not by the Seyfert 1 nucleus).

Lipari (1994) already included Mrk 493 in his IR colour-colour
evolutionary diagram. He found that Mrk 493 is located
at the end of the second sequence of {\bf transition AGNs/NLS1s}
(i.e., close to the power low area).

In addition,  from the study of two sample of 568 and 4037 QSOs
Hu et al. (2008a,b) found that the H$\beta$ emissions,
of almost all these QSOs:
(i) can be decomposed in a broad and intermediate component;
(ii) the shift and width of the intermediate component correlate
with the Fe {\sc ii} emission, but not with the broad one. They also
detected that these broad H$\beta$ emissions are blueshifted of
$\sim$ --400 km s$^{-1}$ in relation with the intermediate H$\beta$
and Fe {\sc ii} emission. They suggested that these results could
be explained by the presence of OF.
L\'{i}pari \&  Terlevich (2006) and in the sub-Section 2.2 (of the present
paper)  already analysed --from the observational and theoretical
point of view-- the  OF process associated with galactic wind as 
one of the main source of the NLR emission, in composite QSOs.

Thus, the GMOS data obtained for the intermediate and narrow  
emission of IRAS 04505-2958, plus the results of similar NLS1 and of
large sample of QSOs, suggest that: it is
important to know more clearly the nature of these emission line
regions before to reach a final conclusion about the mass of the SMBH,
and the host galaxy.
Since at least part of the NLR and ILR in BAL + IR + Fe {\sc ii} QSOs
could be associated with strong OF process + giant-SN/HyN.


\subsection{The Broad Emission at H-Balmer in IRAS 04505-2958
and BAL + IR + Fe {\sc ii} QSOs}

\label{res9.4-Interm+Broad BAL}

In the previous sub-section we have explained that only including a broad,
intermediate and weak-OF components, in the fit of the H-Balmer emission
lines the fitting was correct.
Thus, we found for H$\beta$ and H$\alpha$ emission a broad component with
FWHM-H$\beta$-$_{BROAD}$   of  [2050 $\pm$30] km s$^{-1}$ and 
FWHM-H$\alpha$-$_{BROAD}$   of  [2150 $\pm$30] km s$^{-1}$.
An interesting point is to analyse the possible nature of this component.
Which could be associated  with the SMBH and/or the extreme OF.

According to the extreme OF + explosive + shells composite-scenario for
IRAS 04505-2958, it is interesting to note --from the theoretical point
of view-- that
several authors proposed that at least part of the broad line
{\it emission} region could be associated with OF processes.
In particular, these theoretical works suggest that the BLRs could be
associated with different types of outflow processes. The main models
associated the OF with
ejecta of SN remnants, shocked clouds in nuclear galactic winds,
extended stellar envelopes, accretion disks, jets, etc
Terlevich et al. 1992; Perry 1992; Perry \& Dyson 1992; Dyson, Perry \&
Williams 1992; Scoville \& Norman 1988; Norman \& Miley 1984; see for
a review Sulentic, Marziani, \& Dultzin--Hacyan 2000).
In particular,
Terlevich et al. (1992) showed that all the ELR of the BLR could
be explained in the framework of compact SNR.

L\'{i}pari et al. (2004d) found in all the IR QSOs with OF of their sample
(with more than 50 QSOs) that the H$\beta$ broad line
component is blushifted in relation to the narrow one.
For the broad H$\beta$ component of IRAS 04505-2958
their measured a blushift/OF $\sim$ -1700 km s$^{-1}$.
Which correspond the same velocity of the BAL detected in
C {\sc iv}$\lambda$1550 emission line.
Thus, this result -studied in more detail using GMOS data-
suggest that the optical low ionization BL-emission and the BAL could
be originated in the same OF process, with supershells.

On the other hand, even in the standard model of SMBHs/AGNs the OF
process could play an important role. 
There are two main groups of standard models about the structure and
dynamics of the gas near the core of QSOs and specifically, about  the
broad emission line region (BELR) and broad absorption line region (BALR).
In these models the gas may exist as:

(i) {\it Continuous Winds}:
spectral analysis of Arav et al. (1994, 1997, 1998, 2001, 2005)
seems to show that continuous winds might be better suited to explain
high resolution spectra  of BALR and BELR.

(ii) {\it Discrete Clouds:}
 the idea that gas is partitioned into discrete clouds is the more
 traditional approach to BELR and BALR (see Everett, Konigl,
Kartje 2000; Bottorff et al. 1997).

Therefore, we suggest that at least part of the Fe {\sc ii} emission
and the H-Balmer intermediate + broad width regions could be originated
in OF processes (in the cores of BAL + IR + Fe {\sc ii} QSOs), more
specifically
in warm regions obscured from direct ionizing UV photons. The obscuring
material could be in the form of expanding shells. Giant explosive
events would produce large scale shocks plus shock-heated material.
This scenario is in good agreement with our
recent finding, for the BAL + IR + Fe {\sc ii} QSOs IRAS 07598+6508.
We found that the properties of the BLR of this BAL QSO are consistent
with collisional rather than radiative process (Veron et al. 2006).


\subsection{The Broad Absorption Line systems in IRAS 04505-2958}

\label{res9.5-Narrow BAL}

In general, the GMOS data of
IRAS 04505-2958 and Mrk 231 show  that both QSOs have similar OF processes
and properties,
even both QSOs have  ``relatively narrow" BALs
(L\'{i}pari et al. 2005a, 1994).
In addition, in Paper {\sc i},
we already detected the extended nature of the BAL system {\sc i} of Mrk 231.

Extended BALs were detected in others narrow  BALs, by de-Kool et al.
(2001, 2002).
Using very high resolution Keck spectroscopic data of associated
absorption line (AAL) and BALs QSOs FIRST J104459.6+365605 and
FBQS 0840+3633,
they found that the distances between the AGN and the region where the
OF gas generate the AAL and BAL line are $\sim$700 and $\sim$230 pc,
respectively.

It is interesting to point that
Mrk 231, FIRST J104459.6+365605 and FBQS 0840+3633
(which show extended BAL systems)
are all members of the rare class of low ionization BAL QSOs.
Furthermore, these 3 QSOs are also members of the  ``very" rare sub-class of
Fe {\sc ii} low ionization BAL QSOs with very strong reddening in the
UV continuum. In particular, for Mrk 231 L\'{i}pari et al. (2005a)  found
the presence of strong absorption in the Fe {\sc ii} and Mg {\sc ii} lines;
which are the standard
lines that define the  Fe {\sc ii} low ionization BAL QSO sub-class.
Thus, an interesting alternative that required to be studied in detail is
the possibility that the BAL system of IRAS 04505-2958 is extended.

\clearpage


\section{The QSO as the main source of the ultra luminous IR
emission}

\label{disc-10}

In Sections 1 and 2, we have explained that
the mid and far IR emissions of IRAS 04505-2958 were associated
{\it ``mainly"} with a luminous quasar
(see de Grijp et al. 1987, 1992; Low et al. 1988, 1989; Hutching \&
Neff 1988; L\'{i}pari et al. 2003, 2005a, 2007a,b, 2009; L\'{i}pari \&
Terlevich 2006; Kim et al. 2007; Zhou  et al. 2007; and others).
However, some authors
suggested that the extended object could be a companion/interacting ring
galaxy and also the ``only" source of the ultra luminous IR continuum
emission (Canalizo \& Stockton 2001; Magain et al. 2005; Merrit et al.
2006; and others).

L\'{i}pari et al. (2005a) already proposed 
that the QSO is at least the dominant source of the ultra luminous
IR emission IRAS 04505-2958. Since
this IR source is located in the IR colour-colour diagram exactly in
the sequence of BAL + IR + Fe {\sc ii} QSOs. Furthermore,
the BAL system of this QSO was found using this  IR colours diagram
(i.e. using their mid and far IR emission, which are typical of
 IR QSO; see also de Grijp et al. 1987, 1992; Low et al. 1988, 1989).

In the context of the new GMOS-IFU data, we remark the following
results:

\begin{enumerate}

\item
Regarding the location  of IRAS 04505-2958 in the IR colours diagram,
it is important
to remark that their position is between two standard BAL + IR + Fe {\sc ii}
QSOs: IRAS 07598+6508 and IRAS 17002+5153.
Using new GMOS-IFU data, L\'{i}pari et al. (2008) found new evidences that
IRAS 07598+6508 and IRAS 17002+5153 are also explosive
BAL + IR + Fe {\sc ii} QSOs with supergiant shells (very similar to
IRAS 04505-2958 and Mrk 231).

\item
In the shell S3 the GMOS spectra show young starbursts in the main knot
K1, K2 and K3; with also strong OF processes and  shocks.
Thus the shell could be a second source of IR energy.

\item
For the shell S3
all the GMOS velocity fields show that the kinematics and the multiple
emission line components  ere
related with strong OF process (similar to the kinematics of the
proto-type of expanding external shell in NGC 5514). Thus, the
kinematics of this extended object is not consistent with a interacting
ring galaxy (this is an interesting point since several authors suggested
that the only source of IR emission is a ring galaxy).

Recently using HST-NICMOS and ESO-VLT/VISIR near IR images,
Jahnke et al. (2009) also concluded that S3 is not a collisional ring
galaxy.

\end{enumerate}

Therefore, there are several inconsistencies in the idea that 
the extended object is a ring interacting galaxy and the only
source of the ultra luminous IR energy.
On the other hand,
all the previous enumerated results are in excellent agreement with the
original suggestion that the QSO is the {\it dominant source of
 ultra luminous IR energy}.
However, it is important to note that probably also the 
starburst process detected in the knots K1, K2 and K3 of the shell
S3 could be a second source of  IR emission, which is in
agreement with the detection --in S3-- of CO J $=$ 1-0 line
emission, with a derived mass M(H$_2$) $\sim$ 2 $\times$
10$^{10}$ M$_{\odot}$ and high star formation rate
(Papadopoulos et al. 2008).

Moreover, Jahnke et al. (2009) and Letawe et al. (2009) suggested
that the QSO and the extended companion galaxy are both ultraluminous
IR source. This proposition is in agreement with the scenario propossed
in this paper for IRAS 04505-2958 (QSO + a galaxy in formation).
In addition, we alrady noted that
 Letawe et al. (2009) found --in their IR images-- a point
source close to the knot S3-K5 (strongly obscured, at optical wavelength,
by dust), which they associated
with an AGN, and/or a compact and unusual extremelly brigth starburst.
This last result could be also in agreement with our proposition
that S3 is probably a young galaxy in formation.

Finally, we note that Papdopoulus et al. (2008) suggested a new scenario for
IRAS 04505-2958 and ``some" transition IR QSOs. Specifically, using
their interesting result that the CO emission was detected only in S3
but not in the QSO-host, they proposed that the QSO + S3 (in IRAS 04505-2958)
could be considered an example of gas--poor (elliptical) + gas-rich (spiral)
interaction of galaxies.
In the hyperwind scenario --for IRAS 04505-2958-- the results
of the study of CO in the QSO (absence of CO molecular gas) could be
explained mainly by the ejection of the ISM/CO by  multiple explosive processes
(and the remnant galaxy --of these multiple explosive events-- could be a
dwarf elliptical).

\clearpage


\section{The Host Galaxy of the BAL QSO IRAS 04505-2958}

\label{res11-HostGal}

Recently, several interesting observational and theoretical
studies of the host galaxy of the QSO IRAS 04505-2958 were carried out.
A brief  summary of these results are presented here
(then these results will be compared with the new GMOS-IFU data):

\begin{itemize}

\item
Boyce et al. (1996) --using HST--WFPC2 images-- detected at the
same redshift of the QSO IRAS 04505-2958 a close and extended object.
They associated  this extended object with a companion ring galaxy,
which is probably interacting with the host galaxy of the QSO.

\item
L\'{i}pari et al. (2003, 2005a) --using HST--WFPC2 images and HST/FOS
plus CASLEO spectra-- associated the extended object 
with a hyper shell generated in an extreme OF process (very similar
to those detected in almost all the BAL + IR + Fe {\sc ii} QSOs).

\item
From a detailed study of host galaxies of a sample of 17 QSOs
(using  HST--ACS images, ESO--VLT long-slit spectroscopy),
Magain et al. (2005) found that only in the case of the
QSO HE 0450-2958 the host galaxy was not detected.
This result (the absence of detection of the host) was found in
 high resolution data: the deconvolved HST--ACS images
and the deconvolved ESO--VLT 1D-Spectra.

Magain et al. (2005) proposed that the host galaxy is under
the detection limit could suggests that
the host galaxy is  dark or is absent (i.e., a naked QSO).

\end{itemize}

Therefore,
these HST--WFPC2, HST--ACS images and ESO--VLT, HST--FOS spectra show
two very interesting but controversial results, for the
host galaxy of this BAL IR-QSO.
Specifically:
(1) {\it the host galaxy} --in this bright QSO--  remain undetected,
even using deep HST--ACS images and ESO--VLT spectra (plus using
 deconvolution technique of images and spectra); but
(2) {\it an extended object} --at the same redshift of the
QSO-- was found; which shows a very clear, bright and a knotty
sub-structure.

From the theoretical point of view,
four/five main --and very different--  scenarios were proposed, for
this QSO + host + extended ring object.
In particular:

\begin{enumerate}              

\item
{\bf Interaction of Galaxies.}
Boyce et al. (1996), from the HST--WFPC2 images of IRAS 04505-2958
suggested that this QSO could be the result of an interaction of
galaxies, between the host galaxy of the QSO and a close and very
extended object.

\item
{\bf Explosive BAL + IR + Fe {\sc ii} QSO.}
L\'{i}pari et al. (2005a, 2007a,b, 2009) proposed for
IRAS 04505-2958 an explosive and composite hyperwind scenario.
More specifically, we suggested that:
(a) the close and extended object is a hyper shells, probably forming a
companion/satellite galaxy;
(b) the extreme OF --generated by the composite QSO-- with multiple explosive
process and shells, probably expel a high fraction of the ISM of the host
galaxy.

\item
{\bf Naked QSO.}
Magain et al. (2005) from a study of host galaxy in QSOs (using new
HST--ACS images and ESO--VLT spectra) suggested  that the
host galaxy is absent in their deconvolved data.
They proposed that: the host is dark, or a naked QSO scenario;

\item
{\bf Ejected QSO.}
Haehnlet, Davies \& Rees (2005) analysed theoretically the possibility
that a naked QSO was ejected from the companion ring galaxy-candidate.
Hoffman \& Loeb (2006) discussed the special conditions required for
the theoretical ejected scenario for this QSO.

\item
{\bf Normal Host Galaxy of  NLS1.}
Merritt et al. (2006) suggested that the value of the
black hole mass --of this high luminous version of NLS1 AGN-- is lower
that the value obtained by Magain et al. (2005); and thus, the host galaxy
could be fainter and less massive than the values assumed previously
(by Magain et al. 2005).

\end{enumerate}

In this paper, 
we found new GMOS evidences which are in good agreement with the
explosive and hyperwind model in this 
BAL + IR + Fe {\sc ii} QSO IRAS 04505-2958.
The main new GMOS-IFU evidence (supporting the explosive scenario)
are the following:

\begin{itemize}

\item
Multiple emission line components were detected in the QSO-core, in
the two circumnuclear shell S1, S2 and in the extended hypergiant shell S3.
These components show very high OF velocities (even with $\Delta$V $>$ 2000
km s$^{-1}$)
which could be associated only with extreme and explosive process
(see Sections 2 and 5; and Suchkov et al. 1994).

\item
The kinematics and emission line ratios maps are consistent with an
extreme OF and associated shocks in the QSO and in the shells S1, S2 and S3.
Specifically, the kinematics of the shell S3 show a small scale bipolar OF.

\item
A blue component was detected in all the continuum GMOS map
(3.5$''\times$5$''$ $\sim$ 20$\times$30 kpc), which
is consistent with an extreme galactic wind (associated with the QSO).

\item
The presence of a very extended shell S4 at r $\sim$80 kpc (previously
found by Hutching \& Neff 1988) was discussed. In particular, we found
that this shell is probably associated with a bipolar OF, at
PA $=$ 40$^{\circ}$ and with an opening angle of 95$^{\circ}$ (Fig. 3).

\end{itemize}

Thus, these new GMOS-IFU results are in good agreement with a composite
and explosive hyperwind scenario for the BAL +
IR + Fe {\sc ii} QSOs IRAS 04505-2958.
Furthermore,
this hyperwind model could explain the previous
--apparently-- surprising and controversial results:
               
\begin{enumerate}

\item
{\it The host galaxy}  remain undetected, because the hyperwind
ejected a high fraction of the ISM of the host galaxy; and

\item
{\it The extended, bright and knotty object} is an hyper shell
(similar to those detected in almost all the BAL + IR + Fe {\sc ii}
QSOs) with properties of a shell and also of a companion/satellite galaxy.
The kinematic maps and the spectra of S3 show strong OF, which
suggests that this likely young galaxy is still in the phase of
formation via explosions. Which is in good agreement with the theoretical
explosive models of formation of galaxies (see Section 2 and Ikeuchi 1981;
Ostriker \& Cowie 1981; Berman \& Suchkov 1991).

\end{enumerate}

In conclusion, the hyperwind model
--with multiple extreme explosive events-- 
for BAL + IR Fe {\sc ii} QSOs explain with
a very simple physic, the fact that the extended object plus
the two internal blobs were detected very clearly, but at the same redshift
the host galaxy still remain undetected.

\clearpage


\section{Explosive Model for IRAS 04505$-$2958,
and for Galaxy Formation/End}

\label{disc-12}

For the discussion of the  explosive and composite  hyperwind model
for IRAS 04505$-$2958,
it is important to remark some interesting previous results.
More specifically, \\

\noindent
{\bf From the observational point of view:}

Several surveys of Ly$\alpha$ emitters at high z (Steidel et al.
2000; Keel et al. 1999; Francis et al. 2001;  Matsuda
et al. 2004) have established the existence of extended, highly luminous
Ly$\alpha$ halos (of 50-100 kpc and 1.4 $\times$ 10$^{44}$ erg s$^{-1}$).
In addition, several  extended Ly$\alpha$ halos were
detected in high redshift radio sources (see for references Reuland et al.
2003).

In several  BAL + IR + Fe {\sc ii} QSOs
L\'{i}pari et al. (2003, 2005a, 2007a,b, 2008, 2009) detected very extended
OF processes of 50--100 kpc with giant shells and bubbles. \\


\noindent
{\bf From the theoretical point of view:} 

Dey et al. (1997) and Reuland et al. (2003) proposed
that in high z radio source: starburst and superwinds can generate
extended Ly$\alpha$ nebulae/halos. 
Taniguchi \& Shioya (2000) suggested a starburst hyperwind scenario for
the origin of the high redshift Ly$\alpha$ blobs.

L\'{i}pari et al. (2005a, 2007a,b, 2009); L\'{i}pari \& Terlevich (2006)
proposed a {\it composite and explosive hyperwind scenario} in order to
explain the very extended shells --of $\sim$30--100 kpc-- found in the
BAL QSO IRAS04505-2958. In addition,
they proposed a similar hyperwind for the OF process of 50 kpc in the
BAL + IR + Fe {\sc ii} QSOs Mrk 231 (and also for IRAS 17002+5153 and
IRAS~07598+6508; L\'{i}pari et al. 2008).


\subsection{Explosive Model for the QSO and the shells S1 and S2}

\label{disc-13.1}


The GMOS-IFU data presented in this paper show clear evidences that
an extreme outflow process is present in
the QSO-core of IRAS 04505-2958, and also in the close shells (with
symmetric and circular external--borders) S1 and S2.

From this GMOS study of the OF process (at different scale, from the QSO-core
to the multiple expanding shells systems), we found again that the detection
of multiple OF components and the ELR --for the QSO-core and the shells
S1, S2-- show a good agreement with an extreme galactic wind scenario.
Specifically,
multiple OF emission systems were detected, showing very high
OF velocities. Specifically these OF systems depict 
$\Delta$V from 300 to 3000 km s$^{-1}$.
These very high velocities --of the multiple emission line components--
could be explained  as OF processes. In addition,
the ELR diagrams for the shells S1 and S2 show the typical values associated
with ionization by OF + shocks plus the QSO. 
Thus, our GMOS results suggest the presence of multiple expanding
supergiant shells, which are centered at the location of the QSO
IRAS 04505-2958.


\subsection{Explosive Model for the Shell S3 (and
a new scenario for Galaxy Formation)}

\label{disc-13.2}

Fig. 2 shows  the very extended morphology of the main
hypergiant shell S3.
Again in this plot,
the border of S3 depicts symmetric and circular external--border,
with the centre at the position of the QSO (this result is the
same that the previous one obtained for S1 and S2). Moreover, this
fact is also in agreement with the previous result of Hutching \& Neff
(1988), in the sense that the hyper shell S4 is centered in the QSO. 

Here, we remark some interesting results found for S3, and using
GMOS-3D individual spectra, emission lines maps, and plots of the
kinematics and physical conditions.
All the main knots of the shell S3 show multiple emission lines with
multiple components which could be only associated with an OF process,
in addition these multiple components shows mainly LINER 
properties, which are likely generated by shocks (these shocks are
clearly observed in the [S {\sc ii}]/H$\alpha$ ELR map).

Moreover,
the complex kinematics of S3 --for all the observed emission lines--
show a small scale bipolar OF, with centre in the main knost K4 and K3.
These results show the typical physical properties of an external shell
with extreme explosive OF processes
(for example these results are similar to those obtained for the
kinematics of the prototype of external exploding shell in
NGC 5514; L\'{i}pari et al. 2004c).
In addition, 
the main knots K1, K2, and K3 --of this hyper shell S3--  show a young
starburst, with multiple OF emission line components.

Therefore, {\it S3 exhibits typical properties of a shell in
expansion, but also the properties of a young companion or satellite
galaxy (in formation)}.
More specifically, we have explained that S3 depicts
kinematics of an expanding shell; plus
starburst knots with multiple OF components and ELR typical of
extreme OF process.
A new and strong support for the  young galaxy scenario --for S3--
came from the detection  of CO J $=$ 1-0 line emission in
S3: with a mass M(H$_2$) $\sim$ 2 $\times$ 10$^{10}$ M$_{\odot}$
(Papadopoulos et al. 2008). Which is at least a $\sim$ 30 per cent of
the dynamical mass in the CO-luminous region. This is one of the standard
criteria for the definition of a galaxy in formation.

In conclusion,
the GMOS plus the CO J $=$ 1-0 results are in exellent agreement with
the prediction of theoretical explosive models for the formation of
galaxies/QSOs. Which were already proposed by Ikeuchi (1981),
Ostriker \& Cowie (1981), Berman \& Suchkov (1991). For details of these
explosive models see Section 2.

                                        
\subsection{Hyperwind and Explosive Model for Ly$\alpha$ Blobs}

\label{disc-13.3}

In the introduction of this Section we have explained that
in the last years very extended blobs --specially in Ly$\alpha$--  has been
detected in a variety of high and low redshift objects.
In addition,
the results of the surveys at high z of bright Sub-mm source
(Chapman et al. 2004a; Bower et al. 2004, Swinbank et al. 2005) suggest that
a  high fraction (3/4) of these sources are extended and complex (i.e.,
showing extended and highly luminous Ly$\alpha$ halos;
Chapman et al. 2004a,b).

L\'{i}pari et al. (2004a) already found that
75$\%$ of IR QSOs and mergers (including BAL QSOs) show clear evidence of OF.
Which is the same percent  (75$\%$) found by Chapman et al. (2004b,a) in
their study of Sub-mm sources showing extended and highly luminous
Ly$\alpha$ halos.
Recently, we have started a study of 3D spectroscopic data of high
redshift Sub-mm and Radio BAL-QSOs, using Gemini+GMOS and ESO VLT+VIMOS.
Using these GMOS data we are studying the interesting possibility
(already suggested by Lipari \& Terlevich 2006) that in Sub-mm and
Radio QSOs --at high redshif-- also extreme explosive OF process could play a
main role in their evolution.



\subsection{Explosive Model for the End of the Host Galaxy}

\label{disc-13.4}

Multiple explosive events expelling a high fraction of the
host galaxy could be a probable explanation for interesting and
controversial result the shell S3 is clearly observed,but at the
same redshift the host galaxy of the QSO remain undetected.
Several explosive events can eject a large fraction of the ISM.
Moreover, an extreme galactic wind could strongly change the
kinematics and the physical condition of the ISM in general (and
not only the ejected ISM). Thus, in this way
explosive processes would play a main role in the evolution of the
star formation and therefore in the evolution of the galaxy.
Even extreme OF process could define the mass of the remnant of the
original galaxy. The end product of a multiple
explosive processes was called a galaxy remnant.

We believed that likely in IRAS 04505-2958 we are observing for
the first time a candidatefor a host galaxy  at the end phase
of their evolution, or a galaxy remnant.

Thus, giant QSOs explosions is an interesting process
in order to consider as the base for a  model of galaxy end.
Our observational GMOS-IFU results for BAL + IR + Fe {\sc ii} QSOs,
plus several theoretical works
show a good agreement with explosive models for the end and the formation
of ``some" type of galaxies.

\clearpage


\section{Main Consequences of the Explosive Model:
HyN, CR and Neutrinos/Dark-Matter}

\label{disc-14}

The GMOS-IFU data of IRAS 04505-2958 show new evidence of a multiple
hypergiant symmetric shells,  with centre at the location of the QSO.
These hypershells could be generated only by giant SN and HyN
(Norman \& Ikeuchi 1989; Heiles 1979; Suchkov et al. 1994;
Strickland \&  Stevens 2000; Tenorio-Tagle et al. 1999, 2005, 2006).
In addition, the high resolution spectra of the QSO show multiple
OF components, which could be associated only with OF processes.
Moreover, from our GMOS observational programme and study of nearby
BAL + IR + Fe {\sc ii} QSOs
we found new evidence of super/hyper shells and explosive processes
in four of these objects: Mrk 231, IRAS 04505-2958, IRAS 17002+5153,
IRAS 07598+6508 (Lipari et al. 2007a,b, 2008, 2009).
In the evolutionary, explosive and composite model for
BAL + IR + Fe {\sc ii} QSOs, the presence of multiple shell systems,
 extreme OF, and extreme explosive events are associated
mainly with HyN and giant SNe (Lipari \& Terlevich 2006).

In this paper we show that the OF process of giant type {\sc ii} SN/Hyn
(like 1998E) could explaine the spectra of the very rare class
of NLS1 galaxies and even the spectra of extreme Fe {\sc ii} emitters.
The strong and extreme Fe {\sc ii} emission could not be explained
using standard fotoionization models (see for details Section 2;
Lipari \& Terlevich 2006; Veron et al. 2006).
In addition, the discovery of new giant HyNe (like 2006gy, 2006tf, 2005ap)
confirms our suggestion of the presence of extreme explosive events,
powered by the death of extreme massive stars (like Eta Carinae).
These and similar HyNe in QSOs and galaxies could help to explain several
main themes in Astrophysics.

Specifically, the GMOS results obtained in this paper for
IRAS 04505-2958 are in good agreement with some theoretical studies that
suggest that ultra high energy cosmic rays (CR) and neutrinos are generated
in giant-SN/HyN explosions. Thus, the GMOS results obtained in this paper
could help to discriminate between several theoretical models for the
generation of ultra high energy CR and Neutrinos.
This is probably one of the main astrophysical consequences,  derived
from the study of the explosive model for composite AGNs/QSOs
(see for details Section 2).


\subsection{Diversity of HyperNova associated with GRB} 

\label{disc-14.1}


The GRB-HyN connection (Woosley \& Bloom 2006; see also Colgate 1968)
shows a new type of giant-SN explosion, associated with:
(a) high kinetic energy in the range E$_{SN}$ $\sim$ 10$^{52}$--10$^{53}$ erg,
(b) very broad emission lines, and
(c) strong radio emission indicating
relativistic expansion of $\sim$0.3$\times$c (Kulkarni et al. 1998).
Nomoto et al. (2004, 2006, 2007a,b,c, 2008); Packzinki (1998) studied this
type of giant--SN, that they call HyN.

More specifically, there is strong observational evidence,
that --at least-- some large duration gamma ray burst (GRB) are
associated with giant-SN and starbursts areas.
Five pairs were detected of long duration GRB associated with
confirmed giant-SN/HyN (which show spectra of giant-SN):
HyN 1998bw+GRB 980425, HyN 2003dh+GRB 030329,
HyN 2003lw+GRB 031203, SN 2006aj+XRF 060218 and
SN 2008d/XRT 080109.

For the giant-SN/HyN 2008bw, 2003dh, 2003lw, Nomoto et al. (2007c)
found that these events could be explained
as the collapse to a Black Hole, of the core of massive star of $\sim$40--45
M$_{\odot}$  (and ejected mass of $\sim$4--10 M$_{\odot}$).
For of SN 2006aj Nomoto et al. (2007c) found
that the progenitor  had a smaller mass than the previous
HyN+GRB, with a value of $\sim$20 M$_{\odot}$. This result suggests that
a Neutron star was formed.
For  SN 2008d Tanaka et al. (2008) found for the progenitor  is a
main sequence star with a mass of M$_{MS} =$ 20-25 M$_{\odot}$.
Li (2008), Xu et al. (2008) and Mazzali et al. (2008) considered this
XRT as the lest energetic end of GRBs and XRFs.

Thus,
there are very different types of long duration GRBs and HyNe.
Since even these five pairs/cases of confirmed GRBs associated
with HyNe are very different.
From the study of these five HyN+GRB several interesting consequences
could be derived.
In particular:
(i) For these five pairs of GRB + HyN,
a linear relation was found between the peak of of energy of GRB/XRF
versus peak of bolometric magnitude of the associated giant-SN/HyN
(Li 2008). 
(ii) The detection of the normal type Ibc SN 2008d (associated
with XRT 080109), clearly extends:
{\it the GRB--HyN connection to normal core-collapse SNe}. 
Hence, it has been suggested that probably every core-collapse SN
(type Ib, Ic and II) has a GRB/XRF associated with it (Li 2008). Moreover,
Bloom (2003) proposed that all long duration GRB could be associated with
giant-SN/HyN

Moreover,
even the standard collimated-jet model --for the origin of GRB--
require to be analysed in detail for each GRB/HyN. Since
investigations found that GRBs with softer spectra
tend to have larger jet opening angle: i.e., weakly collimated outflows
(Lamb, Donaghy \& Graziani 2005; Li 2007). Even,
it appear that some GRBs have spherical outflow (see for references
and details Li 2007, 2008).

Therefore, in these very different types of HyN+GRB: the
mildly relativistic ejecta and the relativistic jets  are both
important physical processes; which could generate ultra high
energy emission.


\subsection{Diversity of HyperNova in General} 

\label{disc-14.2}

In addition of HyN associated with GRB, several
types of hypernovae (and giant explosive processes) were
observational and theoretical studied.
Specifically, the following main types of HyN were observed 
and/or theoretically proposed,

\begin{enumerate}

\item
{\bf Radio HyperNovae}

Several years before the discovery of the first HyN (associated with GRB),
Colina \& Perez-Olea (1992, 1995)
suggested that the presence of compact strong Radio-SN remnant
--and strong Radio-SNe-- mean also the existence of HyNe
(that they call Radio-HyN). They proposed that the prototype
of radio hypernova is the Radio-SN/HyN 1979c. We already noted
the  spectra of Radio-HyN 1979c and the BAL + IR + Fe {\sc ii} QSO
are almost identical. In addition,
Weiler et al. (2002) proposed that one of the main process associated
with  the radio emission --in very bright radio-SN/HyN-- can be best
explained
as the interaction of a mildly relativistic shock ($\Gamma$  $\sim$ 1.6)
with a dense pre-explosion stellar wind, in the circumstellar medium.
Which is the same interpretation of the radio emission of HyN+GRB,
suggested by  Kulkarni et al. (1998) and others.

From a survey of SNe at radio emission in Arp 220, 
Lonsdale et al. (2006) reported the important detection of 4 new and
strong radio-SN/HyN in a period of only 12 months, in the two nuclear
regions.
Arp 220 is of one of the prototype of IR Merger with extreme OF +
{\it associated very extended super-giant shells}. This IR Merger
shows two shells with bipolar structure, each one with an extension or
 radius of $\sim$15 kpc  (detected by Heckman et al. 1987, 1990).

\item
{\bf HyperNovae associated with Extreme Massive Stars} \\

An important result, in the HyN field, was the discovery of
the SN 2006gy (in NGC 1260, Smith et al. 2007), that reached a peak
of absolute magnitude of --22, and remain brighter than --21 mag for
about 100 days!.

This SN 2006gy (of type {\sc ii}n) is one of the most luminous SN,
powered by the death of an extremely massive star. This
result confirm one of the main suggestion of the evolutionary and explosive
model for composite AGNs: the existence of giant--SN/HyN explosions,
associated with extreme massive stars, like Eta Carinae (L\'{i}pari et al. 2003,
2005a; L\'{i}pari \& Terlevich 2006).

Very recently, the discovery of new giant-SN or HyN similar to 2006gy
(SN 2006tf and 2005ap; see Smith et al. 2008; Quimby et al. 2007)
confirm the presence of these extreme explosive events.
This type {\sc ii}n HyN and their remnant
could help to explain several main themes in
Astrophysics (see the next sub-Sections).   \\

\item
{\bf HyperNovae associated with Neutron Stars} \\

In accretion disks of AGNs, the star--gas interactions may lead to a
special mode of massive star formation.
Collin \& Zahn (1999) suggest that the residuals of the first
SNe, mainly {\bf neutron stars}, can undergo a new accretion/interaction
phase, with the gas, leading to very powerful SN or hypernova explosions.
We have already suggested that in the core of BAL + IR + Fe {\sc ii} QSOs
some HyN could be generated from the collapse of {\bf neutron stars in
accretion disks}
(these are giant-SN/HyN generated in a second explosive event). \\

\item
{\bf HyperNovae associated with Population {\sc iii} Stars}

It has been suggested that in the Population {\sc iii} star,
extreme massive stars existed  (Abel, Brian, \& Norman 2000;
Bromm, Coppi \& Larson 2002; Nakamura \& Umemura 1999; and others).
Several authors already studied  the collapse --end phases-- of very
and extremely massive Pop. {\sc iii} stars.
Ohkubo et al. (2006) presented a summary of the main theoretical results
obtained for the end phases of extreme massive stars, and for
different ranges of masses:\\
{\it (a)   8 M$_{\odot}$ to  130 M$_{\odot}$}, the stars undergo ONe Fe core
collapse leaving Neutron stars and black hole. \\
{\it (b) 130 M$_{\odot}$ to  300 M$_{\odot}$}, the stars undergo electron
positron pair creation instability, during O burning, releasing more energy
by nuclear burning than the gravitational energy of the star; and thus these
stars disrupt completely as pair-instability SN (PISN).\\
{\it (c) 300 M$_{\odot}$ to  $\sim$1000-10000 M$_{\odot}$}, also these stars
enter in PISN, but continue to collapse (see  Fryer, Woosley, \& Heger 2001;
Ohkubo et al. 2006; Heger et al. 2002; Nomoto et al. 2004, 2006, 2007a,b,c,
2008; and others).

The results of these theoretical SN models suggest that
{\it these very and extreme massive population {\sc iii} (or primordial)
stars} explode as giant--SN/HyN with energies of 10$^{52}$--10$^{53}$ erg
(Fryer et al. 2001; Heger et al. 2002; Ohkubo et al. 2006; Nomoto et al.
2004, 2006, 2007a,b,c, 2008). Moreover,
Collin \& Zahn (1999) suggested that
in the accretion regions of the AGNs, the massive stars could
be similar to extreme massive population {\sc iii} stars.

\end{enumerate}

Finally,
following the results presented in the previous paragraphs we conclude
that very different types of HyNe could be present in composite QSOs/AGNs and
specially in the core of explosive low ionization BAL + IR + Fe {\sc ii}
QSOs. In HyNe  mildly relativistic ejecta  is  probably
the source of ultra high energy (UHE) Cosmic-Rays (UHE-CR) and Neutrinos
(UHE-N).


\subsection{Cosmic Rays associated with Explosive QSOs/AGNs and
HyN}

\label{disc-14.3}

In the last decades, a main astrophysical issue is to understand
the origin of UHE cosmic rays.
Recently, using the Pierre Auger Observatory, Abraham et al. (2007)
found that the extremely high energy CR are generated  with AGNs.
Two different theories and models could explain these P. Auger observations:
(i) Obscured and Collimated AGN/Black-Hole; (ii) Evolutionary, Explosive,
and Composite AGN + starburst Model.

The production of relativistic electrons is in
young SN remnants and it is believed that remnants simultaneously produce
relativistic ions/CRs (see Ellison et al. 2007).
In the evolutionary and composite model for AGNs, HyN explosions  are
a main component; thus we have suggested that giant HyN explosions and
their remnants (RHyN) could be  natural candidates for the origin
--in AGNs-- of UHE CRs (L\'{i}pari et al. 2007b).
In addition,
the large duration and very energetic
gamma ray bursts are associated mainly with HyN explosions.

From the theoretical point of view several groups already analysed
the generation of UHE-CR and UHE-N in: \\

\begin{itemize}

\item
{\bf GRBs in general:} in the fireball blast wave scenario for GRB,
the waves of ejected relativistic plasma that collide with each
other form shocks, which accelerate UHE particles/CR and radiate
high-energy UV photons
(Vietri 1995, 1998a,b, 2003; Waxman 1995; Milgrom \& Usov 1995;
Waxman \& Bahcall 1997, 1999, 2000; Vietri, De Marco, Guetta 2003;
Dai \& Lu 2001; Dermer 2002, 2003, 2007a,b; Dermer \& Atoyan 2006;
Razzaque, Meszaros, Waxman 2004;  Wick et al. 2004;
Fang, Zhang, Wei 2005;
Meszaros \& Razzaque 2006; Murase et al. 2006, 2008; Grupta \& Zhang 2007; 
and others); 

\item
{\bf HyperNovae, associated with neutron stars in GRB:} in the collapse of
neutron star to a black hole (years after the initial SN, that generate the
neutron star), the BH-outflow  interact with the original SN-remnant through
an external shock to form GRB and accelerate UHE particles/CR and radiate
high-energy photons
(Vietri \& Stella 1998, 1999; Dermer \& Mitman 2003)

\item
{\bf HyperNovae, associated with low/sub energetic GRB:} in mildly
relativistic HyN ejecta (similar to HyN 1998bw + GRB980425 and HyN 2003lw
+ GRB031203)   the external shock wave --produced by the ejecta-- could
generate UHE cosmic rays and UHE neutrinos
(Wang et al. 2007; Wang, Razzaque, Meszaros 2008)

\item
{\bf AGN jets:}  the shocks associated with relativistic jets,  in
radio AGNs/QSOs could accelerate UHE particles/CR and radiate high energy
UV-photons
(Berezinky et al. 2006; Dermer et al. 2009); and

\item
{\bf Intergalactic accretion shocks:}
accretion and mergers shocks in massive clusters of galaxies could accelerate
UHE protons/CR, which can give rise neutrinos through {\it pp} interactions
with intercluster gas
(Inoue, Ahoronian \& Sugiyama 2005; Murase, Inoue, Nagataki 2008).

\end{itemize}

Therefore,
in the evolutionary, composite and explosive model for AGNs (and
BAL + IR + Fe {\sc ii} QSOs) the presence of HyNe could generate
UHE-CR and UHE-N, according to the processes and theoretical studies
performed by Wang et al. (2007, 2008); Vietri \& Stella (1998, 1999);
Dermer \& Mitman (2003).
In the core of composite
QSOs and AGNs different types of HyN could be generated in the accretion
regions (Collin \& Zahn 1999), in particular HyNe
associated with: extreme massive stars, and  neutron stars.
In addition,
several works suggested that a high per cent of core collapse SN/HyN are
associated with mildly relativistic ejecta and GRBs.
Thus, in the core of explosive + composite AGNs and BAL + IR +
Fe {\sc ii} QSOs, different types of HyNe are one of the
main candidates for the origin of UHE-CR and UHE-N. \\

\noindent
{\bf Test for the Explosive+HyN AGNs Model (as the source of
UHE-CR)}

An important test for this scenario (that {\it  ``the observed UHE-CR  are
generated by HyN ejecta, in the core of AGNs"})  
is  the following: to detect starburst, giant explosions and the associated
super shells in nearby AGNs.

Recently,
this type of evidence was detected in the prototype of AGN: Centaurus A.
This nearby AGN  is one of sources associated with UHE cosmic rays
by Abraham et al. (2007).
However,
even for Cen A, only very recently and using mid-IR images obtained with
Spitzer Space Telescope, Quillen et al. (2006) found a supergiant nuclear
symmetric+circular shell (at r $\sim$500 pc, from the core of the AGN).
They suggested that this shell is probably associated with a nuclear
starburst (and/or AGN).

Thus, for the nearest AGNs (Cen A) the nuclear starbursts,
explosions and the associated shells were detected only using the
last generation of space telescope (Spitzer Telescope), and in the
mid-IR wavelength range (i.e., in the range of energy  free of
dust absorption).
For more distant AGNs is important the search of
evidences of explosive and OF nuclear processes (i.e.,
supergiant shells, multiple OF emission lines components,
emission line ratios associated with OF + shocks, etc) using 
multi wavelengths data obtained from the last generation of
telescope+instruments.
We expect to find these evidences in the analysis of BAL + IR + Fe {\sc ii}
QSOs (at low and medium redshift), and Sub-mm and Radio BAL QSOs (at high
redshift) through deep Gemini GMOS-IFU spectroscopy plus HST data.



\subsection{HyperNovae as the source of Neutrinos and Dark Matter}

\label{disc-14.4}

In the last sub-Section, we have explained that several groups
already studied theoretically the generation of UHE-N and UHE-CR in GRBs.
More specifically, for the HyN scenario Wang et al. (2007, 2008)
found that in mildly relativistic ejecta of HyNe: the UHE-N
could be generated by the interaction
of the HyN UHE-CR and HyN UV-Optical photons.
In addition, in the HyN scenario
Vietri \& Stella (1998, 1999) also analysed  the generation of UHE-N
associated with the collapse of neutron stars to black holes.

On the other hand,
in the last decades a main issue in Astrophysics is
the search of non-baryonic massive particles which does not interact
strongly with ordinary matter: i.e., weakly interacting 
massive particles (WIMPSs; White 1988).
One of the most attractive candidates for a WIMP are the neutrinos.
Because their thermal motion are so significant, particles like massive
neutrinos are known as Hot Dark Matter (HDM). Very recently,
L\'{i}pari et al. (2007b) proposed that
the discovery of different types of giant and extreme--SNe/HyNe (which
generate UHE-N, UHE-CR, $\gamma$-Rays)
strongly suggest that these neutrinos --generated by HyNe-- are
the probable origin of  dark matter.

From the study of HyN 2006gy,
Smith et al. (2007) proposed that giant--SN/HyN explosions from
extreme massive progenitors could be more numerous --specially, in
Population {\sc iii} stars: i.e., in young objects and in the early
universe-- than previously believed.
{\it Thus,
also the UHE-CR and UHE-N generated by explosion from massive and
extreme massive star are probably more numerous than previously
believed.}

Therefore, it is expected that UHE-CR and UHE-N might have been
generated in the young universe, and also in composite +
explosive QSOs and AGNs (specially, in BAL + IR + Fe {\sc ii} QSOs).

\clearpage


\section{Confirmation of the Explosive Model
(for BAL + IR + Fe {\sc ii} QSOs)}

\label{disc-15}


\subsection{Confirmation 1: Extreme Starburst and [O {\sc ii}]$\lambda$3727
emission in the BAL + Fe {\sc ii} + IR QSO SDSS 143821.40+094623.2}
\label{disc-15a}

Very recently, from a detailed study of the [O {\sc ii}]$\lambda$3727
emission line in QSOs from the very large sample of  Sloan Digital Sky
Survey Data Release 5 (SDSS DR5; Adelman-McCarthy et al. 2007: with
90596 spectra of QSOs),
Lu et al. (2008, 2009, in preparation) reported for the low
ionization  BAL QSO SDSS 143821.40+094623.2 an extreme
[O {\sc ii}]$\lambda$3727 emission, plus large ELR
[O {\sc ii}]$\lambda$3727/[Ne {\sc iii}]$\lambda$3869 and
[O {\sc ii}]$\lambda$3727/[O {\sc iii}]$\lambda$5007.
These results indicate (together with the large far-IR emission)
that the [O {\sc ii}]$\lambda$3727 came from an extreme starburst.
Furthermore, this low ionization BAL QSO is an extreme Fe {\sc ii}
emitter. 
Thus, this is the first BAL + IR + Fe {\sc ii} QSO with extreme
[O {\sc ii}]$\lambda$3727  emission (showing new evidence of
extreme starbursts in this class of QSOs).

These results --detected in the BAL + IR + Fe {\sc ii} QSO SDSS
143821.40+094623.2, at z $\sim$ 0.8--  are an important and independent
confirmation (using a very different method, to that used by us)
of our proposition that
{\it  extreme starbursts and the associated HyN + shells play a main
role in the evolution of low ionization BAL + IR + Fe {\sc ii} QSOs}.


\subsection{Confirmation 2 and 3:
Explosive  BAL + IR + Fe {\sc ii} QSOs IRAS 17002+5153 \& IRAS 07598+6508}
\label{disc-15b}

Very recently,
L\'{i}pari et al. (2008) presented the first results of the study of
the  BAL + IR + Fe {\sc ii} QSOs IRAS 17002+5153 \& IRAS 07598+6508,
using Gemini + GMOS-IFU and HST data.
These data show:

\begin{enumerate}

\item
IRAS 17002+5153:
the 3D spectra in the region of the
shells (L\'{i}pari et al. 2003) show multiple emission line components with
typical properties and ELR of Liners associated with low velocity shocks.\\

\item
IRAS 07598+6508:
the 3D spectra --to the north of the QSO-- in the circumnuclear area where
a possible shell was detected (at r of 2.3$'' \sim$8.0 kpc) show
multiple emission line components also with typical properties and ELR of
Liners/shocks.\\

\end{enumerate}

These Gemini + HST results confirm our previous suggestion that these two
similar low ionization BAL QSOs could be considered as exploding BAL + IR
+ Fe {\sc ii} QSOs (L\'{i}pari 1994). Thus, these GMOS data  are
in good agreement with the explosive model for BAL + IR + Fe {\sc ii} QSOs.


\subsection{Future works and a New Explosive BAL QSO Candidate}

\label{disc-15c}

Our programme of study of BAL QSOs and mergers
with strong OF (using IFU and MOS Spectroscopy) include more than 30
objects already observed. Which are mainly low redshift
BAL + IR + Fe {\sc ii} QSOs, SDSS-Sub$_{mm}$, SDSS-Radio low ionization
BAL QSOs at medium and high redshift (in the range 0.5 $<$ z $<$ 3)
and Ly$\alpha$ emitters at redshift z $\sim$ 5--6.
We are searching for new evidence of extreme explosions and starbursts 
in these systems with extreme OF (similar to those found in IRAS 04505-2958).

Finally we note that using GMOS-IFU spectra  a detailed study
BAL QSO SDSS 030000.56+004828.0 was started.
This BAL QSO shows strong Ca {\sc ii} + Fe {\sc ii} BAL systems,
extreme Fe {\sc ii} emission and a strong fall in the blue continuum.
These are typical spectral features of low ionization BAL + IR + Fe {\sc ii}
QSOs. Moreover,
the  spectrum of this SDSS BAL QSO is  almost the same than
the spectra of the prototype of explosive QSO Mrk 231.
(i.e., a twin of Mrk 231).
Thus SDSS 030000.56+004828.0 is a good candidate --at z = 0.9-- for a new
exploding BAL + Fe {\sc ii} QSO.

\clearpage


\section{Summary and Conclusions}

\label{summary}

In this work we have presented new results  for the BAL QSO IRAS 04505$-$2958
obtained from a study of BAL + IR + Fe {\sc ii} QSOs; based on very deep
Gemini GMOS 3D spectroscopy, and HST images.
We have studied in detail the outflow process and
their associated structures, at two large galactic scales:
two blobs/shells (S1 and S2) at radius r $\sim$ 0.2 and 0.4$''$
($\sim$1.1 and 2.2 kpc);
and  an external hypergiant shell (S3) at r $\sim$  2.0$''$
(11 kpc). The presence of  two external supergiant
shells (S4 and S5) at r $\sim$  10 and 15$''$ ($\sim$55, and 80 kpc)
was discussed.

From this GMOS-IFU study the following main results were obtained:

\begin{enumerate}
             
\item
For the external hypergiant shell S3 at  r $=$ 2.0$''$ (11 kpc) 
the kinematics GMOS maps of the ionized gas ([O {\sc ii}], [Ne {\sc iii}],
[O {\sc iii}], H$\beta$) show a small scale bipolar OF, with characteristics
very very similar to those observed in the prototype of exploding external
super shell in NGC 5514.

\item
The knots K1, K2 and K3 of this hypergiant shells S3 show a stellar
population and emission line ratios consistent with the presence of a
starburst + OF/shocks.

\item
The two internal shells S1 and S2 (at r $\sim$ 1 and 2 kpc)  show
multiple OF components with typical  properties of  nuclear shells.

\item
The shells S1+S2 and S3 are aligned at PA $\sim$ 131$^{\circ}$ with
bipolar OF shape (at $\sim$10--15 kpc scale), and
probably in the blow-out phase.
In addition,
the shells S4 and S5 (at $\sim$60--80 kpc) are aligned at PA $\sim$
40$^{\circ}$, with also bipolar OF shape, which is perpendicular to the more
internal OF.

\item
A strong blue continuum and multiple emission line components were
found in all the observed GMOS field (including the shells, observed
with GMOS: S1, S2 and S3).

\item
Using  optical GMOS and HST data together wit the IR
colour-colour evolutionary diagram for IRAS 04505-2958,
IRAS 07598+6508 and IRAS 17002+5153 observation, 
we have confirmed that the QSO  is  likely the dominant source of
ultra-luminous IR energy associated with IRAS 04505-2958.
However, the starburst detected in the
hyper shell S3 could be also a second source of IR energy.

\end{enumerate}

The nature of the extreme and extended OF process in IRAS 04505$-$2958,
with large and very large scale super/hyper shells (from 1 to $\sim$100 kpc)
was discussed.

Thus, the new GMOS data show a good agreement with an extreme and explosive OF
scenario for IRAS 04505-2958; in which part of the ISM of the host galaxy was
ejected as multiple shells.
This extreme OF process could be also associated with
2 main processes in the evolution of QSOs and their host galaxies:
(i) the formation of young/satellite galaxies by giant explosions; and
(ii) to define the final mass of the host galaxy, and even if the
explosive nuclear outflow is extremely energetic, this process could
disrupt an important fraction (or even all) of the host galaxy.

Finally,
the role of HyNe in BAL + IR + Fe {\sc ii} QSOs and AGNs was analysed.
In particular,
the generation of UHE cosmic rays and neutrino 
--associated with HyNe in BAL + IR + Fe {\sc ii} QSOs-- is discussed.
We sugested that neutrinos associated with HyN could be the source
of dark--matter.


\section*{Acknowledgments}

This research is
based mainly on observations obtained at the Gemini Observatory, which
is  operated
by AURA under cooperative agreement with the NSF-USA on behalf of the Gemini
partnership: NSF-USA, PPARC-UK, NRC-Canada, CONICYT-Chile,
ARC-Australia, CNPq-Brazil and CONICET-Argentina. In addition, in
this work we are using  observations from CASLEO, La Palma, Cala Alto
observatories, and from the archive of the NASA and ESA satellite HST  
(at ESO Garching and STScI--Baltimore).
The authors thank  C. Bornancini, L. Colina, H. Dottori, G. Dubner, J. C.
Forte, D. M., M. Pastoriza and R. Sistero, for stimulating
discussions and help.
We also tank E. Bicca, A. Piatti, J. J. Claria, J. Santos for their
observational templates of stellar populations.
Special tanks to Susan Neff and J. Hutching for their authorization to
adapted our Fig. 3 from their original figure; and also to Lu et al. for
their authorization to comment their results of the BAL QSO SDSS
143821.40+094623.2. We like to thank specially to M. T. Ruiz and N. Suntzeff
for their  spectra of SN 1998E, obtained with the CTIO 4mt telescope.
Finally, we wish to thank the referee for very constructive comments and suggestions, which helped to improve the content,
presentation and discussion of the results of the paper.

\clearpage

%


\begin{table}
\footnotesize \caption{Journal of  observations of IRAS 04505$-$2958 (and
Arp 220)}
\label{tobser}

\begin{tabular}{llllcl}
\hline
\hline
Object&Date & Telescope/ & Spectral region &Expos.\ time   & Comments \\
      &     & instrument &                 & [s]           & \\
\hline

      &     &            &                 &               & \\

I04505-2958
      &2005 Oct 07& 8.1\,m Gemini+GMOS-IFU&R400, $\lambda\lambda$5260--9440 \AA&
1800$\times$2& Seeing-$\langle$FWHM$\rangle$ = 0\farcs9 \\

I04505-2958
      &2005 Dec 25& 8.1\,m Gemini+GMOS-IFU&B600, $\lambda\lambda$4770--7600 \AA&
1800$\times$1& Seeing-$\langle$FWHM$\rangle$ = 0\farcs5 \\

I04505-2958
      &2005 Dec 26& 8.1\,m Gemini+GMOS-IFU&B600, $\lambda\lambda$4770--7600 \AA&
1800$\times$1& Seeing-$\langle$FWHM$\rangle$ = 0\farcs4 \\

I04505-2958
      &2007 Feb 14& 8.1\,m Gemini+GMOS-IFU&R831, $\lambda\lambda$7440--9500 \AA&
1200$\times$1& Seeing-$\langle$FWHM$\rangle$ = 0\farcs5 \\

I04505-2958
      &2007 Feb 14& 8.1\,m Gemini+GMOS-IFU&B600, $\lambda\lambda$3350--6150 \AA&
1200$\times$1& Seeing-$\langle$FWHM$\rangle$ = 0\farcs9 \\

      &     &            &                           &         & \\

I04505-2958
      &1995 Sep 30&{\itshape HST\/}+WFPC2& F702W, $\lambda\lambda$
6895/1389 \AA ($\sim$R)& 1800& Seeing-$\langle$FWHM$\rangle$ = 0\farcs1, archival\\

I04505-2958
      &2004 Oct 01&{\itshape HST\/}+ACS  & F606W, $\lambda
\lambda$5907/2342 \AA ($\sim$V)& 990& Seeing-$\langle$FWHM$\rangle$ = 0\farcs1, archival\\


       &    &            &                           &         & \\
I04505-2958
       &1996 Nov 18&{\itshape HST\/}+FOS& G190H,
$\lambda\lambda$1570--2300 \AA & 1620& archival (specra)\\
       &    &            &                           &         & \\

       &     &            &                           &         & \\

\hline

\end{tabular}

\end{table}

\clearpage


\begin{table}
\footnotesize \caption{Positions of the main knots/areas
in the super shells and selected external regions
(in IRAS 04505$-$2958)}
\label{tknots}

\begin{tabular}{lrrl}
\hline
\hline
Knots/Regions & $\Delta$X    & $\Delta$Y     & D$_{eff}$ \\
              & [$''$]       & [$''$]        & [$''$]   \\
\hline

              &              &               &       \\
{\it Shell S1}&              &               &       \\
Area S1-A1    &  0.00        & -0.20         & 0.15  \\
Area S1-A2    &  0.17        & -0.10         & 0.15  \\
              &              &               &       \\
              &              &               &       \\
{\it Shell S2}&              &               &       \\
Area S2-A1    &  0.00        & -0.40         & 0.15  \\
Area S2-A2    &  0.17        & -0.30         & 0.15  \\
              &              &               &       \\
              &              &               &       \\
{\it Shell S3}&              &               &       \\
Knot S3-K1    & -0.21        &  1.40         & 0.11  \\
Knot S3-K2    &  0.17        &  1.50         & 0.25  \\
Knot S3-K3    & -0.17        &  1.70         & 0.10  \\
Knot S3-K4    & -0.52        &  1.90         & 0.20  \\
Knot S3-K5    & -0.52        &  1.70         & 0.20  \\
              &              &               &       \\
              &              &               &       \\
{\it Regions} &              &               &       \\
region R1     & -1.56        &  0.00         & 0.15  \\
region R2a    &  0.17        & -0.90         & 0.15  \\
region R2b    &  0.17        & -1.10         & 0.15  \\
region R3     &  0.17        &  3.10         & 0.15  \\
region R4     & -0.69        &  0.80         & 0.17  \\
              &              &               &       \\
              &              &               &       \\

\hline

\end{tabular}

Notes:\\
Column 2 and 3: The offset positions of the knots [$\Delta$X,$\Delta$Y]
are given from the QSO-core position (as 0$''$,0$''$).
The Y-axis was aligned at the postion angle PA $=$ 131$^{\circ}$. \\
Column 4:  D$_{eff}$  are the effective diameters of the knots
(derived using the HST-WFPC F702W/R image). \\

\end{table}

\clearpage


\begin{table}
\footnotesize \caption{Main OF Components in the QSO-core and the Shell S3
(of IRAS 04505$-$2958)}
\label{tknots}

\begin{tabular}{lllr}
\hline
\hline
Emission Line Component        & z            & cz            & $\Delta$V(OF) \\
                               &              & [km s$^{-1}$] & [km s$^{-1}$] \\
\hline

                               &              &               &       \\
{\it QSO-core}                 &              &               &       \\
                               &              &               &       \\
Main Component Em.  (MC-EMI)   &  0.28600     & 85800         &  ---   \\
                               &              &               &       \\
Blue OF Component-1 (OF-EB1)   &  0.20300     & 84900         &  -900  \\
Blue OF Component-2 (OF-EB2)   &  0.28033     & 84100         & -1700  \\
Blue OF Component-3 (OF-EB3)   &  0.27700     & 83100         & -2700  \\
                               &              &               &       \\
Red OF Component    (OF-ER )   &  0.29150     & 87450         & +1650  \\
                               &              &               &       \\

                               &              &               &       \\
{\it Shell S3}                 &              &               &       \\
                               &              &               &       \\
Main Component Em. (MC-S3-EMI) &  0.28650     & 85950         &  ---   \\
                               &              &               &       \\
Blue OF Component-1 (OF-S3-EB1)&  0.28483     & 85450         &  -500  \\
Blue OF Component-2 (OF-S3-EB2)&  0.28217     & 84650         & -1300  \\
                               &              &               &        \\
Red OF Component    (OF-S3-ER )&  0.29550     & 87150         & +1200  \\
                               &              &               &       \\
                               &              &               &       \\

\hline

\end{tabular}


\end{table}

\clearpage


\begin{table}
\footnotesize \caption{Emission Lines of the QSO-core (pixel of 0.2$''$)}
\label{flux3dnuc}
\begin{tabular}{llcc}
\hline
\hline

Lines &Component &Fluxes            & FWHM    \\
      &          &QSO-Core          & [km/s]  \\

\hline

[O {\sc ii}]$\lambda$3727
  & MC-EMI           &  8.8      & 480        \\
  & OF-EB1           &  1.0      & 170        \\
  &                  &           &            \\

H$_{11}$$\lambda$3771
  & MC-EMI ${Interm}$&  (0.8)    & (830)      \\
  &                  &           &            \\

H$_{10}$$\lambda$3798
  & MC-EMI ${Interm}$&   1.2     &   790       \\
  &                  &           &            \\

H$_{9}$$\lambda$3835
  & MC-EMI ${Broad}$ &  (0.3)    & (2150)       \\
  & MC-EMI ${Interm}$&   2.0     &   700       \\
  &                  &           &            \\
  & MC-EMI 1-Compon. &   2.4     &  880       \\
  &                  &           &            \\

[Ne {\sc iii}]$\lambda$3869
  & MC-EMI           &  6.8      &  540        \\
  & OF-EB1           &  0.9      &  180        \\
  &                  &           &            \\

H$_8$$\lambda$3889
  & MC-EMI ${Broad}$ & (0.5)     & (2100)      \\
  & MC-EMI ${Interm}$&  4.7      &   650       \\
  &                  &           &            \\
  &MC-EMI 1-Compon+OF&  5.5      &  940     \\
  &                  &           &            \\

H$\epsilon\lambda$3970
  & MC-EMI ${Broad}$ &   6.0     &  2020       \\
  & MC-EMI ${Interm}$&  10.0     &   740       \\
  & OF-EB1           &   0.5     &   210       \\
  & OF-EB2           &   0.4     &   190       \\
  &                  &           &            \\
  &MC-EMI 1-Compon+OF&  17.0     &  1090       \\
  &                  &           &            \\

H$\delta\lambda$4102
  & MC-EMI ${Broad}$ &  10.0     &  2490      \\
  & MC-EMI ${Interm}$&  12.0     &   700       \\
  & OF-EB1           &   0.6     &   250      \\
  & OF-EB2           &   0.5     &   230       \\
  &                  &           &            \\
  &MC-EMI 1-Compon+OF&  22.0     &  1120      \\
  &MC-EMI 1-Compon.  &  25.0     &  1350      \\
  &                  &           &            \\

[Fe {\sc v}]$\lambda$4181
  & MC-EMI           &   5.9     &   980      \\
  &                  &           &            \\

H$\gamma\lambda$4340

  & MC-EMI ${Broad}$ &  11.0     &  2200       \\
  & MC-EMI ${Interm}$&  20.0     &   730       \\
  & OF-EB1           &   0.7     &   240       \\
  & OF-EB2           &   0.6     &   210       \\
  & OF-EB3           &   0.6     &   200       \\

  &                  &           &            \\
  &MC-EMI 1-Compon+OF&  29.0     &  1070       \\
  &MC-EMI 1-Compon.  &  32.0     &  1330       \\
  &                  &           &            \\

Fe {\sc ii}$\lambda$4489+91
  & MC-EMI           &  3.5      &  820      \\
Fe {\sc ii}$\lambda$4523
  & MC-EMI           &  5.3      &  800      \\
Fe {\sc ii}$\lambda$4556
  & MC-EMI           &  4.8      &  780      \\
Fe {\sc ii}$\lambda$4583
  & MC-EMI           &  5.2      &  790      \\
Fe {\sc ii}$\lambda$4629
  & MC-EMI           &  5.4      &  810      \\
Fe {\sc ii}$\lambda$4661
  & MC-EMI           &  5.6      &  blend      \\


  &                  &           &            \\

\hline

\end{tabular}

\end{table}

\clearpage


\begin{table}
\addtocounter{table}{-1}
\footnotesize \caption{Continuation}
\label{flux3dnucc}
\begin{tabular}{llcc}
\hline
\hline

Lines &Component &Fluxes            & FWHM    \\
      &          &QSO-Core          & [km/s]  \\

\hline

H$\beta\lambda$4861
  & MC-EMI ${Broad}$ &  20.0     &  2050       \\
  & MC-EMI ${Interm}$&  24.0     &   780       \\
  & MC-EMI ${Narrow}$& ( 4.0)    &  (230)       \\
  & OF-EB1           &   0.9     &   260        \\
  & OF-EB2           &   0.7     &   235       \\
  & OF-EB3           &   0.7     &   210       \\
  &                  &           &            \\
  &MC-EMI 1-Compon+OF&  44.0     &  1050       \\
  &MC-EMI 1-Compon.  &  48.0     &  1300       \\
  &                  &           &            \\

Fe {\sc ii} (42)$\lambda$4925
  & MC-EMI           &   3.1     &  830      \\
  &                  &           &            \\

[O {\sc iii}]$\lambda$5007
  & MC-EMI ${Interm}$&  13.0     &   610       \\
  & MC-EMI ${Narrow}$&   5.0     &   280       \\

  &                  &           &            \\
  & MC-EMI 1-Compon. &  19.1     &   530      \\
  &                  &           &            \\

Fe {\sc ii}$\lambda$5159
  & MC-EMI           &  2.5      &  780      \\
Fe {\sc ii}$\lambda$5169
  & MC-EMI           &  3.0      &  730      \\
Fe {\sc ii}$\lambda$5198
  & MC-EMI           &  2.8      &  720      \\
Fe {\sc ii}$\lambda$5220
  & MC-EMI           &  4.4      &  810      \\
Fe {\sc ii}$\lambda$5235
  & MC-EMI           &  3.6      &  690      \\
Fe {\sc ii}$\lambda$5276
  & MC-EMI           &  4.2      &  660      \\
Fe {\sc ii}$\lambda$5317
  & MC-EMI           &  3.8      &  650      \\
Fe {\sc ii}$\lambda$5362
  & MC-EMI           &  3.0      &  620      \\
Fe {\sc ii}$\lambda$5385
  & MC-EMI           &  4.0      &  800      \\
                     &           &            \\

[O {\sc i}]$\lambda$6300
  & MC-EMI           & ---       &  ---         \\
  &                  &           &            \\

H$\alpha\lambda$6563
  & MC-EMI ${Broad}$ &  73.0     &  2150      \\
  & MC-EMI ${Interm}$&  72.0     &   800       \\
  & OF-EB1           &   2.3     &   210       \\
  & OF-EB2           &   2.1     &   200       \\
  & OF-EB3           &   2.2     &   210       \\
  & OF-ER3           &   1.0     &   220       \\
  &                  &           &            \\
  &MC-EMI 1-Compon+OF& 140.0     &  1020       \\
  &MC-EMI 1-Compon.  & 151.0     &  1240       \\
  &                  &           &            \\

[N {\sc ii}]$\lambda$6583
  & MC-EMI           &   ---     & ---        \\
  &                  &           &            \\

[S {\sc ii}]$\lambda$6717/31
  & MC-EMI           &  (1.6)    & (280)      \\
  &                  &           &            \\

[S {\sc ii}]$\lambda$6731
  & MC-EMI           &  (2.0)    & (300)       \\
  &                  &           &             \\


H$\alpha$/H$\beta$ 
  &MC-EMI ${Interm}$& 3.00      &             \\

Fe {\sc ii}$\lambda$4570/H$\beta$
  &MC-EMI ${Interm}$& 1.24      &             \\

[O {\sc iii}]$\lambda$5007/H$\beta$
  &MC-EMI ${Interm}$& 0.54      &             \\

\hline

\end{tabular}

\noindent
Column 2: emission line components (see Section 5). \\
Column 3: the fluxes are given in units of 10$^{-15}$ erg cm$^{-2}$ s$^{-1}$ \\
The H-Balmer lines show the results of 3 fitting processes,
using: 
[1] one H-Balmer main component;
[2] one H-Balmer main component plus OFs; and
[3] Broad, intermediate, and narrow H-Balmer components, plus OFs. \\
All the H-Balmer broad components show a blueshift of $\sim$ 500 km s$^{-1}$,
in relation to the corresponding H-Balmer intermediate components. \\
The Fe {\sc ii} emission lines are at the same redshift of the H-Balmer
intermediate components (and also they show the same FWHM). \\
The errors/$\sigma$ in the fluxes and FWHM are less than 10$\%$. \\
The values between parentheses are data with low S/N.

\end{table}

\clearpage


\begin{table}
\footnotesize \caption{Emission lines of the main areas of the
shells S1 and S2 (GMOS-B600)}
\label{flux3ds12}
\begin{tabular}{llcccc}
\hline
\hline

Lines &Compon&         &Fluxes      &        &                  \\
      &     &Area S1-A1 &Area S1-A2 &Area S2-A1 &Area S2-A2      \\
      &     &[0.0$''$,-0.2$''$] &[0.2$''$,-0.1$''$] &
             [0.0$''$,-0.4$''$] &[0.2$''$,-0.3$''$]  \\

\hline

  &         &         &         &         &        \\

  &         &         &         &         &            \\

H$\beta\lambda4861$
  & MC-EMI  & 6.30    & 8.30    & 6.00    & 3.50      \\
  & OF-EB1  & 0.3     & 0.3     & 0.2     & 0.2       \\
  & OF-EB2  & 0.2     & 0.3     & 0.2     & 0.1      \\
  &         &         &         &         &            \\


[O{\sc iii}]$\lambda5007$
  & MC-EMI  & 46.0    & 55.1    & 38.0    & 27.4      \\
  &         &         &         &         &           \\

[O {\sc i}]$\lambda6300$
  & MC-EMI  & 4.00    & 3.90    & 3.00    &  2.70      \\
  &         &         &         &         &            \\

H$\alpha\lambda6563$
  & MC-EMI  & 18.3    & 25.0    & 17.1    &  15.0      \\
  & OF-EB1  &  0.8    &  0.9    &  0.7    &   0.5      \\
  & OF-EB2  &  0.6    &  0.8    &  0.6    &   0.4     \\
  & OF-EB3  &  0.5    &  0.6    & ---     &  ---      \\
  &         &         &         &         &           \\


[N {\sc ii}]$\lambda6583$
  & MC-EMI  &  6.00   & 11.0    & 7.00    &  7.00    \\
  &         &         &         &         &           \\

[S {\sc ii}]$\lambda6717$
  & MC-EMI  & 3.70    & 2.80    & 6.50    &  2.40     \\
  &         &         &         &         &          \\

[S {\sc ii}]$\lambda6731$
  & MC-EMI  & 6.00    & 3.20    & 3.50    &  3.50   \\
  &         &         &         &         &          \\
  &         &         &         &         &          \\

H$\alpha$/H$\beta$
  & MC-EMI  & 2.9     & 3.0     & 2.9     &  4.3      \\
  &         &         &         &         &          \\

FWHM H$\alpha$
  & MC-EMI  &  390    & 380     & 370     & 360      \\
FWHM [O {\sc iii}]
  & MC-EMI  &  360    & 350     & 355     & 340      \\

  &         &         &         &         &          \\

\hline

\end{tabular}

\noindent
The fluxes are given in units of 10$^{-16}$ erg cm$^{-2}$ s$^{-1}$.\\
Column 2: emission line components (see Section 5).
In particular,
MC-EMI means the Main Component of the emission line, and
OF-EB1 the outflow emission line of the blue component-1. \\
Line 3: the X and Y offset (from the QSO-core, as 0,0)
for each GMOS spectrum, in each knot (see Table 2).
The GMOS Y-axis was positioned at PA $=$ 131$^{\circ}$. \\
The FWHM are given in unit of Km/s.\\
The errors/$\sigma$ in the fluxes and FWHM are less than 10$\%$.\\
The values between parentheses are data with low S/N.

\end{table}

\clearpage


\begin{table}
\footnotesize \caption{Emission  lines of the
main knots of the shells S3 }
\label{flux3ds3}
\begin{tabular}{llccccc}
\hline
\hline

Lines &Component&     &         &Fluxes         &      &              \\
      &     &Knot S3-K1 &Knot S3-K2 &Knot S3-K3 &Knot S3-K4&Knot S3-K5 \\
      &     &[-0.2$''$,1.4$''$] &[ 0.2$''$,1.5$''$] &[-0.2$''$,1.7$''$] &
             [-0.5$''$,1.9$''$] &[-0.5$''$,1.7$''$] \\

\hline

  &         &         &         &         &        &              \\

[O{\sc ii}]$\lambda$3727
  & MC-EMI  &  3.70   &  5.00   & 4.60    & 1.00   & 1.60         \\
  & OF-EB   &  ---    &  0.60   & 2.00    & 0.40   & ---          \\
  &         &         &         &         &        &              \\

H$_{11}\lambda$3771
  & MC-EMI  &  0.14   &  0.10   &  0.30   & 0.08   &  0.08         \\
  &         &         &         &         &        &              \\

H$_{10}\lambda$3798
  & MC-EMI  &  0.10   &  0.10   &  0.10   & 0.07   &  0.08         \\
  &         &         &         &         &        &              \\

H$_{9}\lambda$3835
  & MC-EMI  &  0.20   &  0.20   &  0.20   & 0.09   &  0.15         \\
  &         &         &         &         &        &              \\

H$_8$$\lambda$3889
  & MC-EMI  &  0.30   &  0.10   &  0.05   & 0.08   &  0.10         \\
  &         &         &         &         &        &              \\

H$\epsilon\lambda$3970
  & MC-EMI  &  0.10   &  0.20   &  0.10   & 0.05   &  0.10         \\
  &         &         &         &         &        &              \\

H$\delta\lambda$4102
  & MC-EMI  &  ---    &  ---    &  weak   & 0.10   &  0.20         \\
  &         &         &         &         &        &              \\

H$\gamma\lambda$4340
  & MC-EMI  &  ---    &  ---    & ---     & weak   &  ---         \\
  &         &         &         &         &        &              \\

H$\beta\lambda$4861
  & MC-EMI  & 1.10    & 1.30    & 1.20    & 0.28   & 0.75         \\
  & OF-EB1  & 0.40    & 0.60    & 0.50    & 0.10   & 0.40         \\
  & OF-EB2  & 0.50    & ---     & 0.60    & 0.10   & 0.40         \\
  &         &         &         &         &        &              \\

[O{\sc iii}]$\lambda$5007
  & MC-EMI  & 1.11    & 1.40    & 1.17    & 0.60   & 0.65         \\
  & OF-EB1  & 0.20    & 0.30    & 0.20    & 0.10   & ---          \\
  & OF-EB2  & 0.20    & ---     & ---     & ---    & ---          \\
  &         &         &         &         &        &              \\

[O {\sc i}]$\lambda$6300
  & MC-EMI  & 0.24    & 0.35    & 0.28    &  0.35  & 0.50         \\
  & OF-EB1  & ---     & 0.30    & ---     &  ---   & ---          \\
  &         &         &         &         &        &              \\

H$\alpha\lambda6563$
  & MC-EMI  & 2.80    & 5.50    & 3.60    &  1.33  & 2.20         \\
  &         &         &         &         &        &              \\

[N {\sc ii}]$\lambda$6583
  & MC-EMI  &  1.40   & 2.40    & 1.80    &  0.56  & 0.90         \\
  &         &         &         &         &        &              \\

[S {\sc ii}]$\lambda$6717
  & MC-EMI  & 0.35    & 1.30    & 0.60    &  0.40  & 0.50         \\
  &         &         &         &         &        &              \\

[S {\sc ii}]$\lambda$6731
  & MC-EMI  & 0.24    & 1.10    & 0.70    &  0.40  & 0.50          \\
  &         &         &         &         &        &               \\
  &         &         &         &         &        &               \\

H$\alpha$/H$\beta$ 
  & MC-EMI  & 2.7     & 3.9     & 3.0     &  4.6   & (3.2)          \\
  &         &         &         &         &        &                \\

FWHM H$\alpha$ 
  & MC-EMI  &  400    & 380     & 405     & 390    & 380            \\
FWHM [O {\sc iii}]$\lambda$5007  
  & MC-EMI  &  380    & 375     & 395     & 380    & 385            \\
FWHM [O {\sc ii}]$\lambda$3727 
  & MC-EMI  &  295    & 290     & 300     & 270    & 260         \\

  &         &         &         &         &        &               \\

\hline

\end{tabular}

\noindent
The fluxes are given in units of 10$^{-16}$ erg cm$^{-2}$ s$^{-1}$.\\
Column 2: emission line components (see Section 5).
In particular,
MC-EMI means the Main Component of the emission line, and
OF-EB1 the outflow emission line of the blue component-1. \\
Line 3: the X and Y offset (from the QSO-core, as 0,0)
for each GMOS spectrum, in each knot (see Table 2).
The GMOS  Y-axis was aligned at the position angle PA $=$ 131$^{\circ}$. \\
The FWHM are given in unit of Km/s.\\
The errors/$\sigma$ in the fluxes and FWHM are less than 15$\%$. \\
The values between parentheses are data with low S/N.

\end{table}

\clearpage


\begin{table}
\footnotesize \caption{Absorption lines of the
main knots of the shells S3 }
\label{flux3ds3}
\begin{tabular}{llccccc}
\hline
\hline

Lines &Component&     &         &EqW [\AA]      &          &           \\
      &     &Knot S3-K1 &Knot S3-K2 &Knot S3-K3 &Knot S3-K4&Knot S3-K5 \\
      &     &[-0.2$''$,1.4$''$] &[ 0.2$''$,1.5$''$] &[-0.2$''$,1.7$''$] &
             [-0.5$''$,1.9$''$] &[-0.5$''$,1.7$''$] \\

\hline

  &         &         &         &         &        &              \\
  &         &         &         &         &        &              \\

H$_{11}\lambda$3771
  & MC-ABS  &  3.8    &  5.8    &  5.7    &  4.0   &  6.0         \\
  &         &         &         &         &        &              \\

H$_{10}\lambda$3798
  & MC-ABS  &  8.7    &  8.5    & 12.0    & 11.8   &  8.8         \\
  &         &         &         &         &        &              \\

H$_{9}\lambda$3835
  & MC-ABS  &  8.5    & 15.5    & 11.3    & 10.8   & 10.2         \\
  &         &         &         &         &        &              \\

H$_8$+He I$\lambda$3889
  & MC-ABS  & 10.2    & 13.0    &  7.8    & 12.8   & 11.0         \\
  &         &         &         &         &        &              \\

Ca II-H$\lambda$3933
  & MC-ABS  &  4.8    &  4.5    &  5.1    &  7.1   &  6.0         \\
  &         &         &         &         &        &              \\

H$\epsilon\lambda$3970+Ca {\sc ii}K
  & MC-ABS  &   9.3   & 10.5    &  9.1    & 13.7   & 11.5         \\
  &         &         &         &         &        &              \\

H$\delta\lambda$4102
  & MC-ABS  &   6.5   &  3.5    &  5.8    & 11.2   &  10.0         \\
  &         &         &         &         &        &              \\

H$\gamma\lambda$4340
  & MC-ABS  &  ---    &  ---    & ---     &  7.0   &   8.0        \\
  &         &         &         &         &        &              \\

H$\beta\lambda$4861
  & MC-ABS  & ---     & ---     & ---     &  5.0   &   4.0         \\

  &         &         &         &         &        &              \\
  &         &         &         &         &        &              \\
FWH-Min. H$\epsilon\lambda$3970  
  & MC-EMI  &  495    & 485     & 460     & 570    & 590          \\

  &         &         &         &         &        &               \\
  &         &         &         &         &        &               \\

\hline

\end{tabular}

\noindent
The Equivalent Widths (EqW) are given in units \AA \\ 
Column 2: absorption components (see Section 5).
In particular,
MC-ABS means the Main Component of the absorption line. \\
Line 3: the X and Y offset (from the QSO-core, as 0,0)
for each GMOS spectrum, in each knot (see Table 2).
The GMOS  Y-axis was aligned at the position angle PA $=$ 131$^{\circ}$. \\
The FWHM are given in unit of Km/s.\\
The errors/$\sigma$ in the EqW are less than 13$\%$. 

\end{table}

\clearpage


\begin{table}
\footnotesize \caption{Emission Lines of main/several external Regions
(of the IRAS 04505-2958, in the GMOS field)}
\label{flux3ds3}
\begin{tabular}{llccccc}
\hline
\hline

Lines &Compon&        &Fluxes      &      &          &                \\
      &     & Region R1& Region R2a& Region R2b& Region R3& Region R4      \\
      &     &[-1.6$''$,0.0$''$] &[ 0.2$''$,-0.9$''$]  &[0.2$''$,-1.1$''$]
            &[ 0.2$''$,3.1$''$] &[-0.7$''$, 0.8$''$]\\

\hline

      &     &      &       &       &        &          \\

[O{\sc ii}]$\lambda3727$
  & MC-EMI  & 1.50 & 3.00  & 2.10  & ---    & 2.00    \\
  & OF-EB1  & 0.60 & ---   & 0.70  & ---    & 0.30    \\
  &         &      &       &       &        &          \\

H$\gamma\lambda4340$
  & MC-EMI  & ---  & 0.60  & ---    & ---   & 0.65    \\

H$\beta\lambda4861$
  & MC-EMI  & 0.50 & 1.30  & 0.45   &(0.23) & 2.00      \\
  & OF-EB1  & 0.40 & ---   & ---    & ---   & 0.40     \\
  & OF-EB2  & 0.20 & ---   & ---    & ---   & 0.50     \\
  &         &      &       &        &       &         \\

[O{\sc iii}]$\lambda5007$
  & MC-EMI  & 0.50 & 2.50  & 0.35   &(0.24) & 1.82   \\
  & OF-EB1  & 0.30 & ---   & ---    & ---   & 0.20   \\
  & OF-EB2  & 0.20 & ---   & ---    & ---   & ---   \\
  &         &      &       &        &       &         \\

[O {\sc i}]$\lambda6300$
  & MC-EMI  & 0.15 & 1.40  & 0.75   & 0.14  & 0.40    \\
  &         &      &       &        &       &          \\

H$\alpha\lambda6563$
  & MC-EMI  & 1.00 & 3.70  & 1.75   & 0.60  & 5.90     \\
  &         &      &       &        &       &          \\

[N {\sc ii}]$\lambda6583$
  & MC-EMI  & 0.35 & 3.58  & 4.17   & 0.68  & 2.20    \\
  &         &      &       &        &       &        \\

[S {\sc ii}]$\lambda6717$
  & MC-EMI  & 0.30 & 1.68  & 1.95   & 0.45  & 0.70     \\
  &         &      &       &        &       &          \\

[S {\sc ii}]$\lambda6731$
  & MC-EMI  & 0.30 & 1.00  & 1.00   & 0.26  & 0.70  \\
  &         &      &       &        &       &       \\
  &         &      &       &        &       &       \\

H$\alpha$/H$\beta$
  & MC-EMI  & (2.0)& 3.0   & 3.8    &  2.8  & 3.3      \\
  &         &      &       &        &       &           \\

FWHM H$\alpha$
  & MC-EMI  & 400  &  380  &  440   &  460  & 420      \\
FWHM [O {\sc iii}]$\lambda$5007
  & MC-EMI  & 360  &  340  &  370   &  380  & 395      \\
FWHM [O {\sc ii}]$\lambda$3727
  & MC-EMI  & 280  &  250  &  260   &  ---  & 265     \\

  &         &      &       &        &       &           \\

\hline

\end{tabular}

\noindent
The fluxes are given in units of 10$^{-16}$ erg cm$^{-2}$ s$^{-1}$
(from GMOS/IFU R831, B600 and R400 spectroscopy).\\
Column 2: emission line components (see Section 5).\\
Line 3: the X and Y offset (from the QSO-core, as 0,0)
for each GMOS spectrum, in each knot (see Table 2).
The GMOS  Y-axis was positioned at PA $=$ 131$^{\circ}$.  \\
The FWHM are given in unit of Km/s.\\
The errors/$\sigma$ in the fluxes and FWHM are less than 15$\%$. \\
The values between parentheses are data with low S/N.

\end{table}

\clearpage


\begin{table}
\footnotesize \caption{Emission Line Ratios of the QSO-core, main knots of
the shells S1, S2 and S3 plus several external regions}
\label{elrs1}
\begin{tabular}{lcccccc}
\hline \hline

                 &       &       &       &       &      &        \\
Knots/Regions    &Compon & log[O{\sc iii}]/H$\beta$ &log[O{\sc i}]/H$\alpha$ 
& log[N{\sc ii}]/H$\alpha$  & log[S{\sc ii}s]/H$\alpha$ & Spectral Type \\
\hline
                  &       &       &       &       &       &       \\

{\it QSO}         &       &       &       &       &       &       \\
QSO-Core (0\farcs2)&MC-EMI& -0.36 &  ---  & ---   & ---   & ---   \\
                  &       &       &       &       &       &       \\
                  &       &       &       &       &       &       \\

{\it Shell S1}    &       &       &       &       &       &       \\

Area S1-A1        &MC-EMI &  0.86 & -0.66 & -0.48 & -0.26 & SHOCKS + AGN\\

Area S1-A2        &MC-EMI &  0.87 & -0.77 & -0.36 & -0.62 & SHOCKS + AGN + H {\sc ii}\\

                  &       &       &       &       &       &       \\
                  &       &       &       &       &       &       \\
{\it Shell S2}    &       &       &       &       &       &       \\

Area S2-A1        &MC-EMI &  0.89 & -0.76 & -0.39 & -0.23 & SHOCKS + AGN\\

Area S2-A2        &MC-EMI &  0.89 & -0.75 & -0.33 & -0.40 & SHOCKS + AGN\\

                  &       &       &       &       &       &       \\
                  &       &       &       &       &       &       \\
{\it Shell S3}    &       &       &       &       &       &       \\

Knot S3-K1        &MC-EMI &  0.00 & -1.07 & -0.30 & -0.68 & LINER + H {\sc ii} \\

Knot S3-K2        &MC-EMI &  0.07 & -1.20 & -0.36 & -0.36 & LINER + H {\sc ii} \\

Knot S3-K3        &MC-EMI & -0.01 & -1.08 & -0.30 & -0.51 & LINER + H {\sc ii}  \\

Knot S3-K4        &MC-EMI &  0.08 & -0.58 & -0.38 & -0.22 & LINER \\

Knot S3-K5        &MC-EMI & -0.04 & -0.60 & -0.37 & -0.32 & LINER \\

                  &       &       &       &       &       &        \\
                  &       &       &       &       &       &        \\
{\it Regions}     &       &       &       &       &       &       \\

Region R1         &MC-EMI &  0.00 & -0.82 & -0.45 & -0.22 & LINER   \\

Region R2a        &MC-EMI &  0.19 & -0.42 &  0.00 & -0.15 & LINER  \\

Region R2b        &MC-EMI & -0.11 & -0.37 &  0.38 &  0.23 & LINER  \\

Region R3         &MC-EMI & -0.02 & -0.59 &  0.09 &  0.11 & LINER  \\

Region R4         &MC-EMI & -0.04 & -1.20 & -0.43 & -0.62 & LINER + H {\sc ii} \\
                  
                  &       &       &       &       &       &       \\
                  &       &       &       &       &       &       \\

\hline

\end{tabular}

\noindent
The wavelengths of the main used lines are [O {\sc iii}]$\lambda5007$;
[O {\sc i}]$\lambda6300$; [N {\sc ii}]$\lambda6583$;
[S {\sc ii}s]$\lambda\lambda$6716+6731.\\
Column 2: emission line components (Compon),
as in Table 3.\\
Column 7: spectral type,  using mainly the diagrams
log  log [S {\sc ii}]/H$\alpha$ vs log [O {\sc i}]/H$\alpha$,
log  log [O {\sc iii}]/H$\beta$ vs log [S {\sc ii}]/H$\alpha$, 
log  log [O {\sc iii}]/H$\beta$ vs log [O {\sc i}]/H$\alpha$
(from Heckman et al. 1990: Figure 14; L\'{i}pari et al. 2004d).\\
The errors/$\sigma$ in the emission lines ratios are less than 15$\%$. 

\end{table}




\clearpage

\begin{figure*}
\vspace{12.0 cm}
\begin{tabular}{cc}
\includegraphics{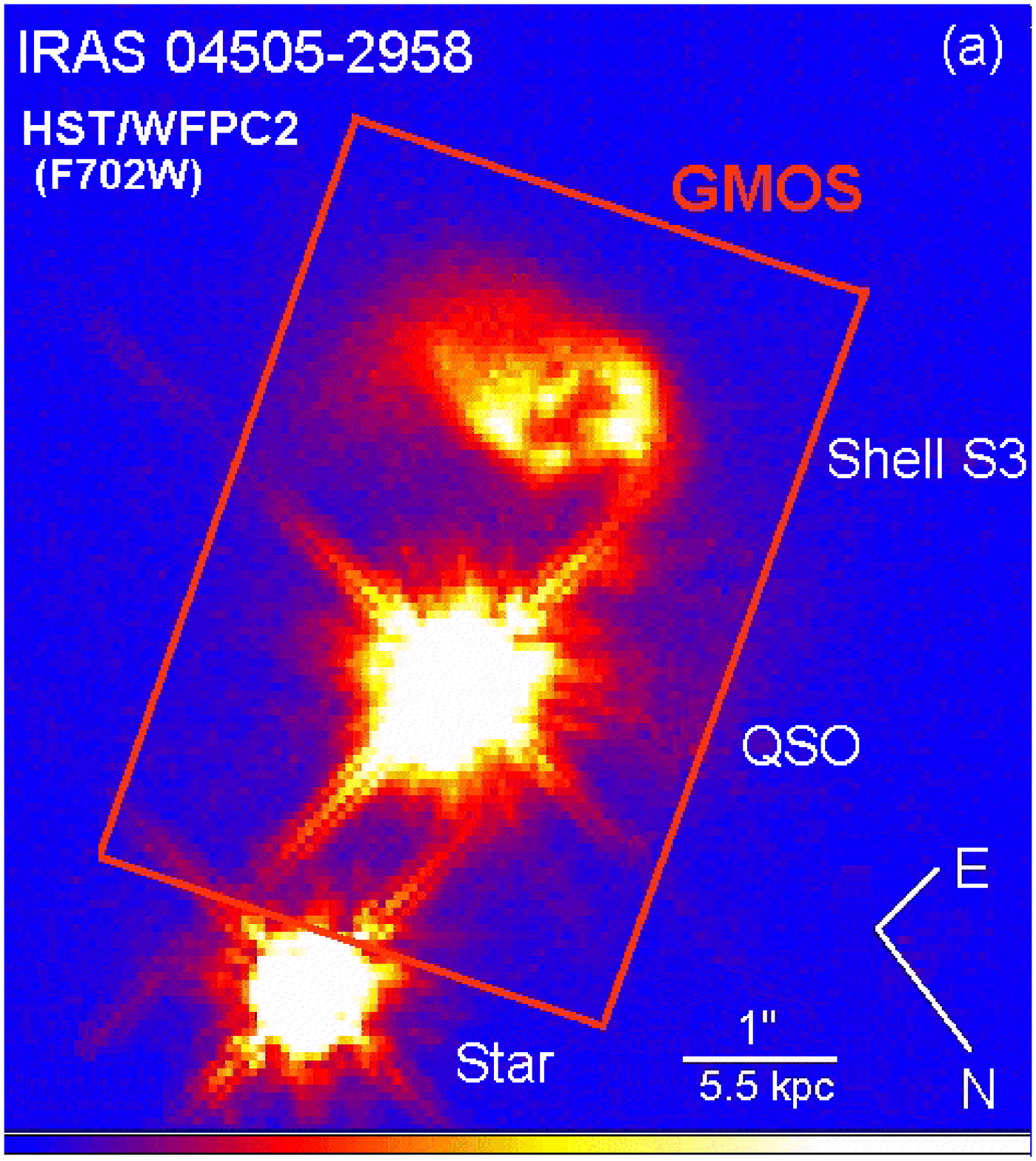}&
\includegraphics{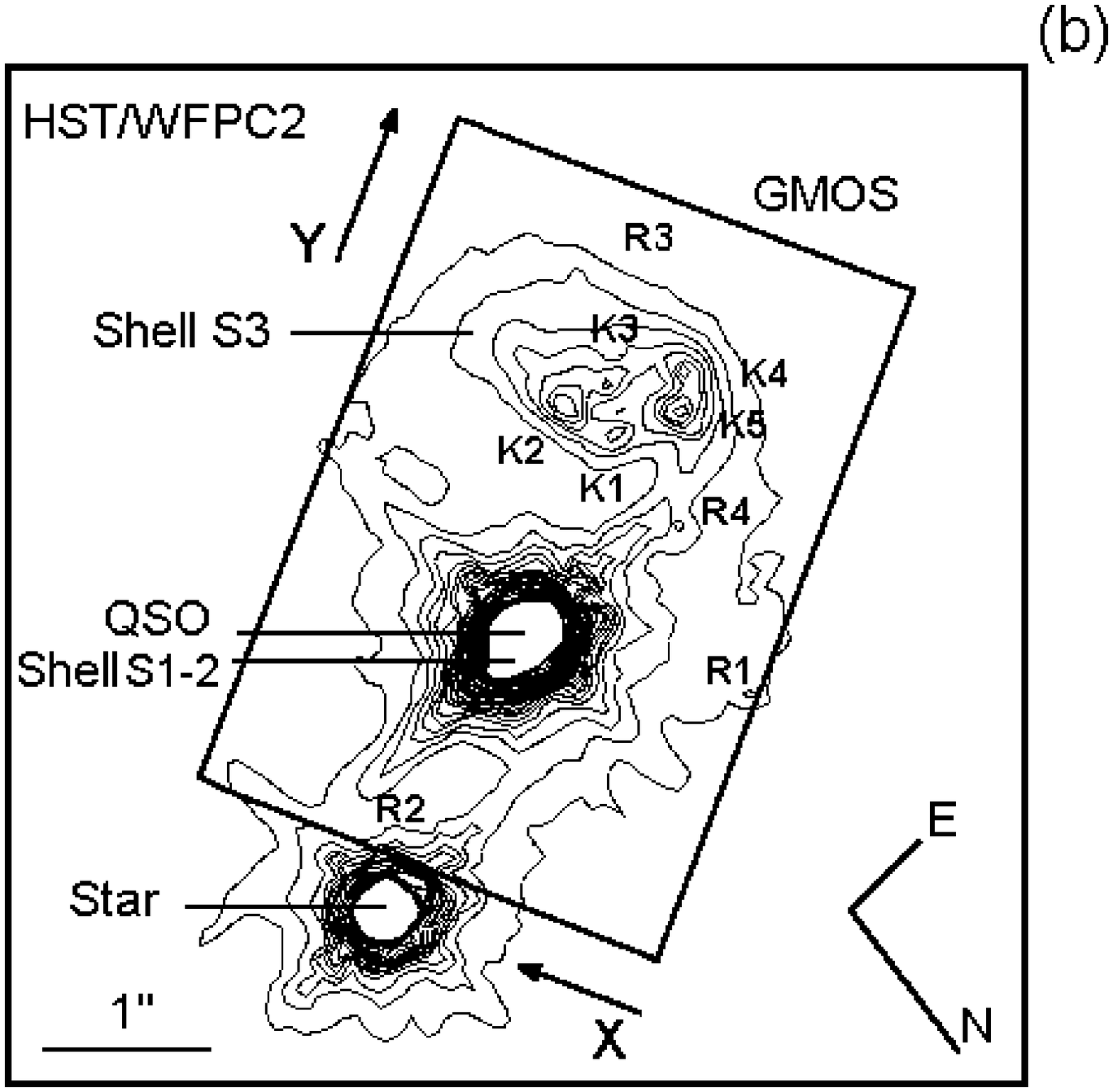} \cr
\includegraphics{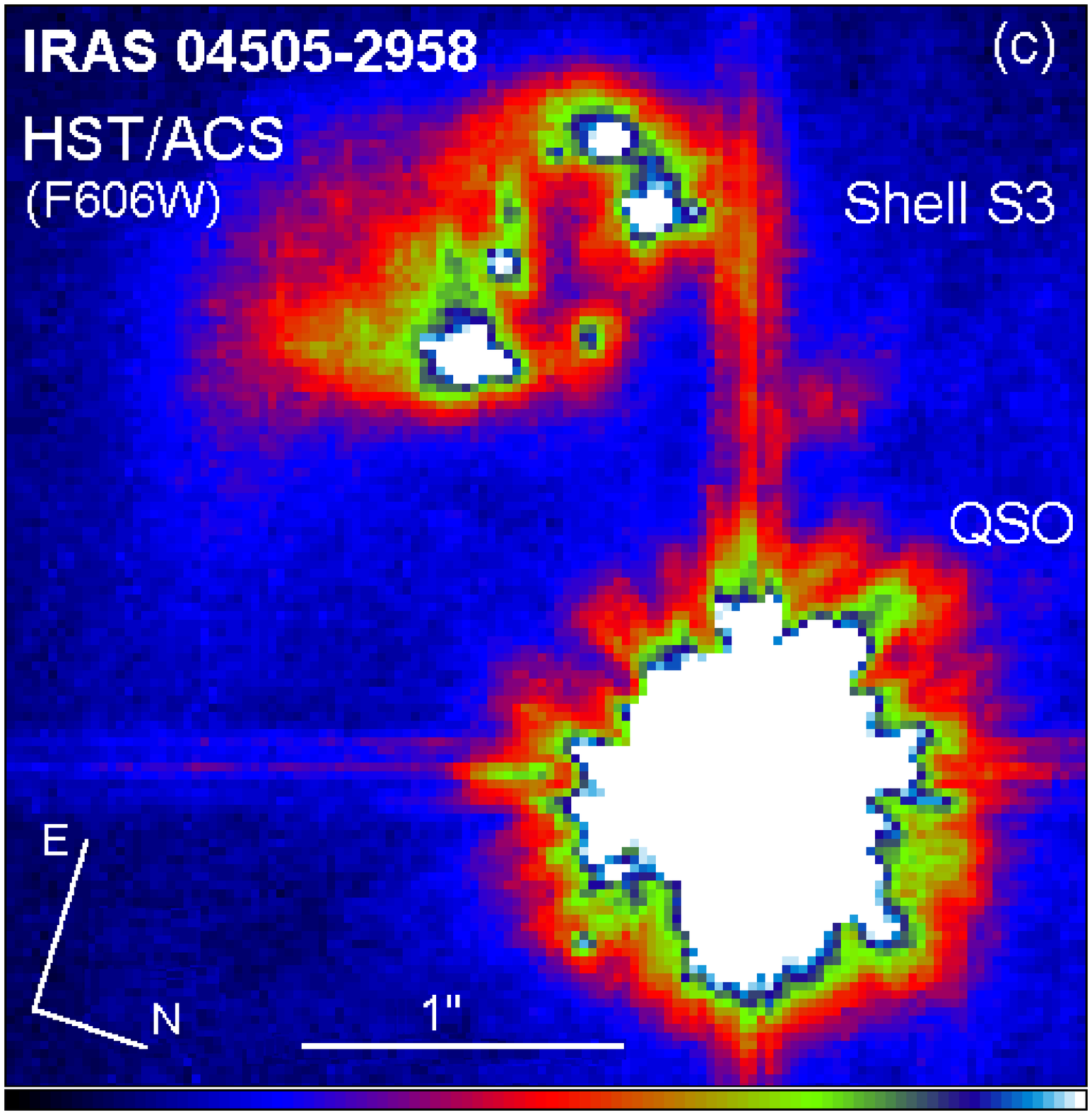}&
\includegraphics{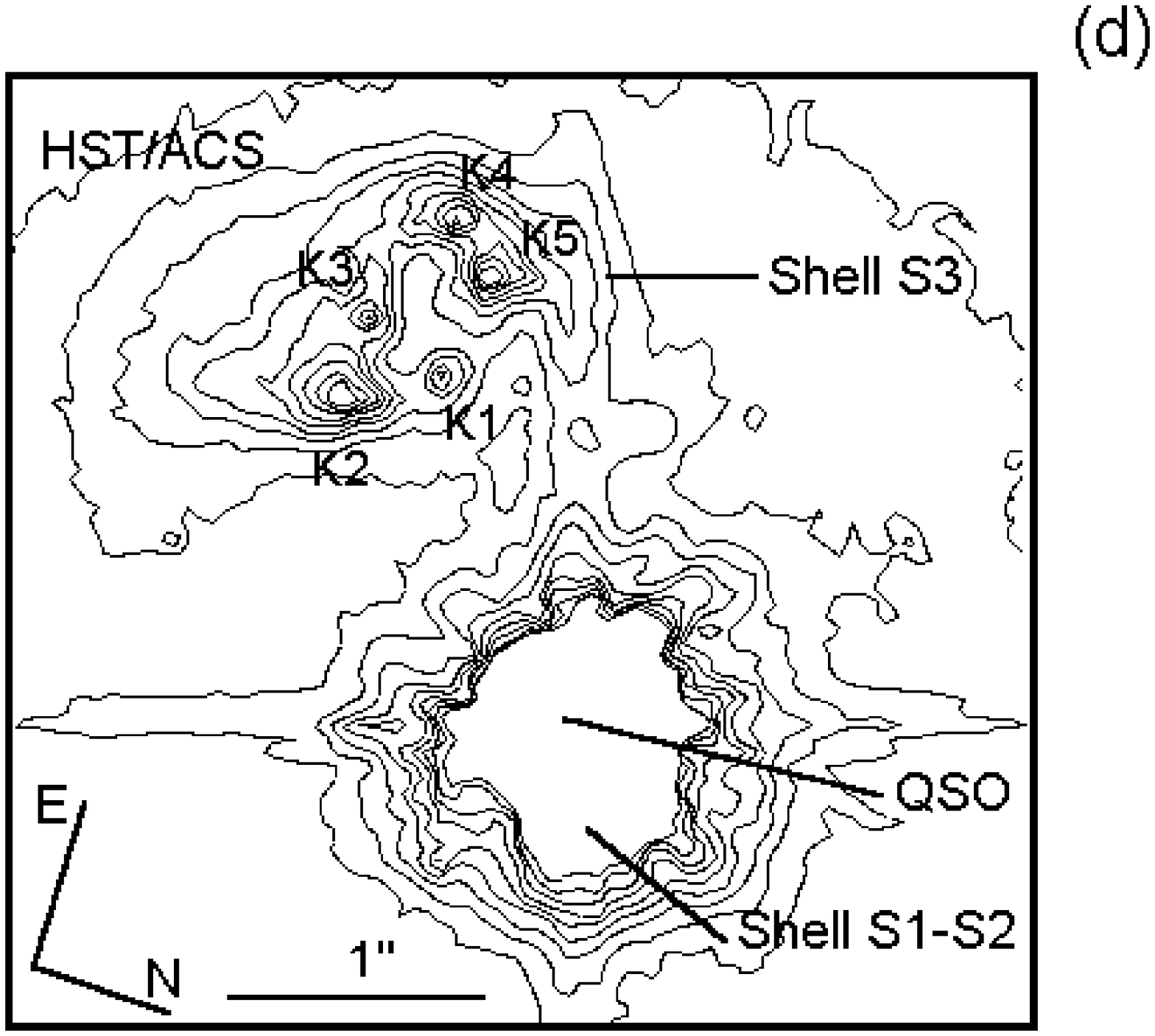} \cr
\end{tabular}
\vspace{7.0 cm}
\caption {
HST WFPC2+F702W ($\sim$R) and ACS+F606W ($\sim$V) high resolution images
(a, c) and  contour--images (b, d)  are depicted, for IRAS 04505-2958.
Which show the main shells and their knots (see for details the text).
The GMOS observed field is shown in orange color.
The GMOS  Y-axis was located at the position angle PA $=$ 131$^{\circ}$.
}
\label{fig1}
\end{figure*}

\clearpage

\begin{figure*}
\vspace{12.0 cm}
\begin{tabular}{c}
\includegraphics{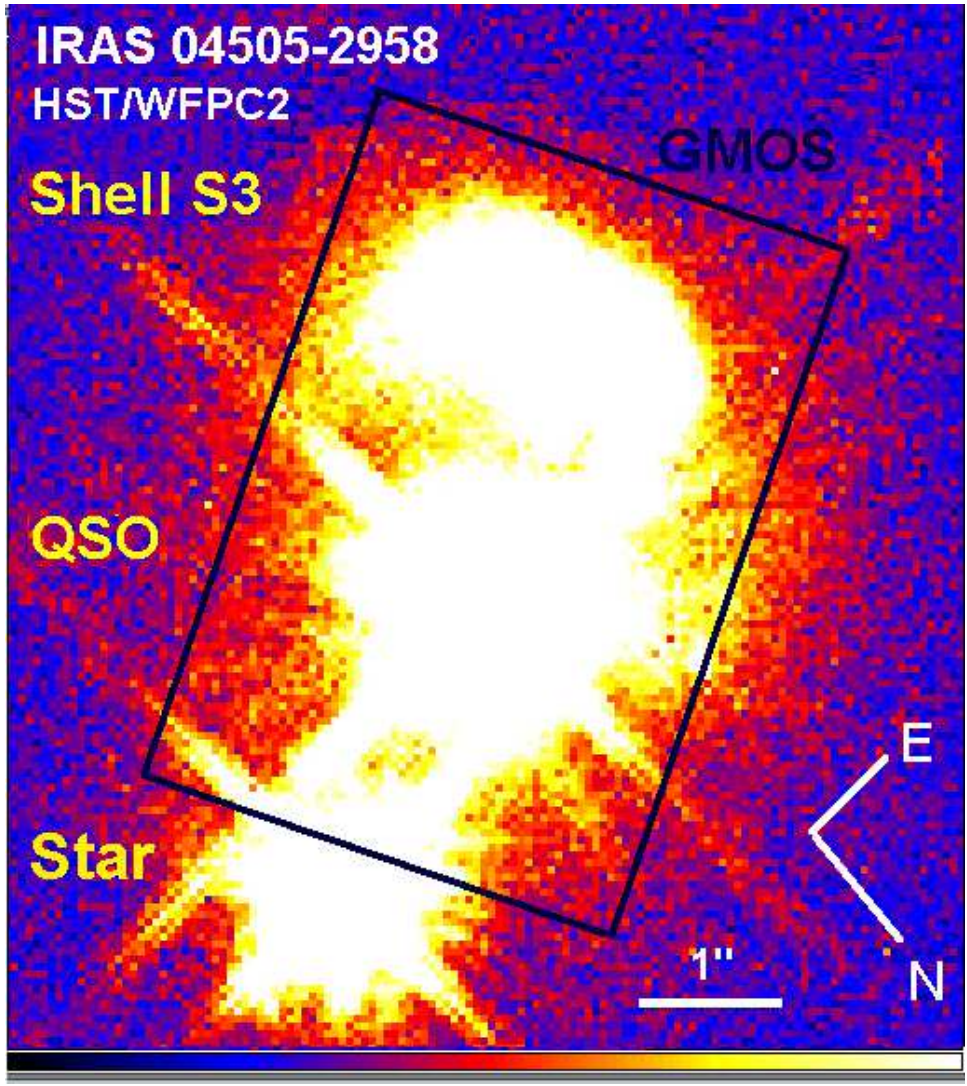}\cr
\end{tabular}
\vspace{8.0 cm}
\caption {
Deep HST--WFPC2+F702W ($\sim$R) 
high resolution broad band image of IRAS04505-2958
showing all the extension of the shell S3.
The GMOS  Y-axis was positioned at PA $=$ 131$^{\circ}$. 
}
\label{f2}
\end{figure*}

\clearpage

\begin{figure*}
\vspace{12.0 cm}
\begin{tabular}{c}
\includegraphics{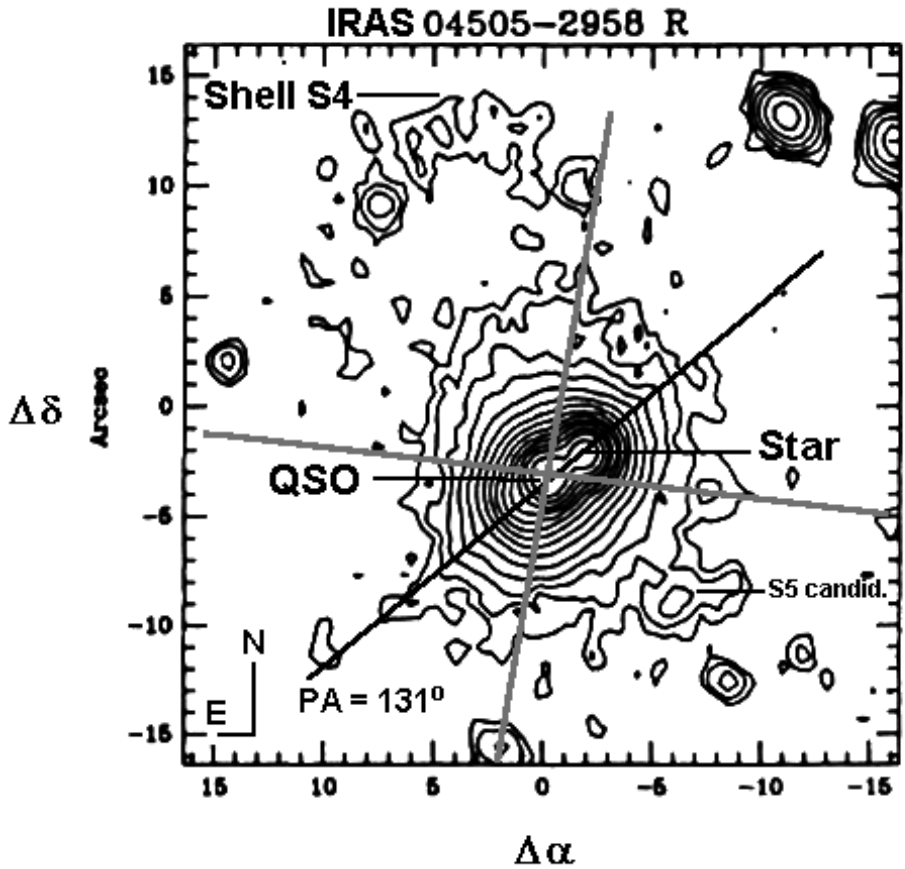}\cr
\end{tabular}
\vspace{8.0 cm}
\caption {
Wide field
CFHT broad band R-image of the QSO IRAS 04505-2958 showing the external
shell S4 (adapted from Hutching \& Neff 1988; their Fig. 1).
The gray lines show a possible opening angle for the more external out
flow process, probably associated with the shells S4 and S5 (at
PA  = 40$^{\circ}$; see the text).
}
\label{fig3}
\end{figure*}

\clearpage

\begin{figure*}
\vspace{12.0 cm}
\begin{tabular}{c}
\includegraphics{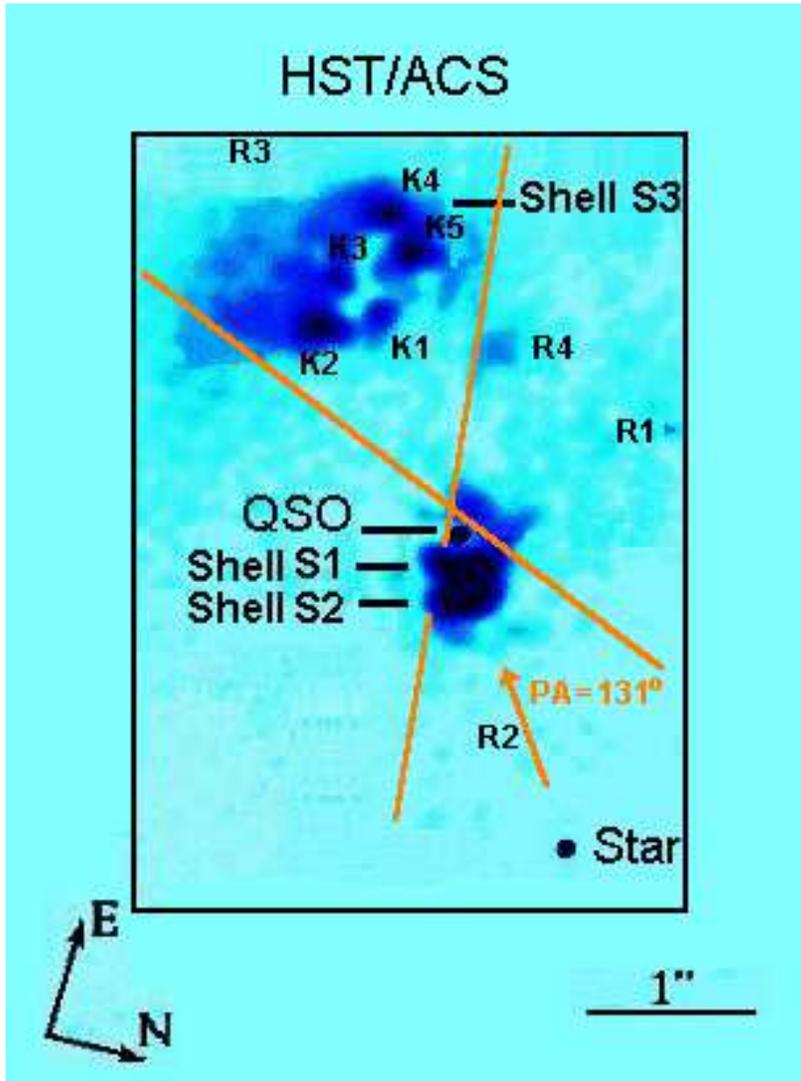}\cr
\end{tabular}
\vspace{8.0 cm}
\caption {
HST--ACS+F606W ($\sim$V) high resolution  image of the QSO IRAS 04505-2958
and the shells system S1, S2 and S3 (adapted from
Magain et al. 2005, and ESO Press Release 23/05, 14 September 2005).
The  lines show a possible opening angle for the more internal out
flow process, associated with the shells S1+S2 and S3 (at PA = 131$^{\circ}$;
which is perpendicular to the PA of the more external OF; see the text).
}
\label{fig4}
\end{figure*}

\clearpage

\begin{figure*}
\vspace{12.0 cm}
\begin{tabular}{cc}
\includegraphics{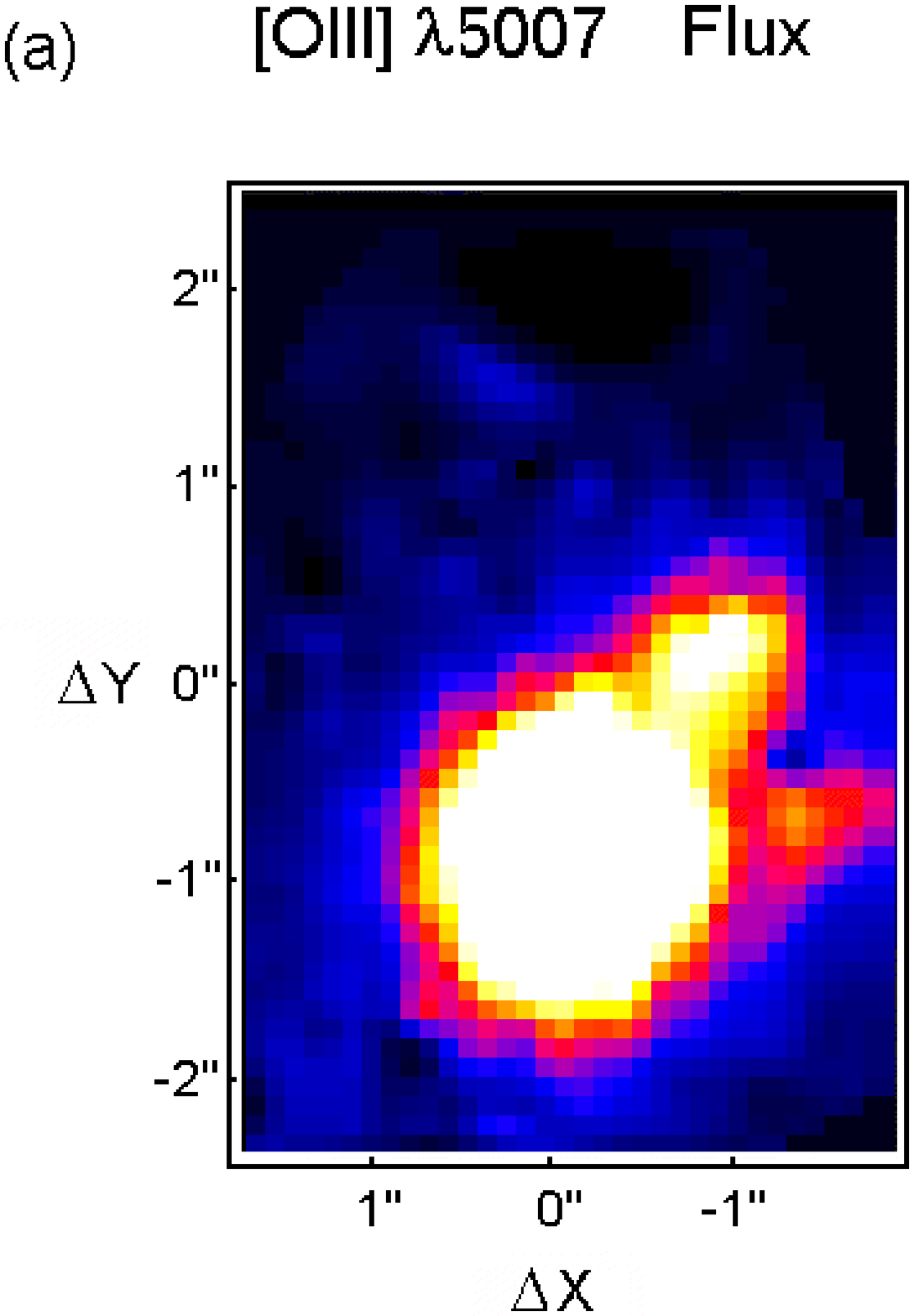}&
\includegraphics{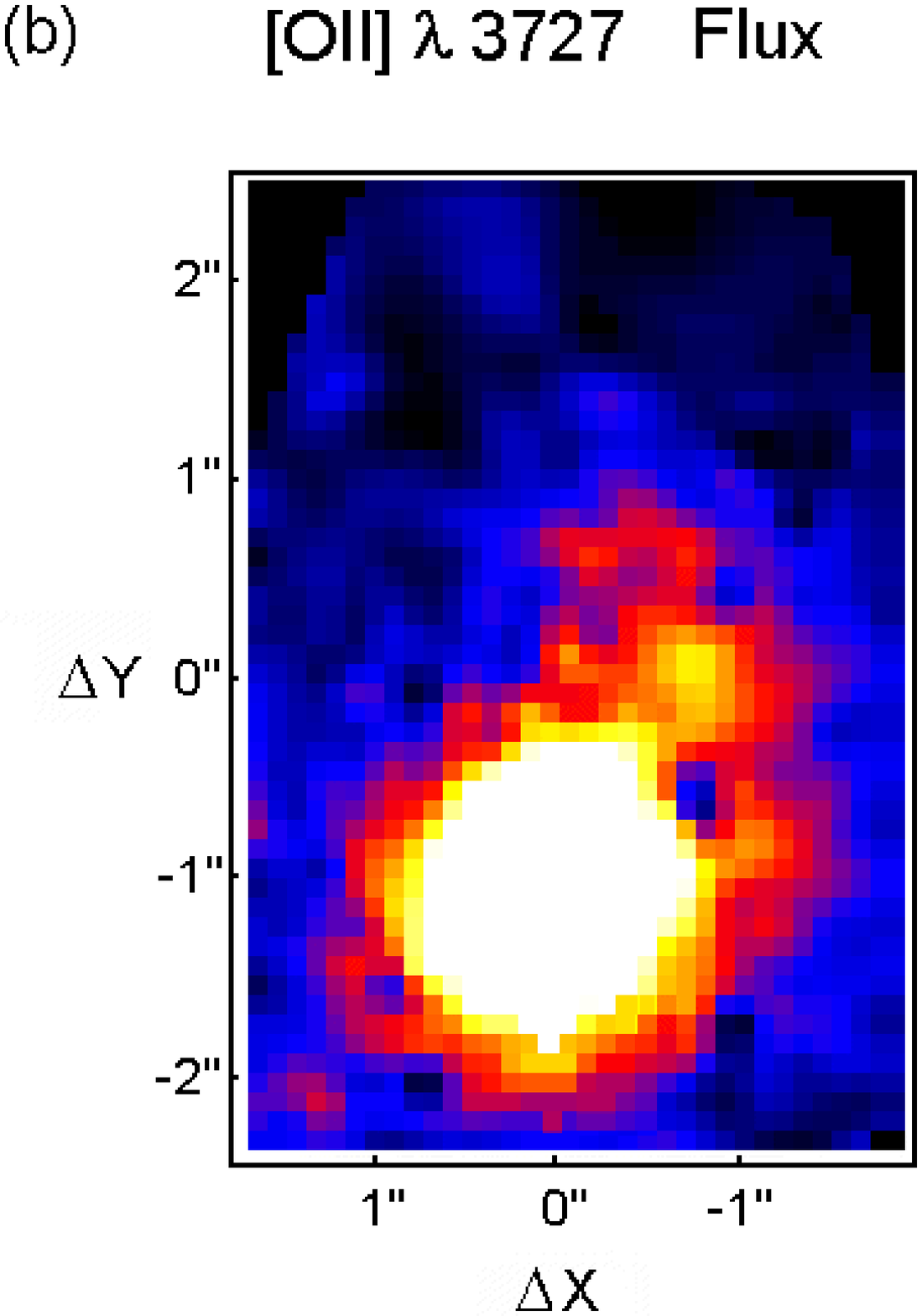} \cr
\includegraphics{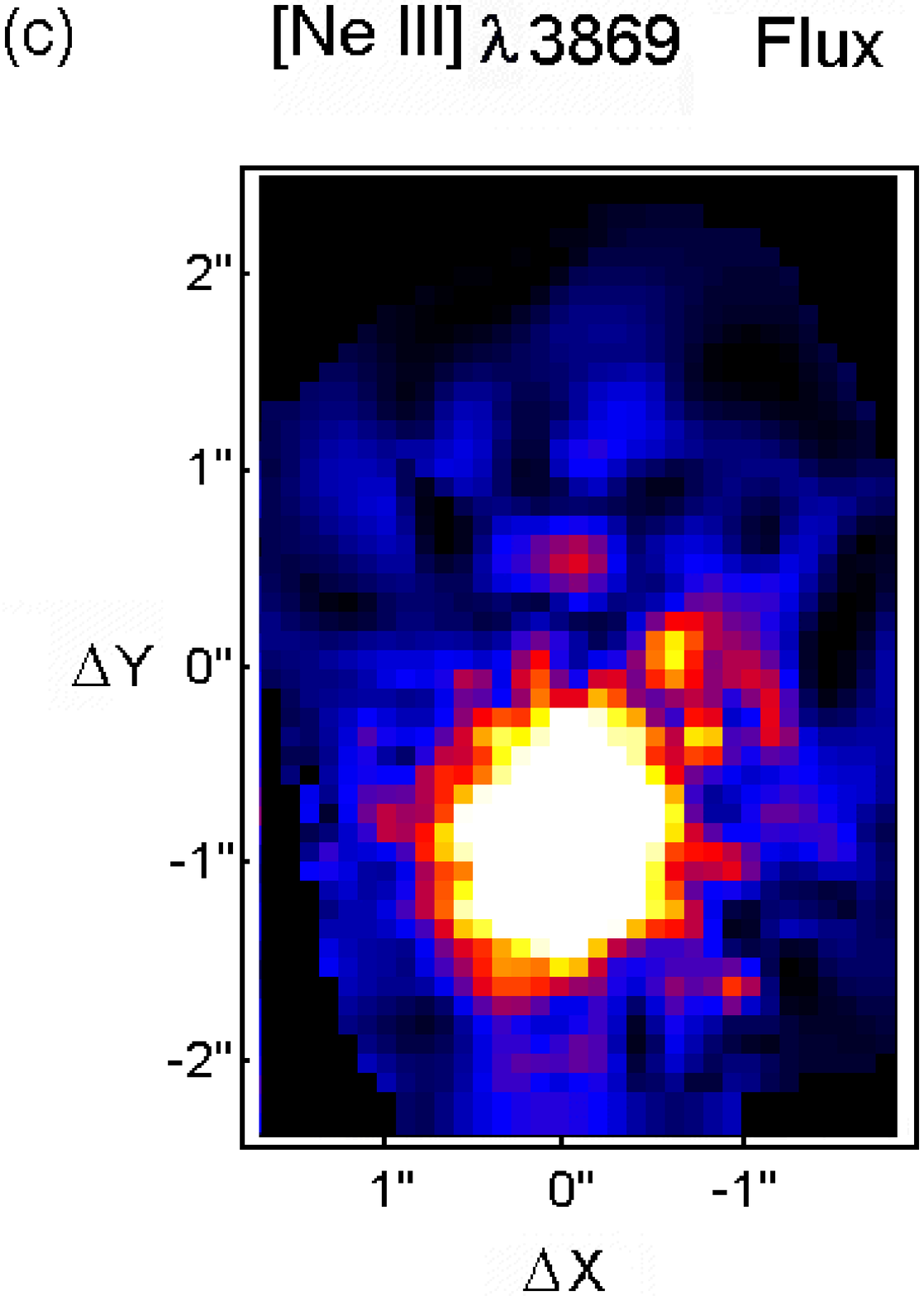}&
\includegraphics{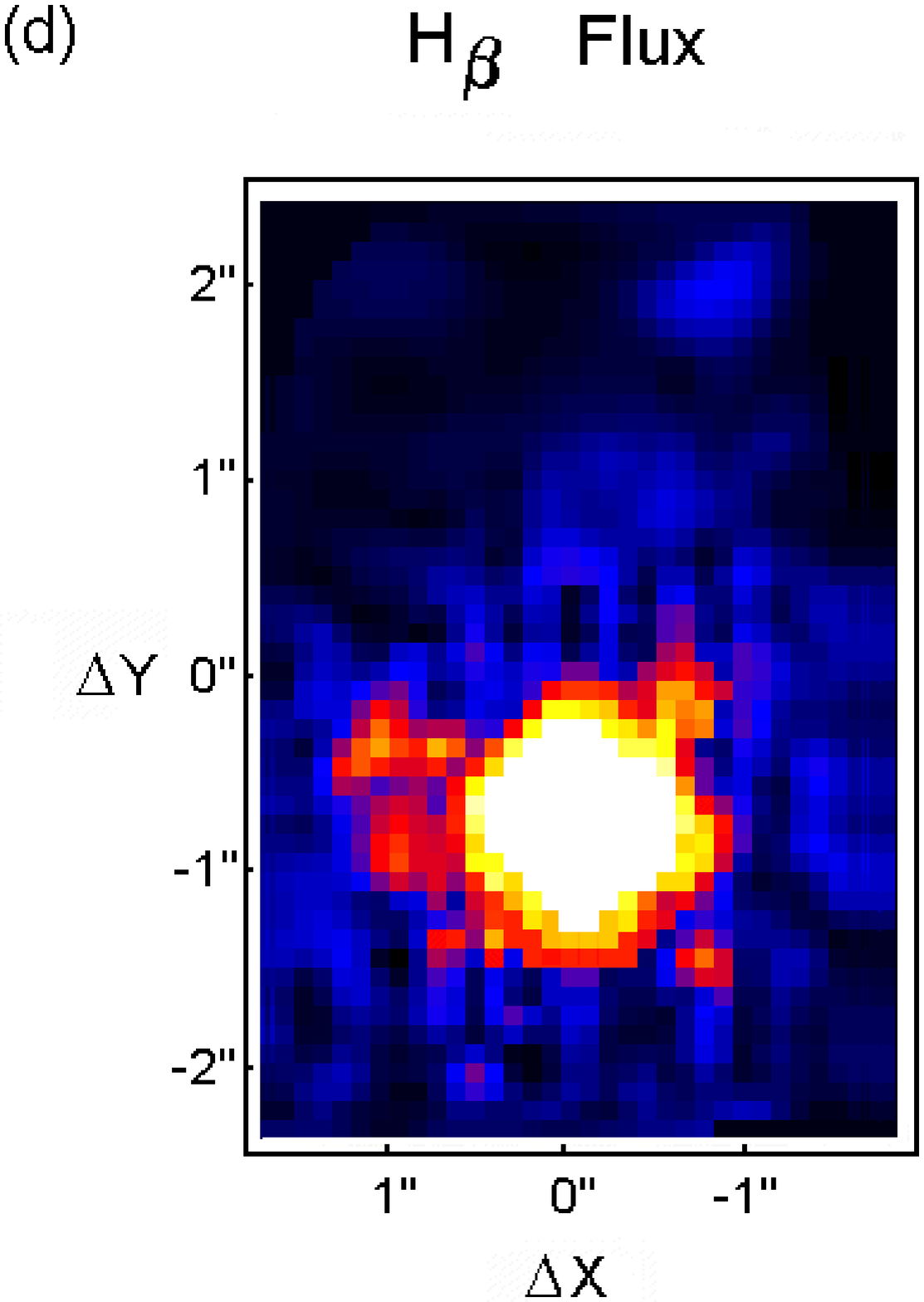} \cr
\end{tabular}
\vspace{8.0 cm}
\caption {
Gemini GMOS-IFU maps (3.5$'' \times$ 5$''$) of the emission lines:
[O {\sc iii}]$\lambda$5007, [O {\sc ii}]$\lambda$3727,
[Ne {\sc iii}]$\lambda$3869 and H$\beta$.
The QSO-core (in each GMOS maps) is positioned at
$\Delta X \sim$ 0.0$''$, and $\Delta Y \sim$ -1.0$''$.
All the GMOS field were observed at  the position angle 
PA = 131$^{\circ}$.
}
\label{fig5}
\end{figure*}


\clearpage

\begin{figure*}
\vspace{12.0 cm}
\begin{tabular}{c}
\includegraphics{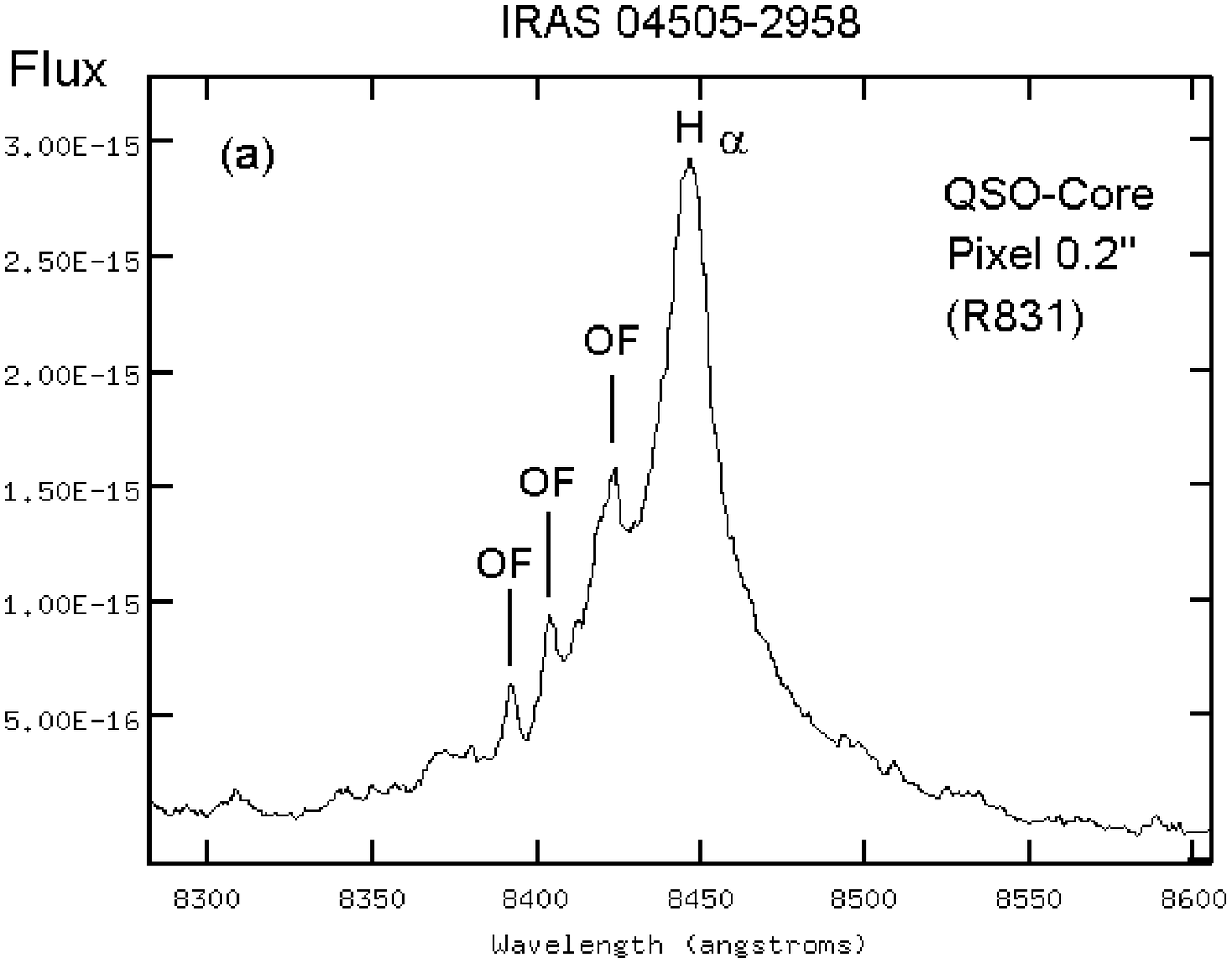}   \cr
\includegraphics{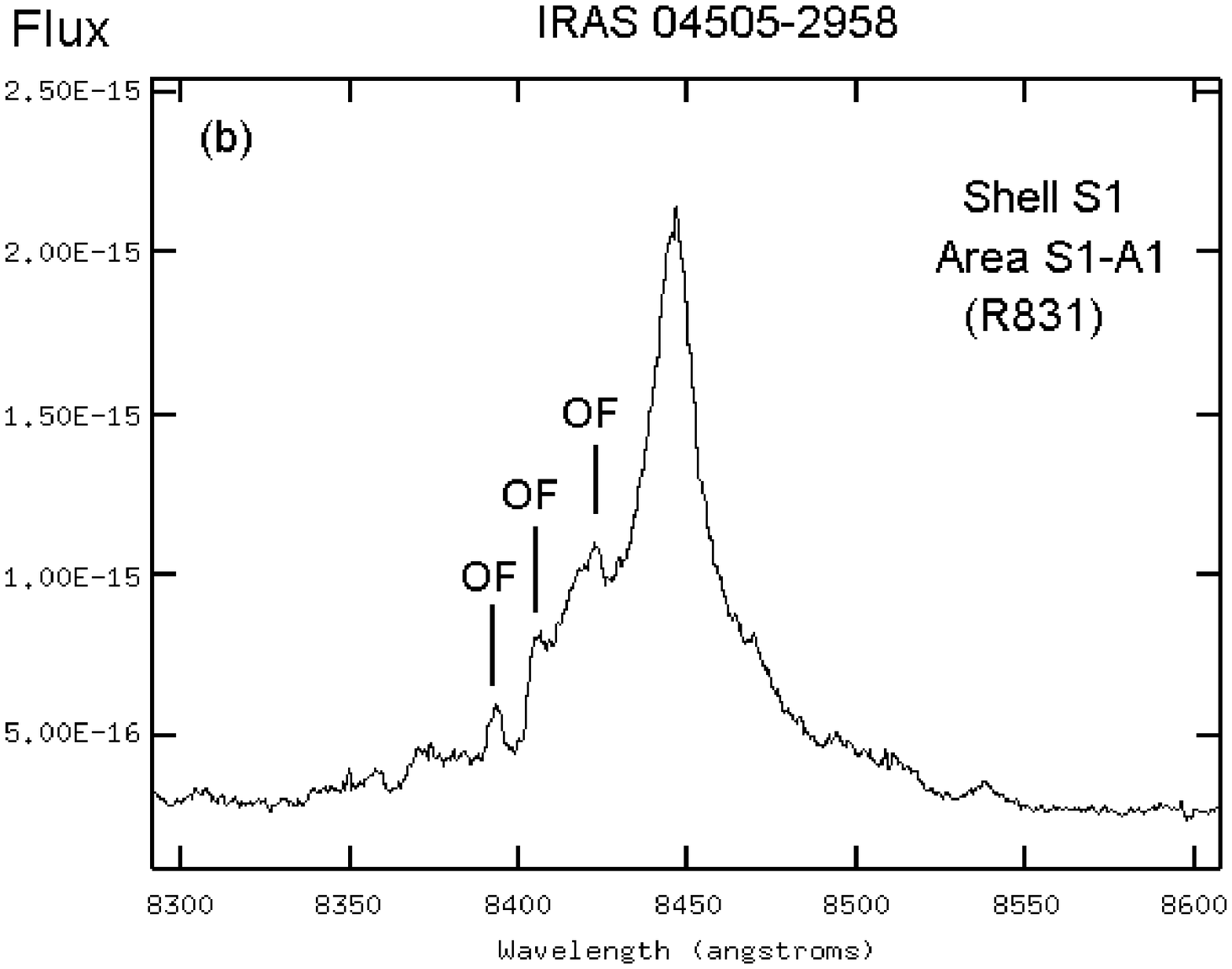} \cr
\includegraphics{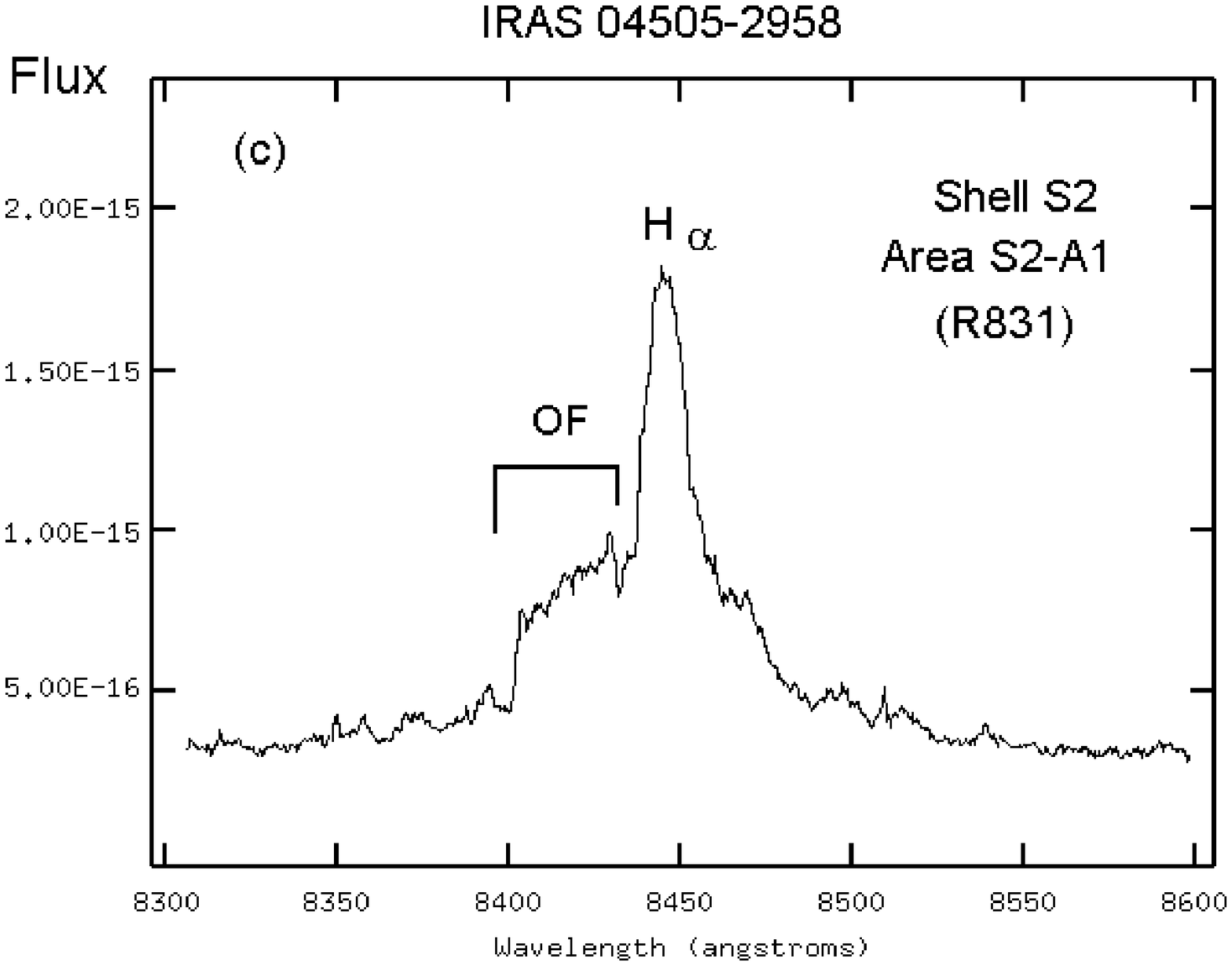} \cr
\end{tabular}
\vspace{8.0 cm}
\caption {
GMOS high resolution spectra (R831) for the QSO-core, and
the shells S1 and S2, of the
QSO IRAS 04505-2958. For the  wavelength range of H$\alpha$.
These GMOS spectra show the 3 main blue OF-components. 
}
\label{fig6}
\end{figure*}


\clearpage

\begin{figure*}
\vspace{12.0 cm}
\begin{tabular}{cc}
\includegraphics{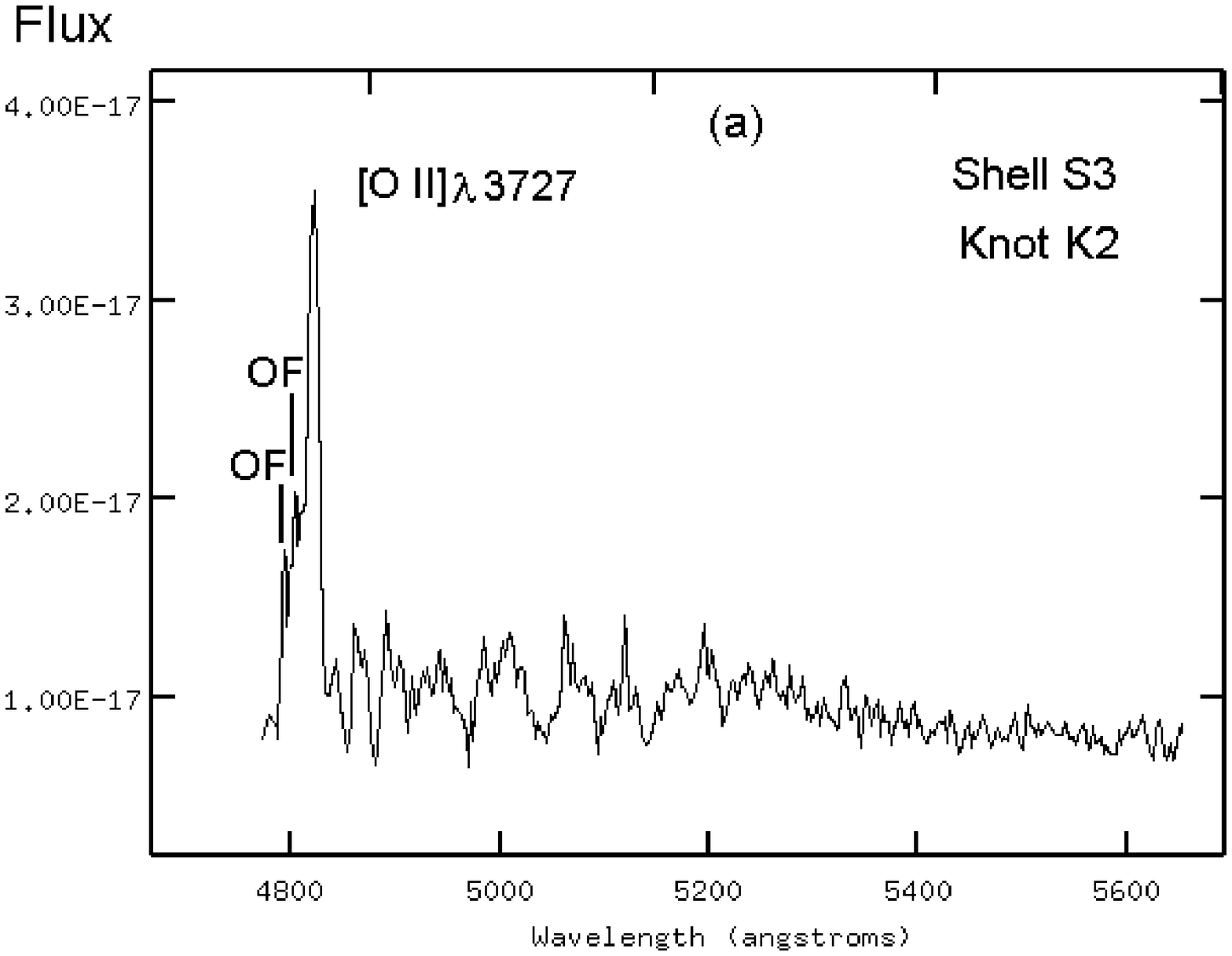}&
\includegraphics{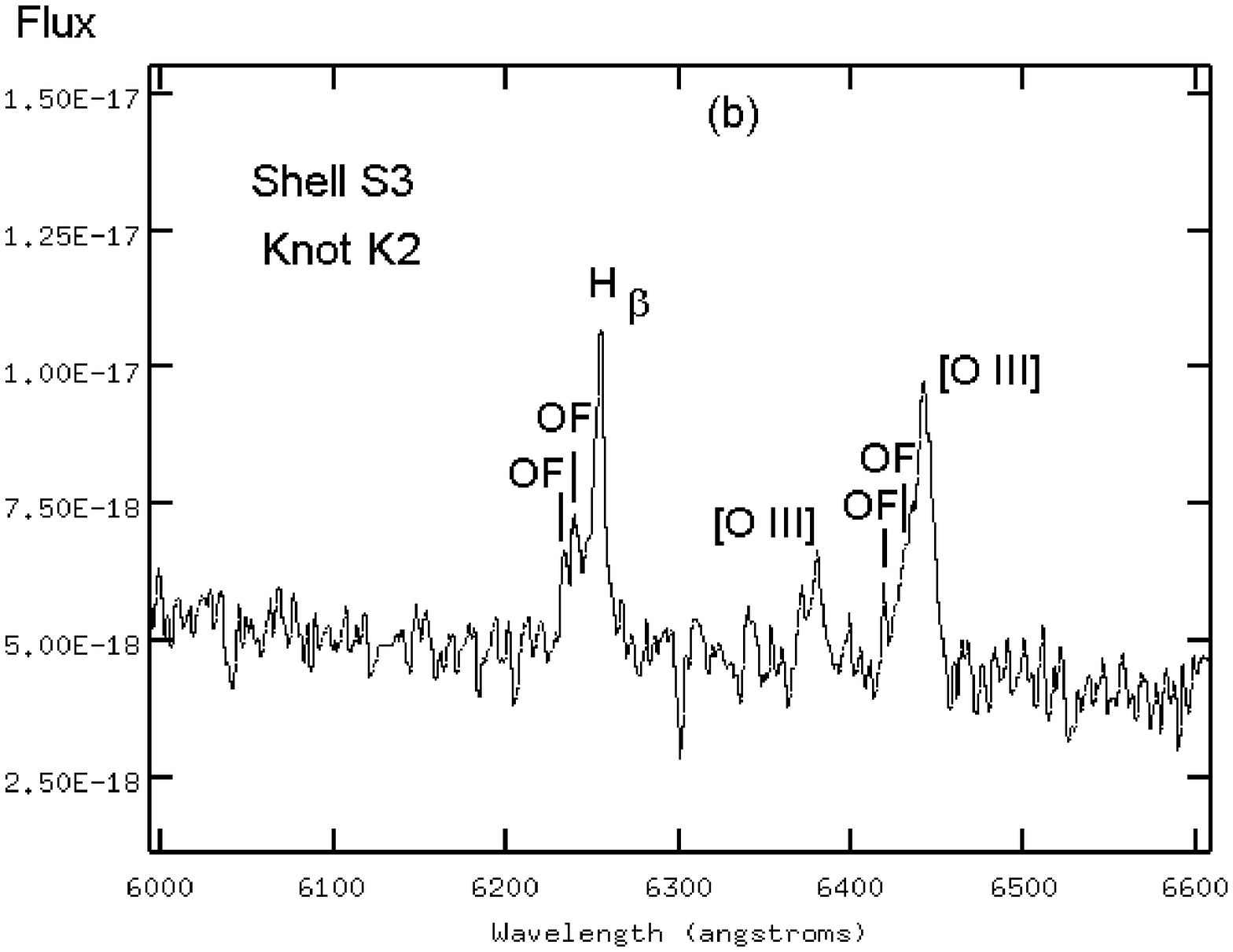} \cr
\includegraphics{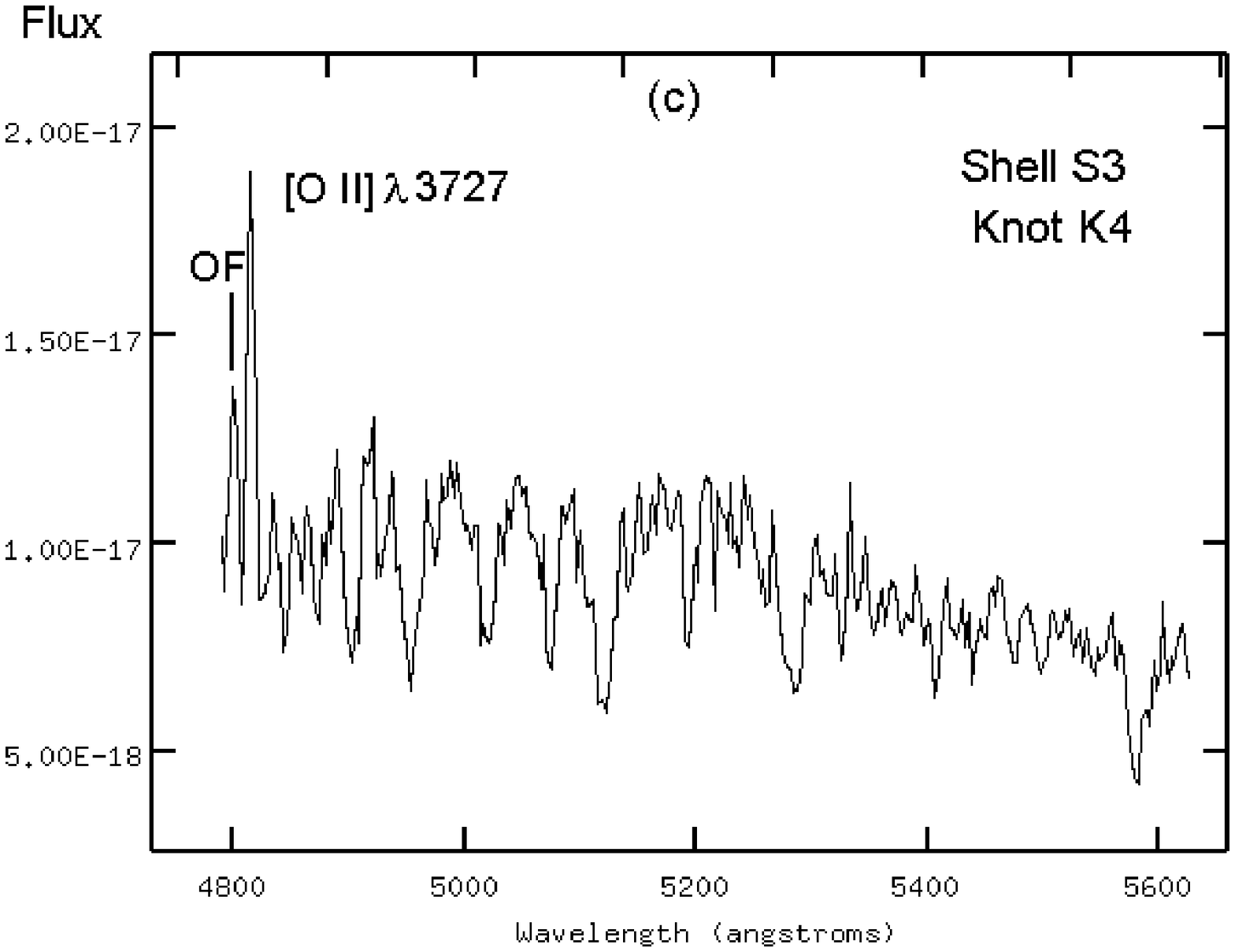}&
\end{tabular}
\vspace{8.0 cm}
\caption {
GMOS-IFU B600 spectra  for knots of the Shell S3.
These GMOS spectra show the main blue OF-components. 
For the  wavelength range of H$\beta$ and [O {\sc ii}]$\lambda$3727.
}
\label{fig7}
\end{figure*}


\clearpage

\begin{figure*}
\vspace{12.0 cm}
\begin{tabular}{c}
\includegraphics{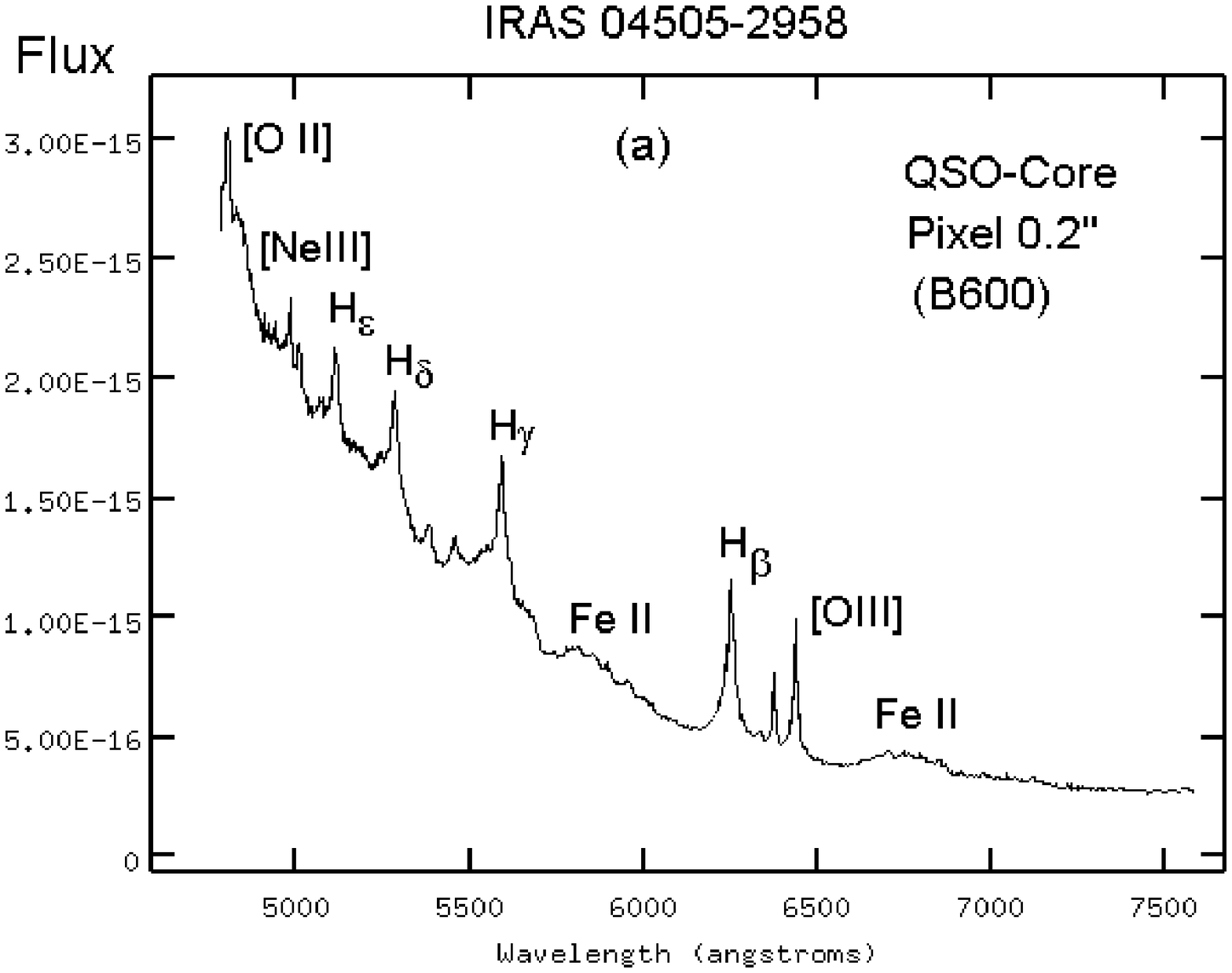}   \cr
\includegraphics{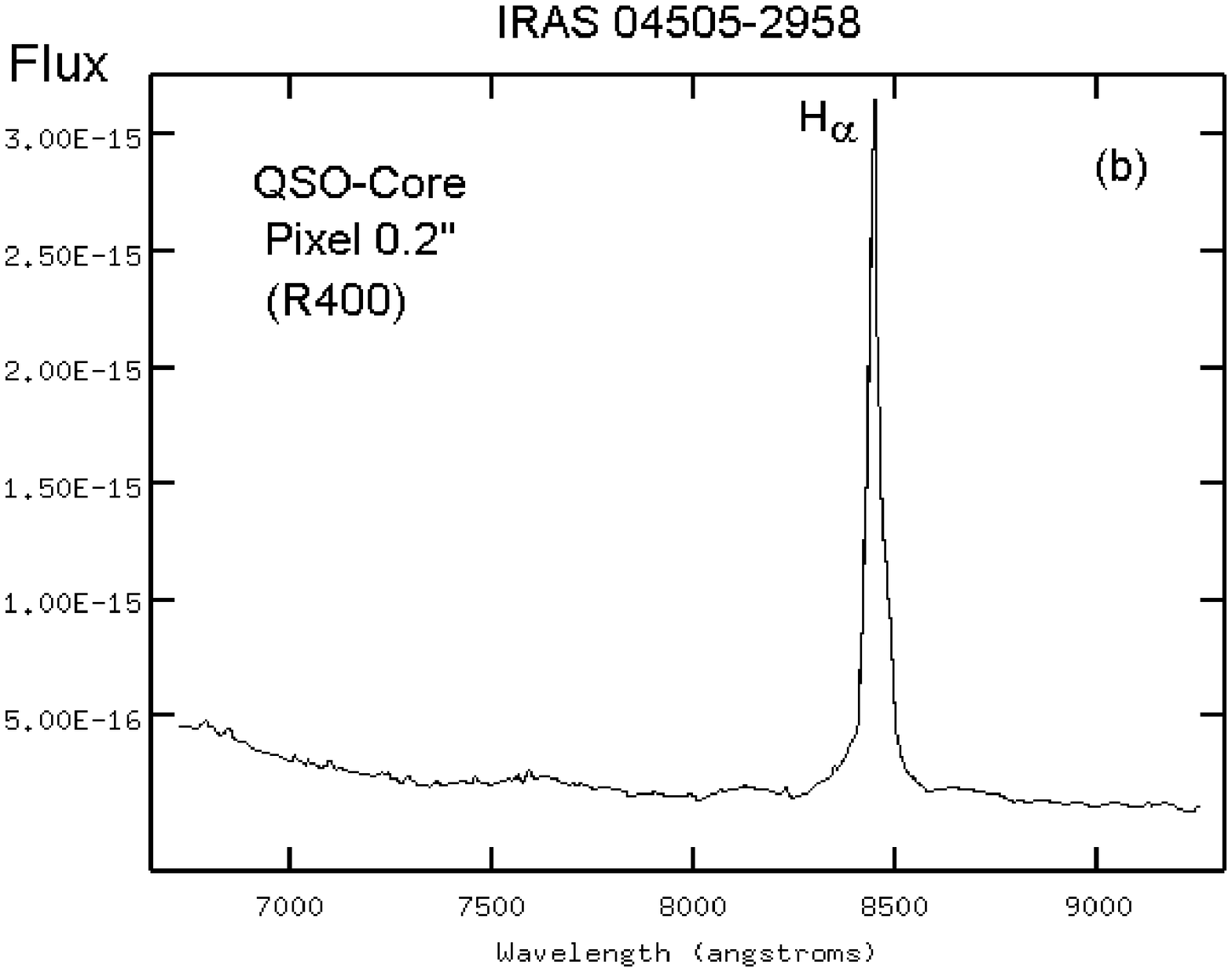} \cr
\end{tabular}
\vspace{8.0 cm}
\caption {
GMOS-IFU optical spectra  of the QSO-core, for a pixel of 0.2$''$ (a and b).
HST--FOS UV spectra of the QSO-core,  showing the BAL system at
C IV$\lambda$1550 (panel c)
}
\label{fig8}
\end{figure*}

\clearpage

\begin{figure*}
\vspace{12.0 cm}
\begin{tabular}{c}
\includegraphics{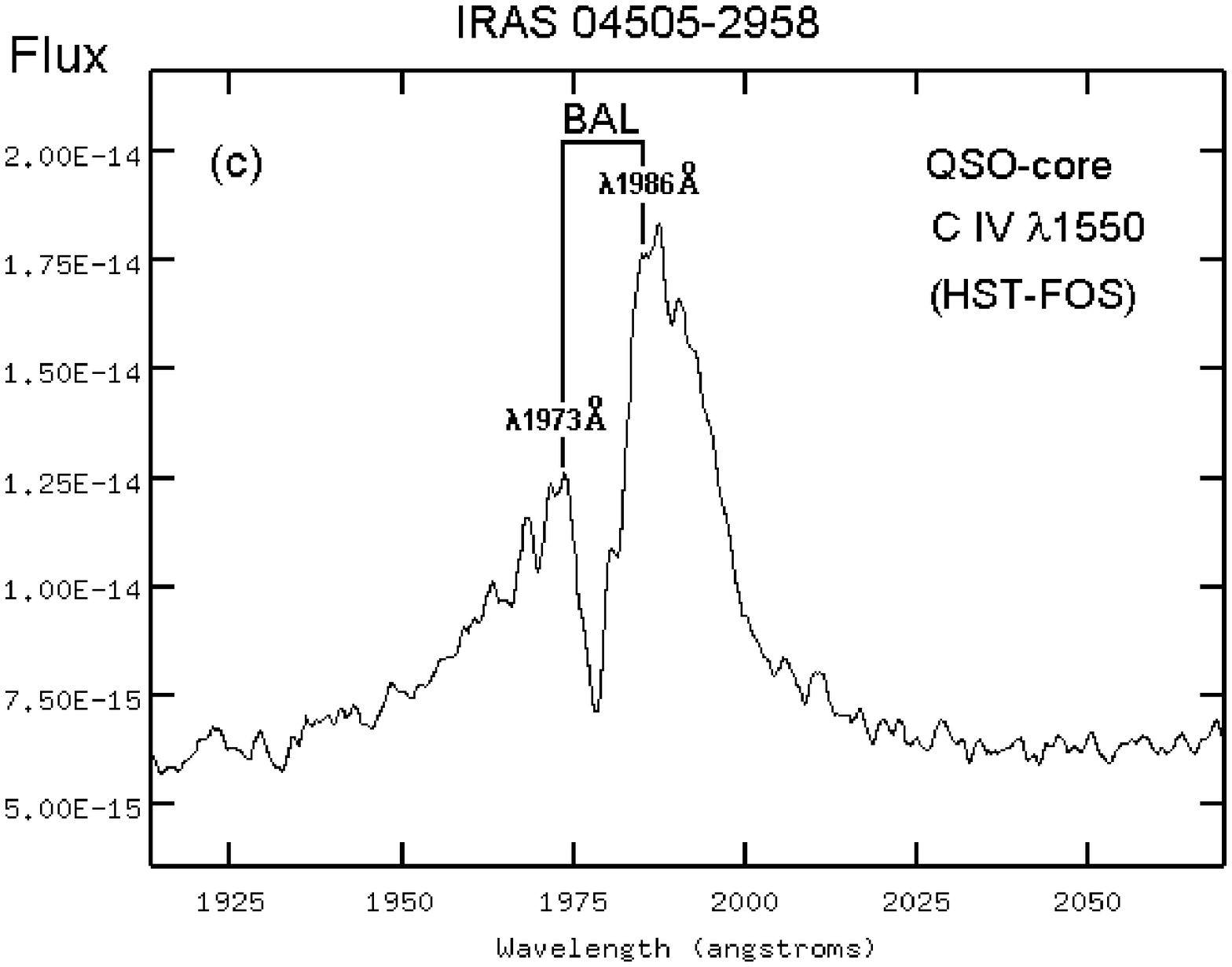}\cr
\end{tabular}
\vspace{8.0 cm}
\addtocounter{figure}{-1}
\caption {Cont.
}
\label{fig8c}
\end{figure*}


\clearpage

\begin{figure*}
\vspace{12.0 cm}
\begin{tabular}{cc}
\includegraphics{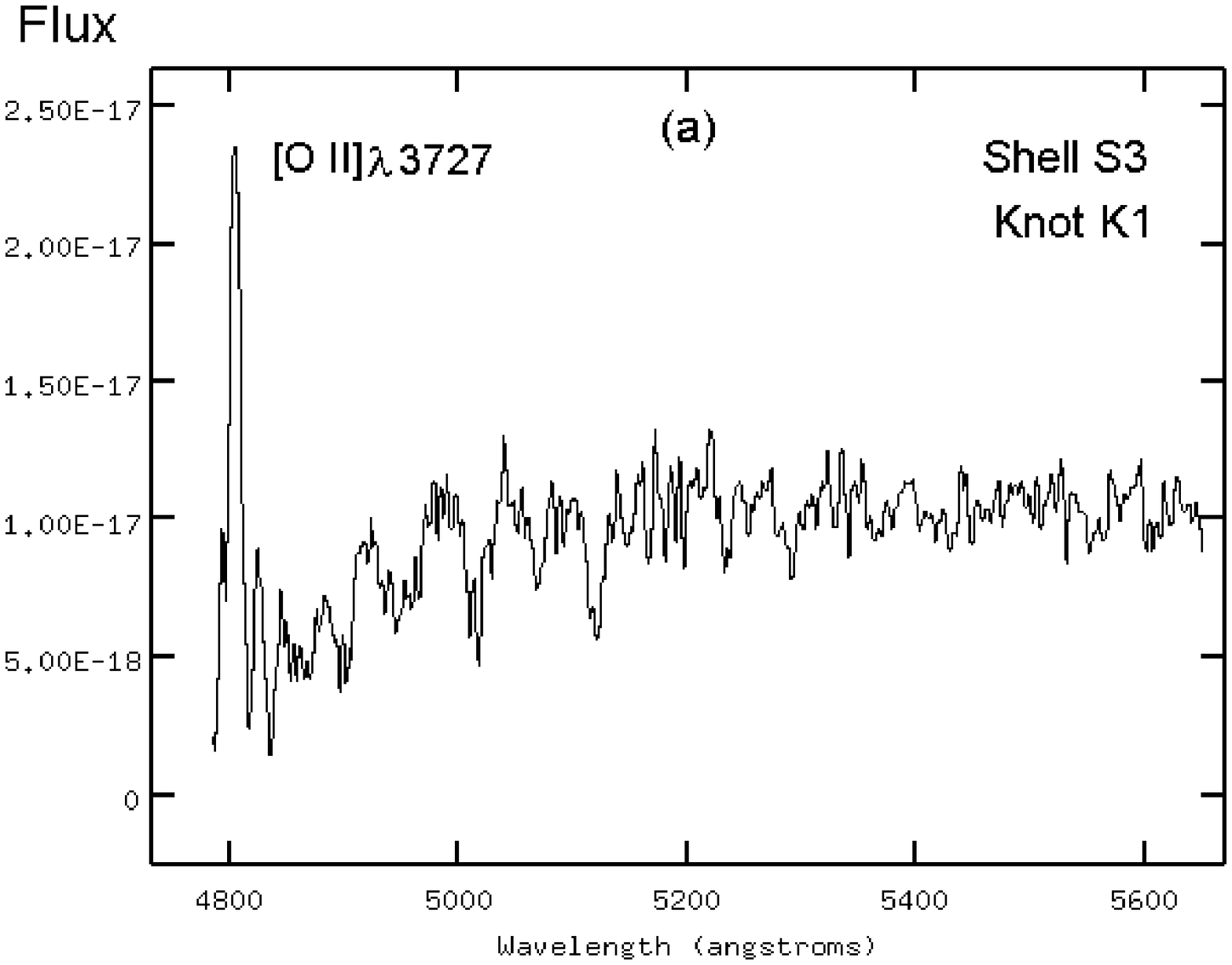}& 
\includegraphics{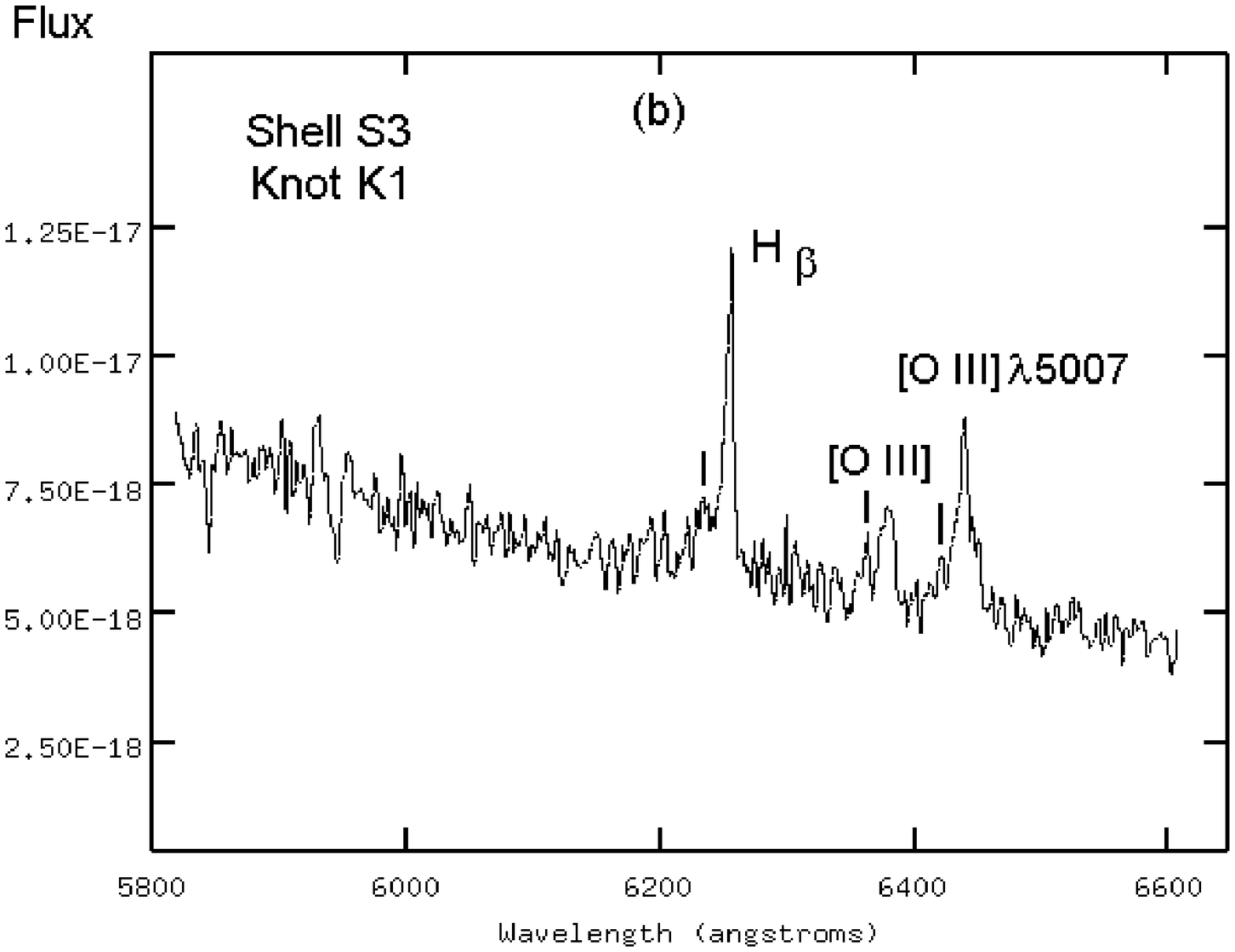} \cr
\includegraphics{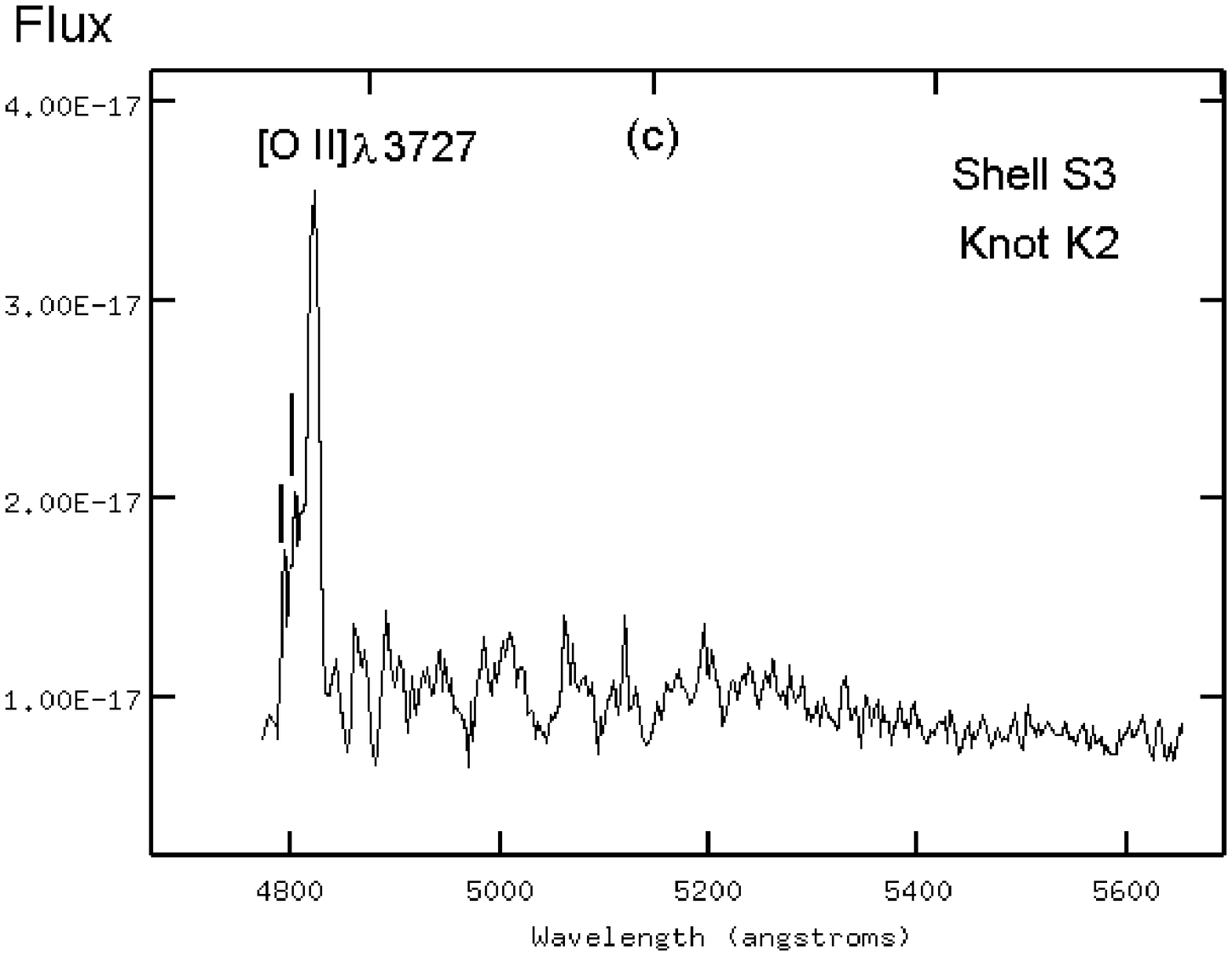}&
\includegraphics{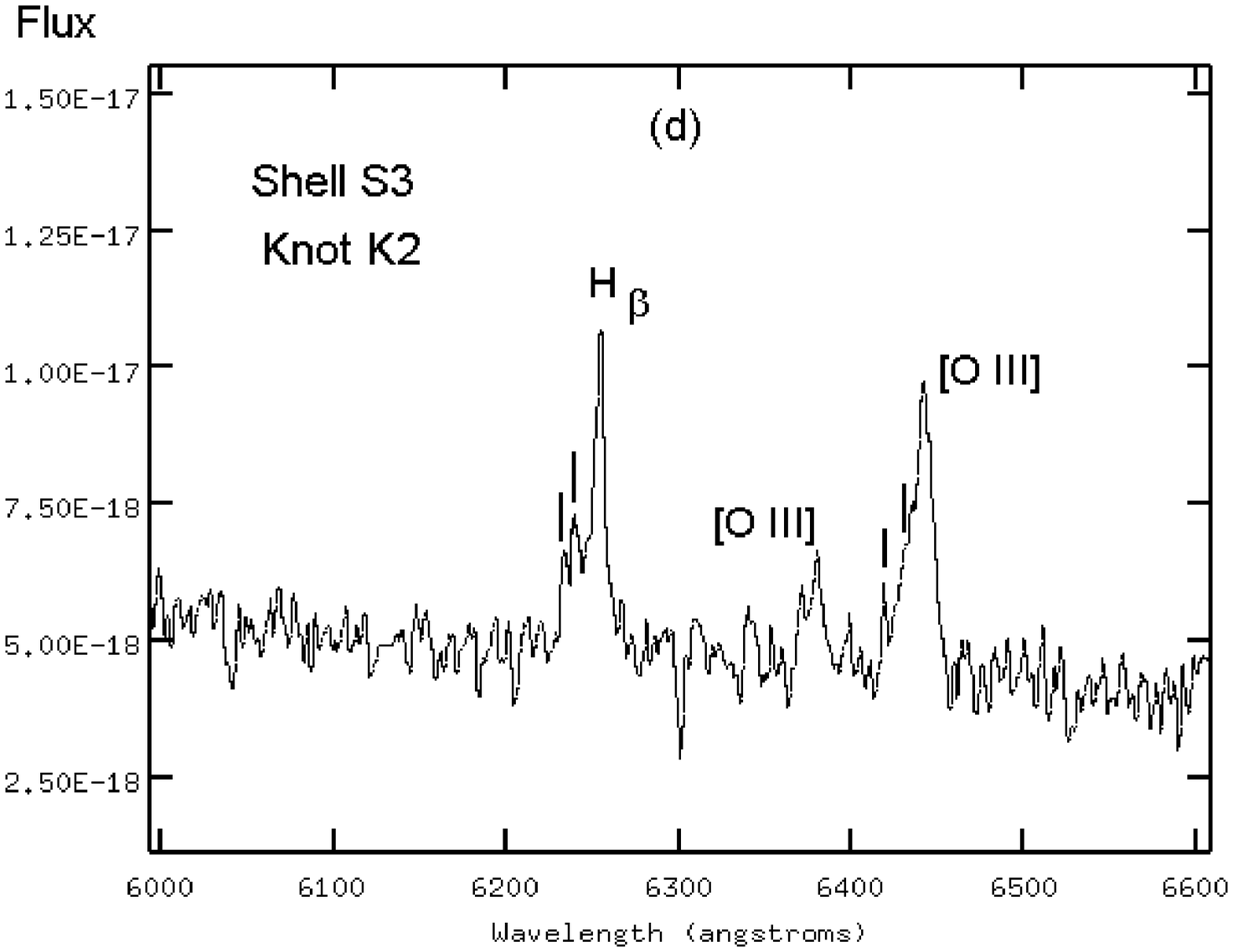} \cr
\includegraphics{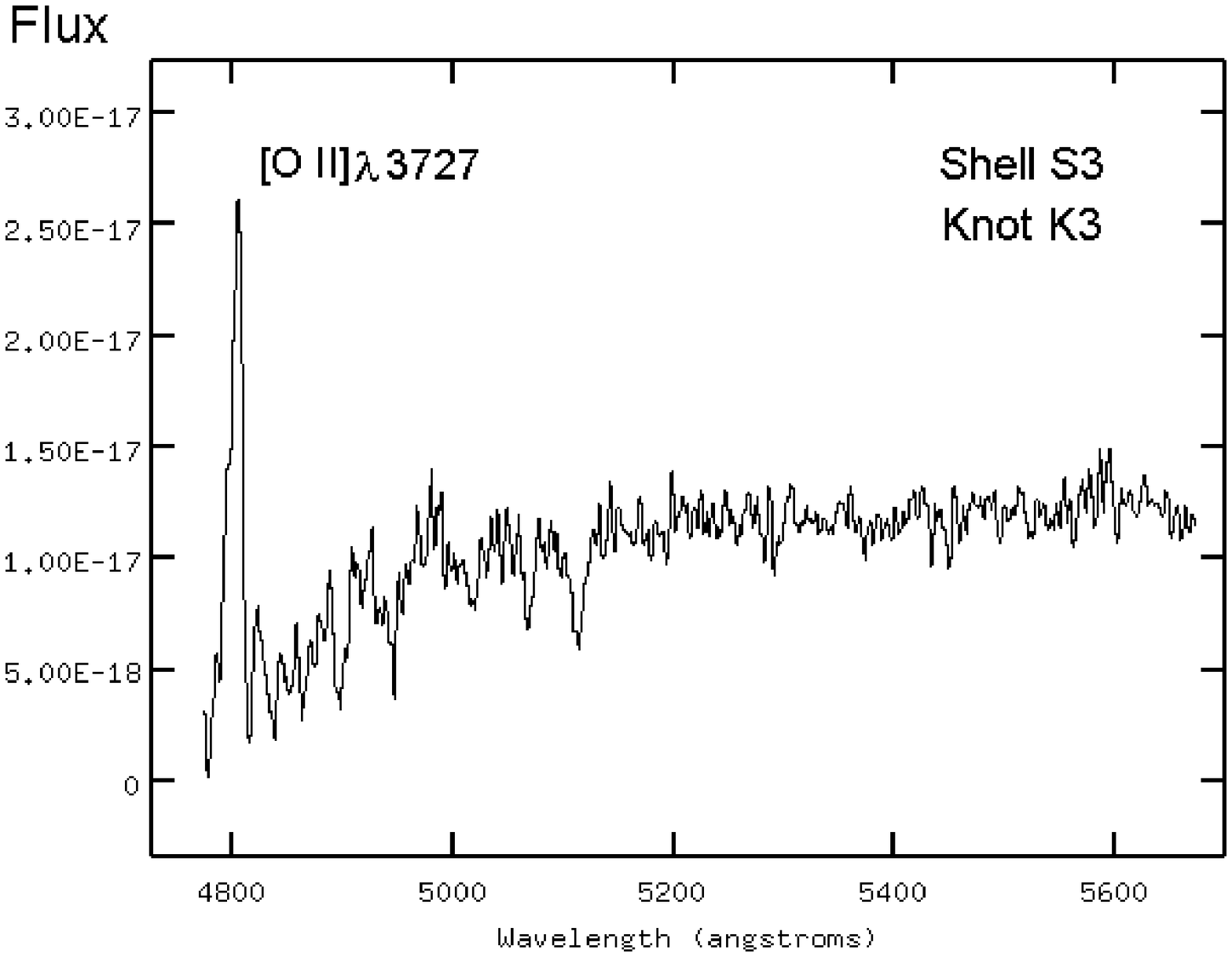}& 
\includegraphics{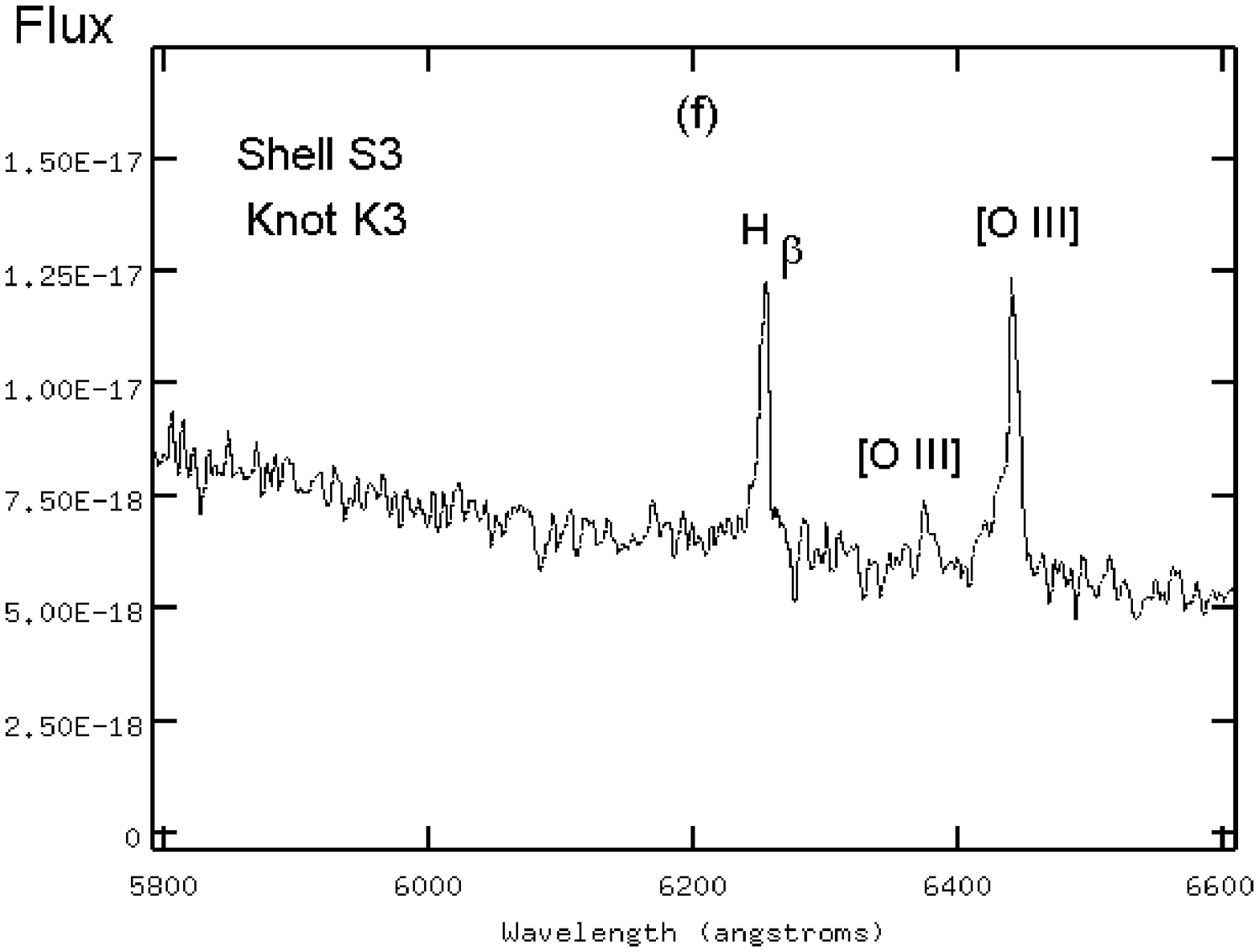} \cr
\end{tabular}
\vspace{8.0 cm}
\caption {
GMOS-IFU spectra --at the [O {\sc ii}]$\lambda$3727--H$\gamma$, and
H$\beta$ + [O {\sc iii}]$\lambda$5007 + Fe {\sc ii} wavelength ranges--
of the main knots of the shell S3, of IRAS 04505-2958.
}
\label{fig9}
\end{figure*}

\clearpage

\begin{figure*}
\vspace{12.0 cm}
\begin{tabular}{cc}
\includegraphics{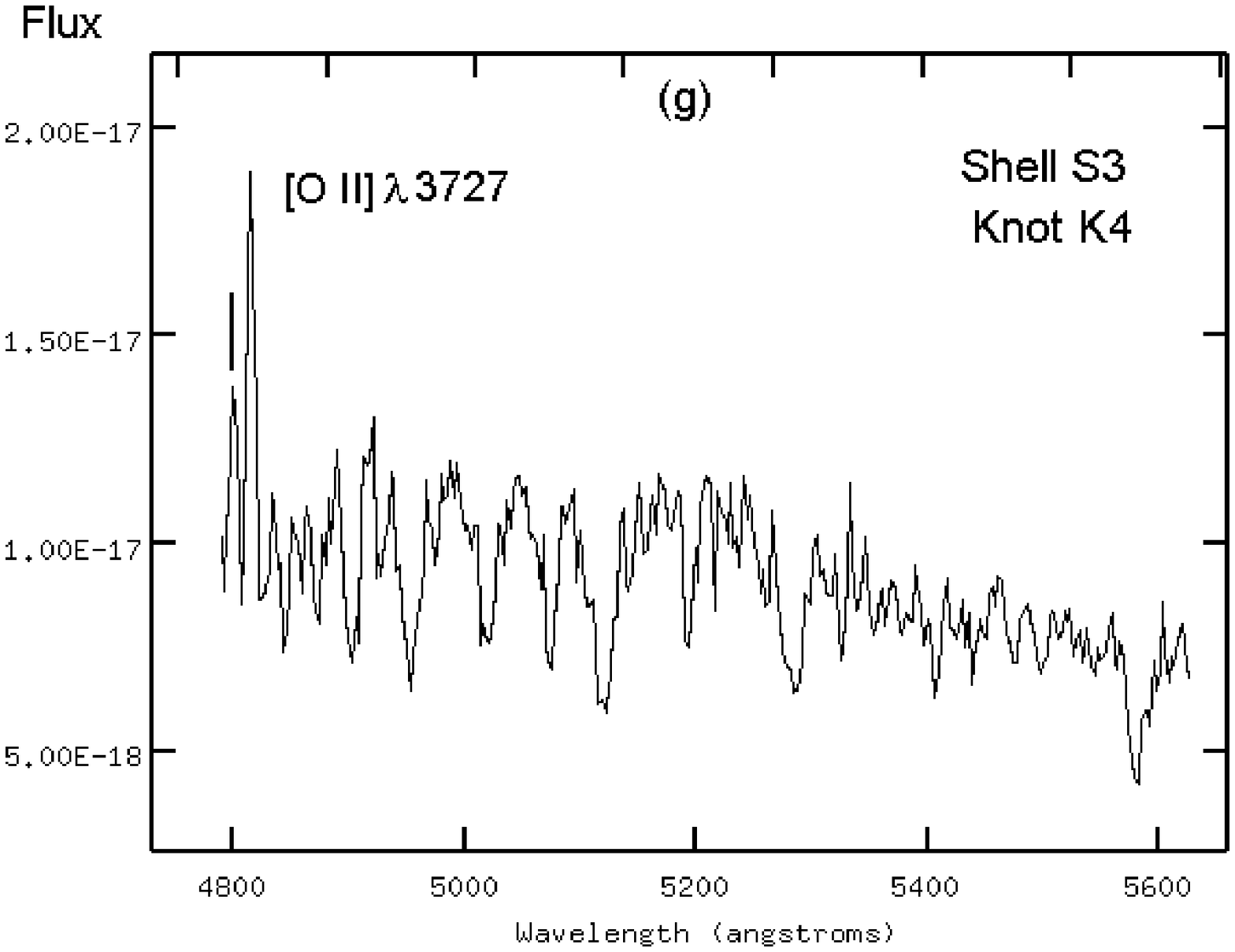}& 
\includegraphics{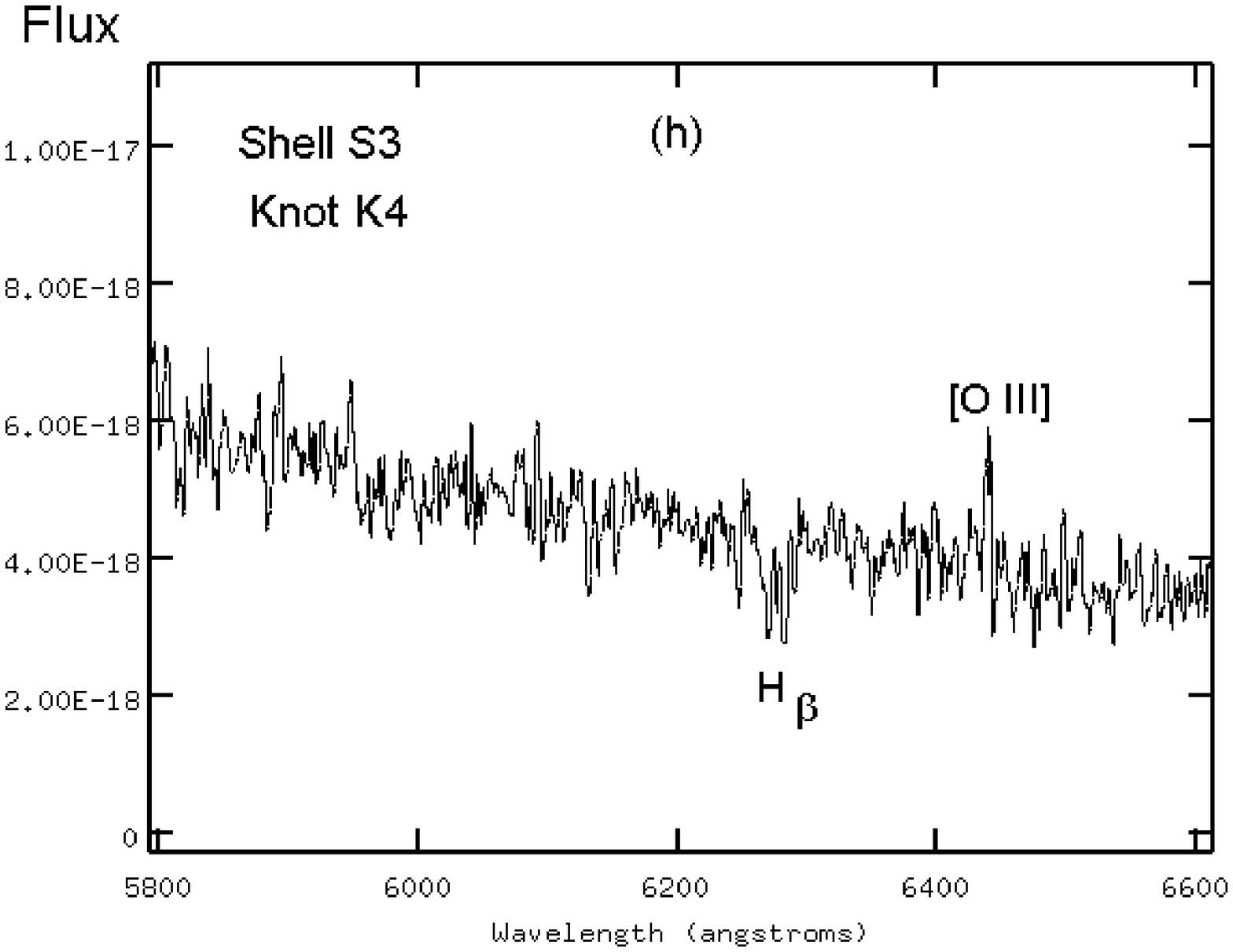} \cr
\includegraphics{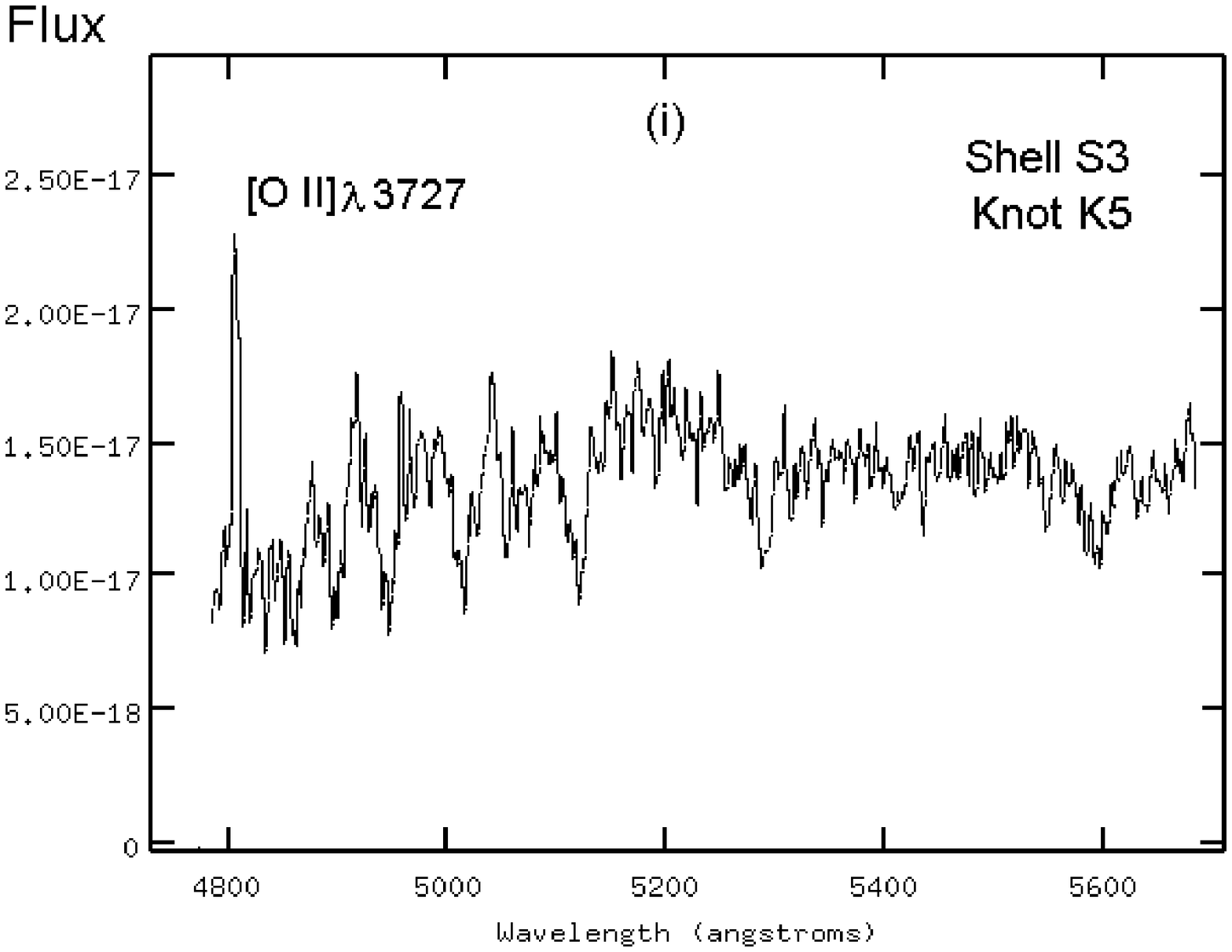}&
\includegraphics{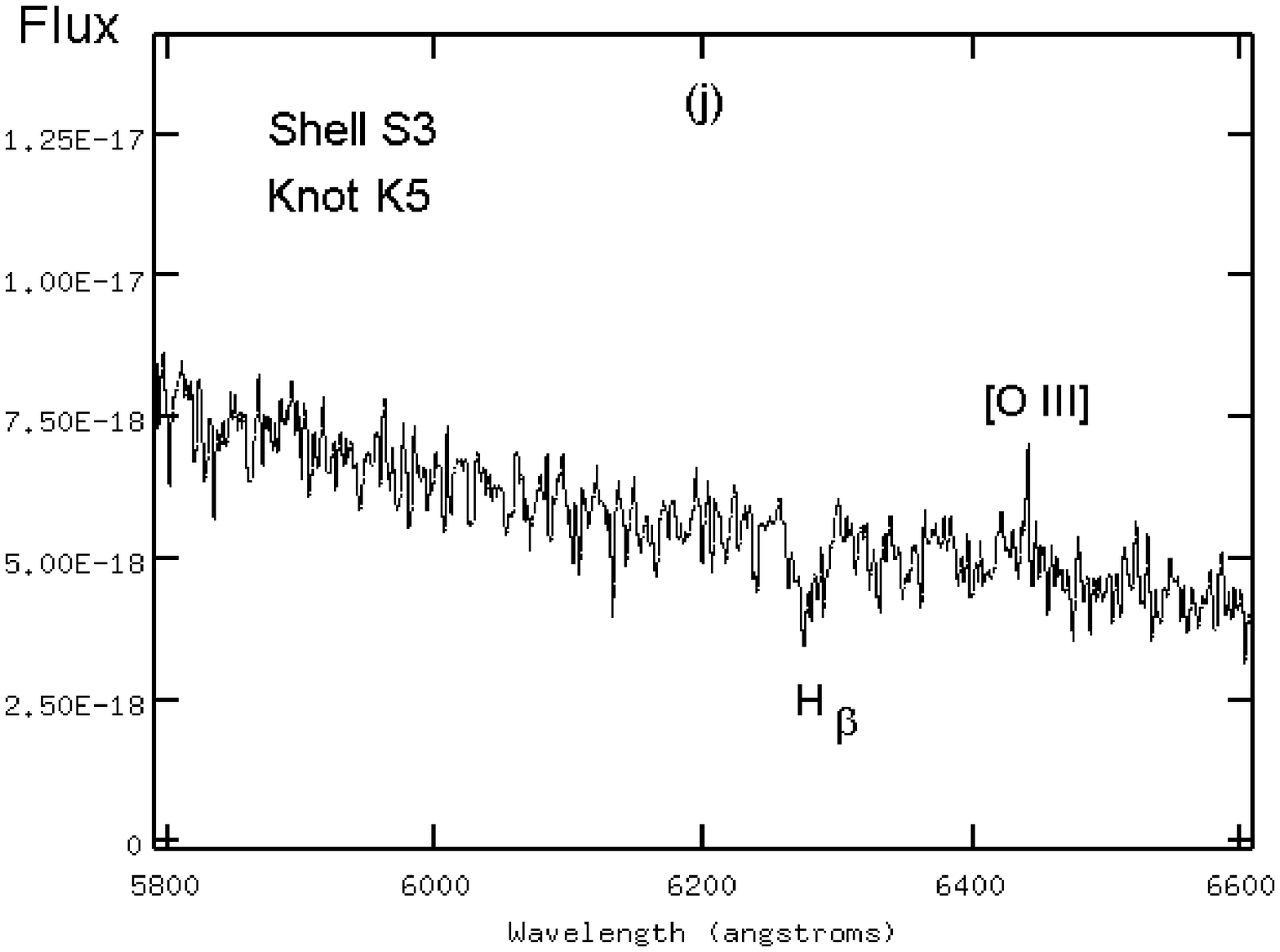} \cr
\end{tabular}
\vspace{8.0 cm}
\addtocounter{figure}{-1}
\caption {Contin.
 }
\label{fig9c}
\end{figure*}


\clearpage

\begin{figure*}
\vspace{12.0 cm}
\begin{tabular}{cc}
\includegraphics{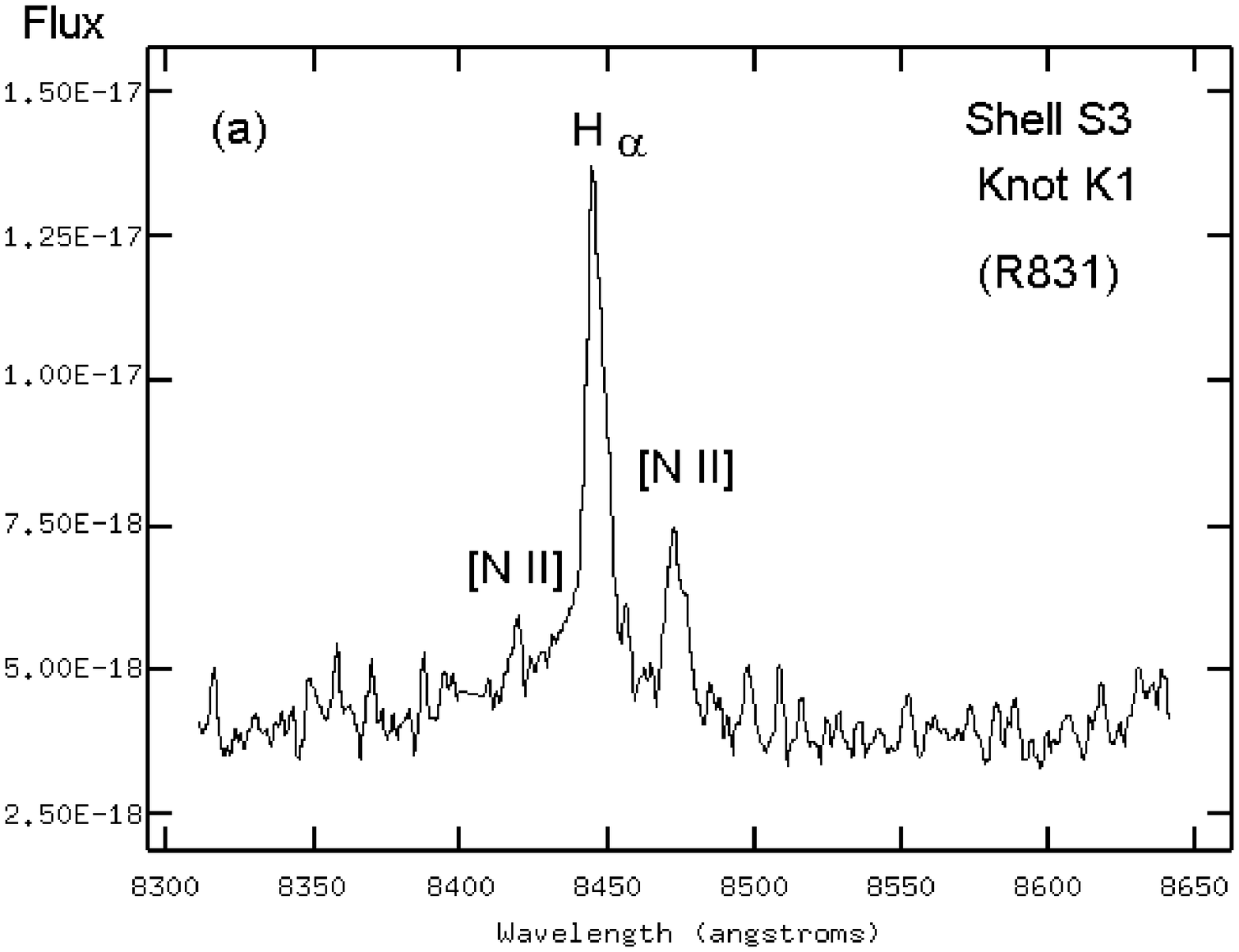}& 
\includegraphics{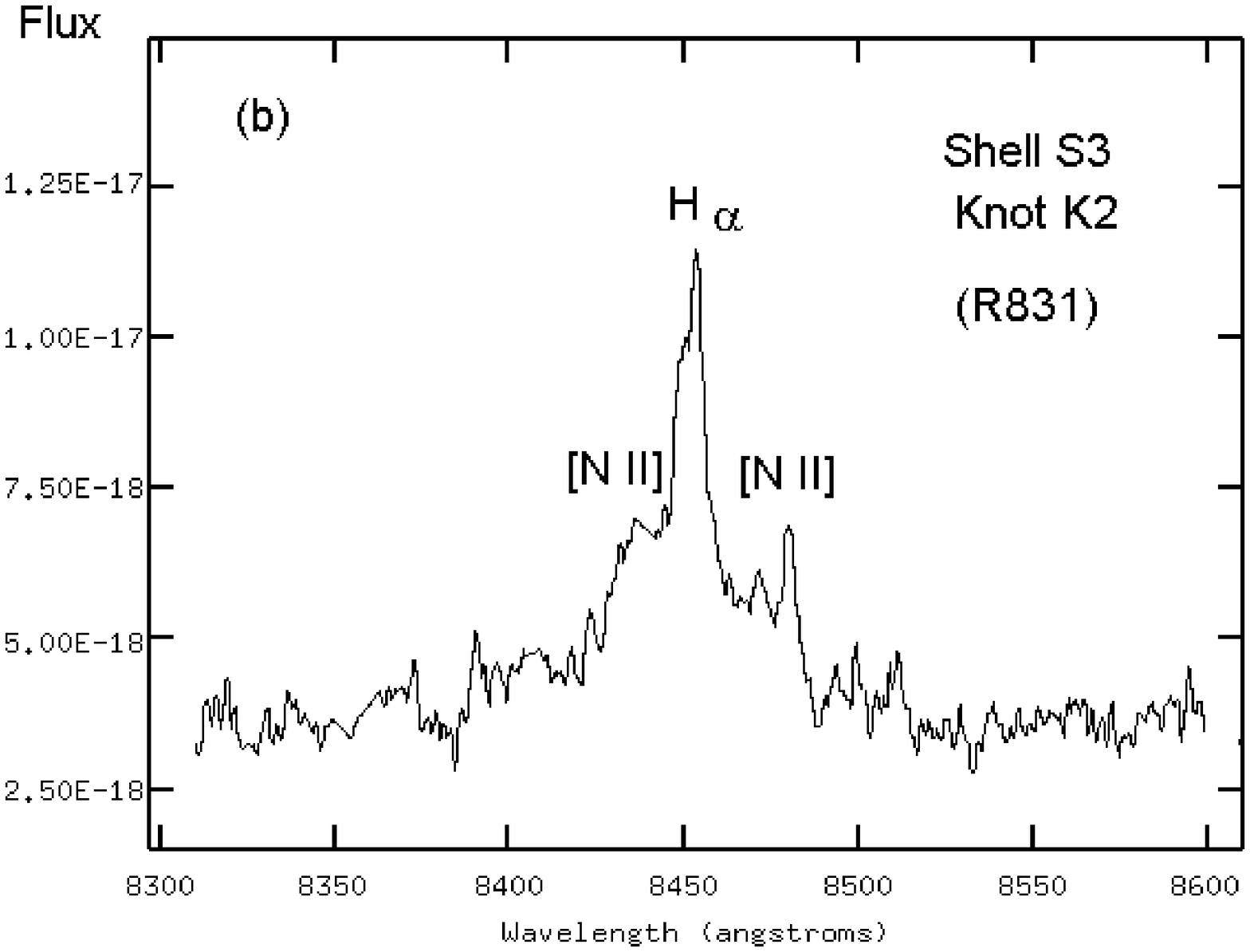} \cr
\includegraphics{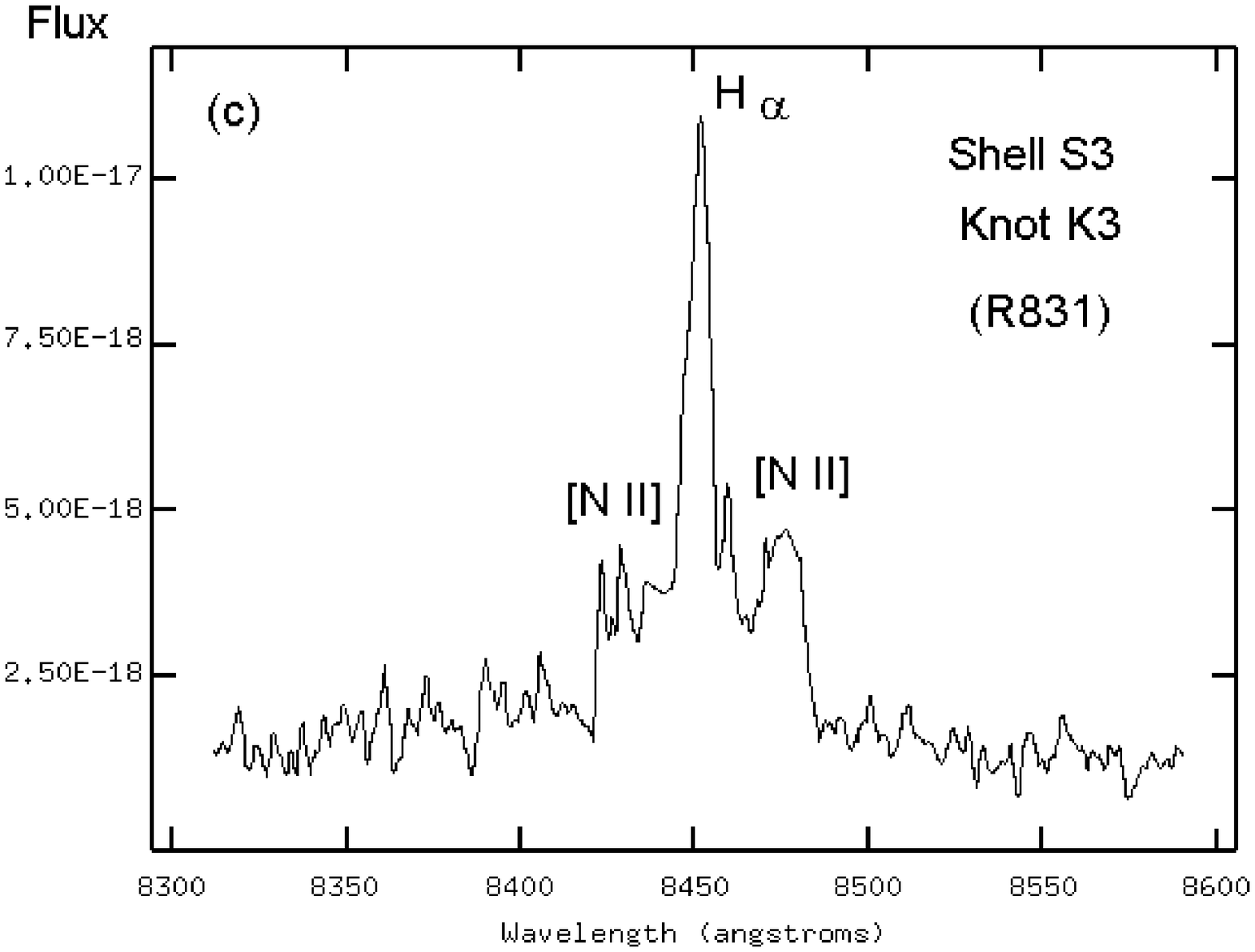}&
\includegraphics{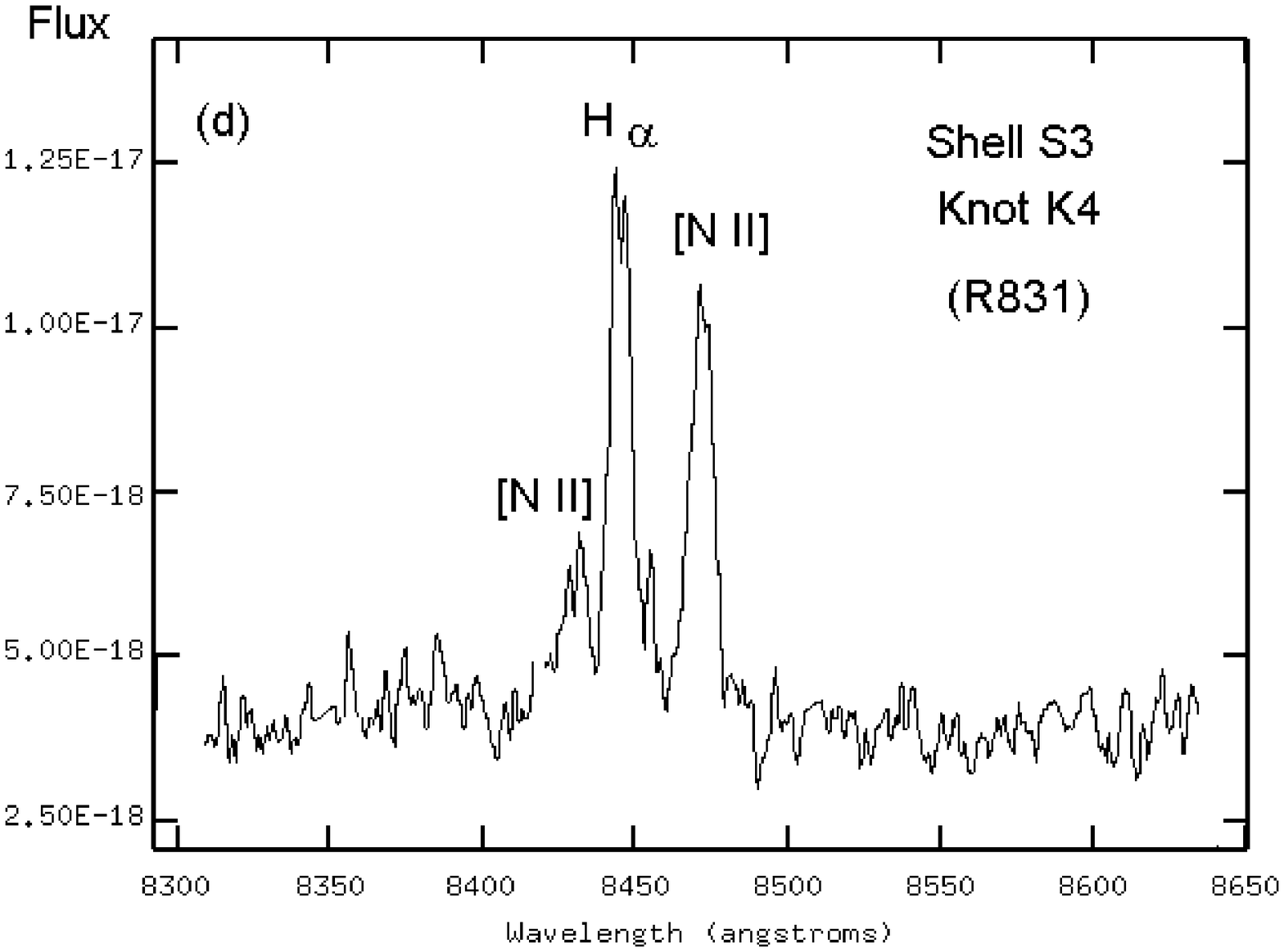} \cr
\includegraphics{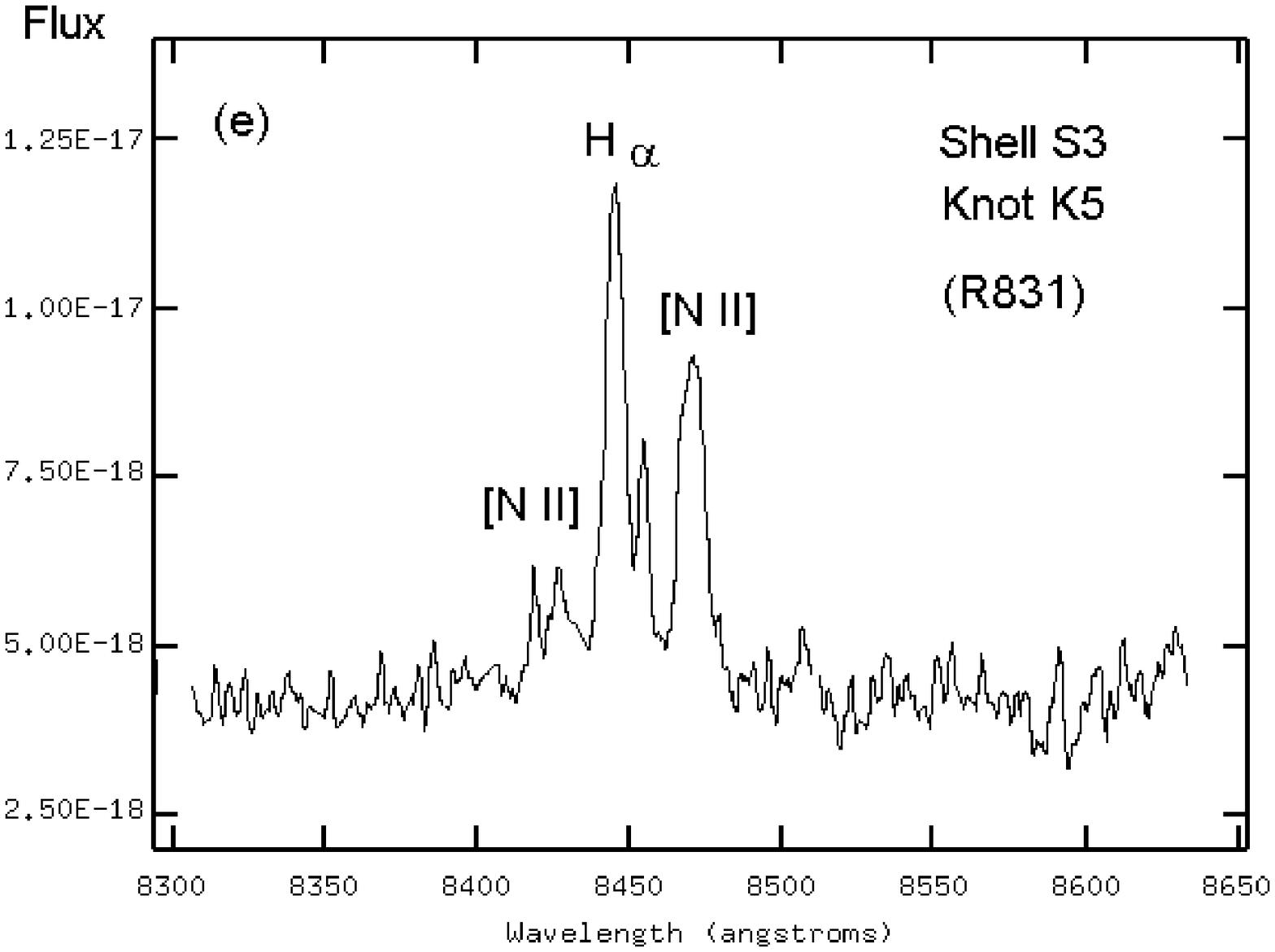}& 
\end{tabular}
\vspace{8.0 cm}
\caption {
GMOS-IFU high resolution spectra (R831) of the main knots of the
shells S3. For the wavelength range of H$\alpha$.
 }
\label{fig10}
\end{figure*}


\clearpage

\begin{figure*}
\vspace{12.0 cm}
\begin{tabular}{cc}
\includegraphics{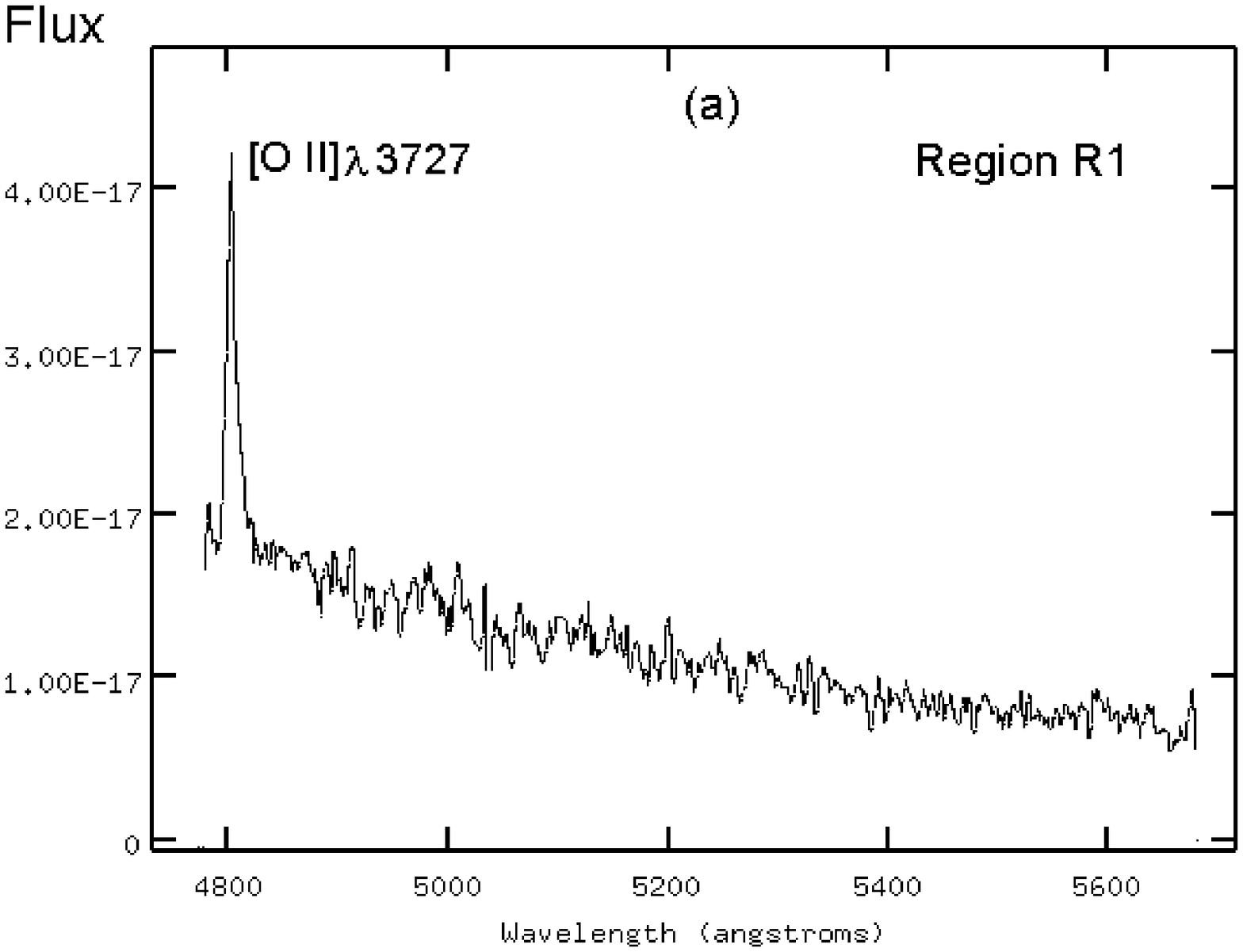}& 
\includegraphics{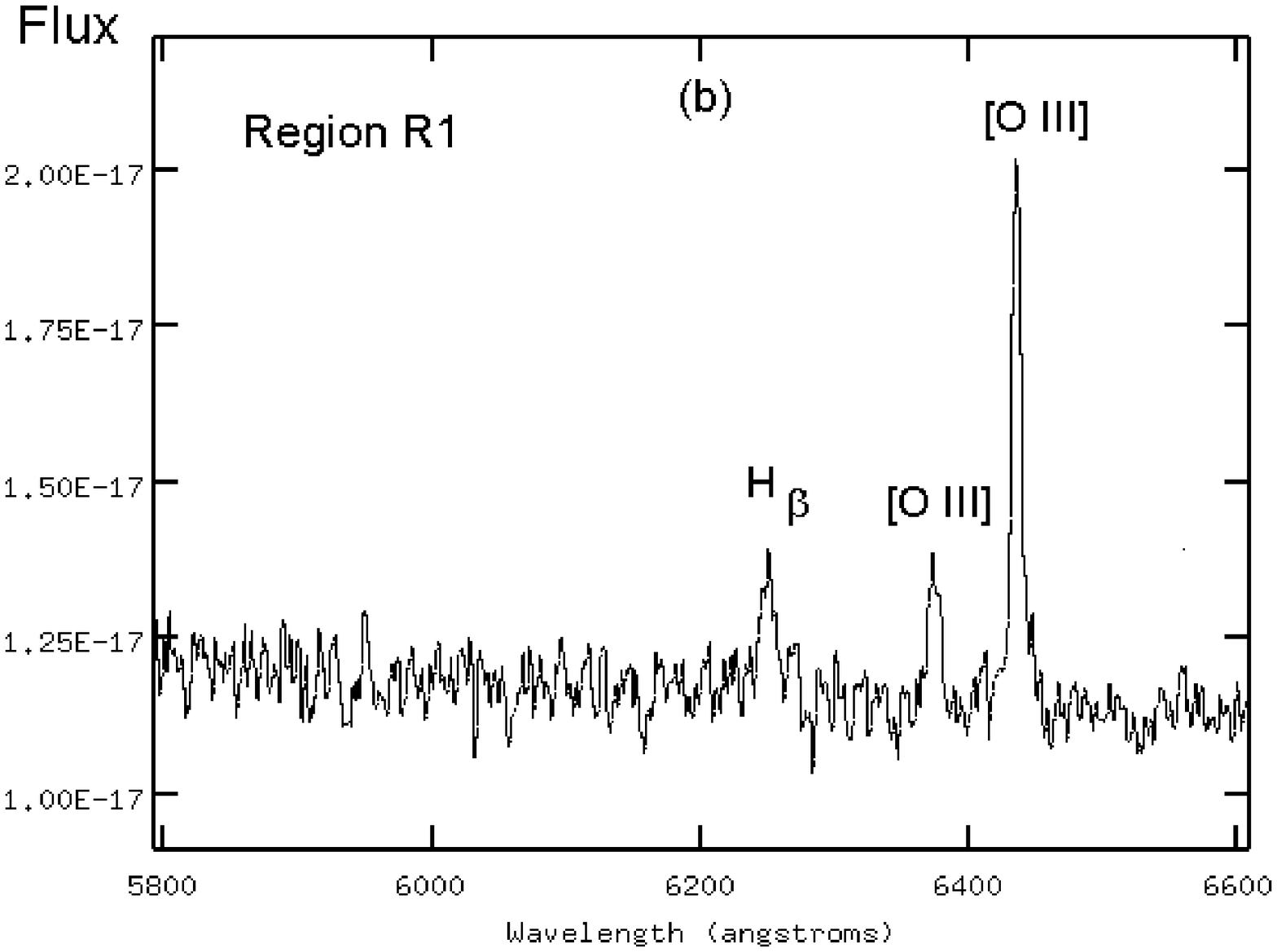} \cr
\includegraphics{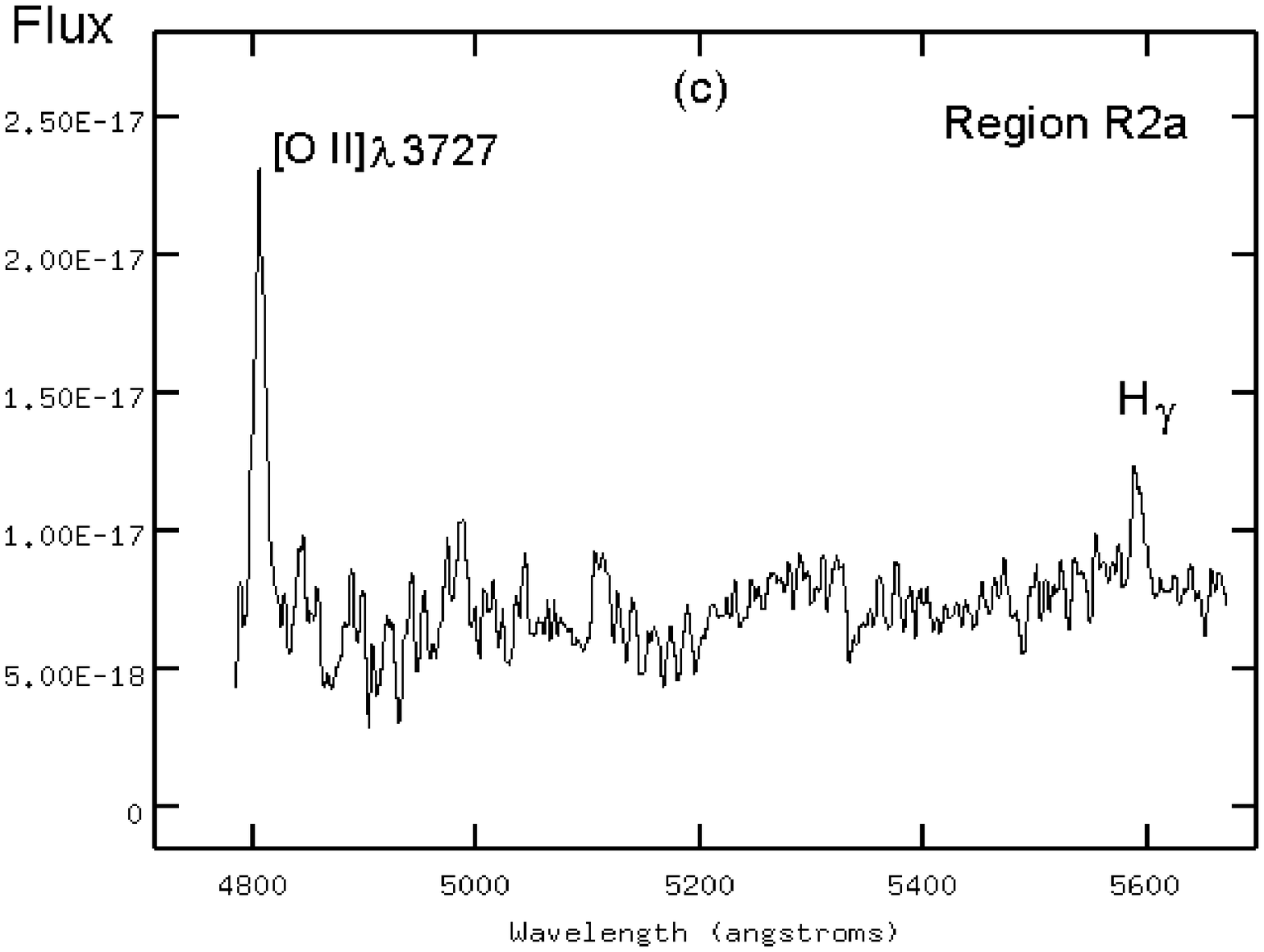}&
\includegraphics{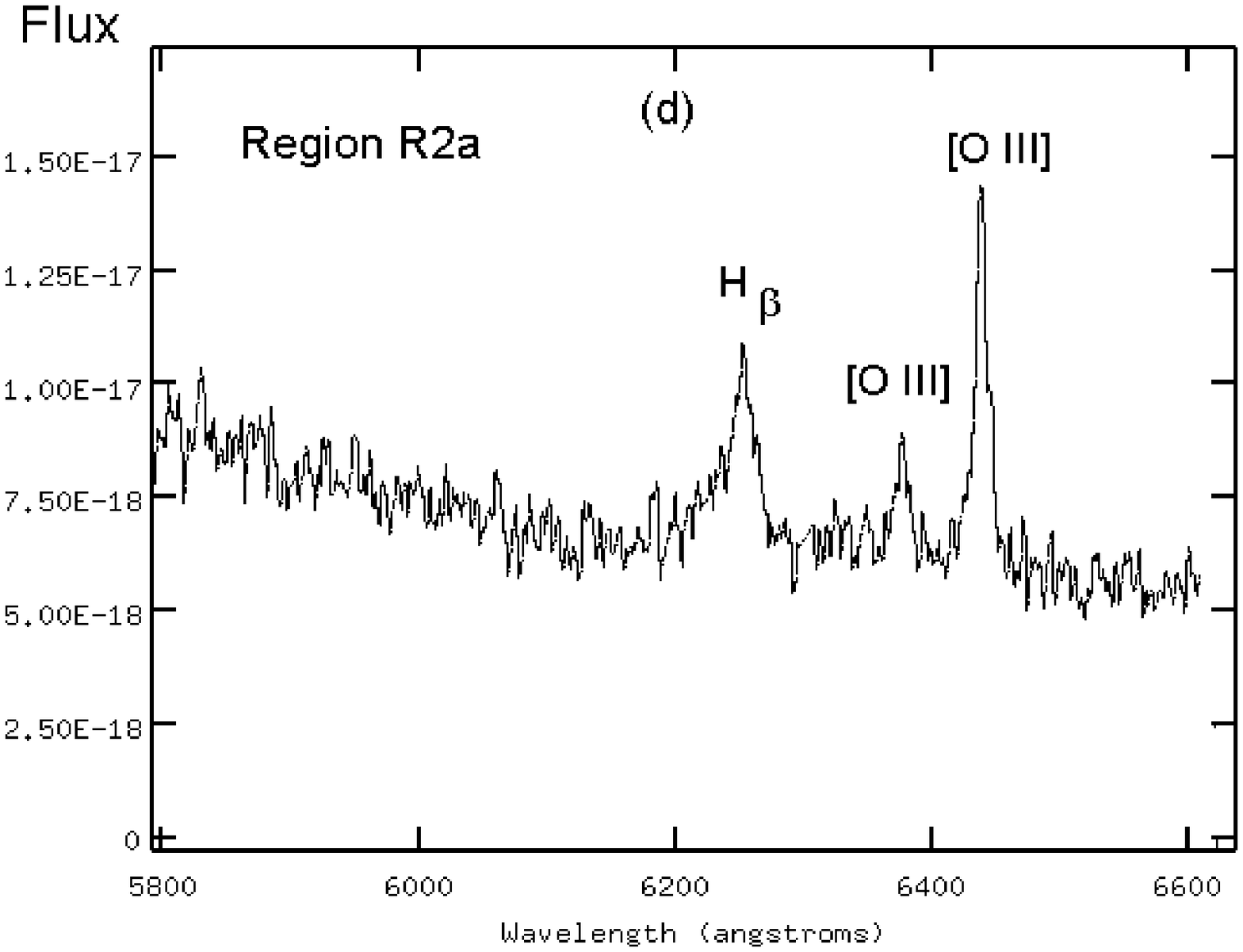} \cr
\includegraphics{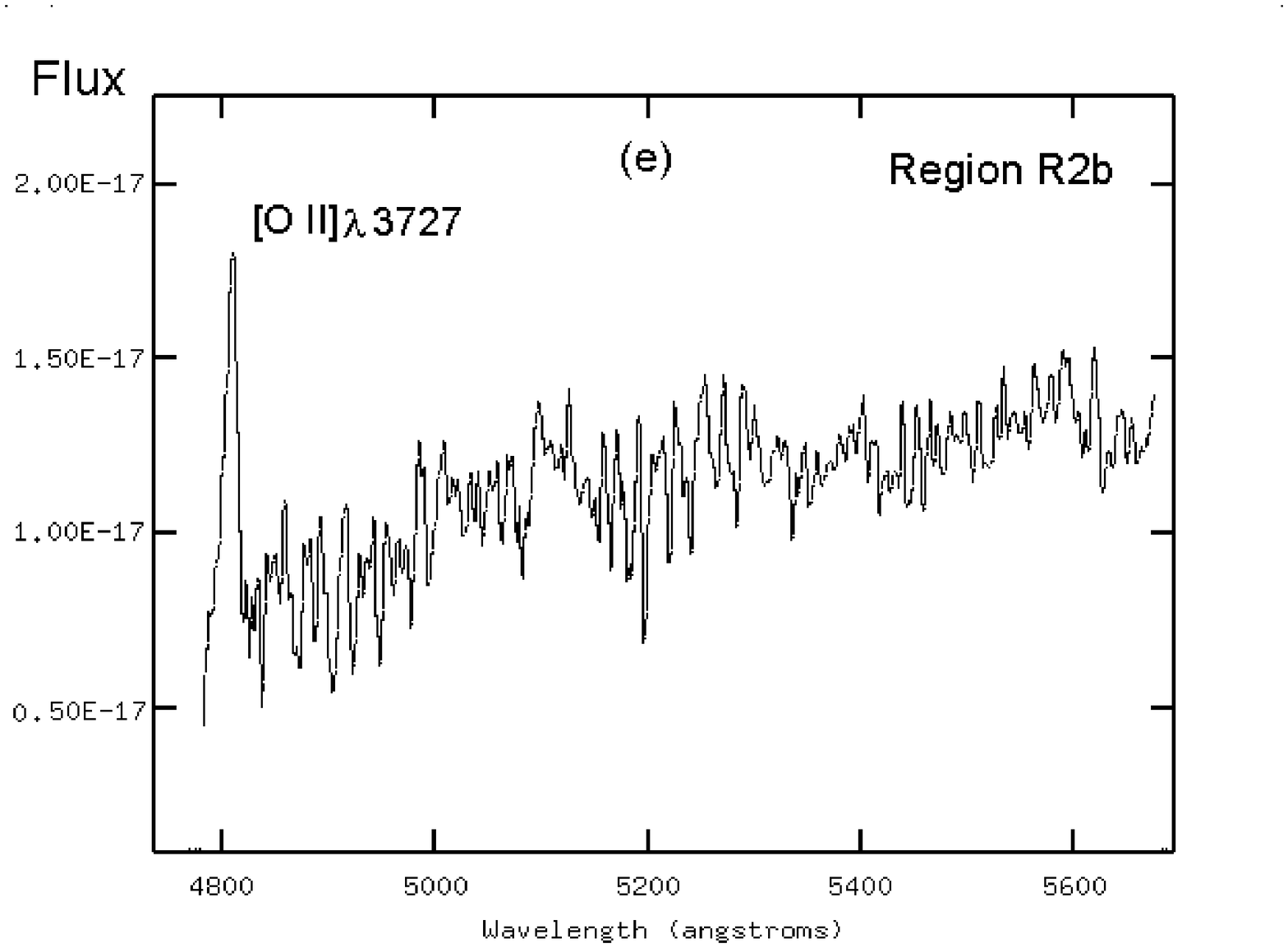}& 
\includegraphics{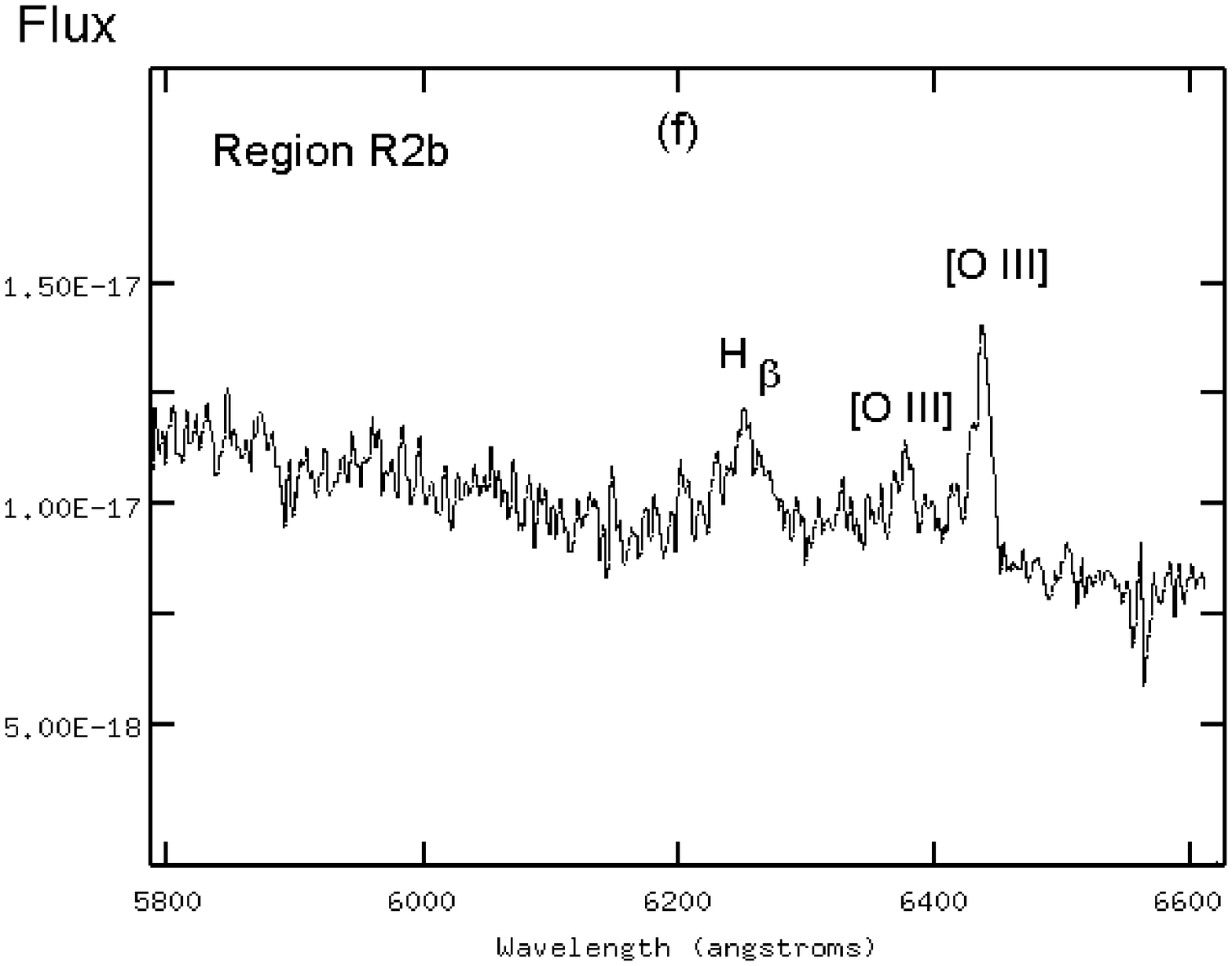} \cr
\end{tabular}
\vspace{8.0 cm}
\caption {
GMOS-IFU spectra of selected external regions R1, R2a, R2b, R3, and R4
(see the text) in the field of the QSO IRAS 04505-2958. For the
wavelength ranges:
[O {\sc ii}]$\lambda$3727--H$\gamma$ and
H$\beta$ + [O {\sc iii}]$\lambda$5007 + Fe {\sc ii}.
}
\label{fig11}
\end{figure*}

\clearpage

\begin{figure*}
\vspace{12.0 cm}
\begin{tabular}{cc}
\includegraphics{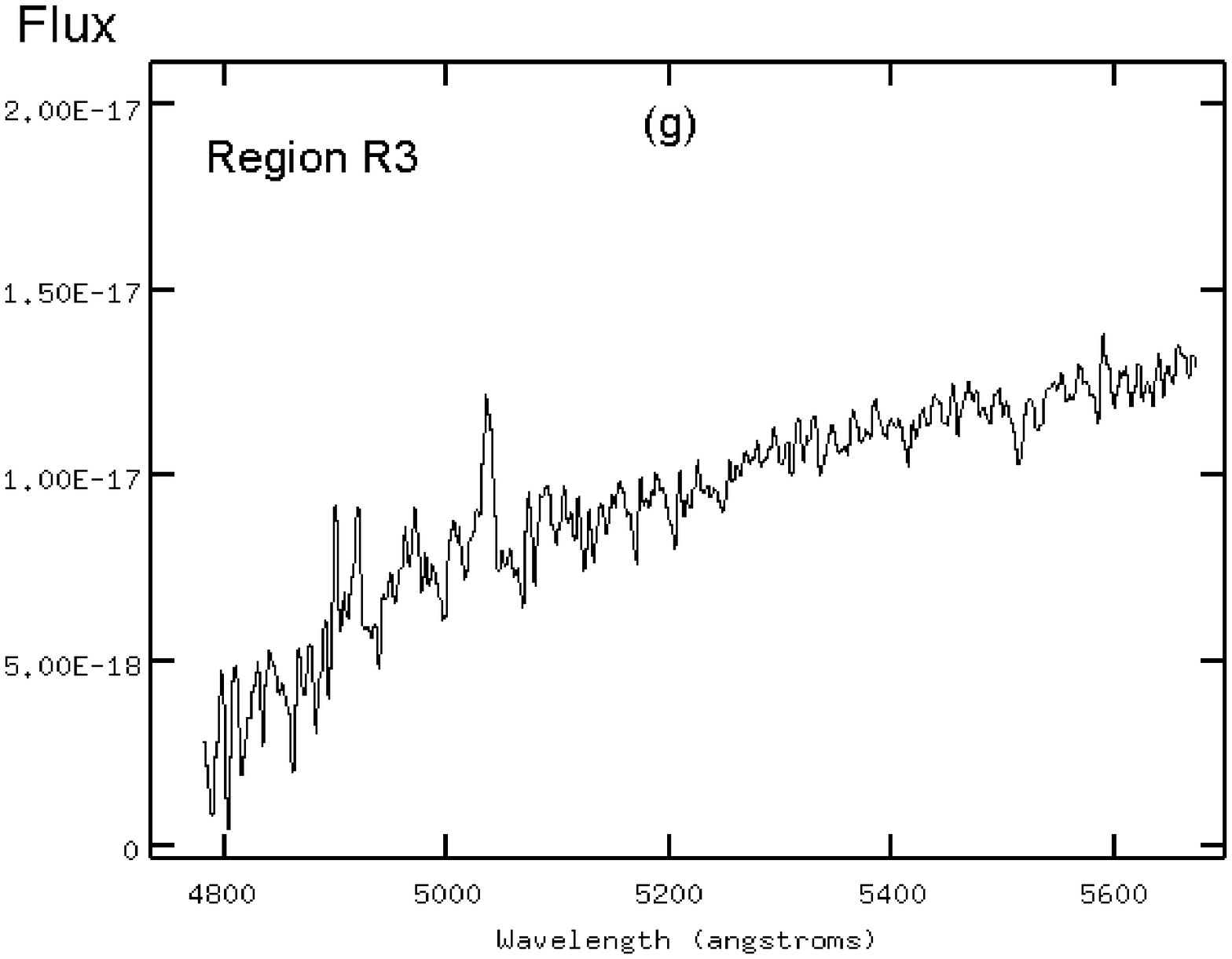}& 
\includegraphics{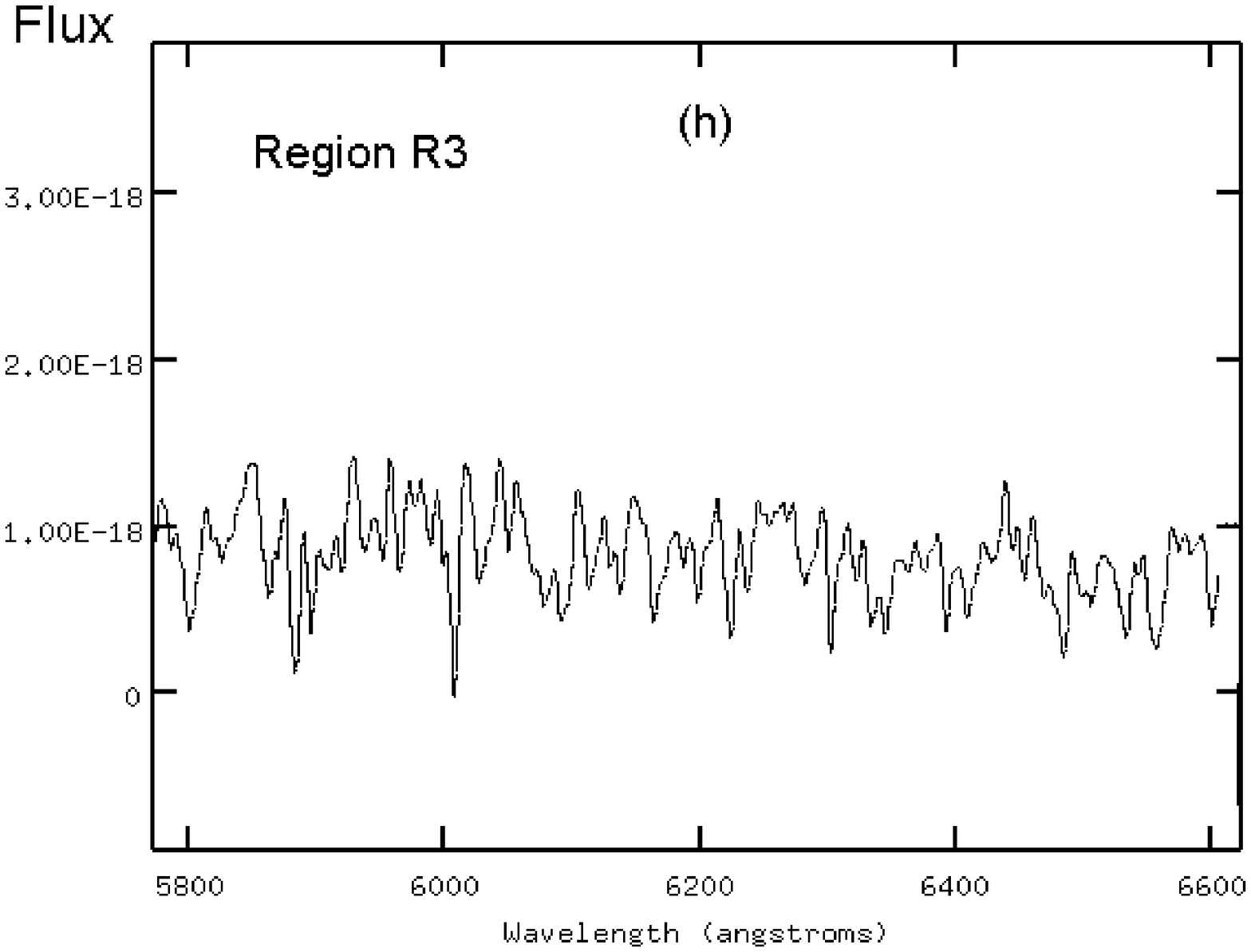} \cr
\includegraphics{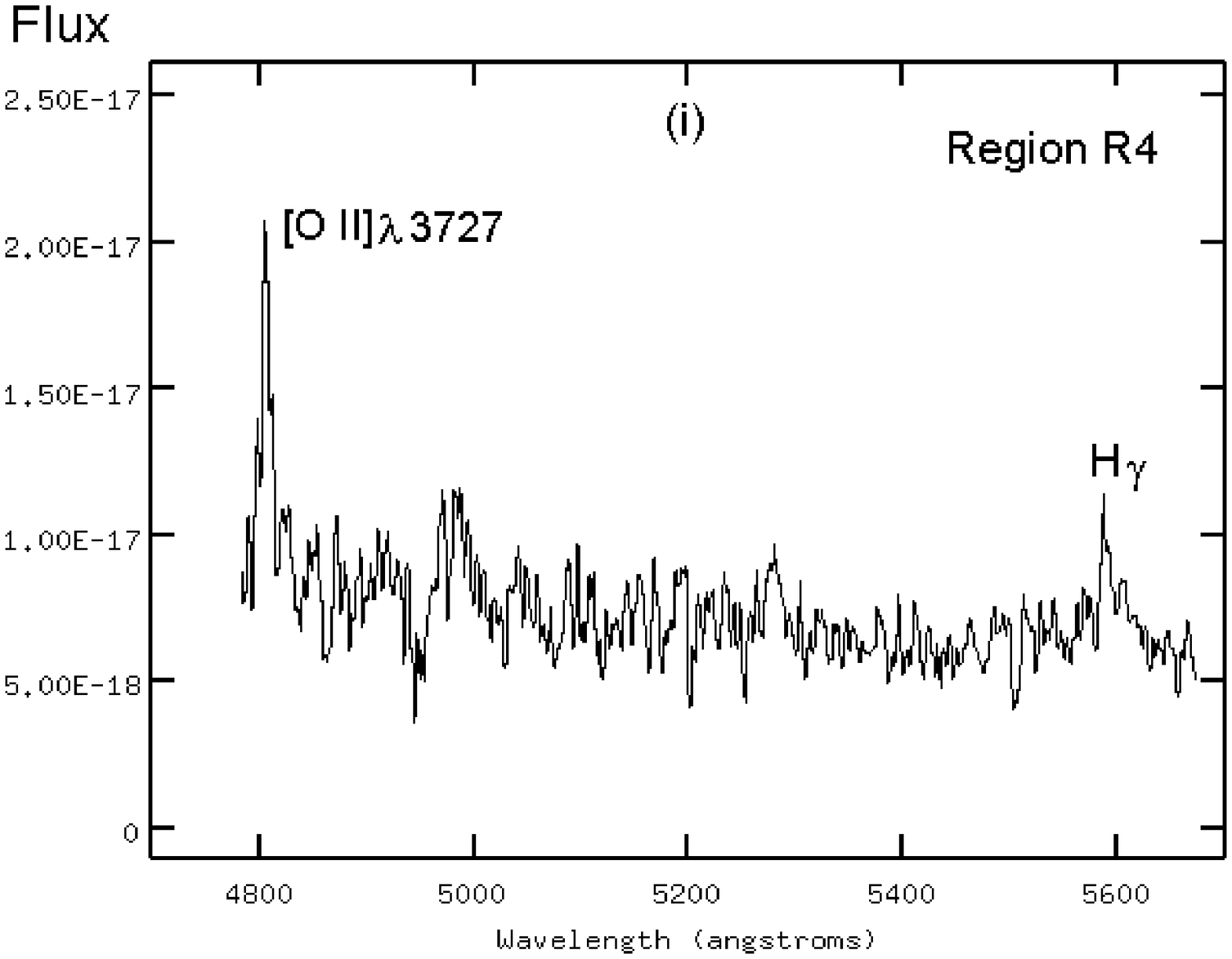}&
\includegraphics{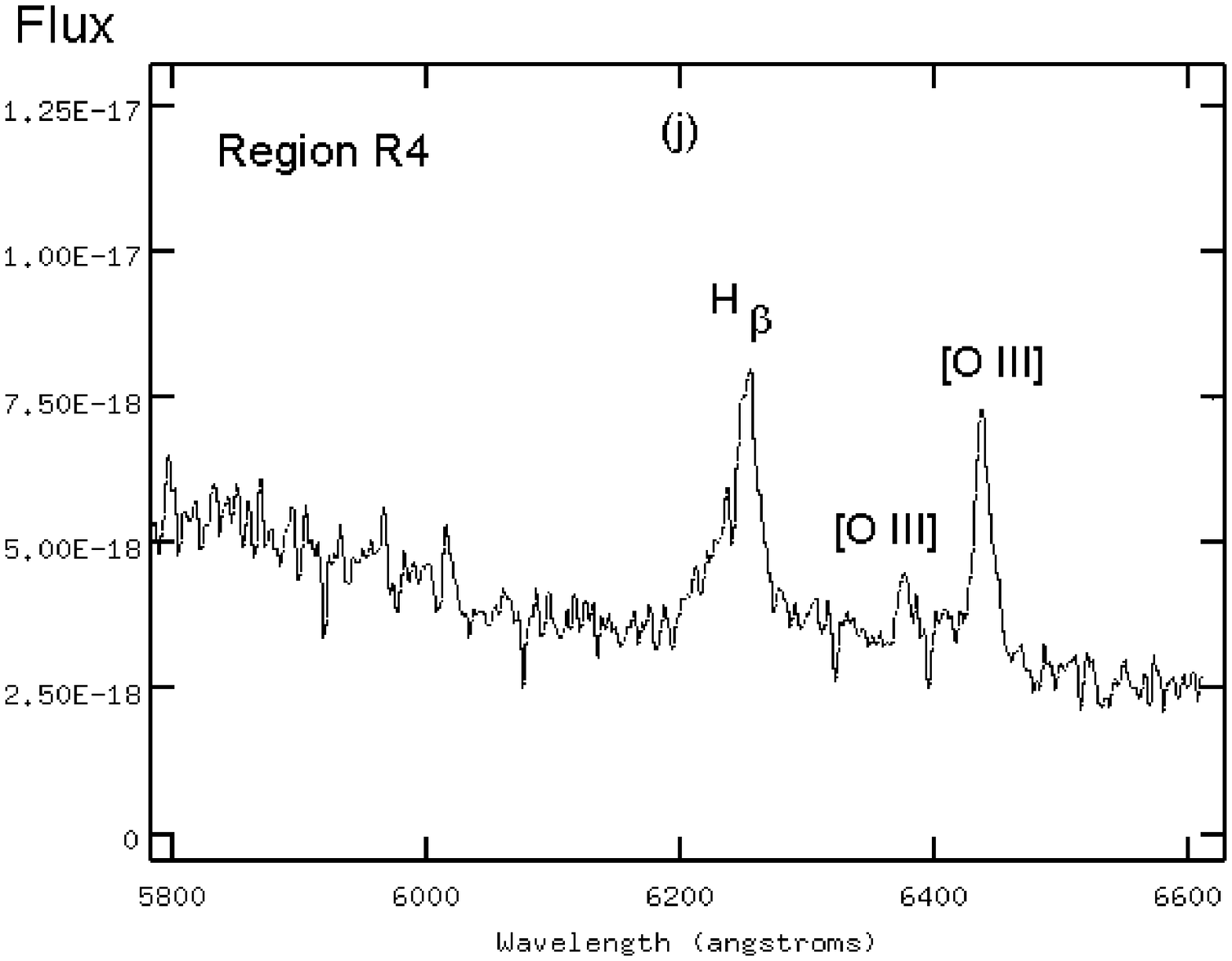} \cr
\end{tabular}
\vspace{8.0 cm}
\addtocounter{figure}{-1}
\caption {
Contin.
}
\label{fig11c}
\end{figure*}


\clearpage

\begin{figure*}
\vspace{12.0 cm}
\begin{tabular}{cc}
\includegraphics{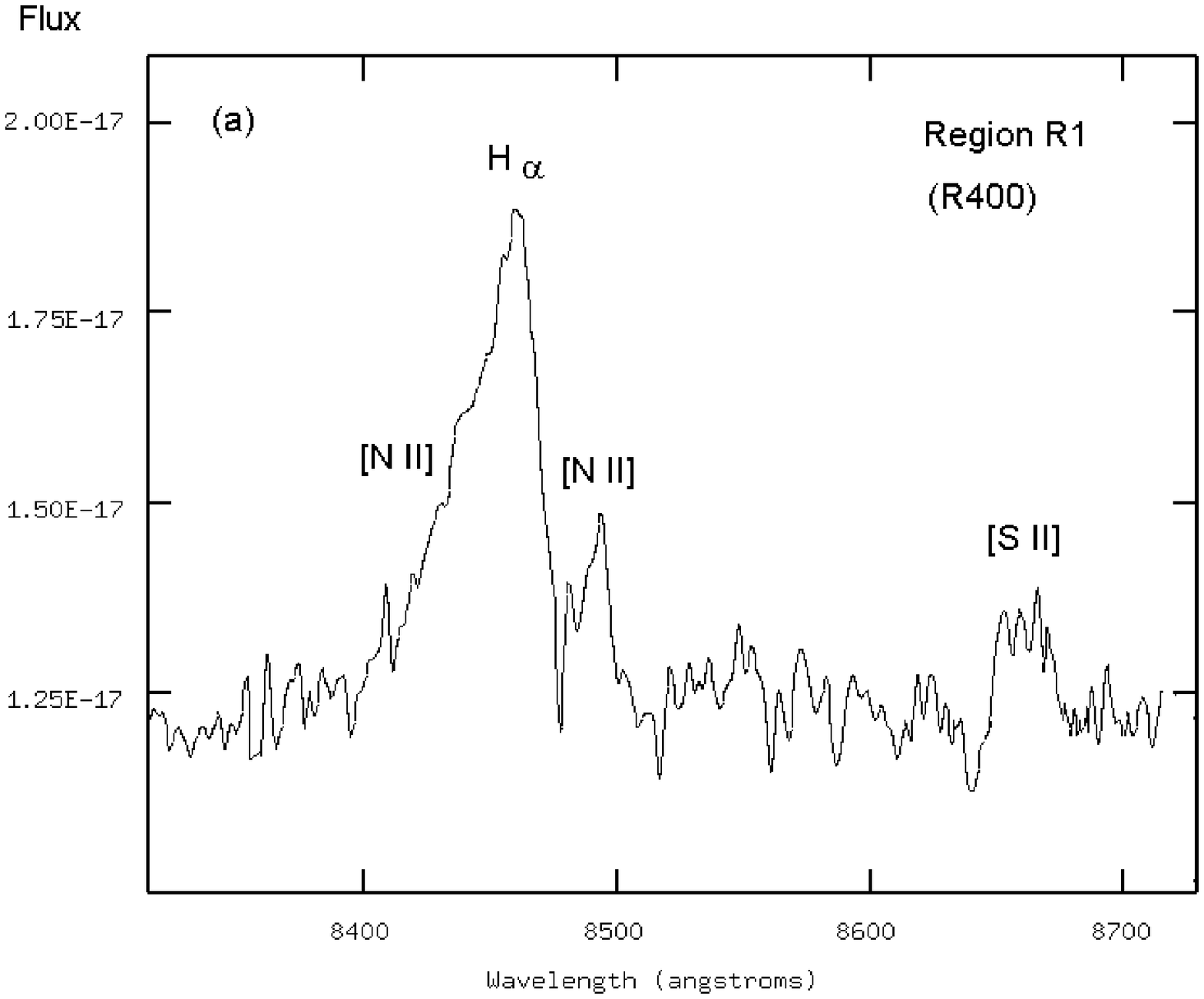}& 
\includegraphics{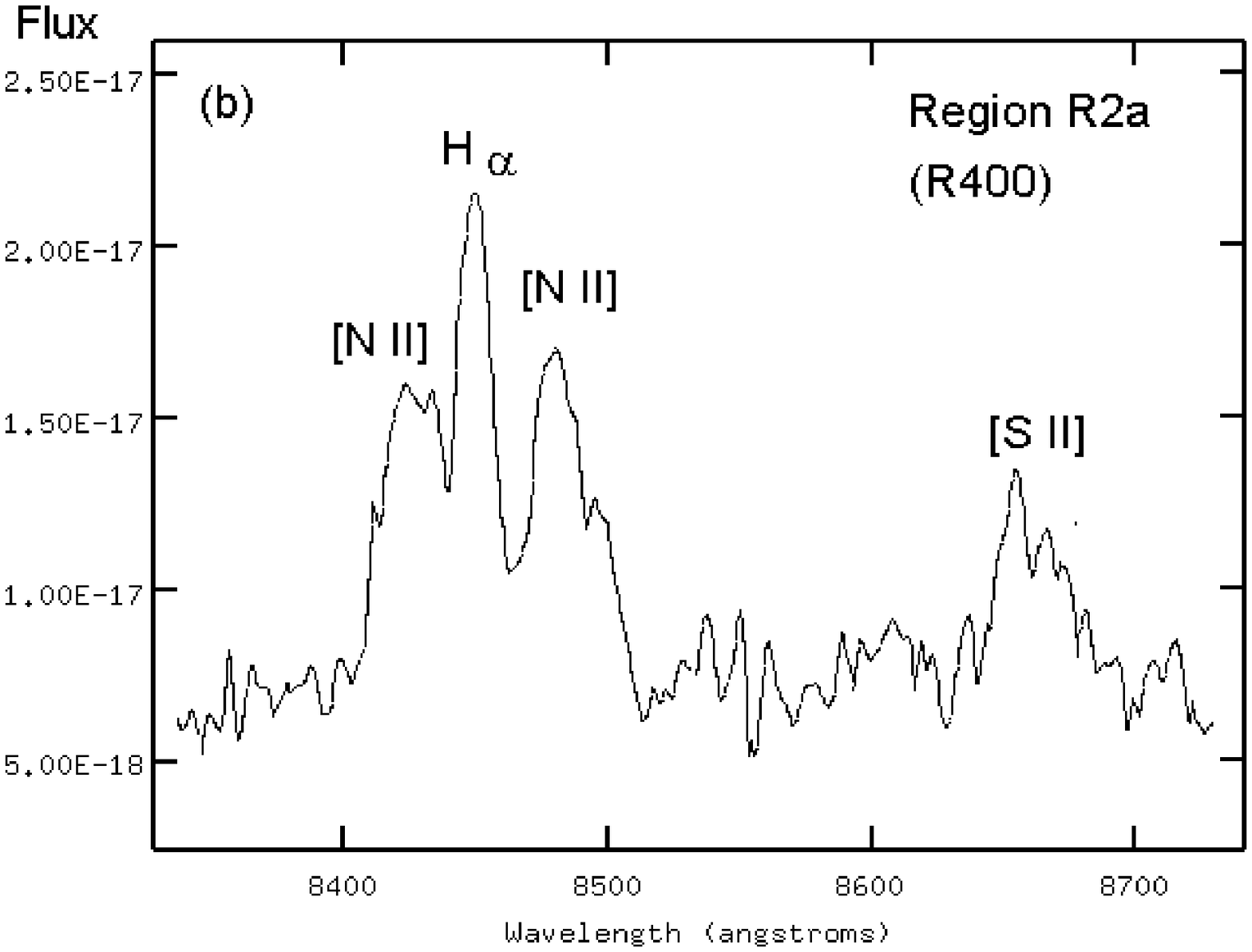} \cr
\includegraphics{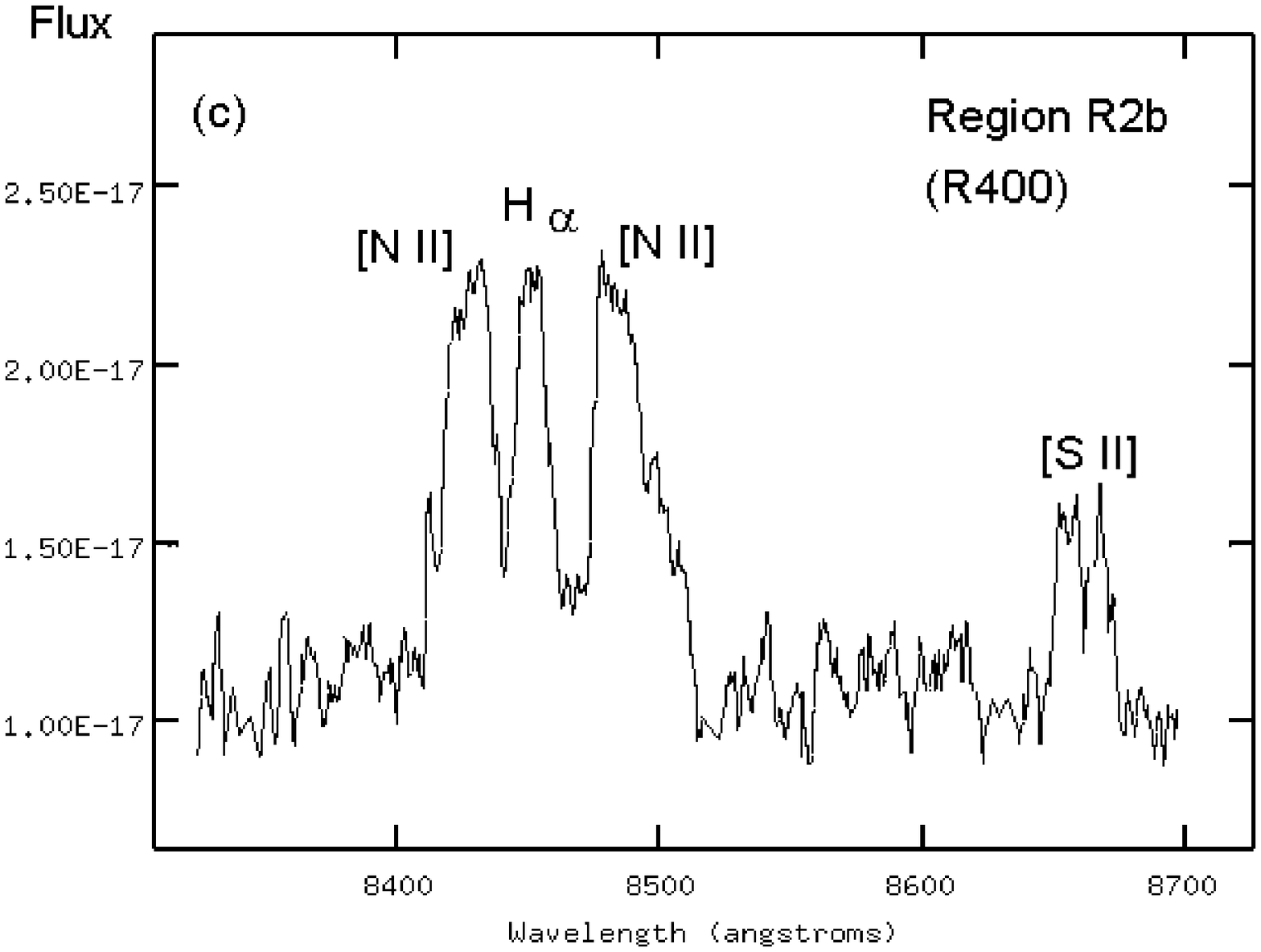}&
\includegraphics{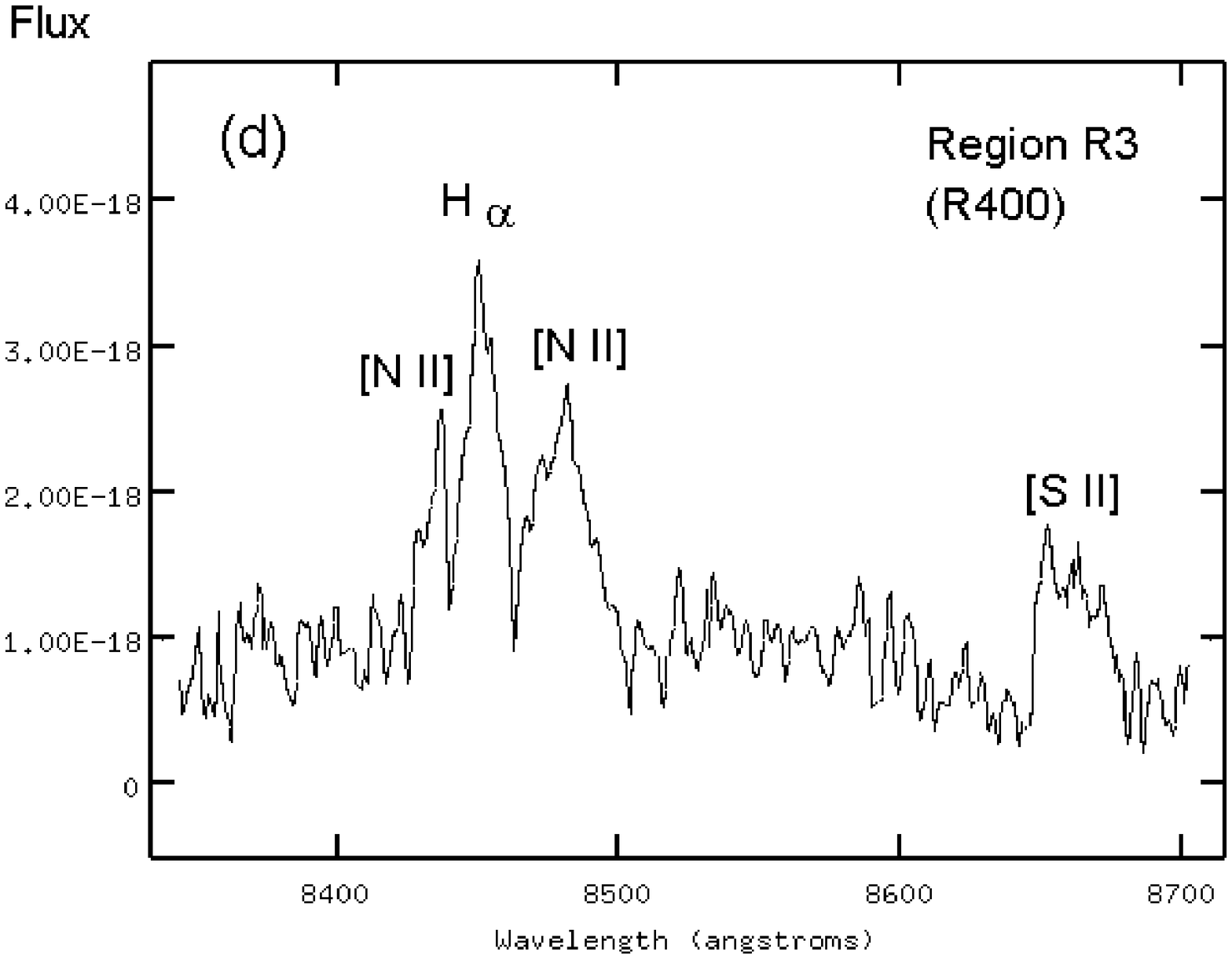} \cr
\includegraphics{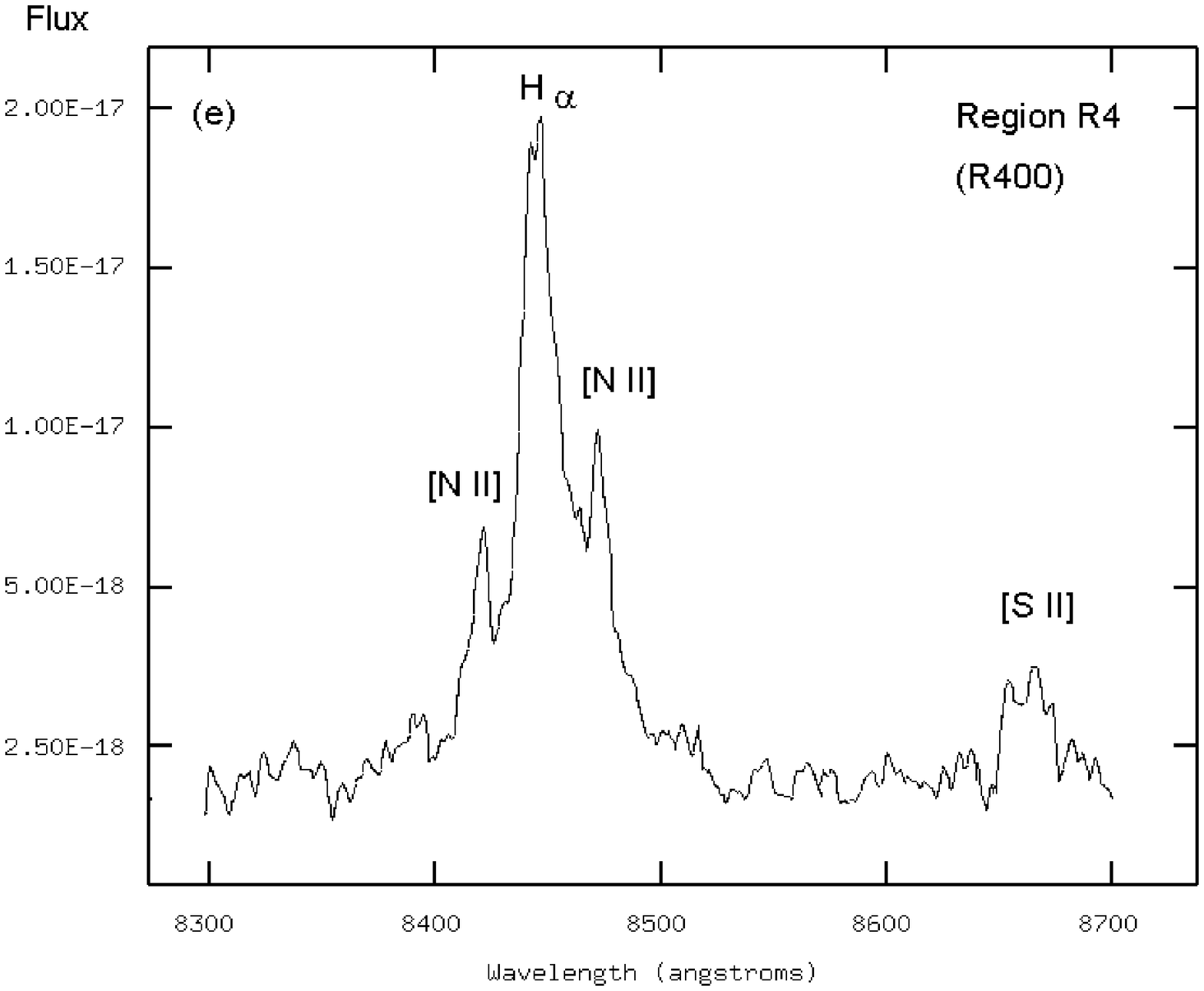}& 
\end{tabular}
\vspace{8.0 cm}
\caption {
GMOS-IFU spectra of the selected external regions R1, R2a, R2b, R3, and R4.
For the  wavelength range of H$\alpha$.
}
\label{fig12}
\end{figure*}


\clearpage

\begin{figure*}
\vspace{12.0 cm}
\begin{tabular}{c}
\includegraphics{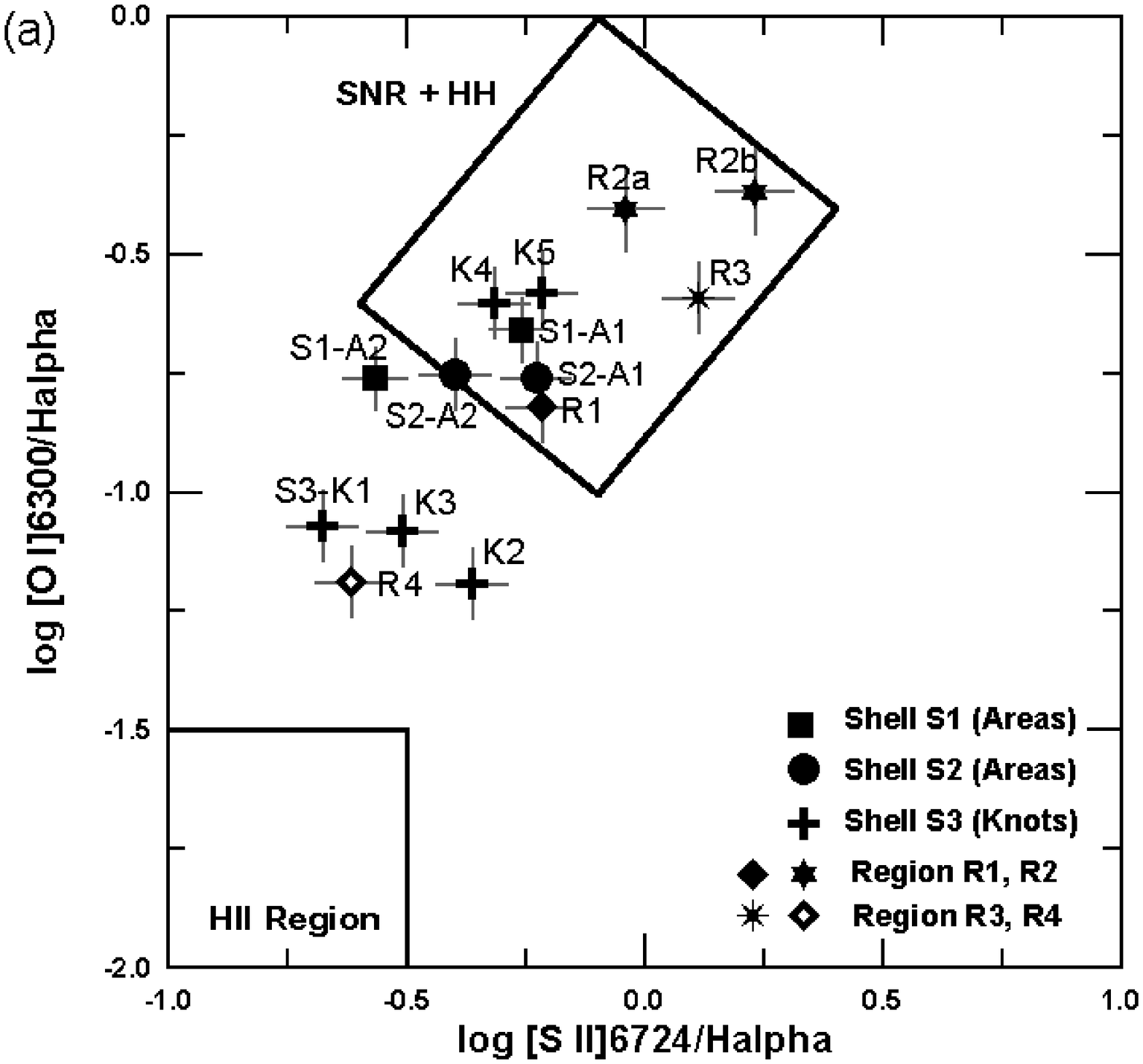}\cr
\end{tabular}
\vspace{8.0 cm}
\caption {
Emission line ratios diagrams:
(a) [S {\sc ii}]/H$\alpha$ vs, [O {\sc i}]/H$\alpha$,
(b) [S {\sc ii}]/H$\alpha$ vs, [O {\sc iii}]$\lambda$5007/H$\beta$
(for the shells S1, S2, S3 and several external regions).
}
\label{fig13}
\end{figure*}

\clearpage

\begin{figure*}
\vspace{12.0 cm}
\begin{tabular}{c}
\includegraphics{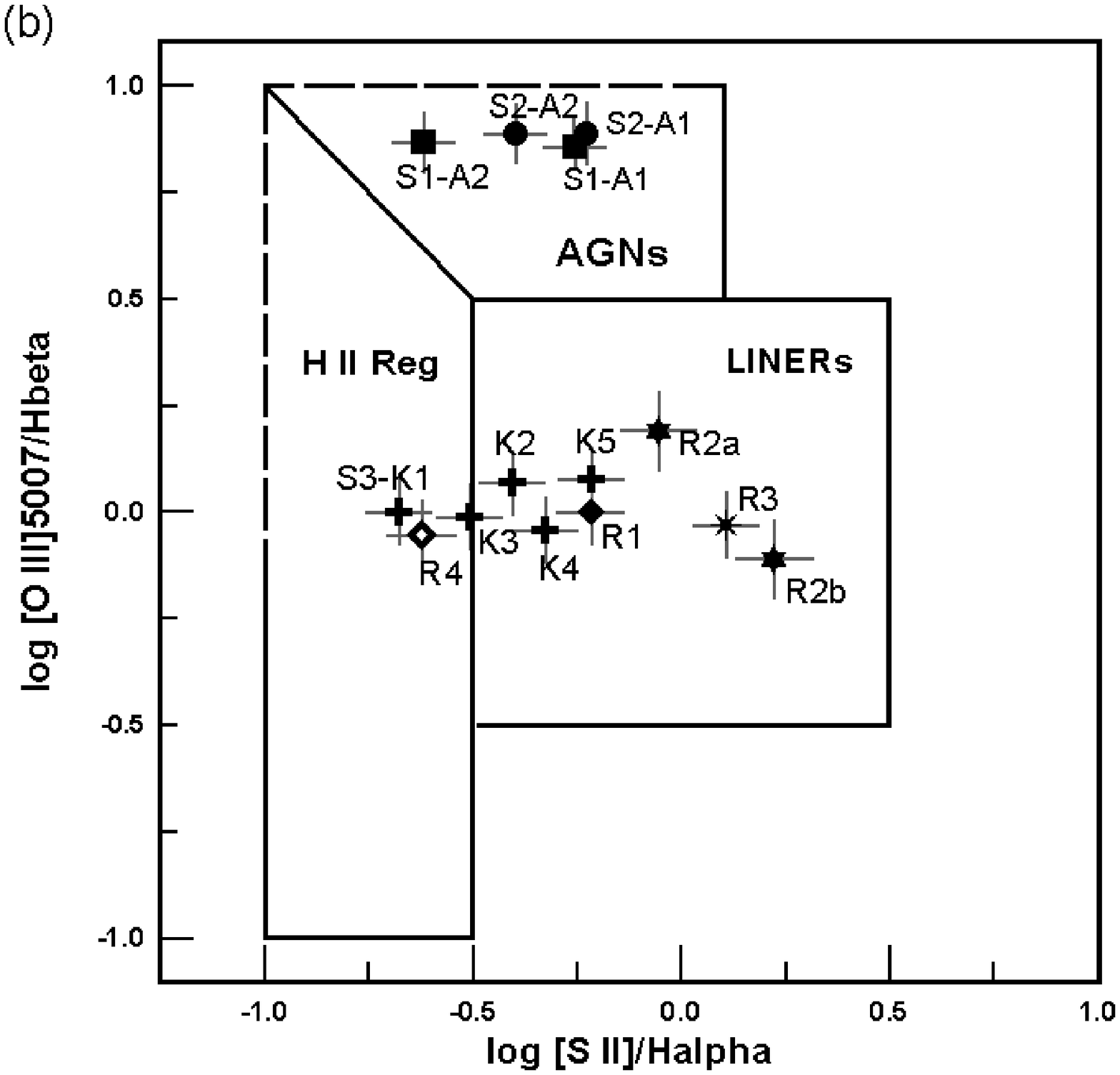}\cr
\end{tabular}
\vspace{8.0 cm}
\addtocounter{figure}{-1}
\caption {Contin.
}
\label{fig13c}
\end{figure*}


\clearpage

\begin{figure*}
\vspace{12.0 cm}
\begin{tabular}{cc}
\includegraphics{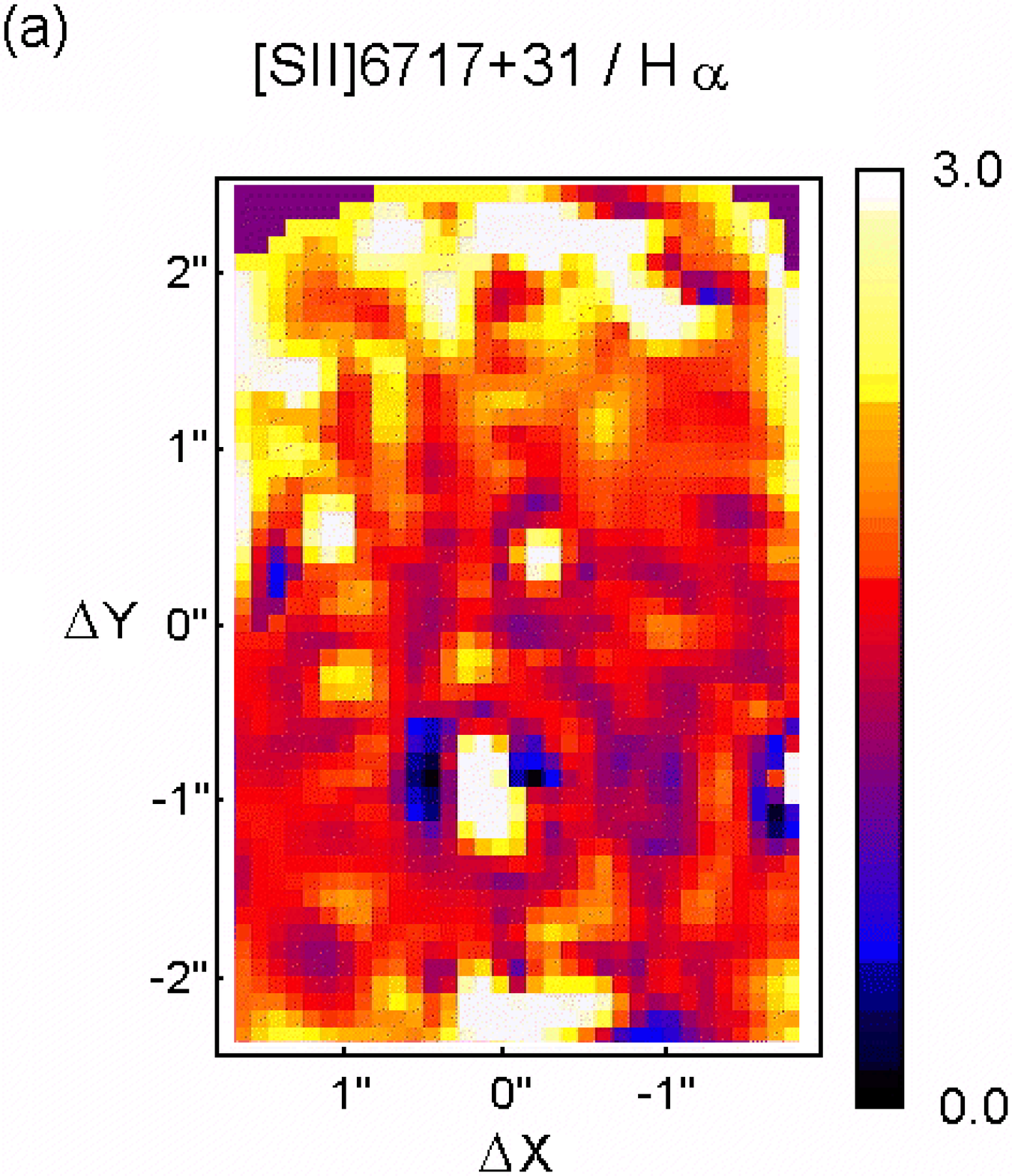}&
\includegraphics{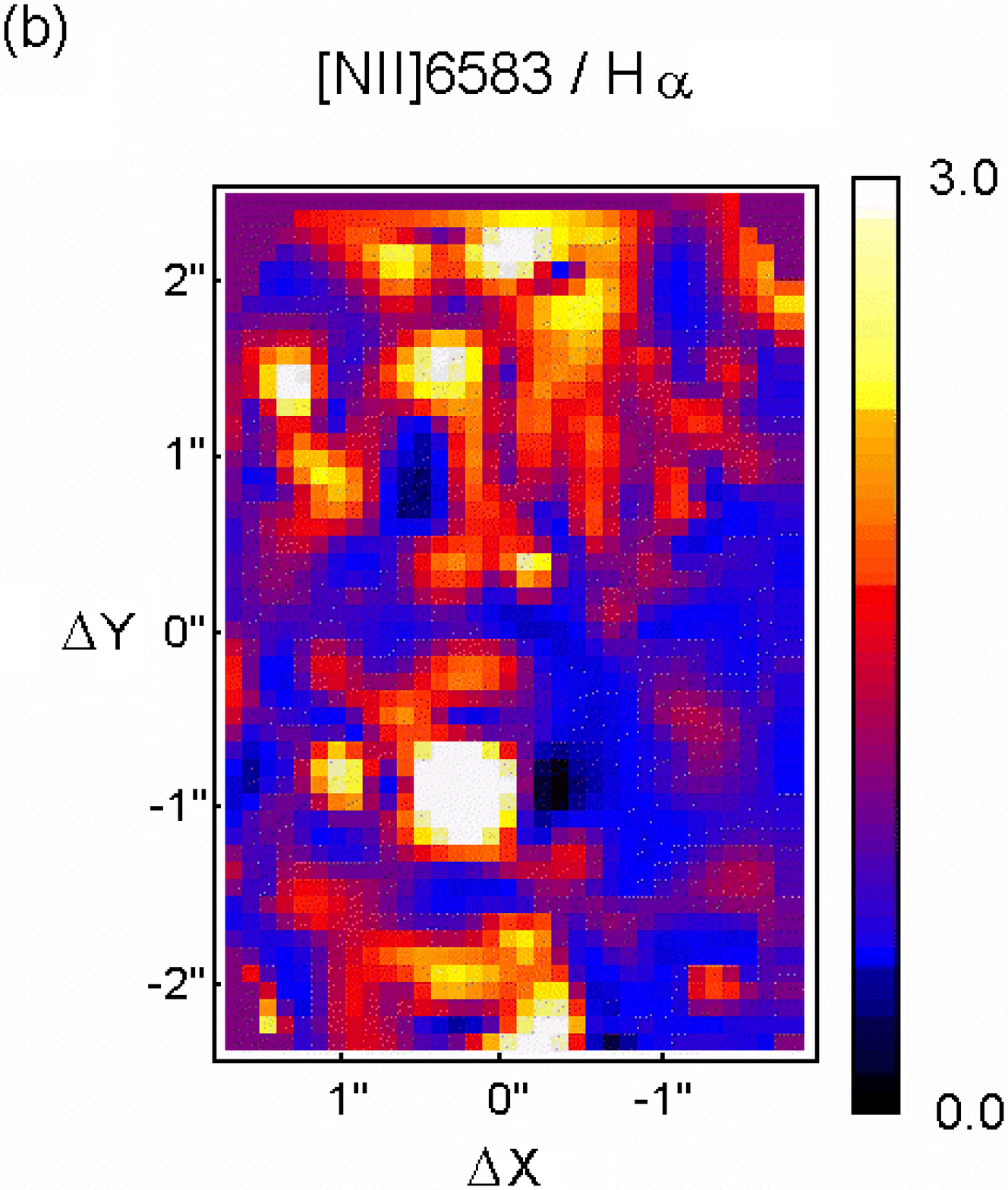} \cr
\includegraphics{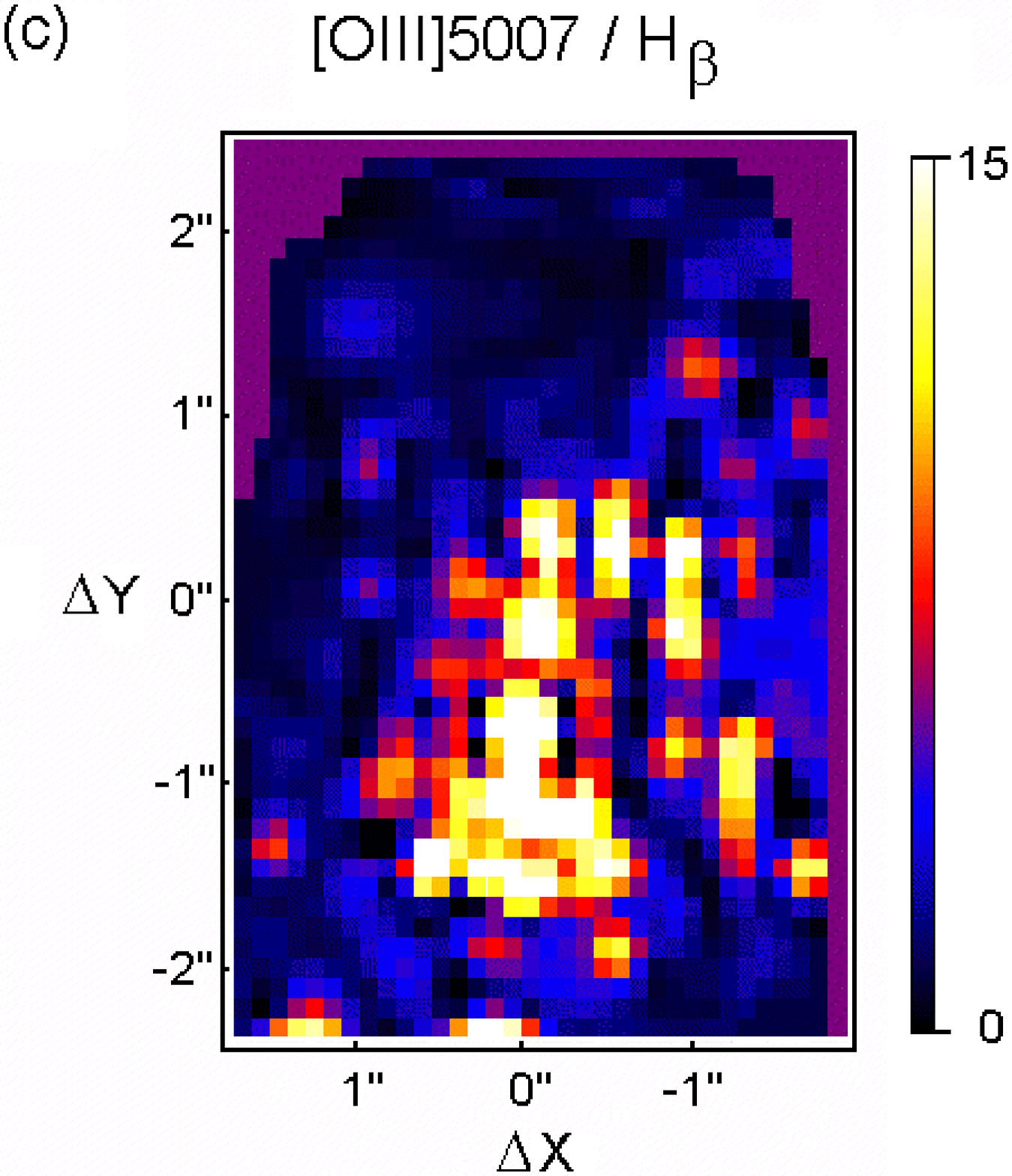}&
\includegraphics{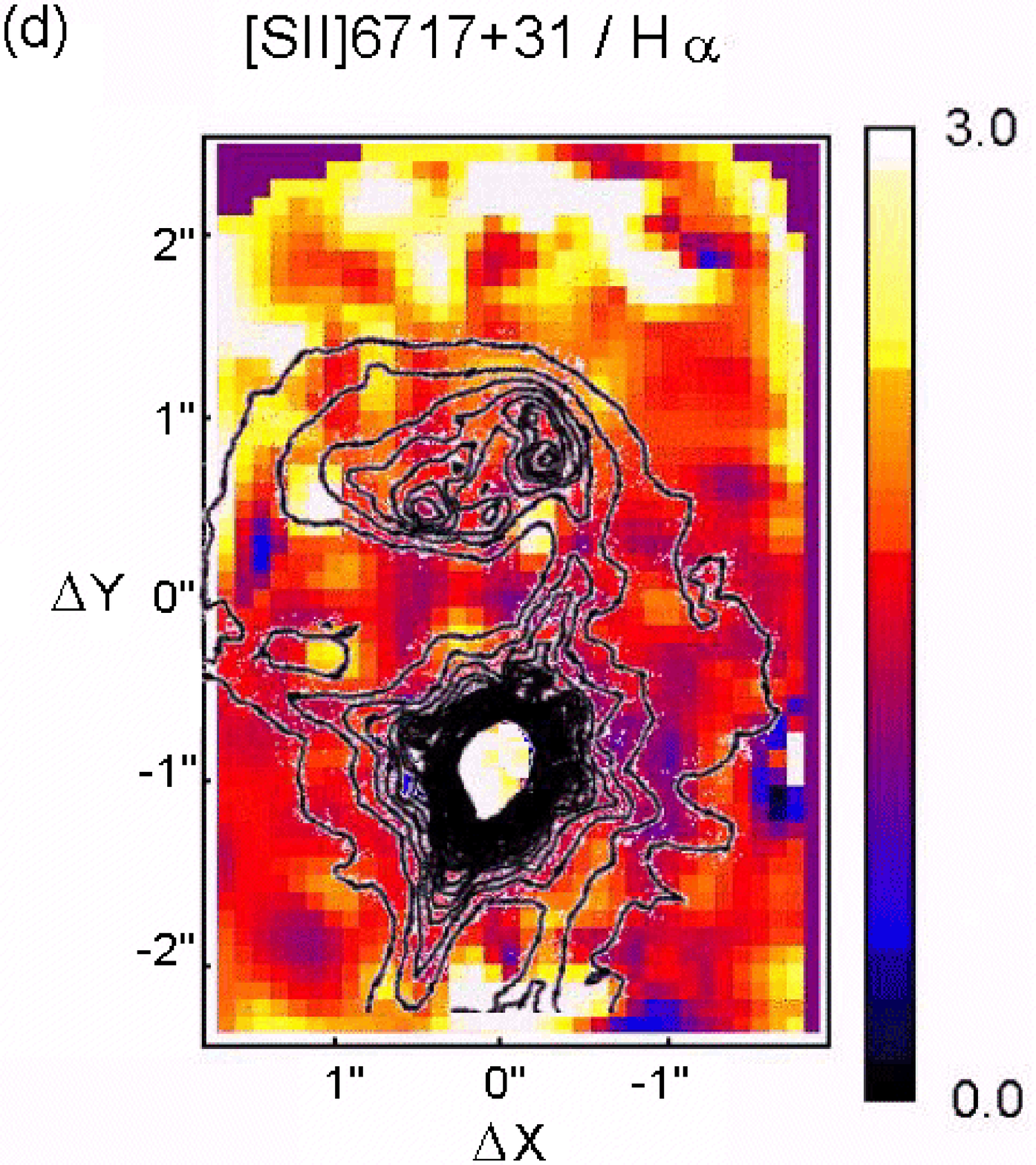} \cr
\end{tabular}
\vspace{8.0 cm}
\caption {
GMOS maps of the emission line ratios:
[S {\sc ii}]$\lambda$6717+31/H$\alpha$,
[N {\sc ii}]$\lambda$6583/H$\alpha$ and
[O {\sc iii}]$\lambda$5007/H$\beta$.
The QSO (in each GMOS maps) is positioned at
$\Delta X \sim$ 0.0$''$, and $\Delta Y \sim$ -1.0$''$. 
Panel (d) shows the superposition of the
[S {\sc ii}]/H$\alpha$ map and the HST--WFPC2-R  contour image.
}
\label{fig14}
\end{figure*}


\clearpage

\begin{figure*}
\vspace{12.0 cm}
\begin{tabular}{cc}
\includegraphics{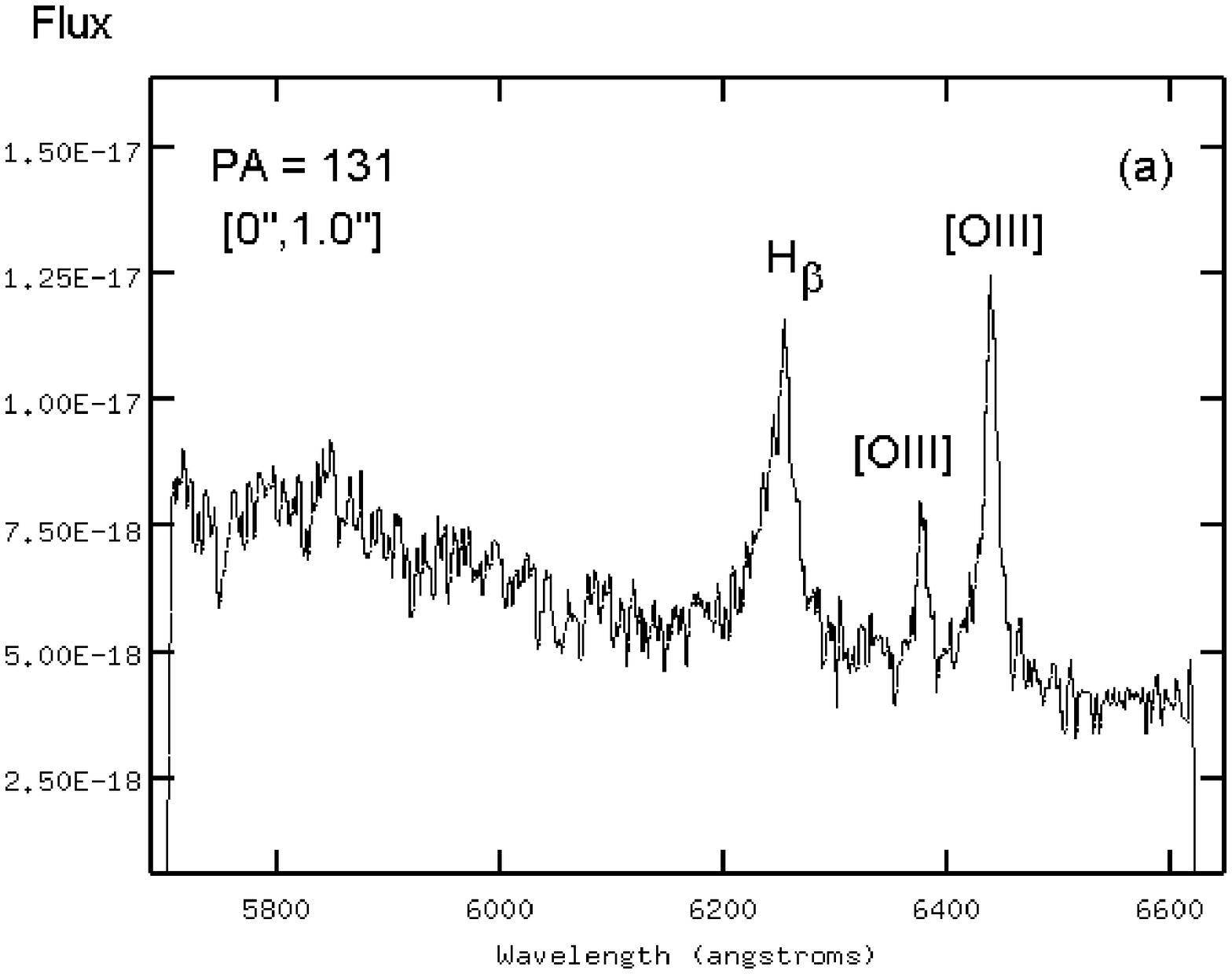}& 
\includegraphics{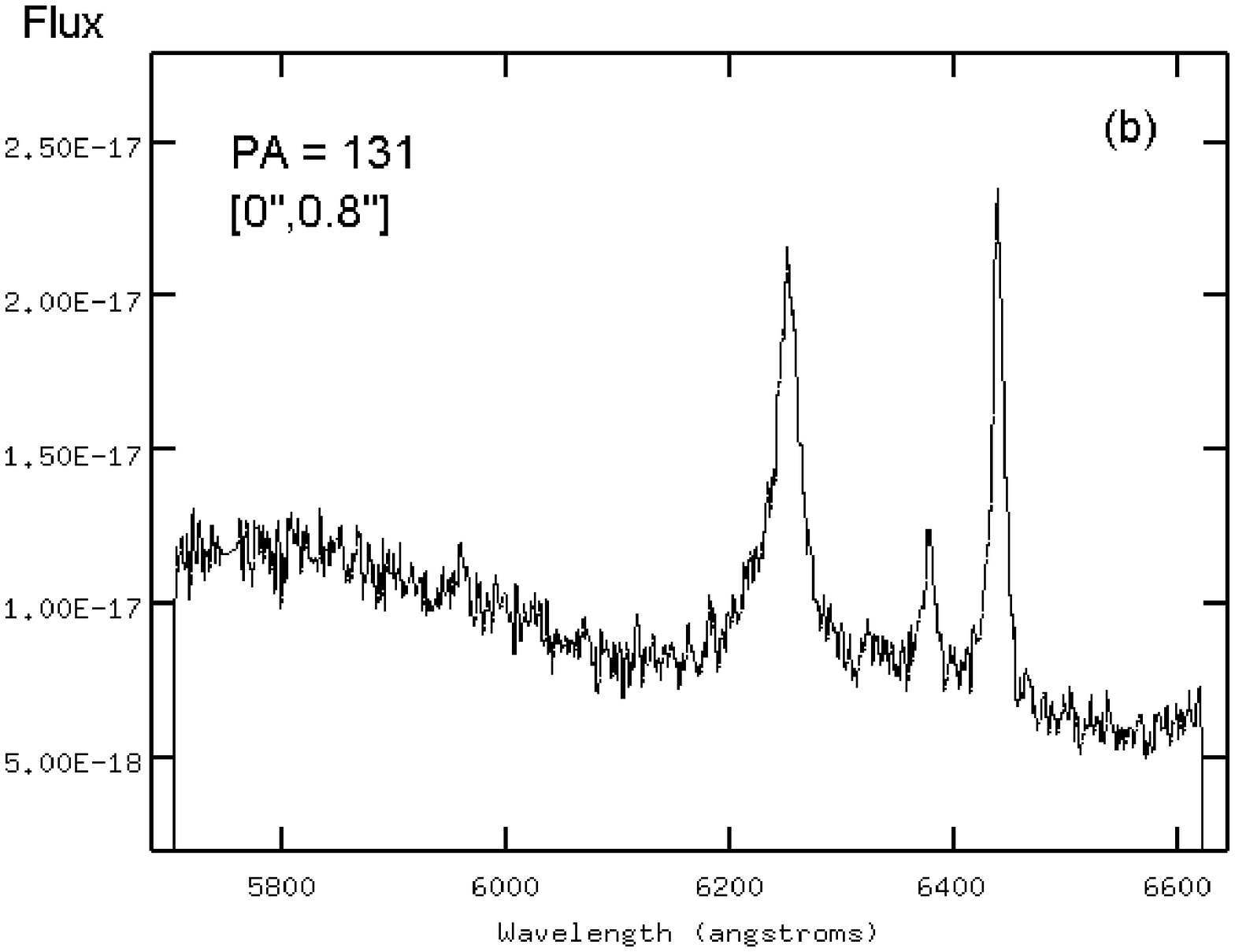} \cr
\includegraphics{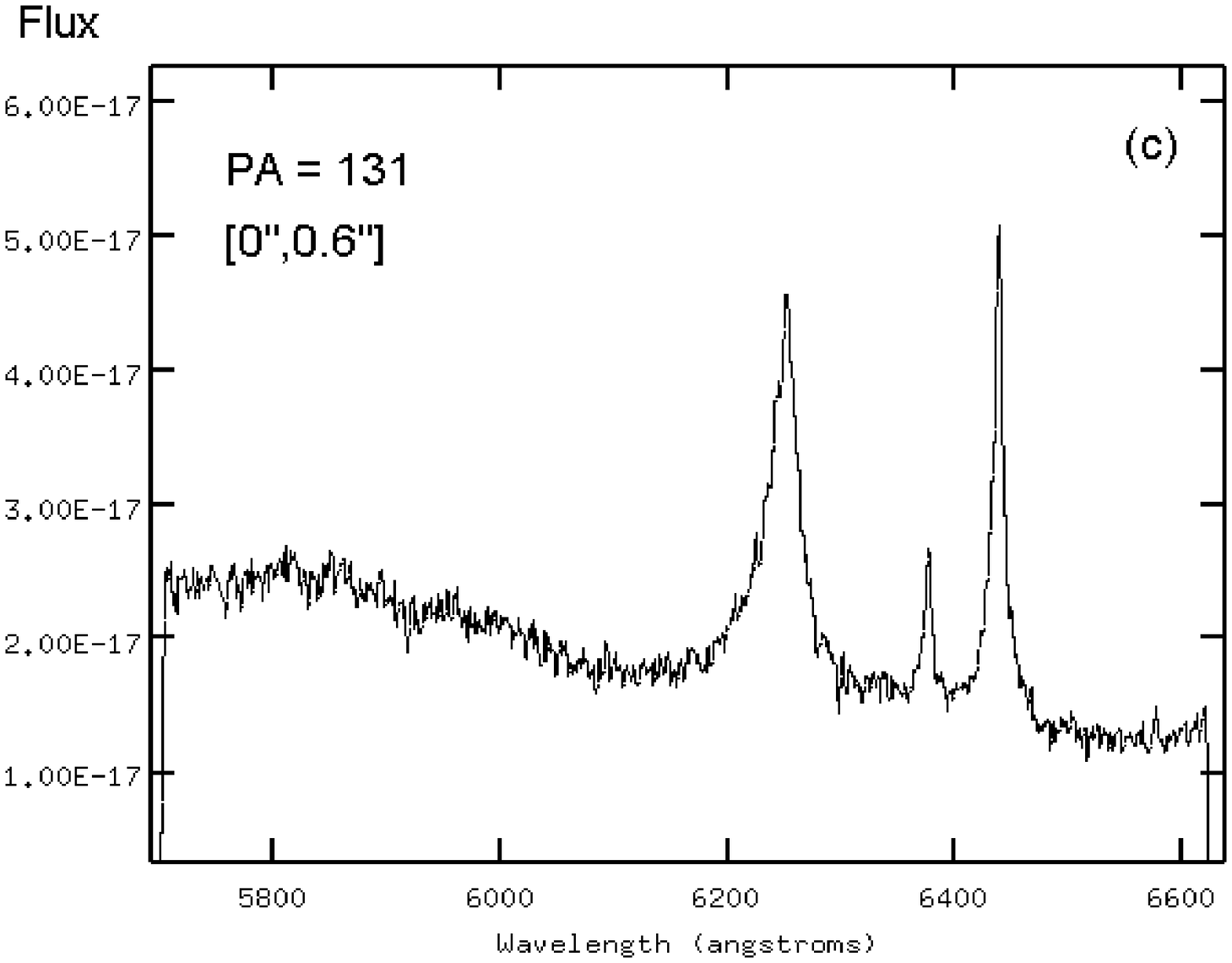}&
\includegraphics{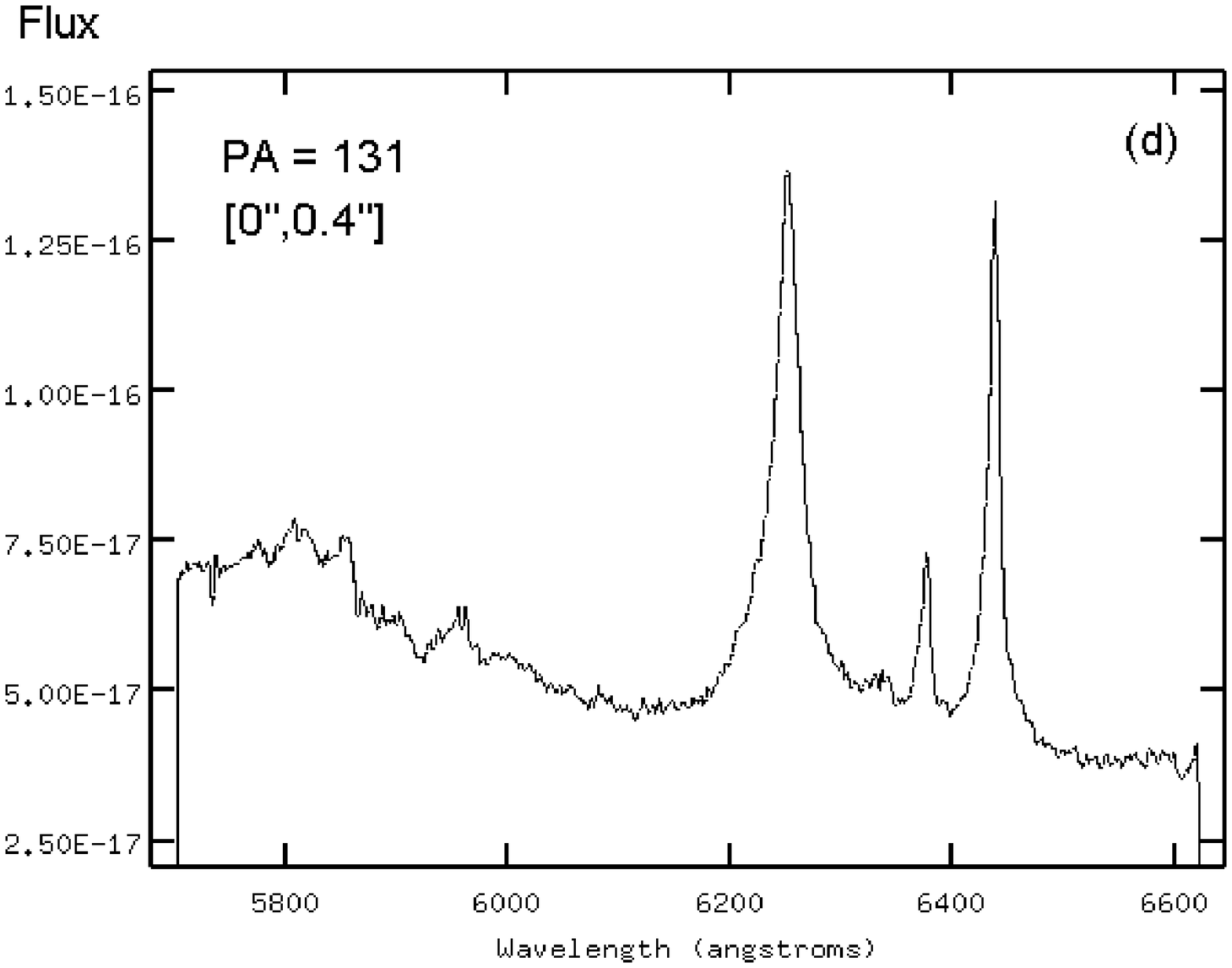} \cr
\includegraphics{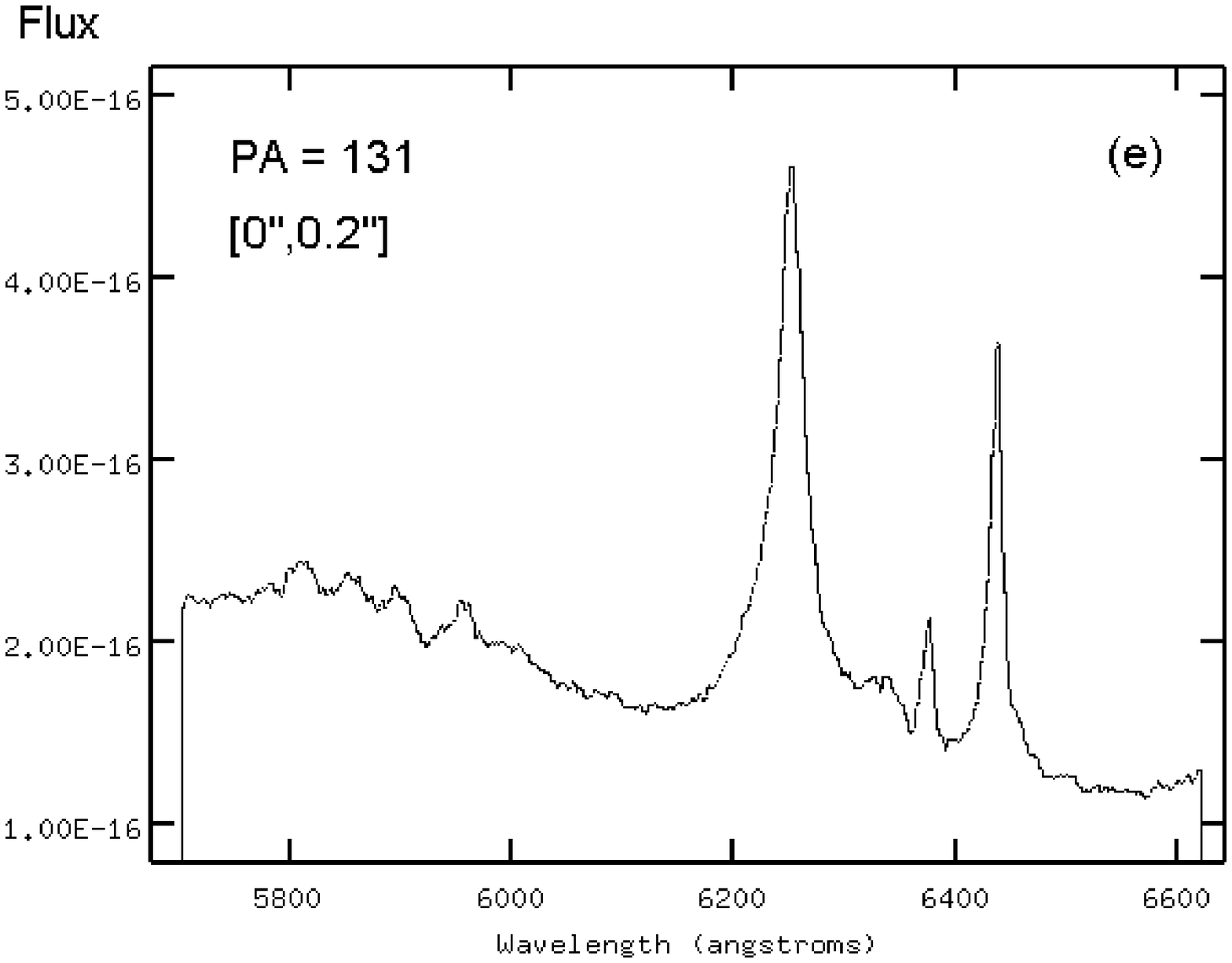}& 
\includegraphics{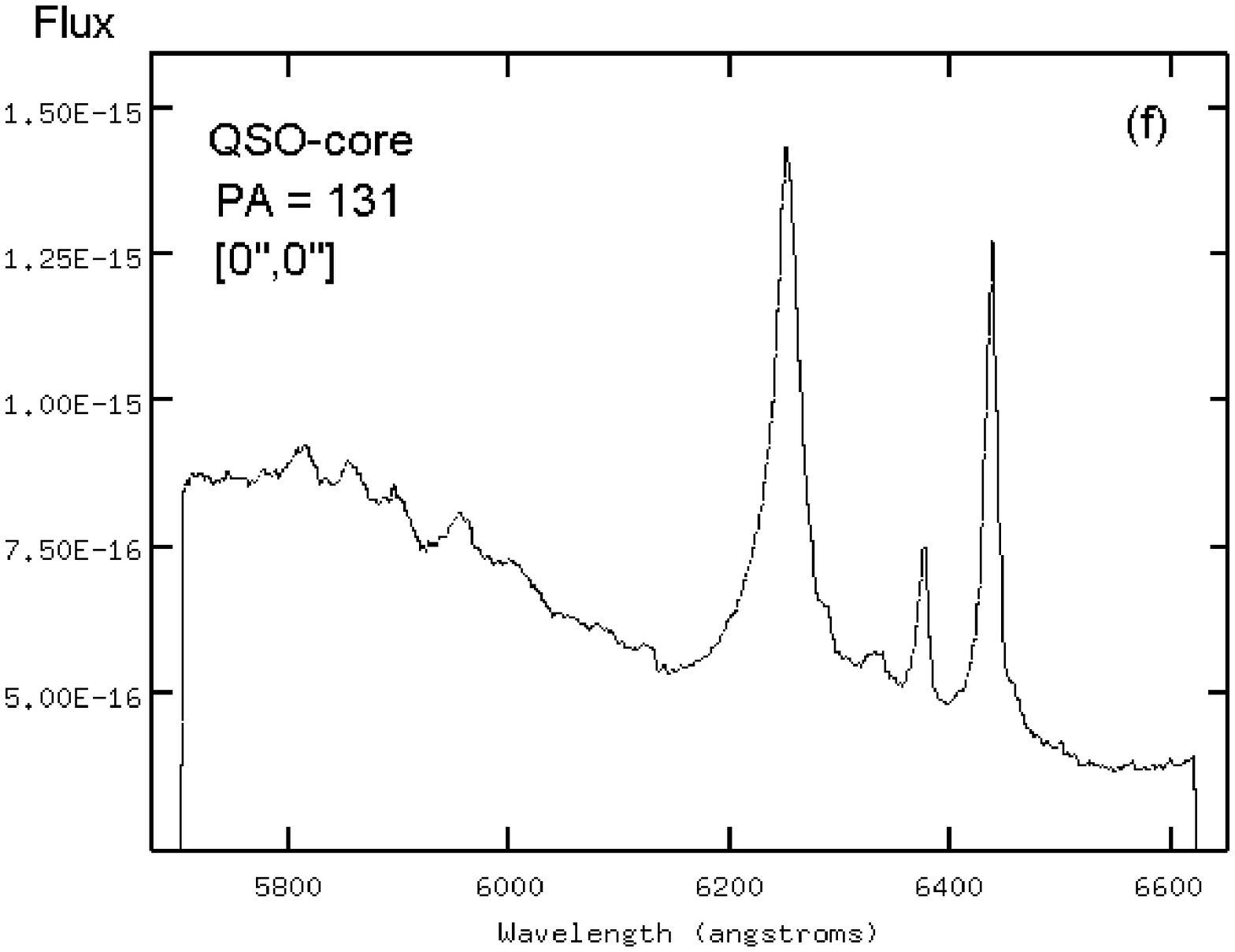} \cr
\end{tabular}
\vspace{8.0 cm}
\caption {
Sequence of individual GMOS spectra at position angle PA = 131$^{\circ}$ and
for the wavelength range of H$\beta$ + [O {\sc iii}]$\lambda$5007 +
Fe {\sc ii}  showing the presence of the strong blue continuum.
The blue continuum component was found in all the GMOS field.
The offset positions are from the QSO-core, and
in the GMOS  X and Y-axis (the Y-axis was located at PA $=$ 131$^{\circ}$). 
}
\label{fig15}
\end{figure*}

\clearpage

\begin{figure*}
\vspace{12.0 cm}
\begin{tabular}{cc}
\includegraphics{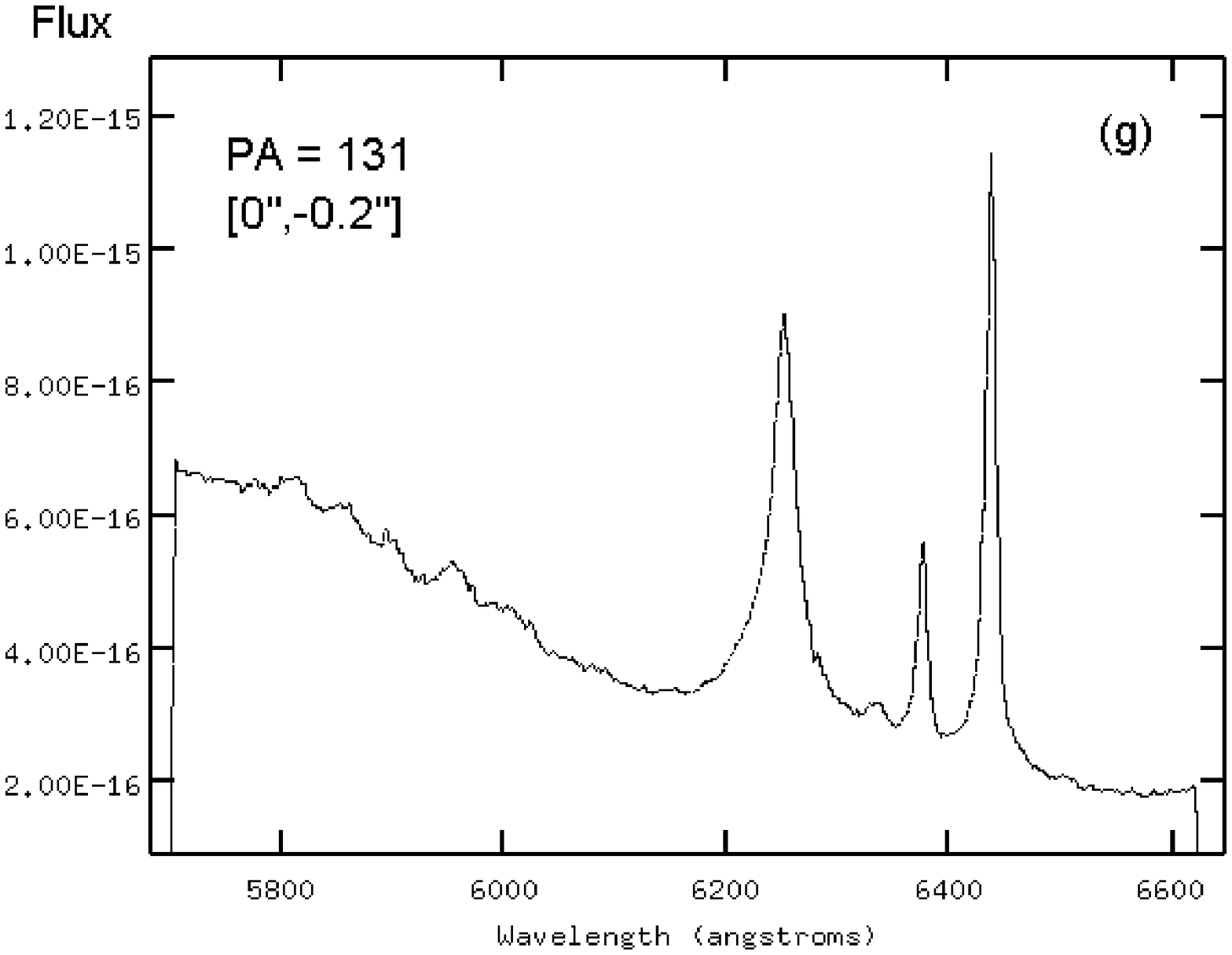}& 
\includegraphics{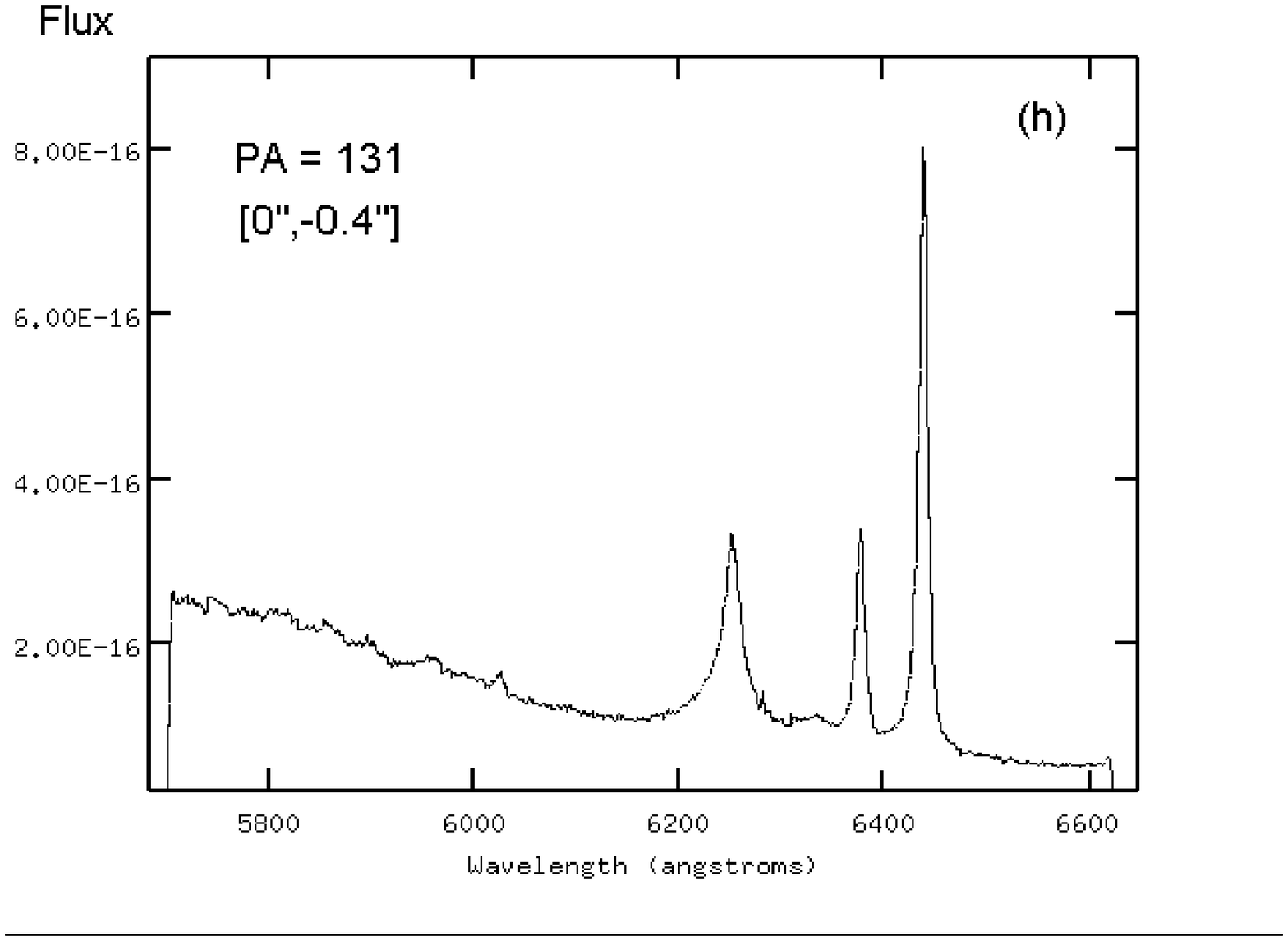} \cr
\includegraphics{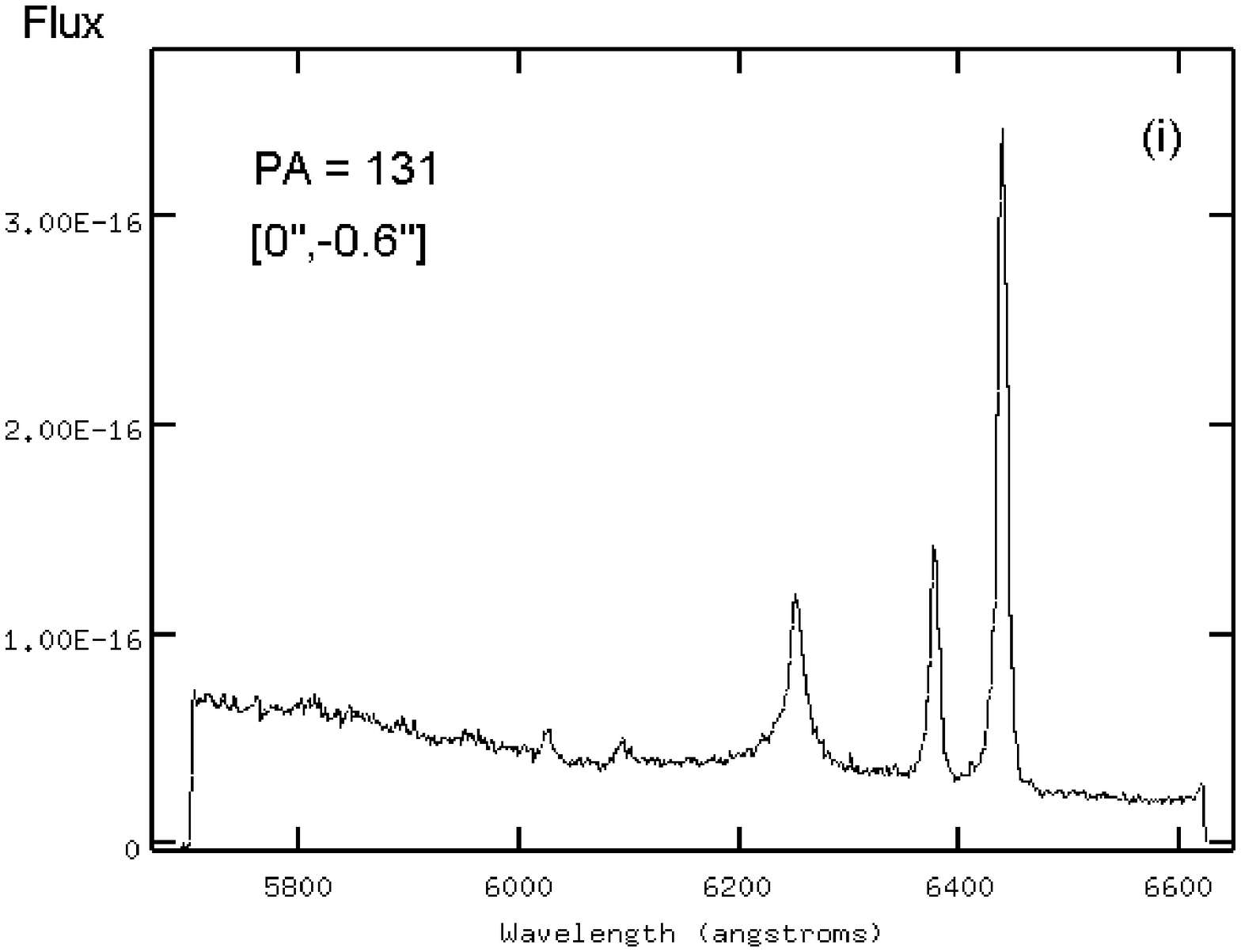}&
\includegraphics{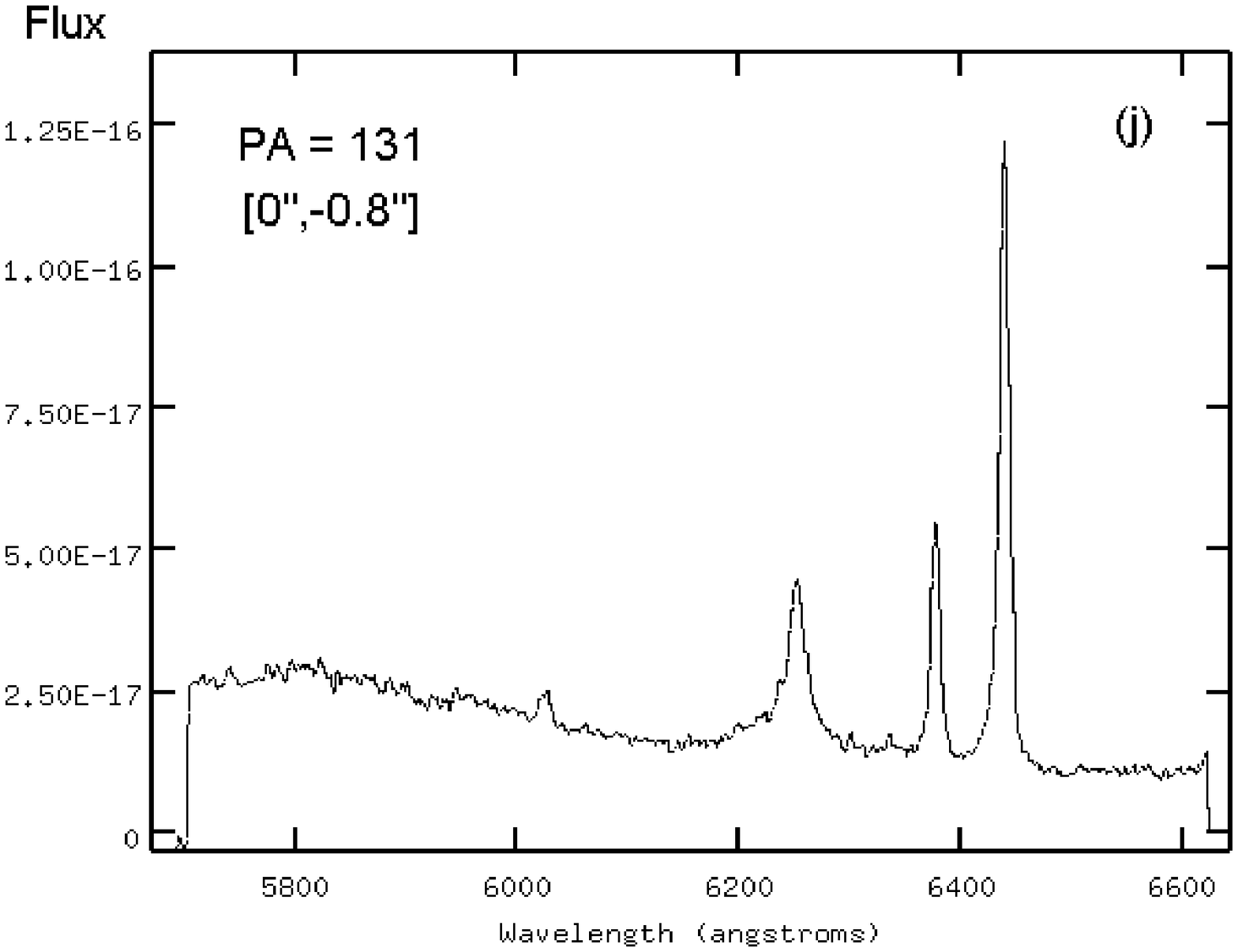} \cr
\includegraphics{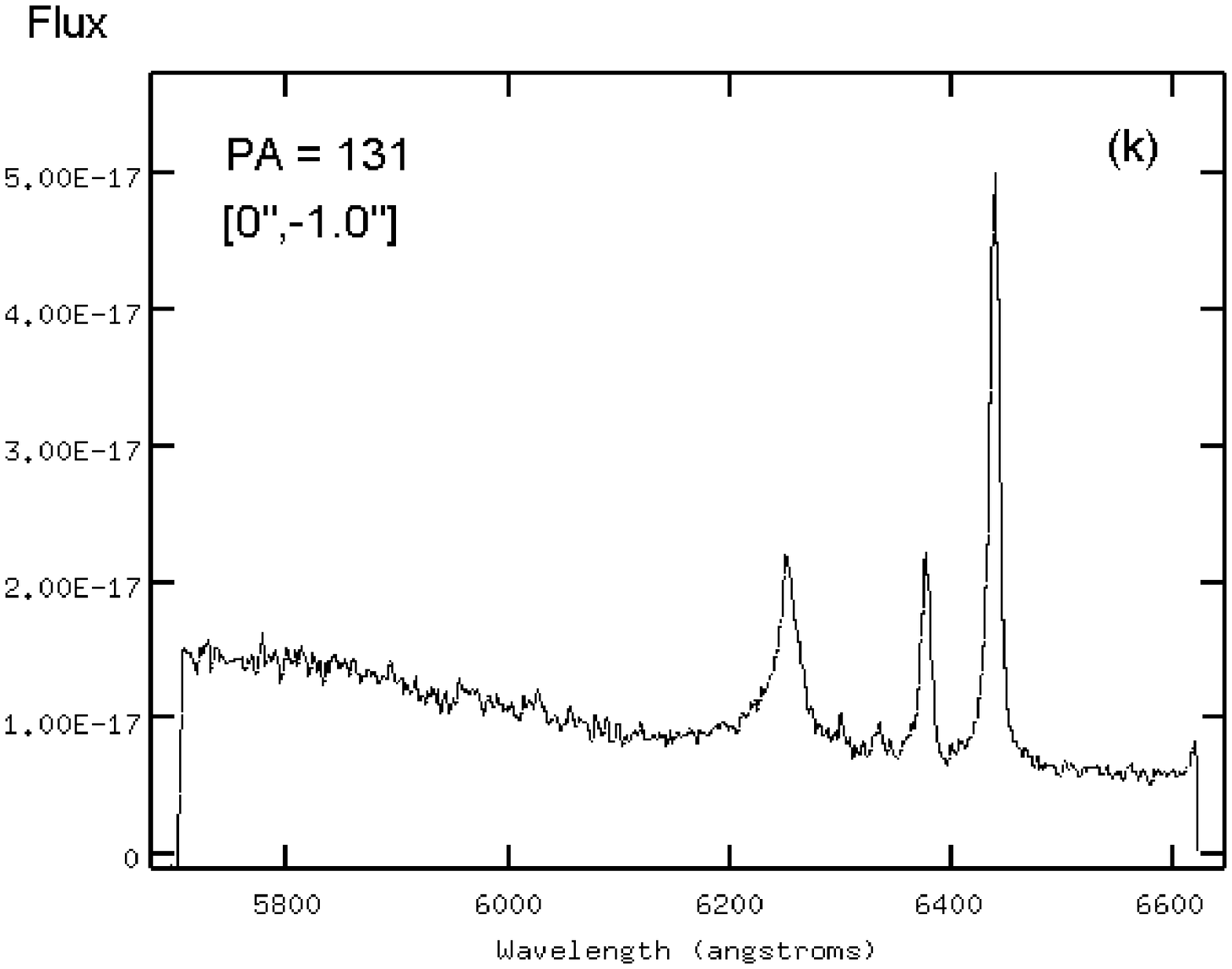}& 
\includegraphics{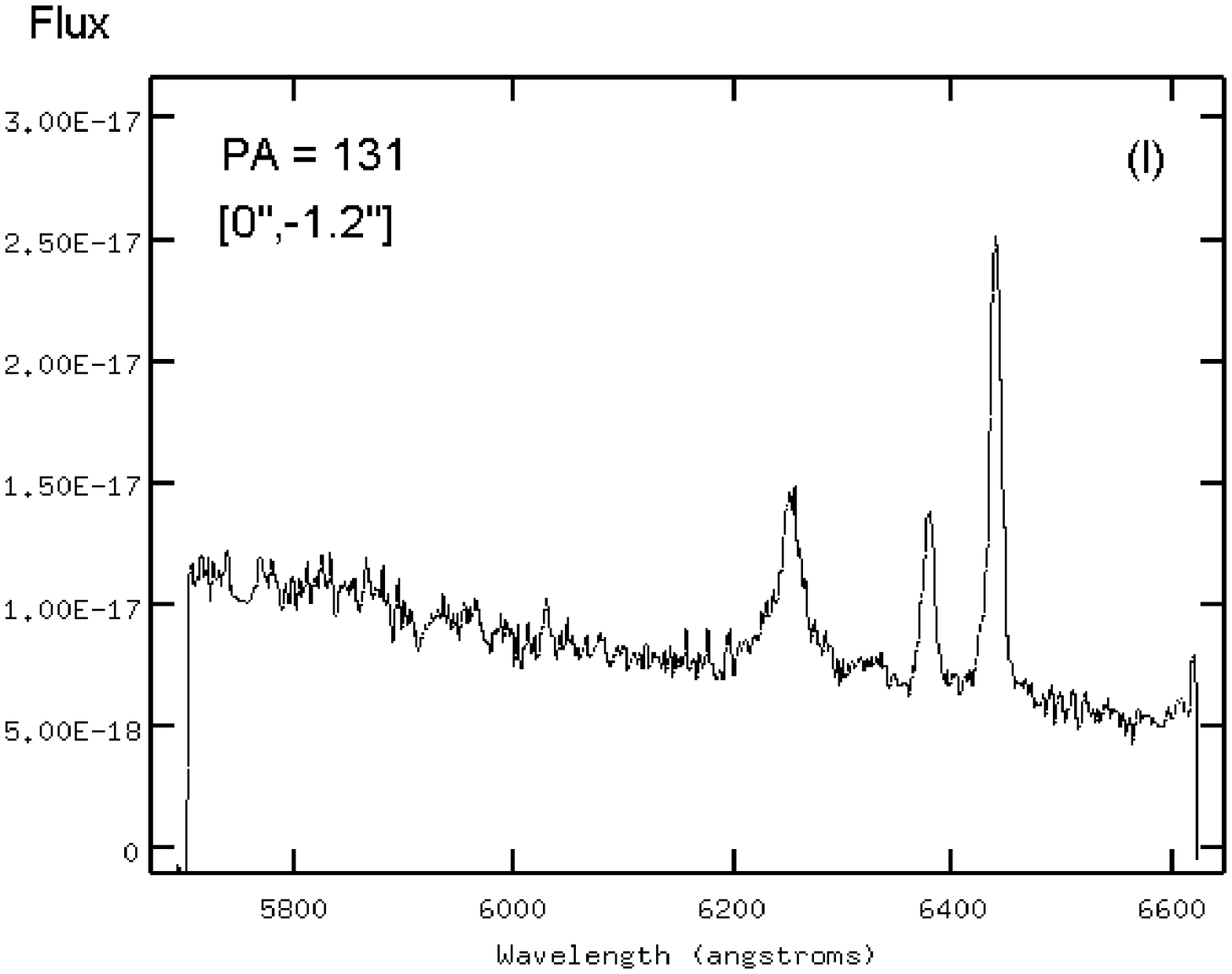} \cr
\end{tabular}
\vspace{8.0 cm}
\addtocounter{figure}{-1}
\caption {Contin.
 }
\label{fig15c}
\end{figure*}


\clearpage

\begin{figure*}
\vspace{12.0 cm}
\begin{tabular}{c}
\includegraphics{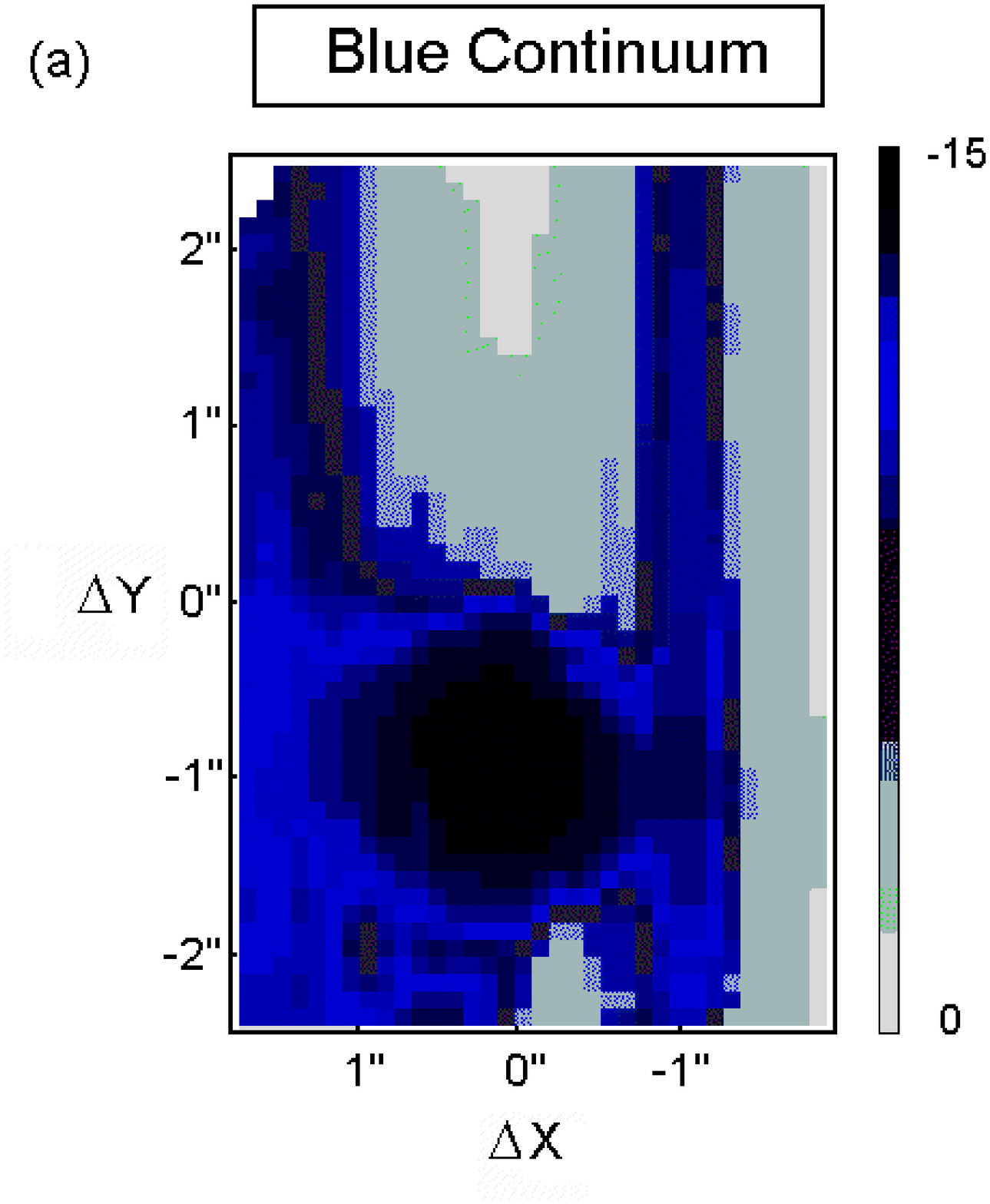}\cr
\end{tabular}
\vspace{8.0 cm}
\caption {
GMOS-IFU map of the Continuum Colour, for IRAS 04505-2958 (panel a):
showing in all the field a strong blue component. Panel (b) shows the
superposition of the GMOS colour map and the HST-WFPC2 R contours.
For details see the text.
}
\label{fig16}
\end{figure*}

\begin{figure*}
\vspace{12.0 cm}
\begin{tabular}{c}
\includegraphics{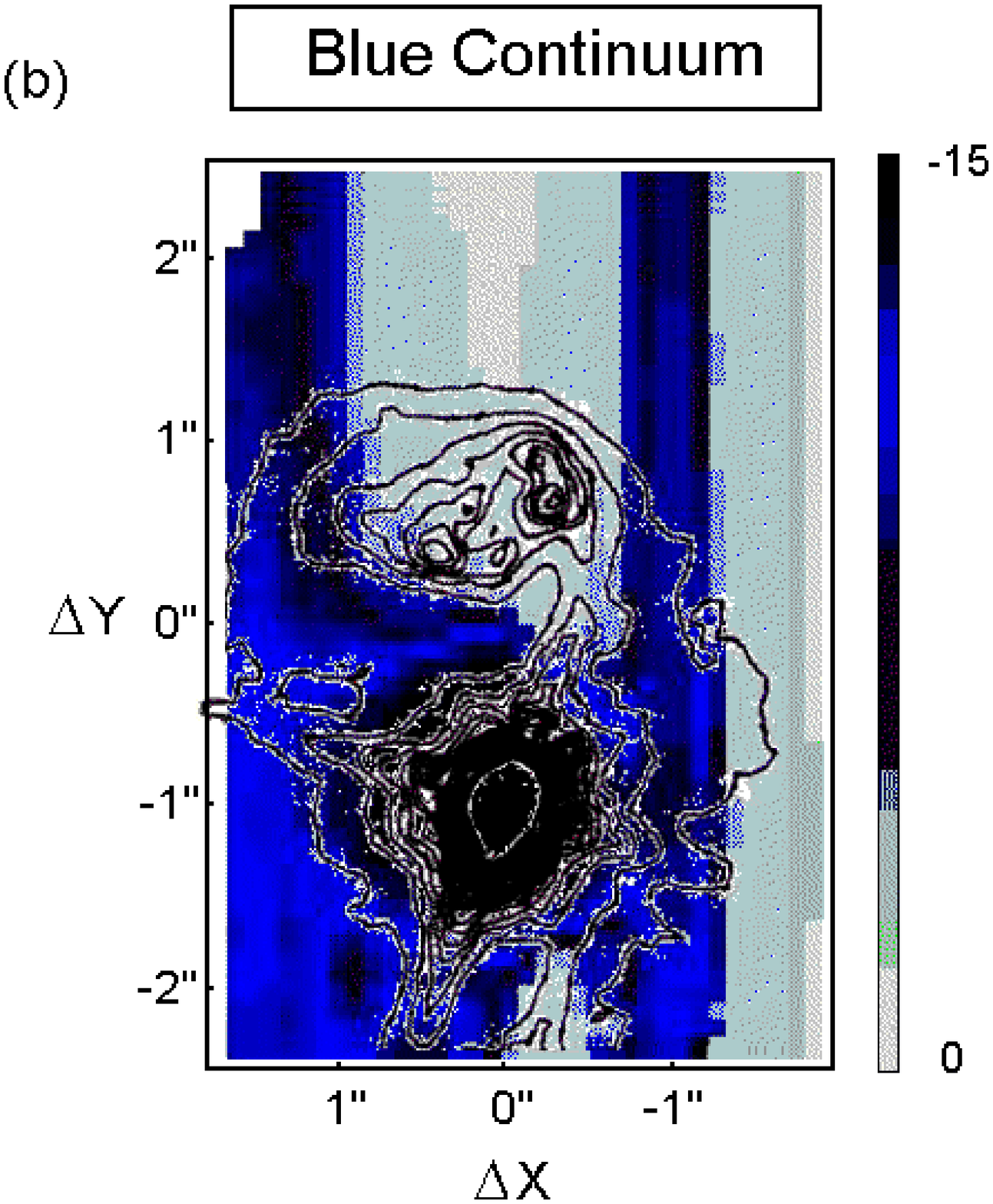}\cr
\end{tabular}
\vspace{8.0 cm}
\addtocounter{figure}{-1}
\caption {Contin.
}
\label{fig16c}
\end{figure*}


\clearpage

\begin{figure*}
\vspace{12.0 cm}
\begin{tabular}{cc}
\includegraphics{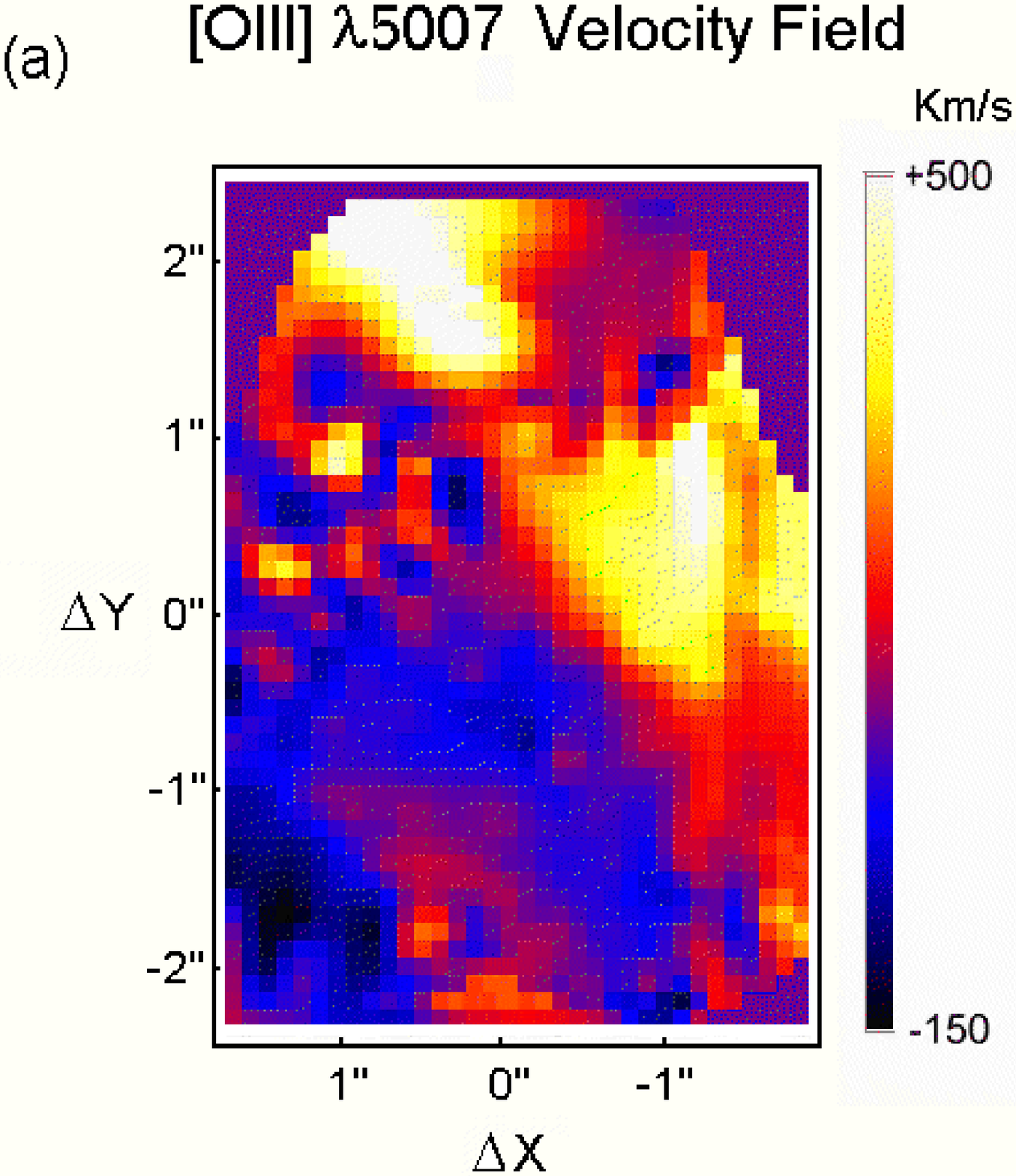}&
\includegraphics{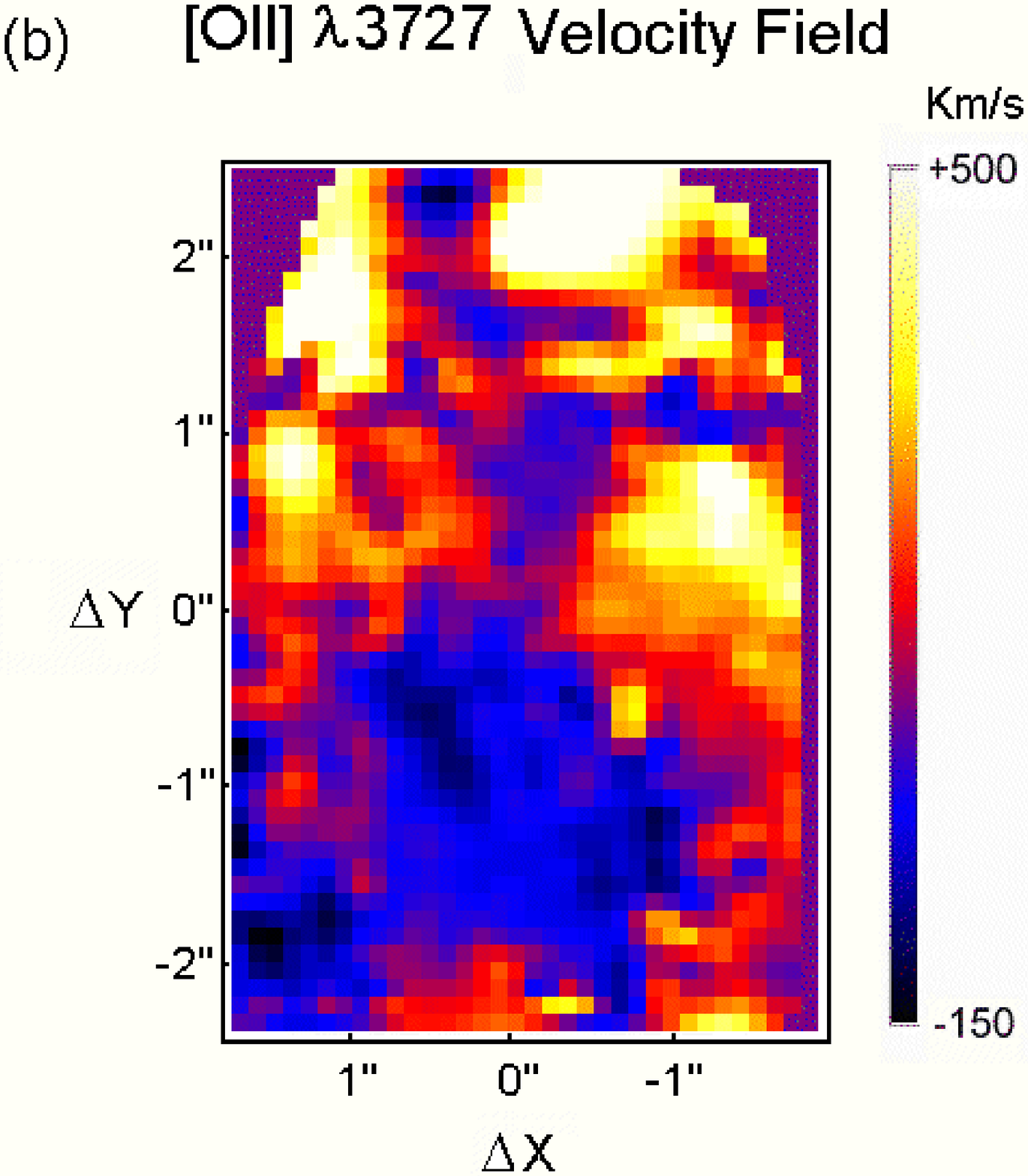} \cr
\includegraphics{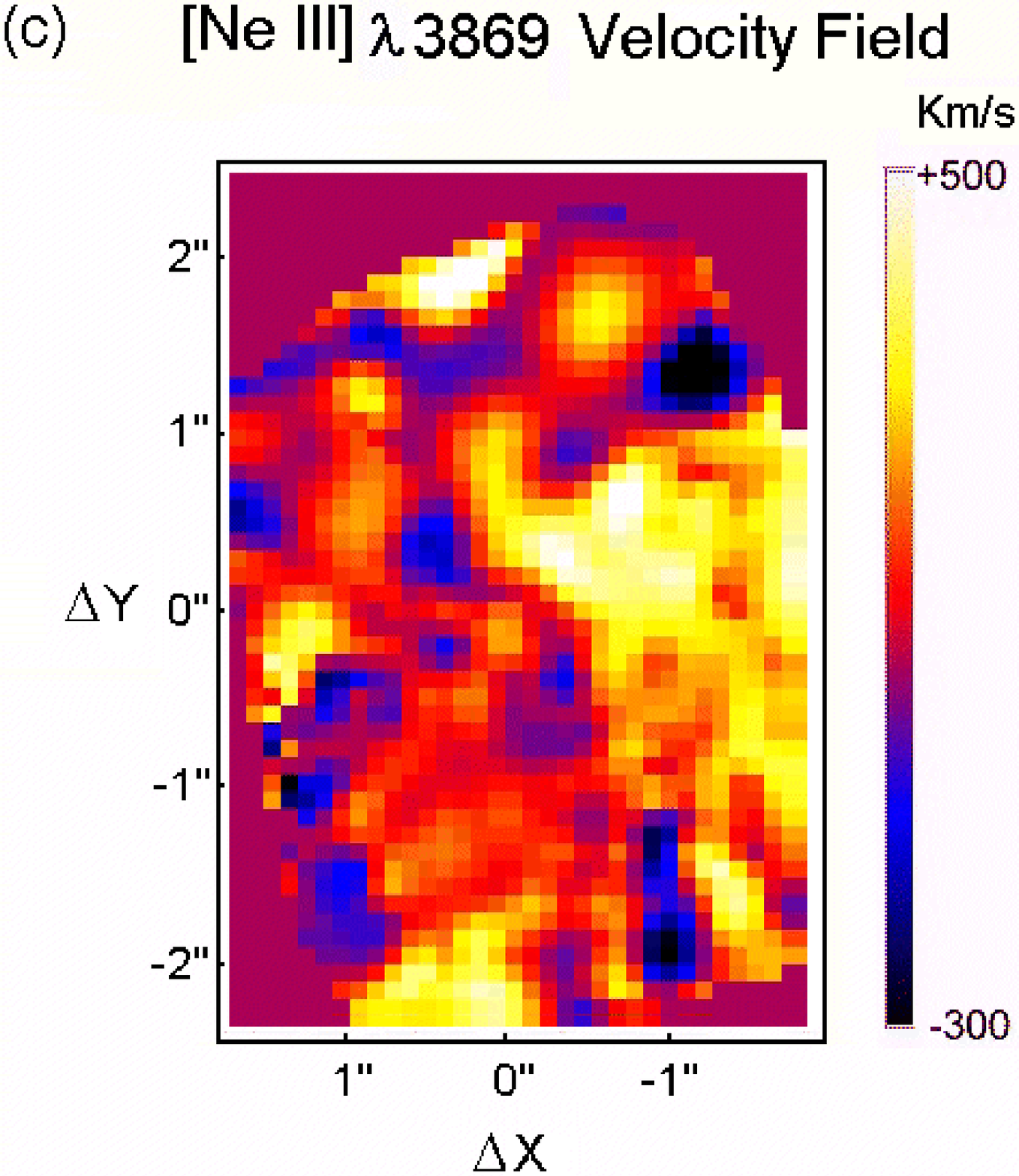}&
\includegraphics{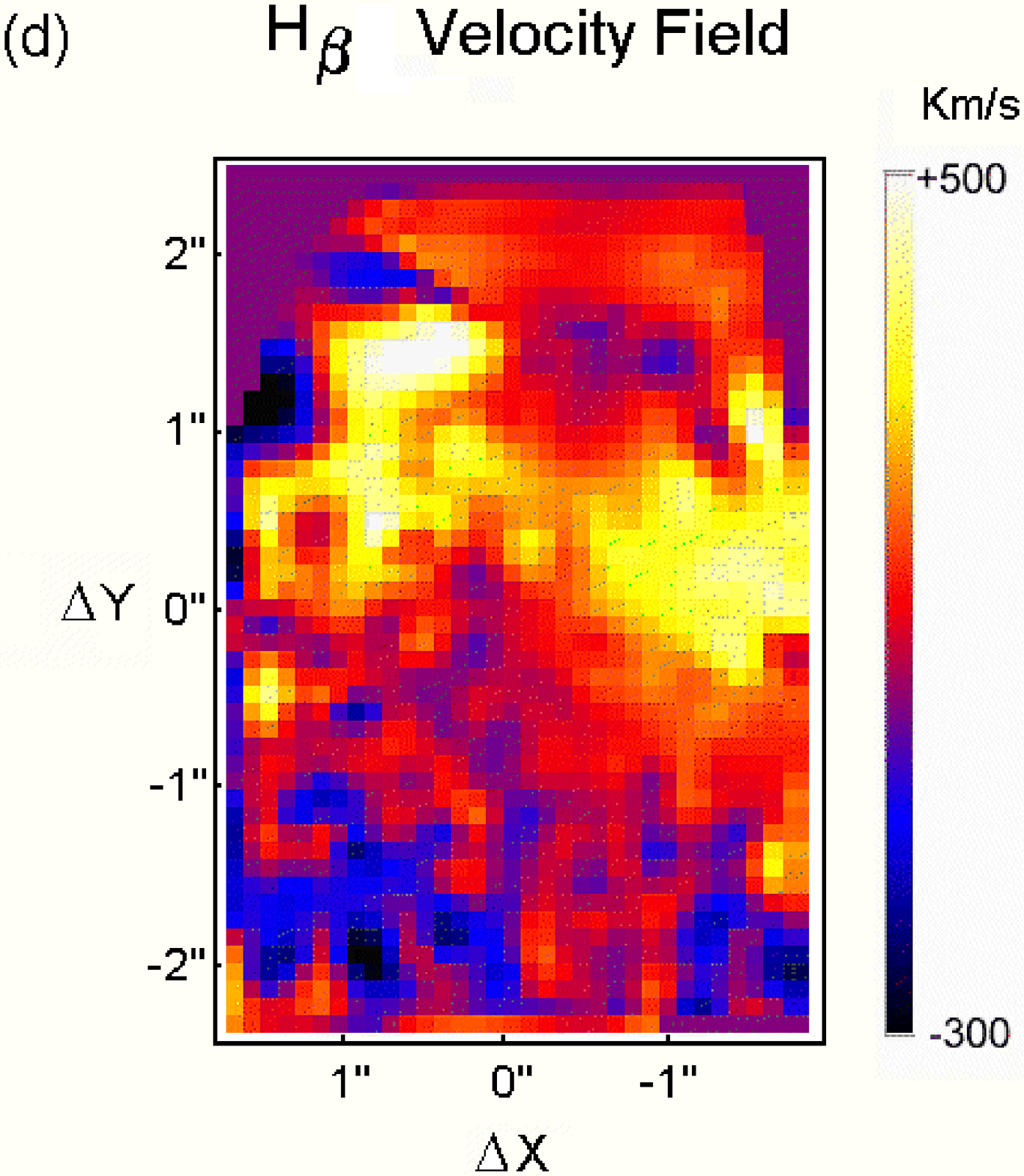} \cr
\end{tabular}
\vspace{8.0 cm}
\caption {
GMOS velocity field maps (a, b, c, d) for the emission lines:
[O {\sc iii}]$\lambda$5007, [O {\sc ii}]$\lambda$3727,
[Ne {\sc iii}]$\lambda$3869 and H$\beta$.
Panel (e) shows the superposition of the [O {\sc iii}] VF and the
HST--WFPC2  contour image (with the R filter).
In the GMOS maps the QSO-core is positioned at
$\Delta X \sim$ 0.0$''$, and $\Delta Y \sim$ -1.0$''$.
}
\label{fig17}
\end{figure*}

\clearpage

\begin{figure*}
\vspace{12.0 cm}
\begin{tabular}{c}
\includegraphics{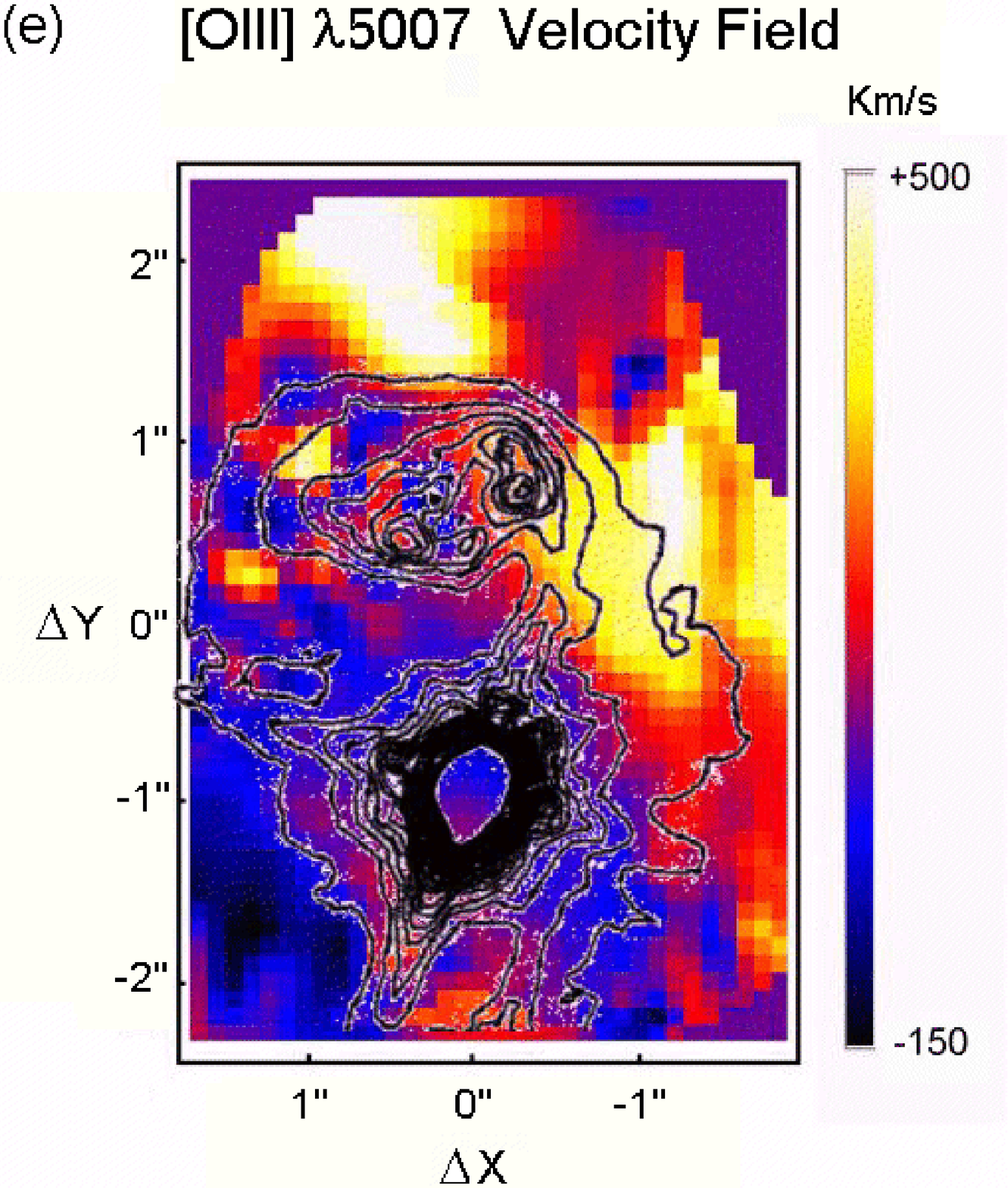}\cr
\end{tabular}
\vspace{8.0 cm}
\addtocounter{figure}{-1}
\caption {Contin.
}
\label{fig17c}
\end{figure*}


\clearpage

\begin{figure*}
\vspace{12.0 cm}
\begin{tabular}{c}
\includegraphics{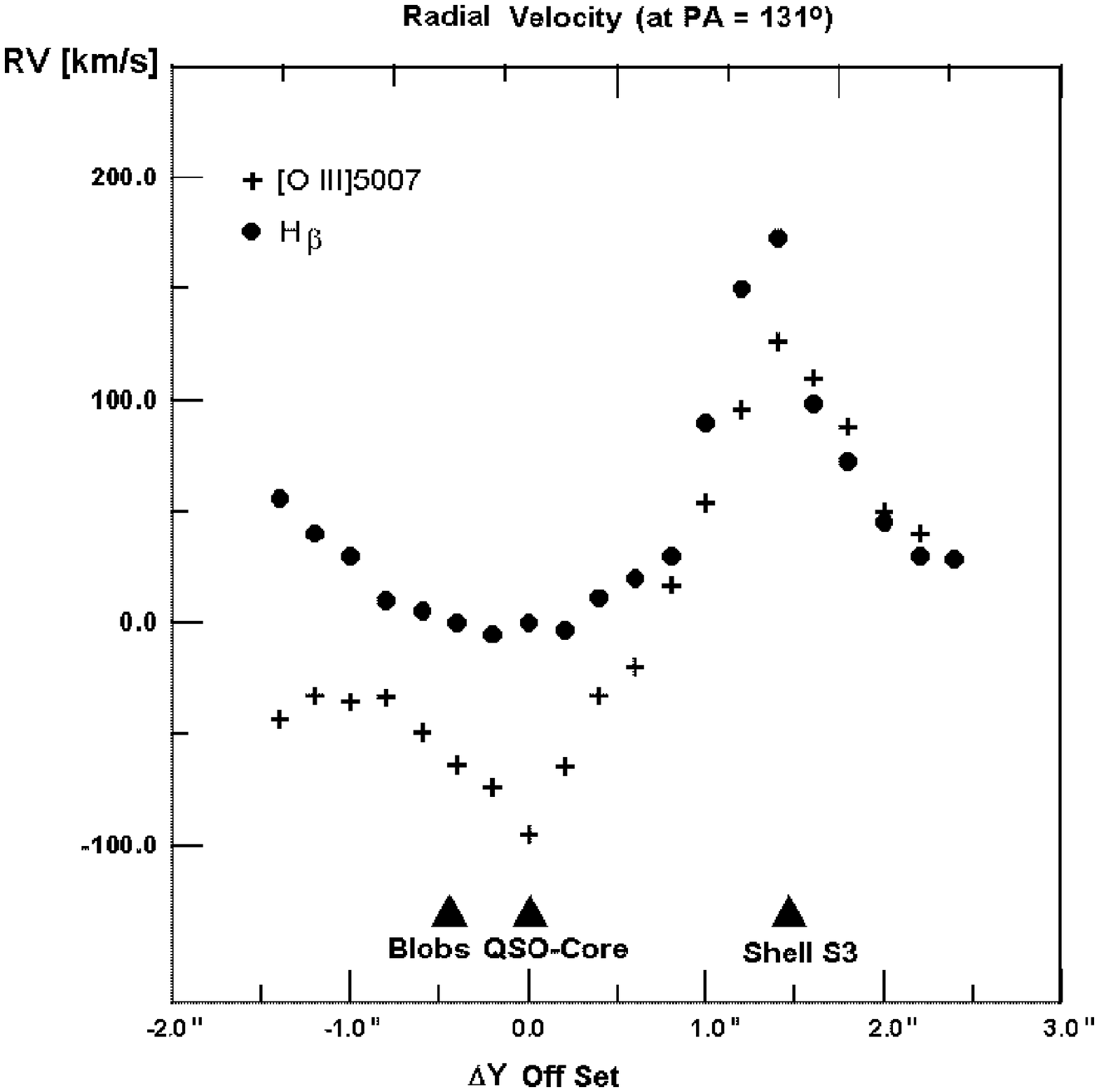}\cr
\end{tabular}
\vspace{8.0 cm}
\caption {
Radial velocity profile/variation along the position angle
PA $=$ 131$^{\circ}$,
for the emission lines H$\beta$ and [O {\sc iii}]$\lambda$5007.
}
\label{fig18}
\end{figure*}


\clearpage

\begin{figure*}
\vspace{12.0 cm}
\begin{tabular}{c}
\includegraphics{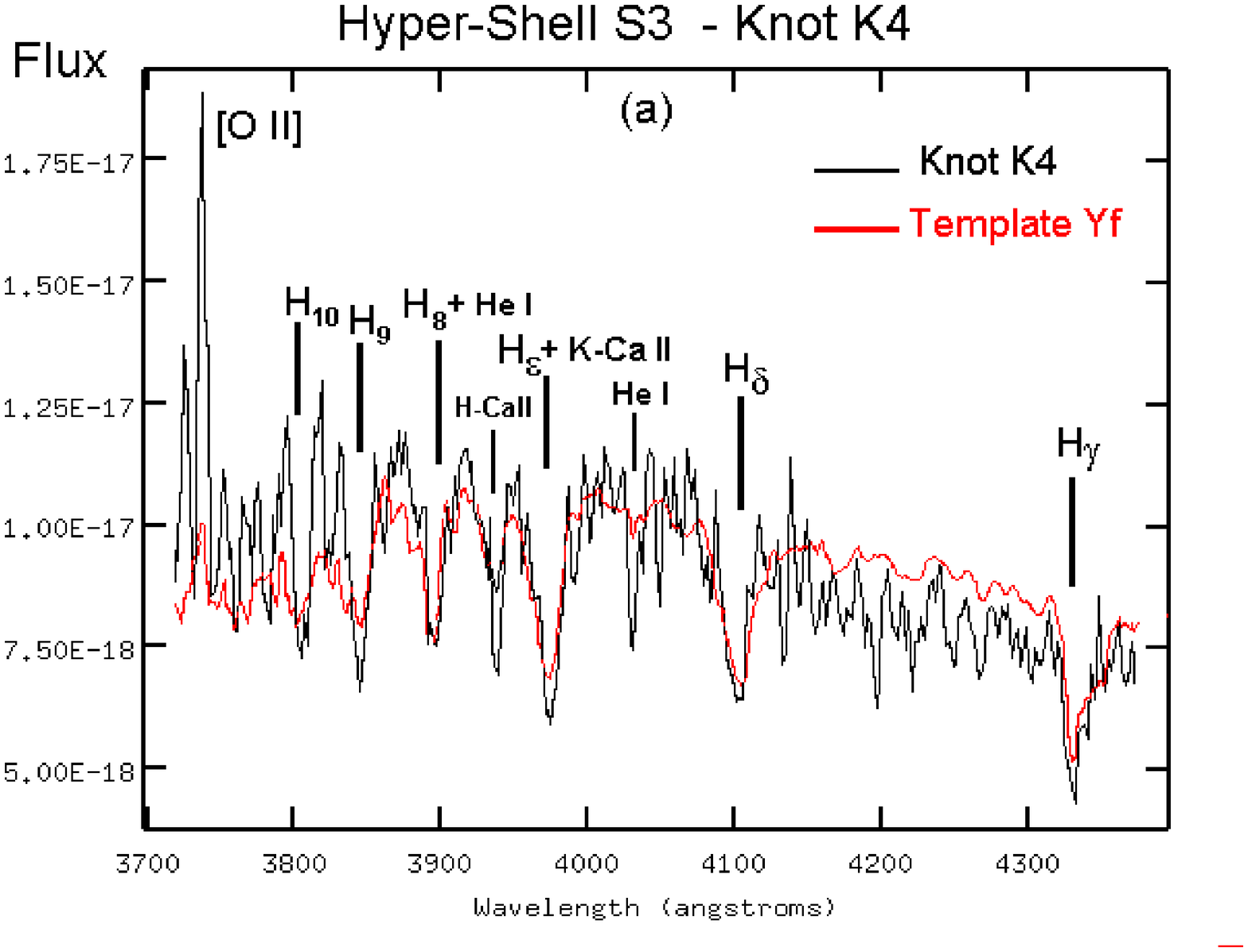}   \cr
\includegraphics{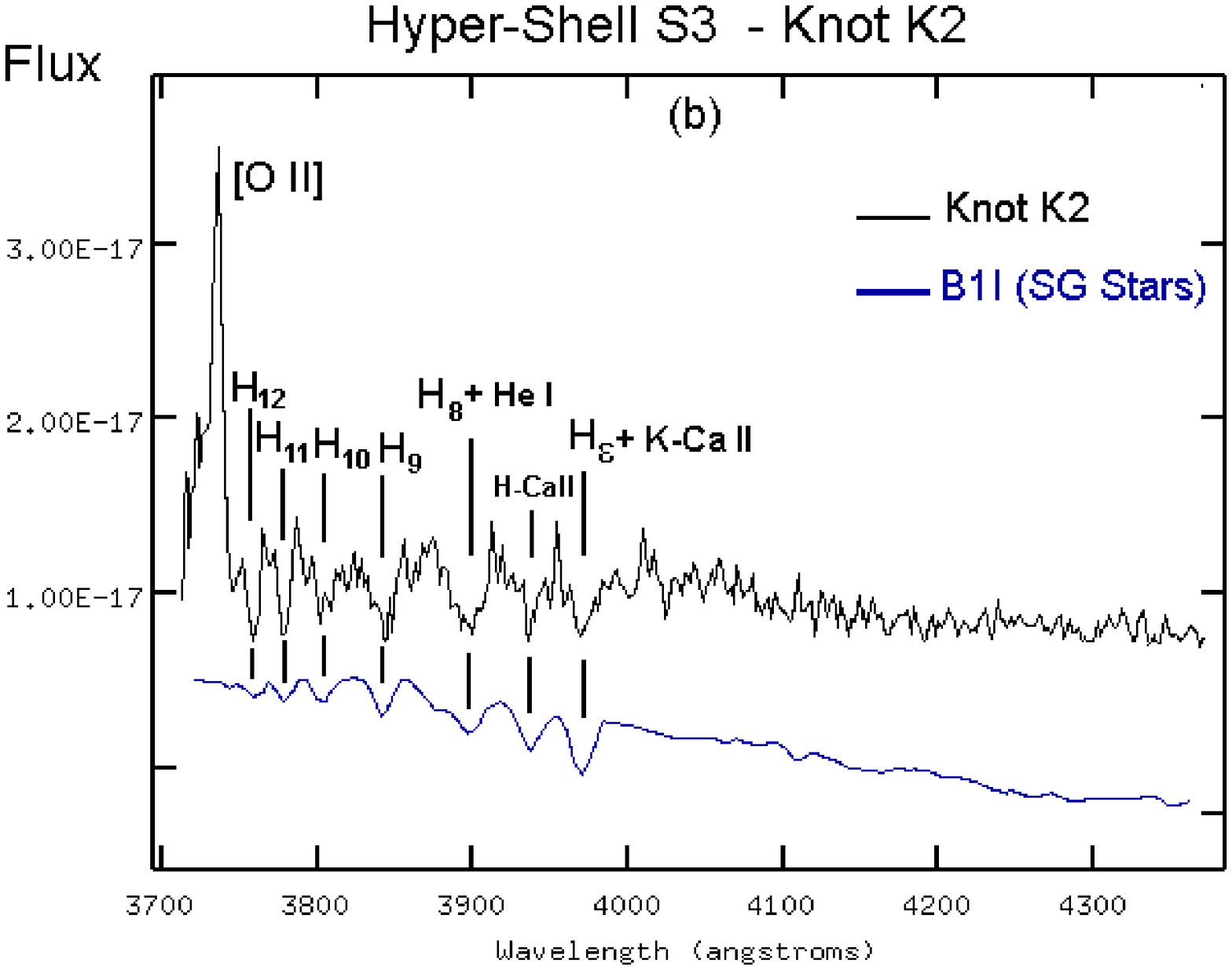} \cr
\end{tabular}
\vspace{8.0 cm}
\caption {
Result of the study of the Stelar Population in the main knots of the
shell S3.
The panel (a) shows the superposition of the GMOS spectra of the knot K4 and
the Templates of Stellar Clusters, of 125 Myr.(from Piatti
et al. 2002; Bica 1988).
The panel (b) depicts the spectra of the knot K2 and the best fit of
B1I supergiant stars (from the library of Silva \& Cornell 1992).
}
\label{fig19}
\end{figure*}

%


\clearpage

\begin{figure*}
\vspace{12.0 cm}
\begin{tabular}{cc}
\includegraphics{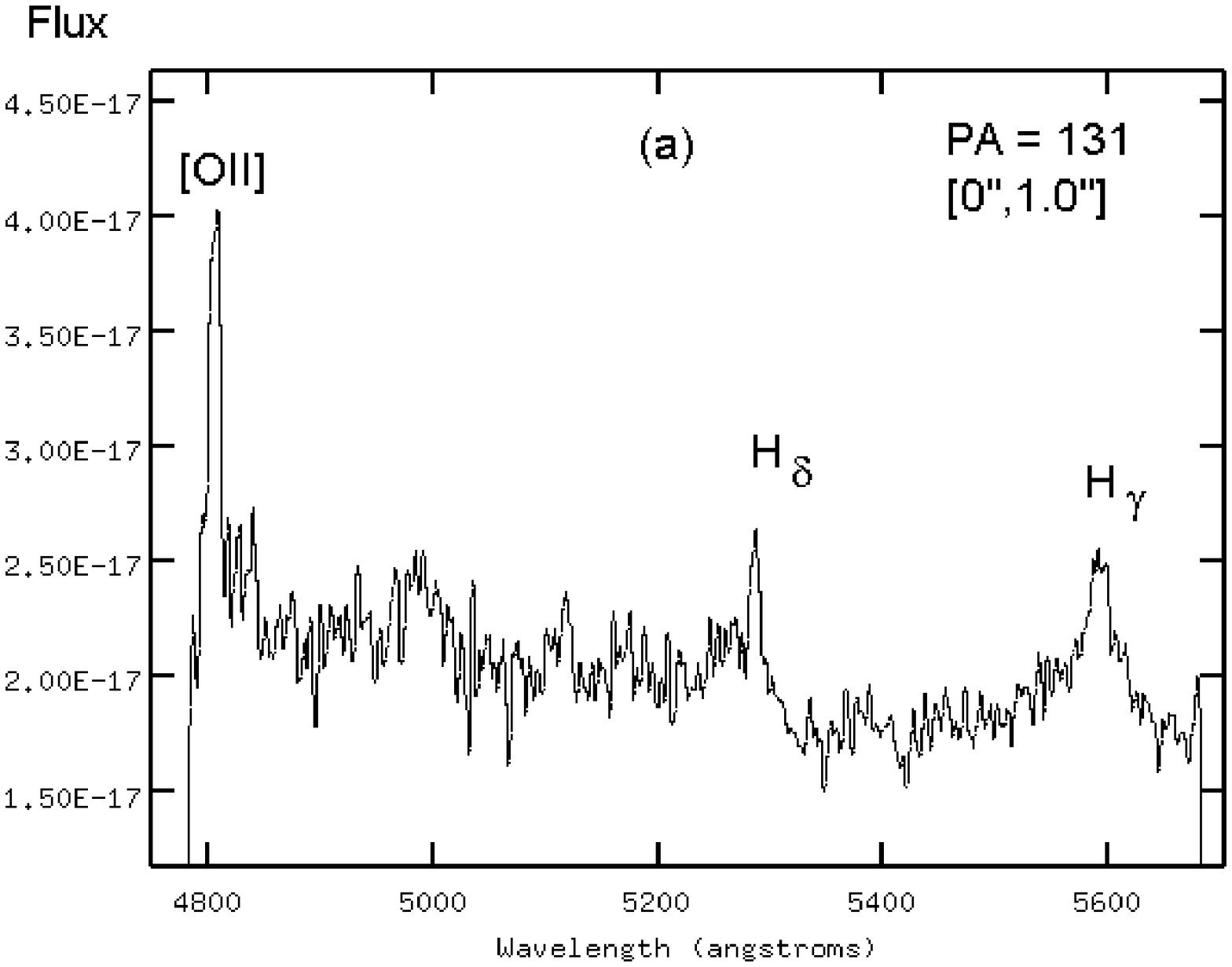}& 
\includegraphics{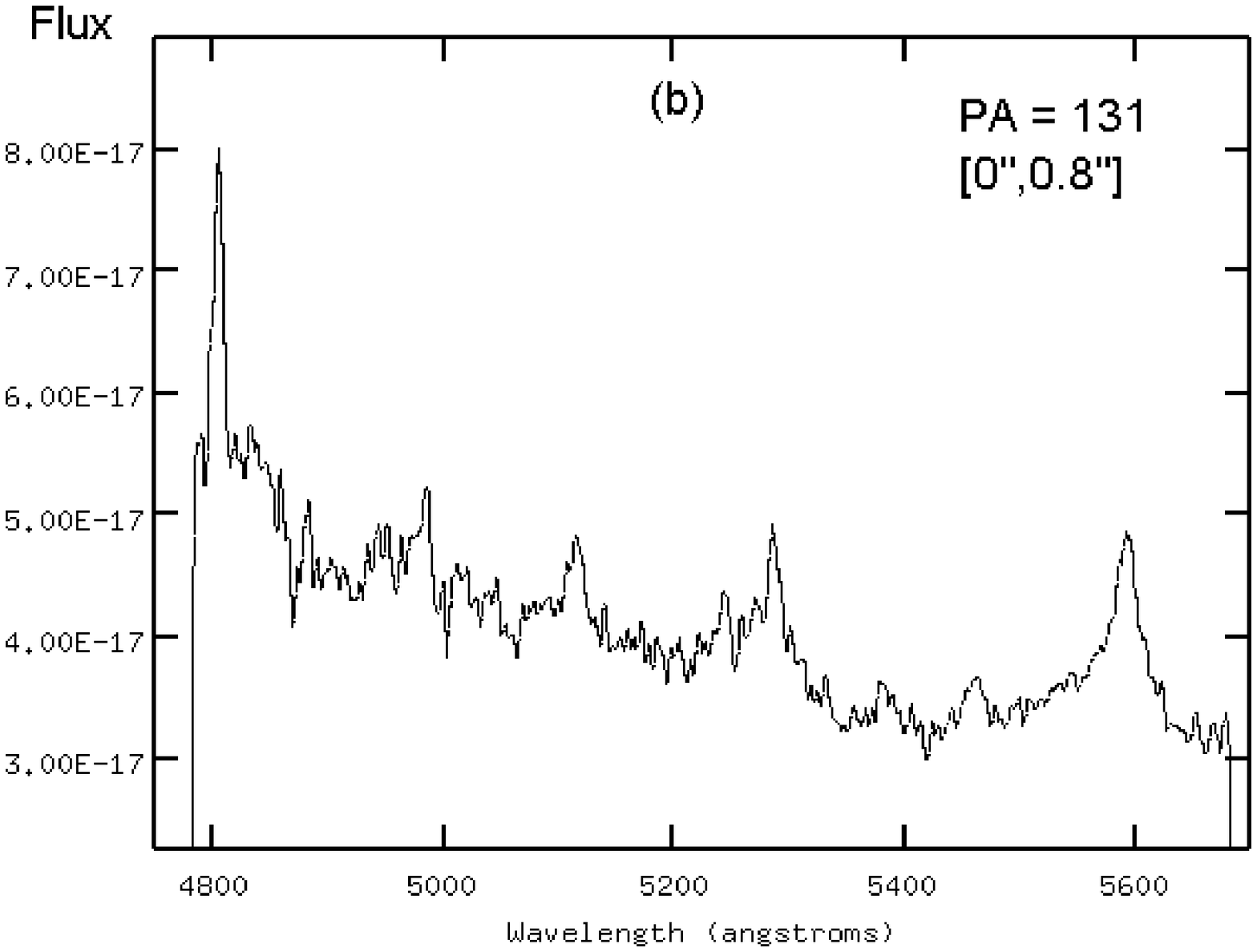} \cr
\includegraphics{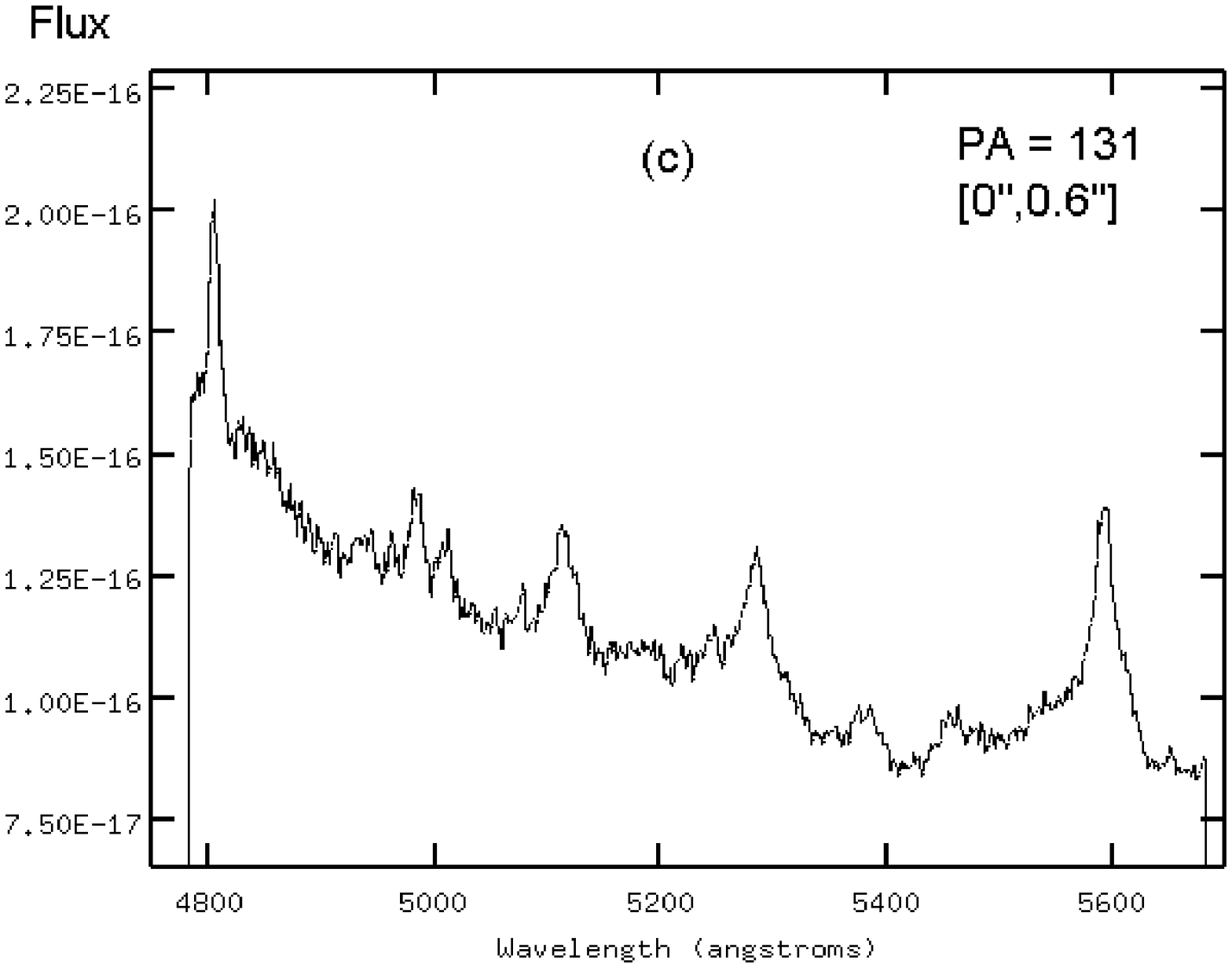}&
\includegraphics{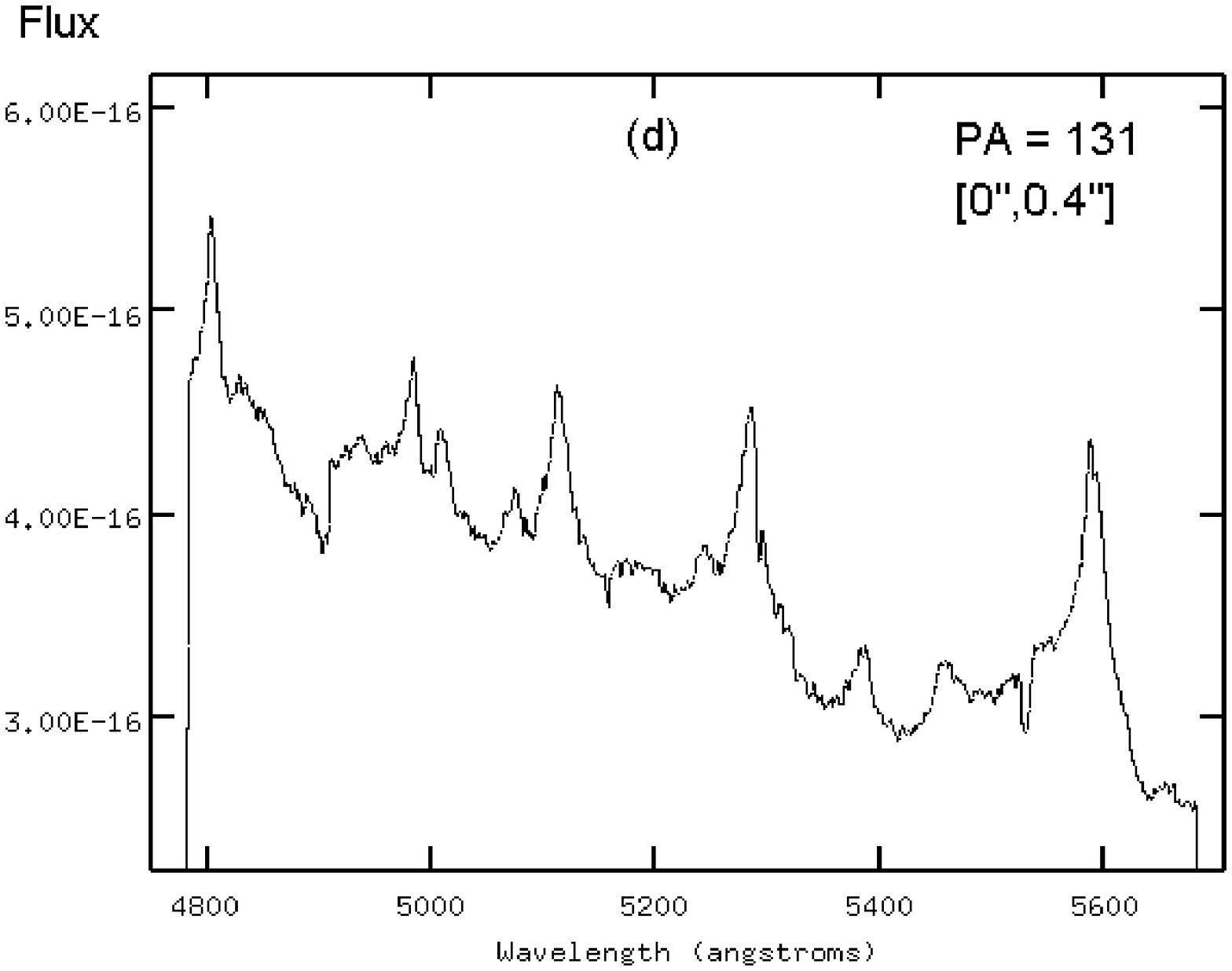} \cr
\includegraphics{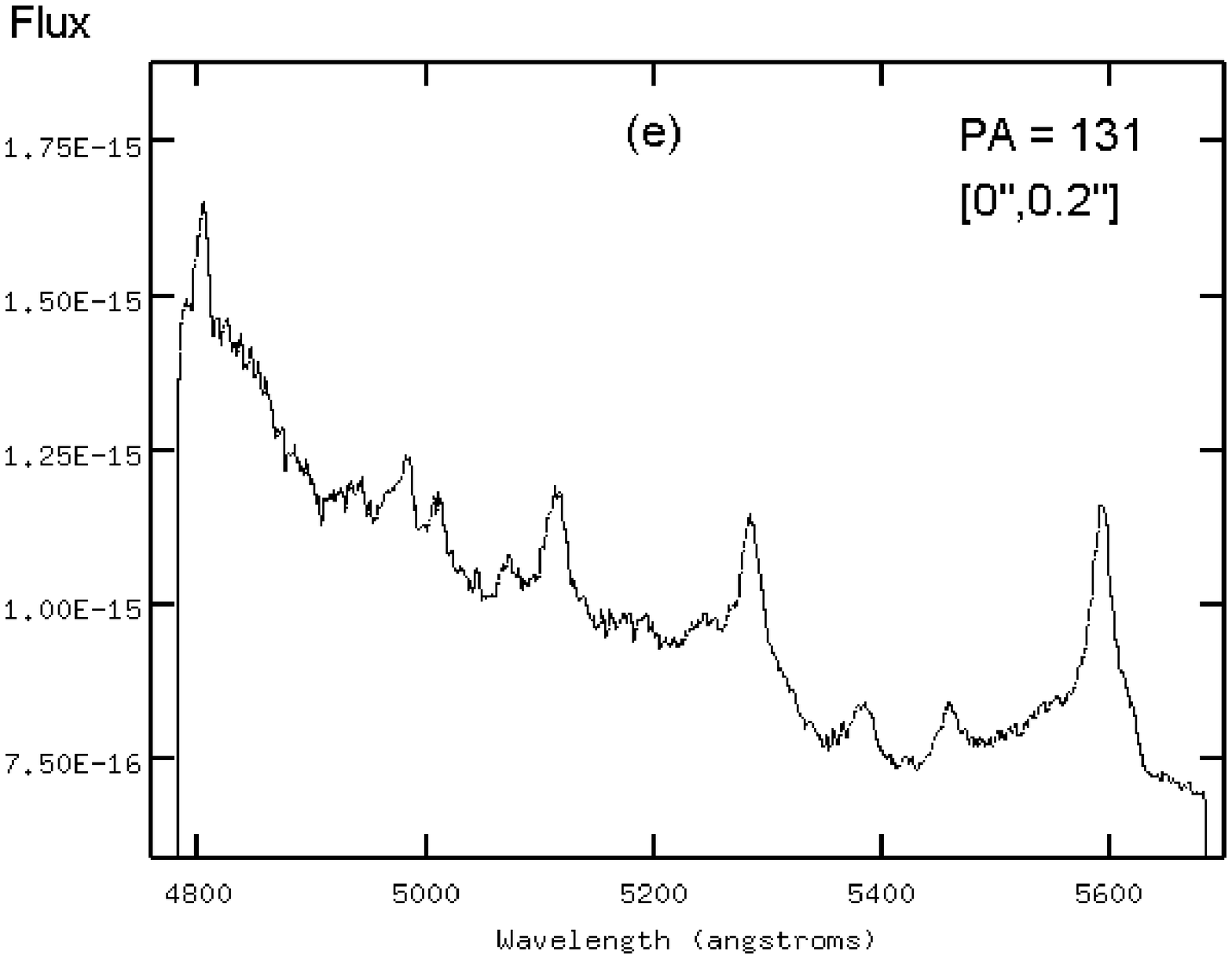}& 
\includegraphics{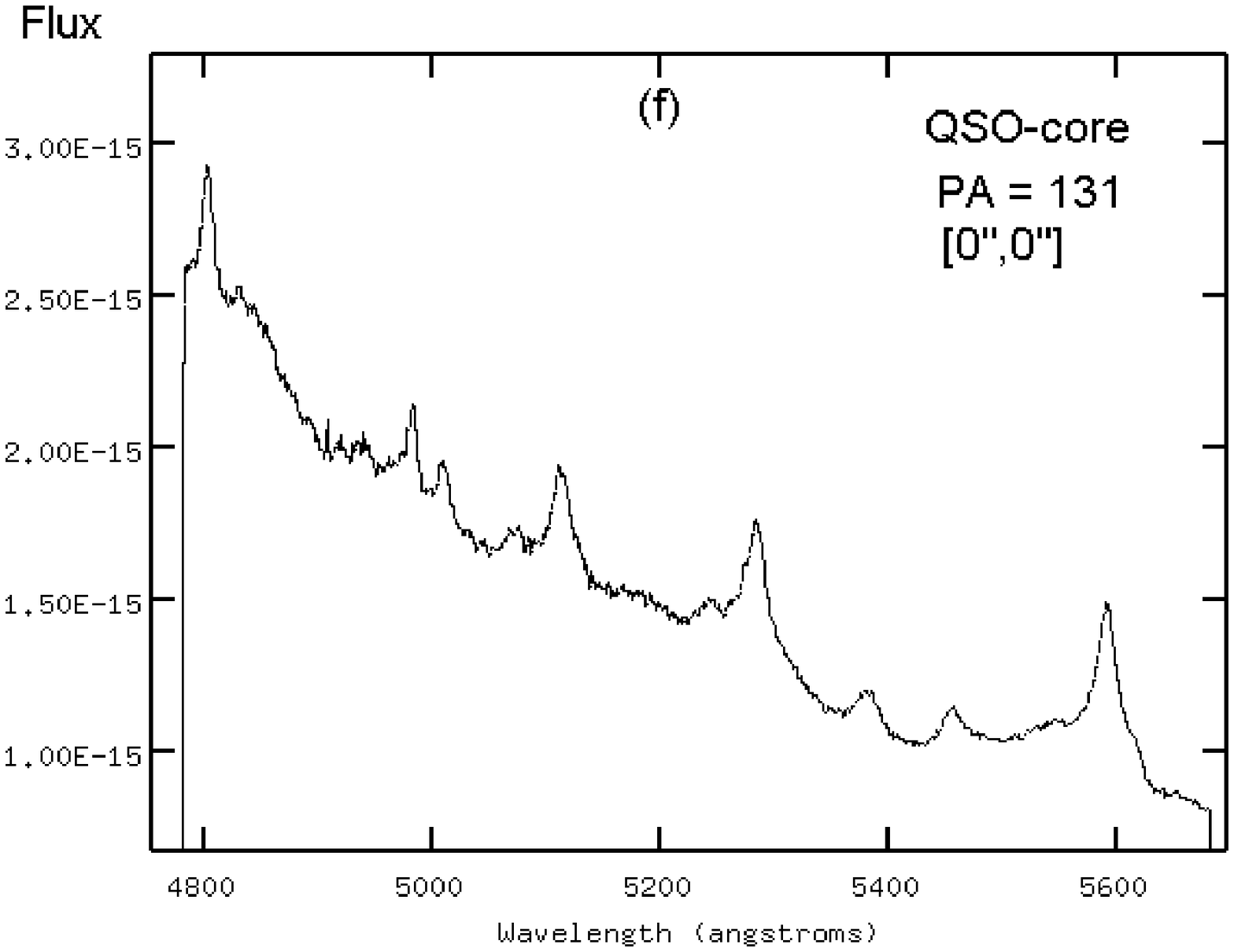} \cr
\end{tabular}
\vspace{8.0 cm}
\caption {
Sequence of individual GMOS-IFU spectra at PA = 131$^{\circ}$ and for the
wavelength range  of [O {\sc ii}]$\lambda$3727--H$\gamma$ 
showing interesting variations.
The offset positions are from the QSO-core, and
in the GMOS  X and Y-axis (the Y-axis was located at PA $=$ 131$^{\circ}$). 
 }
\label{fig20}
\end{figure*}

\clearpage

\begin{figure*}
\vspace{12.0 cm}
\begin{tabular}{cc}
\includegraphics{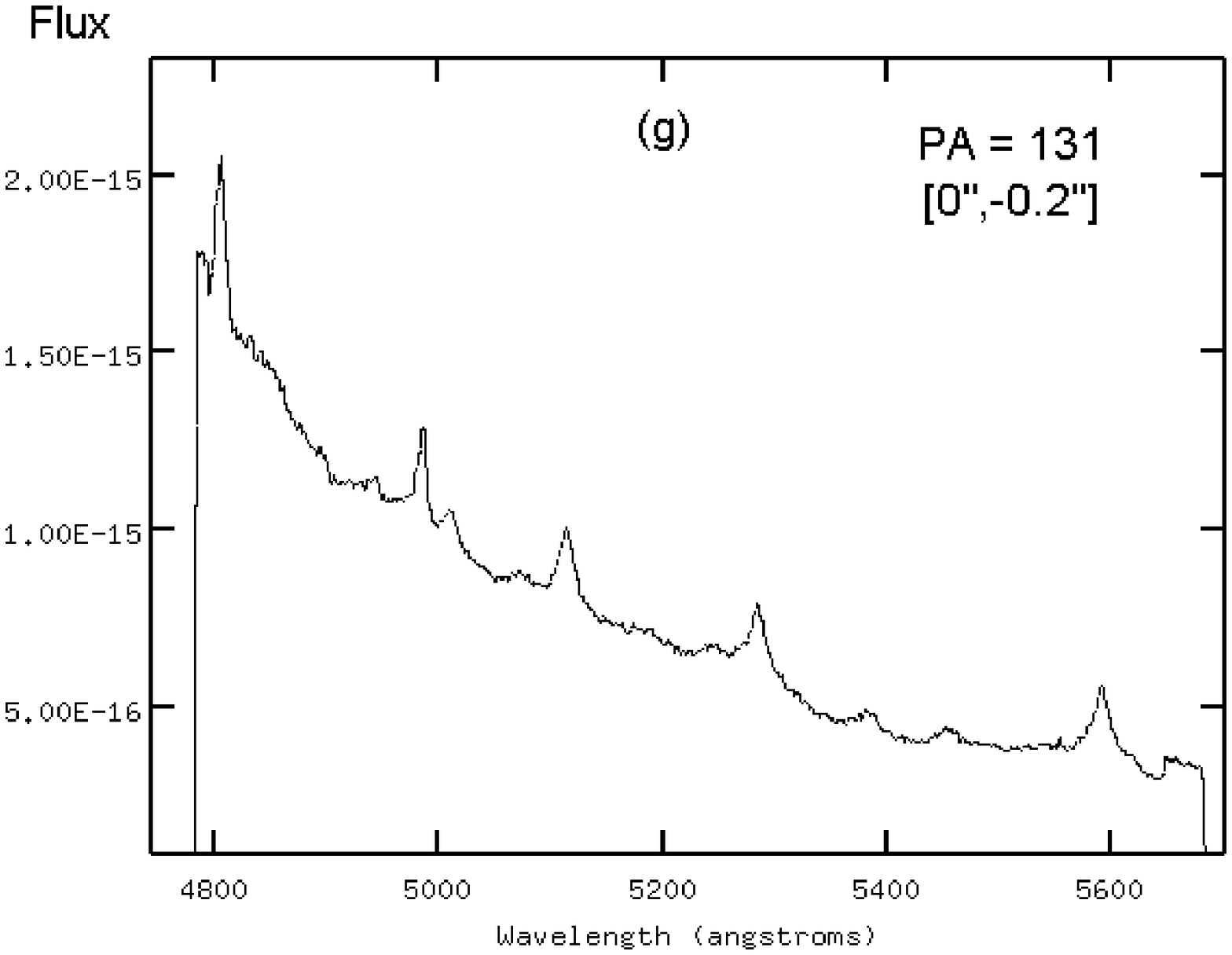}& 
\includegraphics{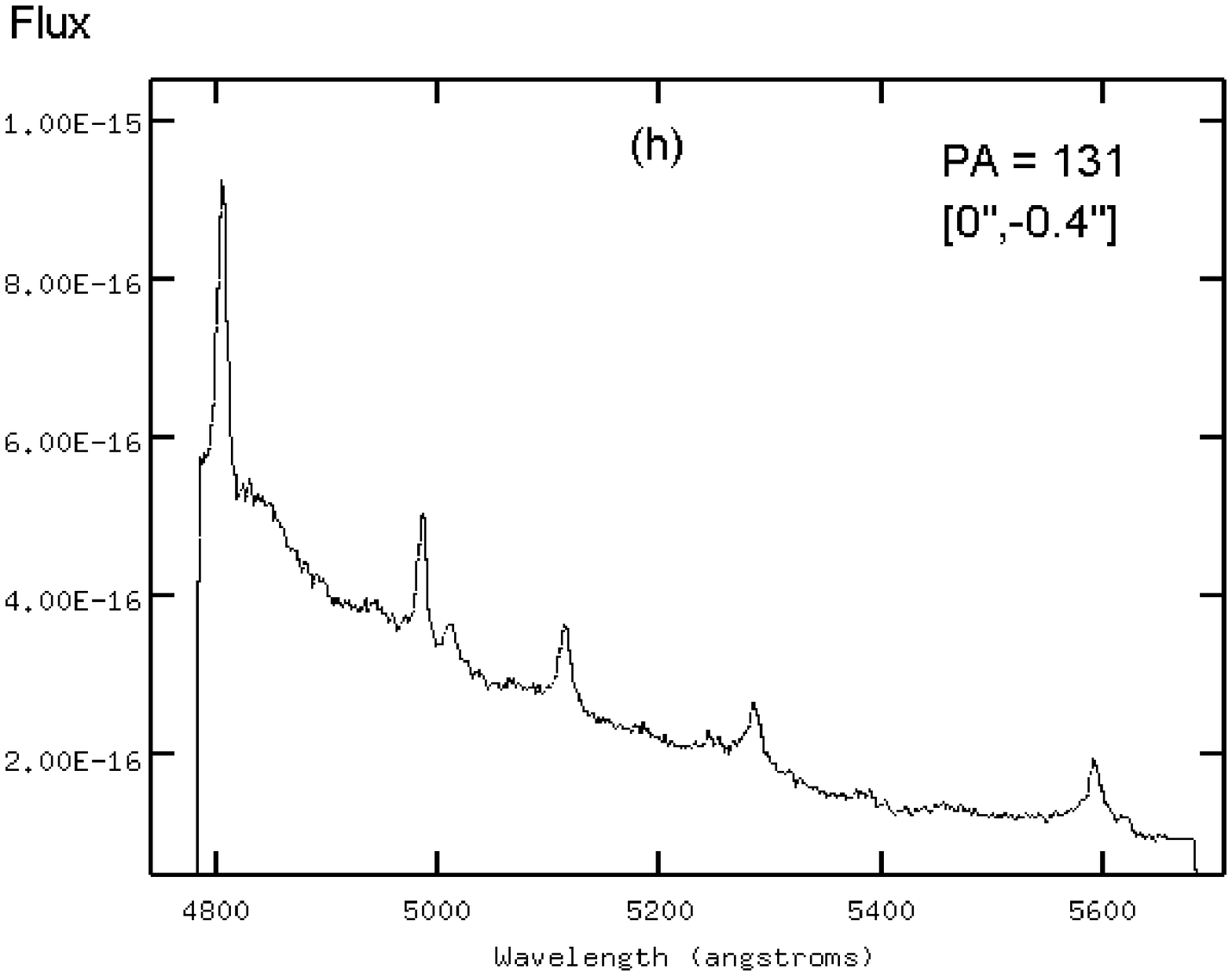} \cr
\includegraphics{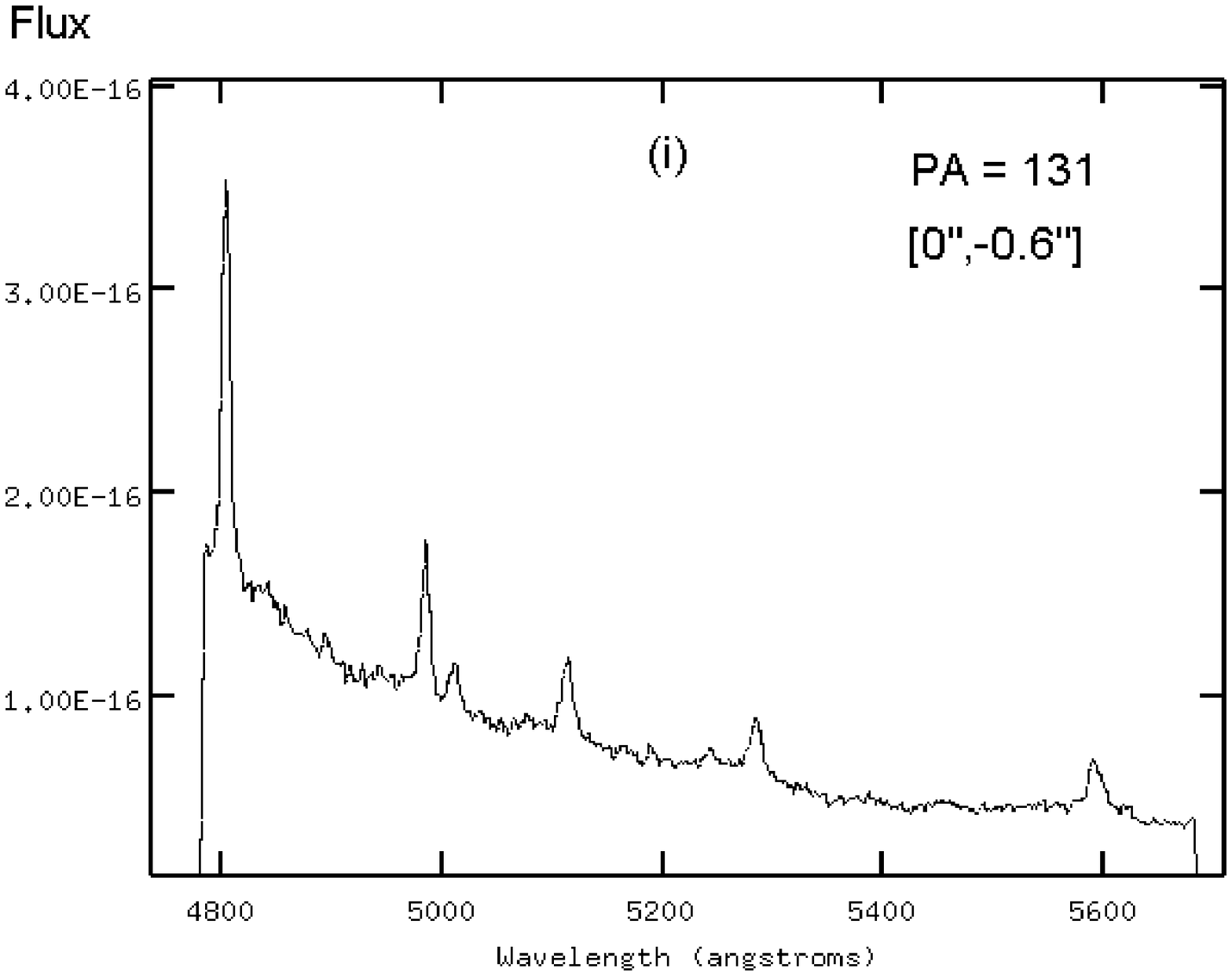}&
\includegraphics{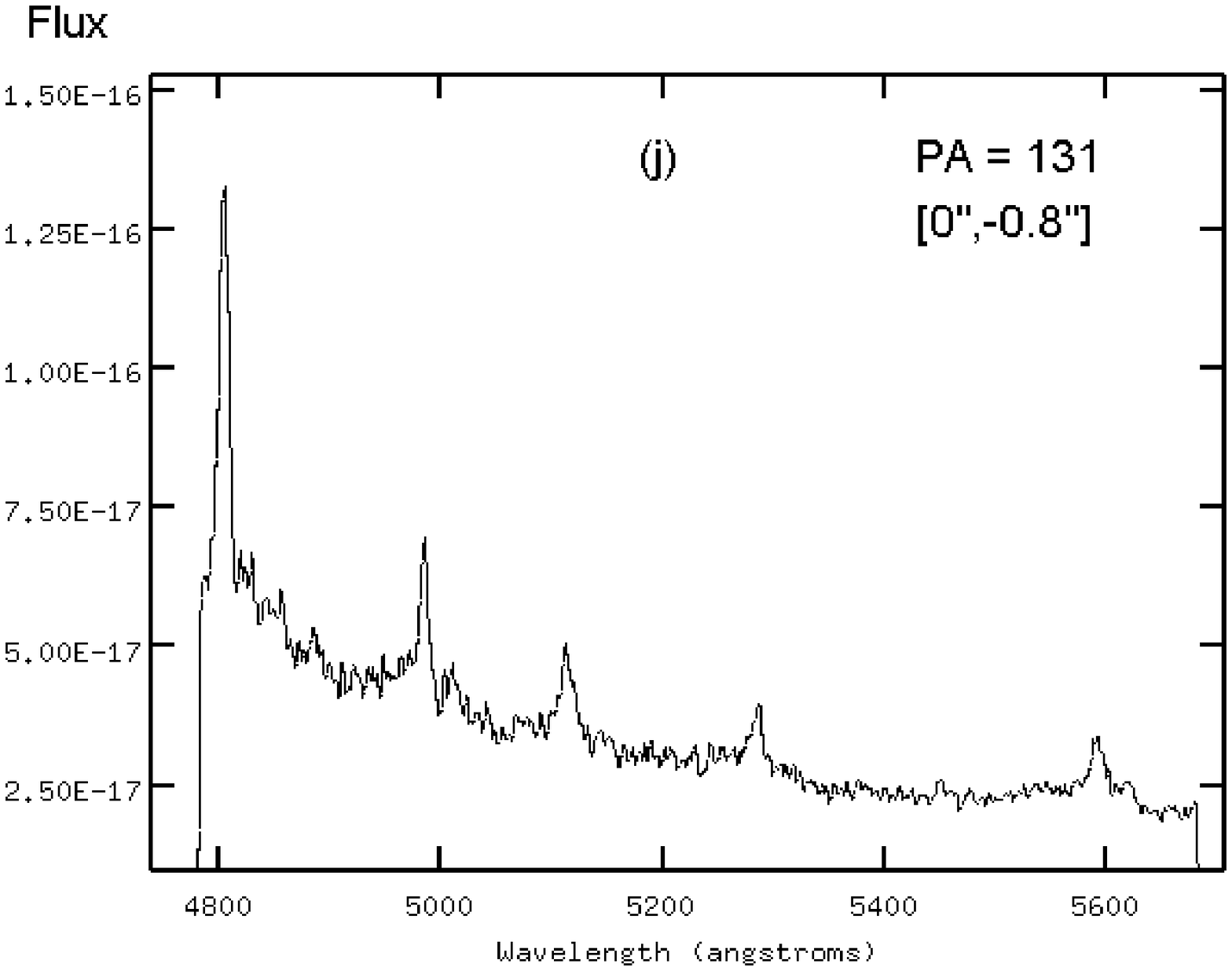} \cr
\includegraphics{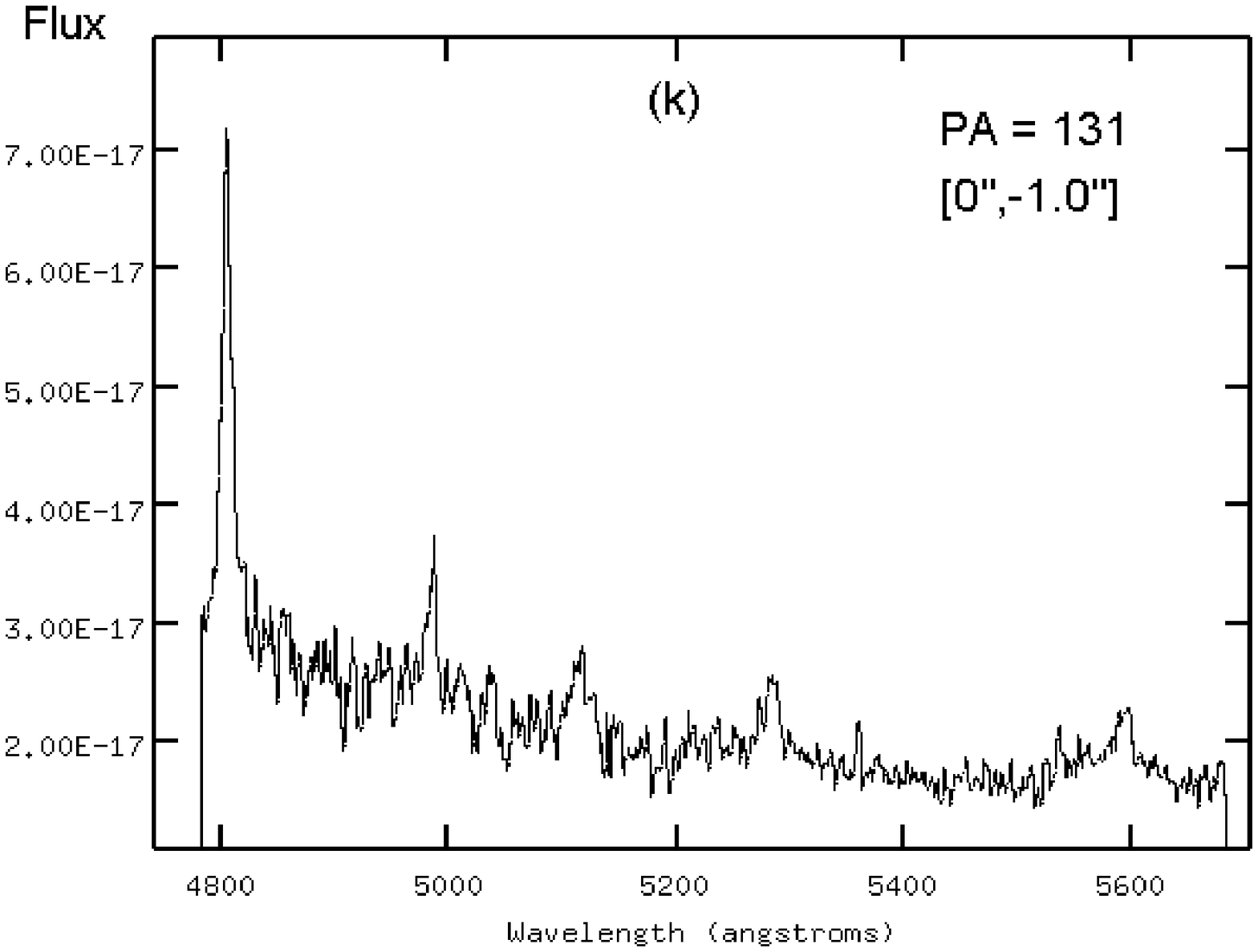}& 
\includegraphics{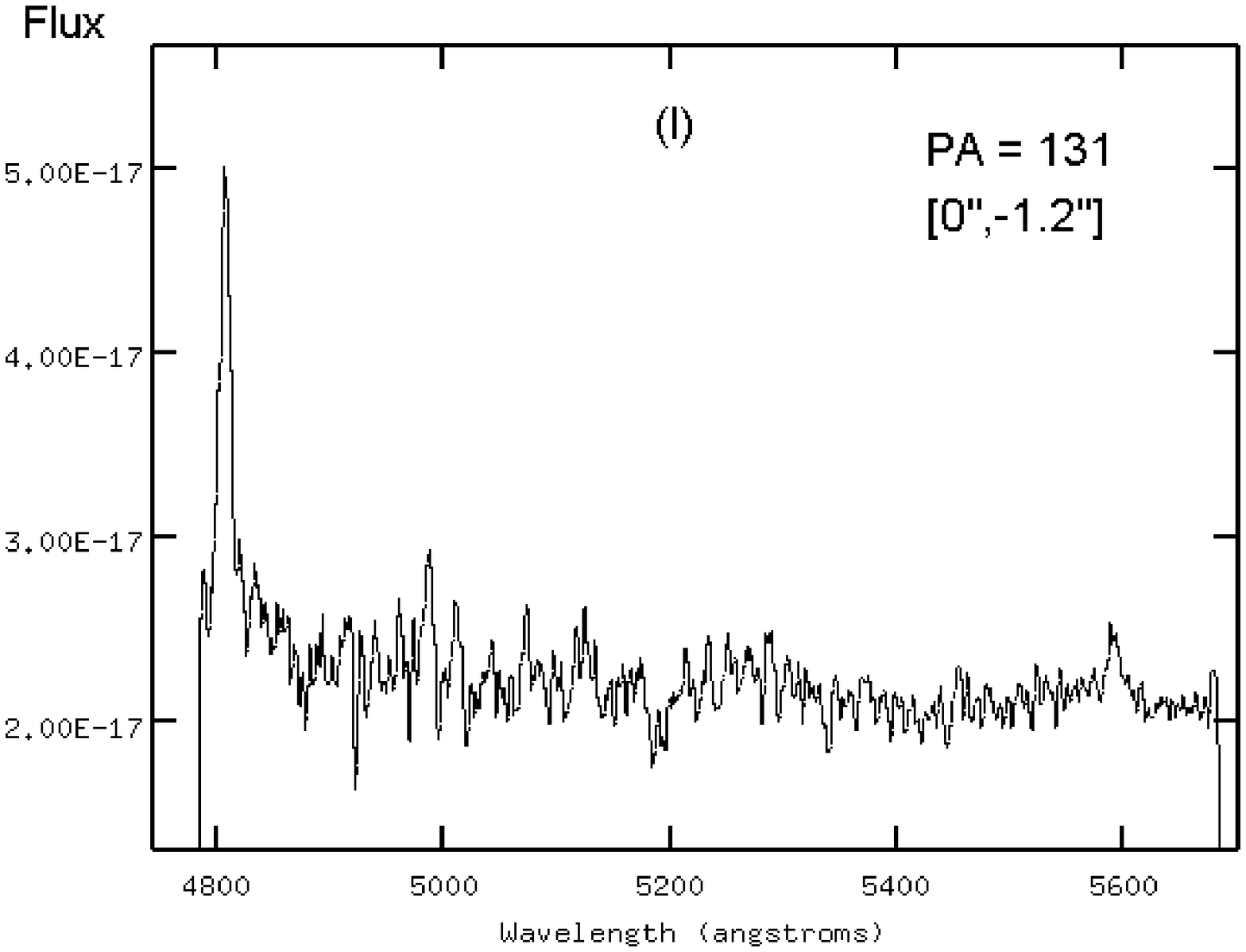} \cr
\end{tabular}
\vspace{8.0 cm}
\addtocounter{figure}{-1}
\caption {Contin.
 }
\label{fig20c}
\end{figure*}


\clearpage

\begin{figure*}
\vspace{12.0 cm}
\begin{tabular}{cc}
\includegraphics{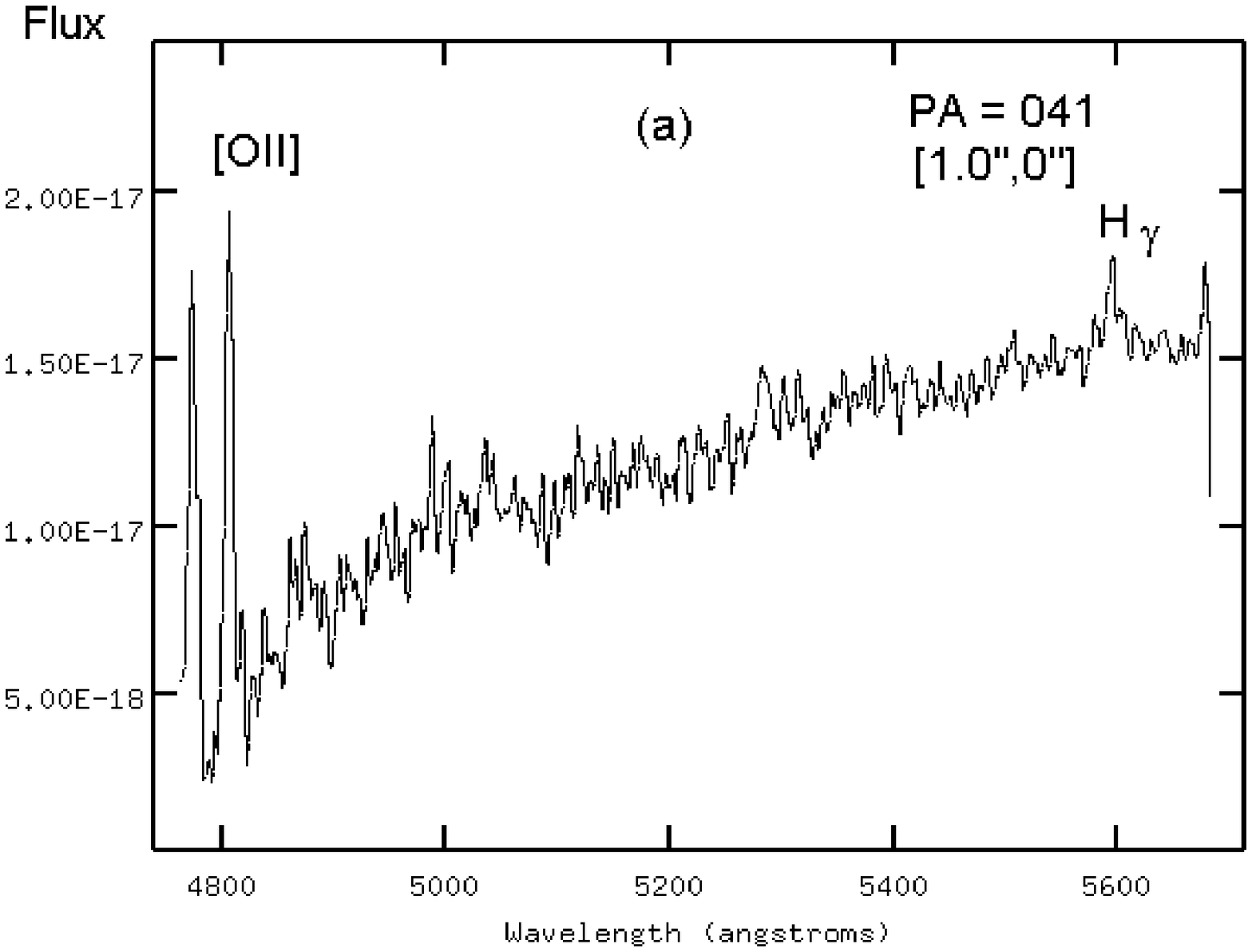}& 
\includegraphics{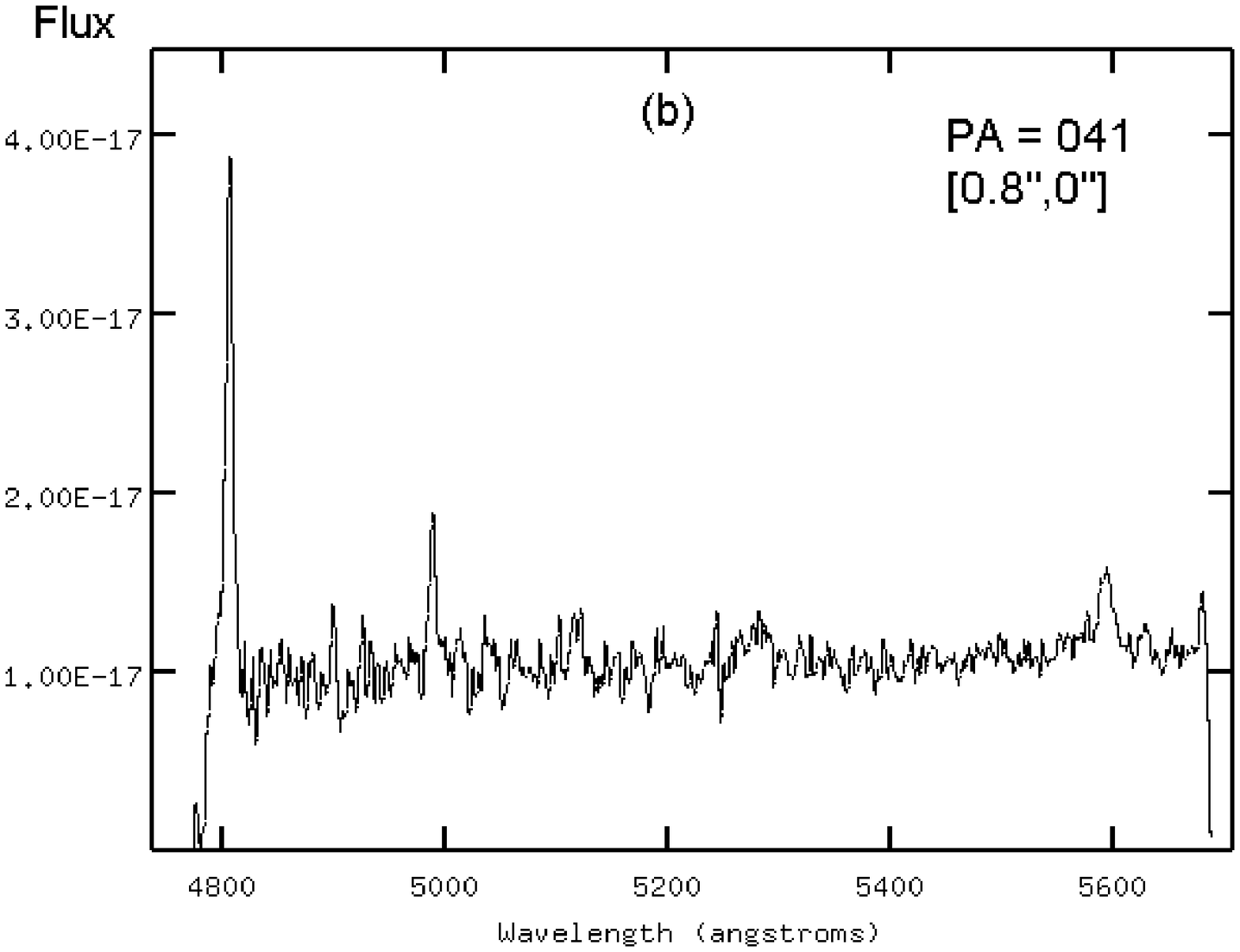} \cr
\includegraphics{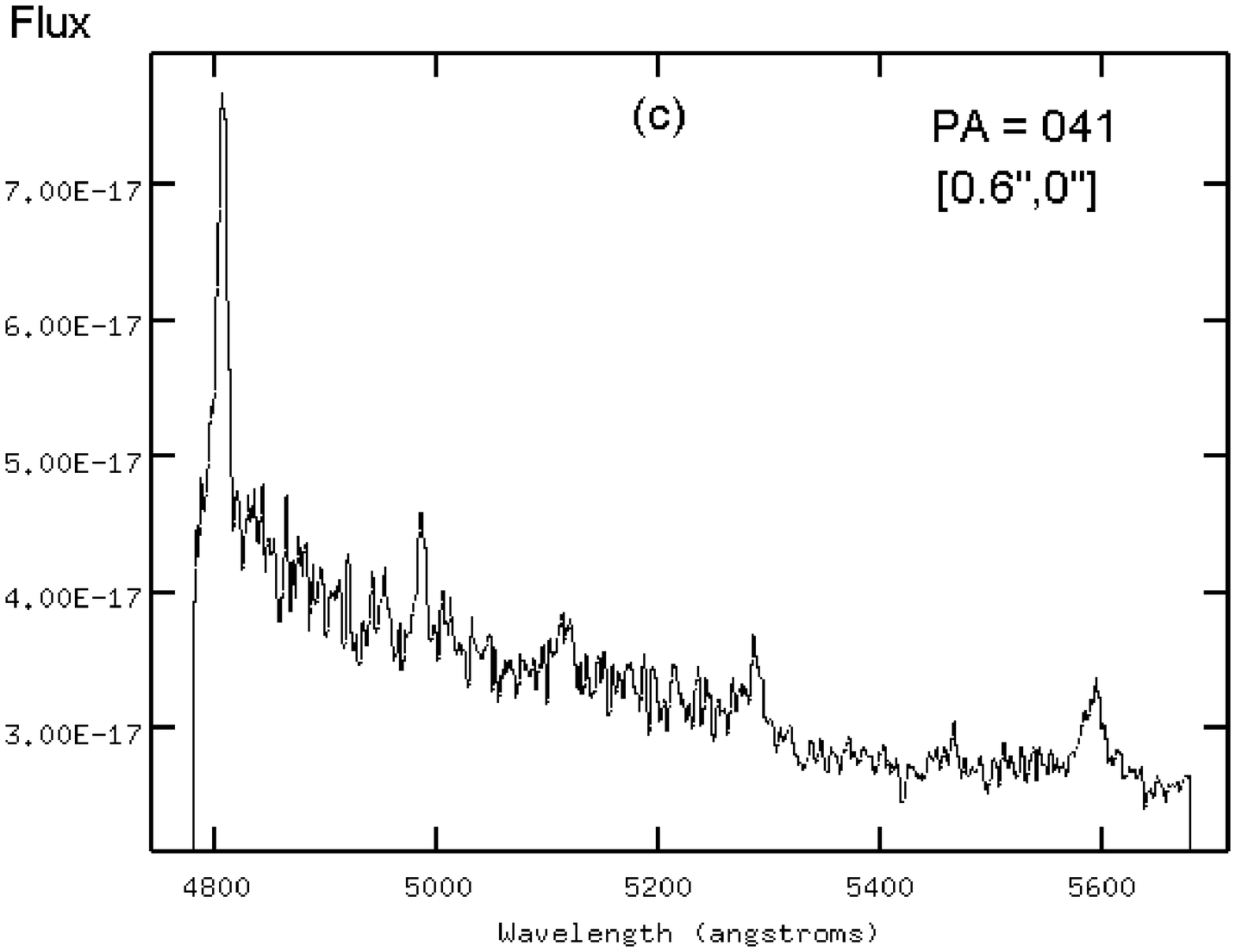}&
\includegraphics{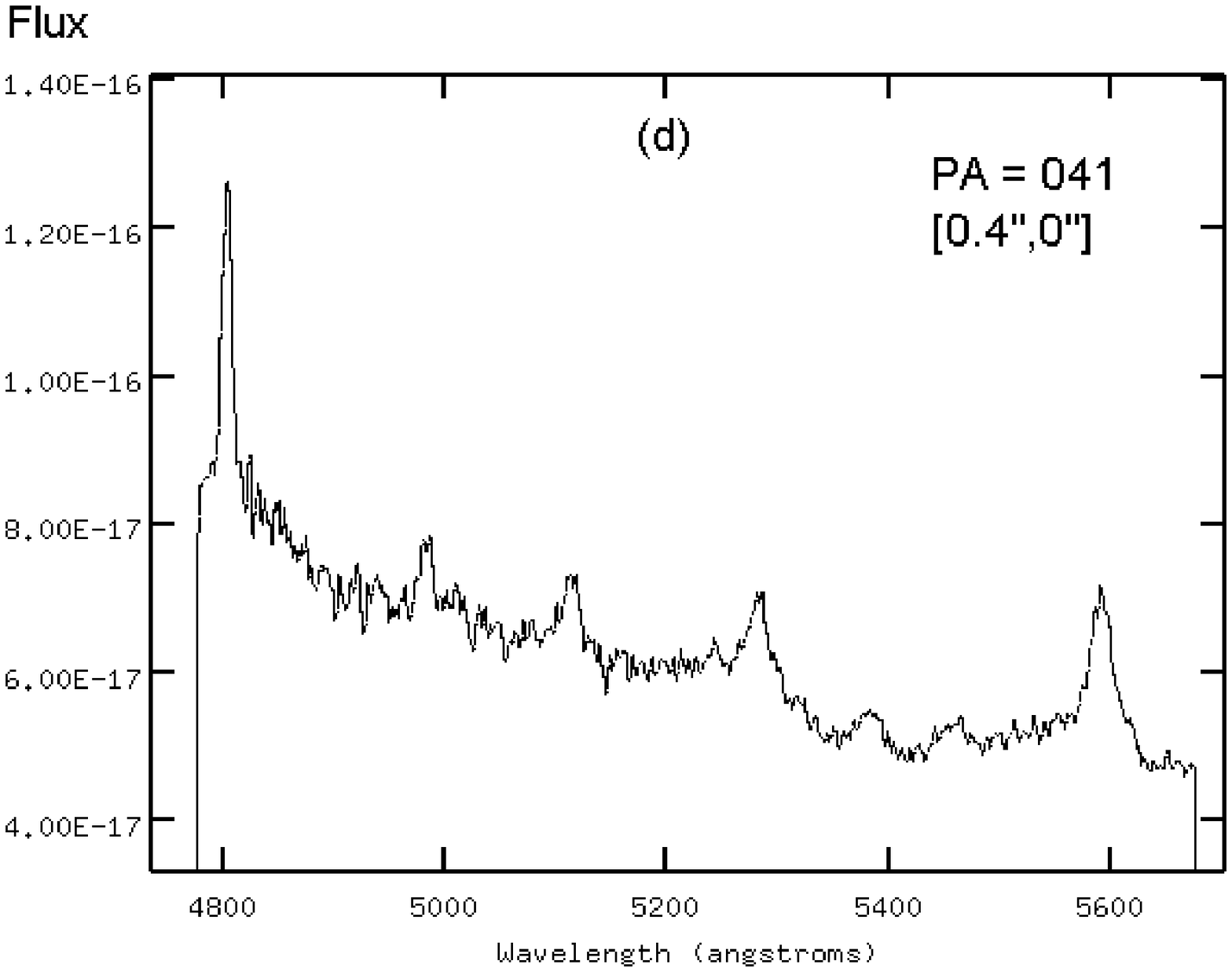} \cr
\includegraphics{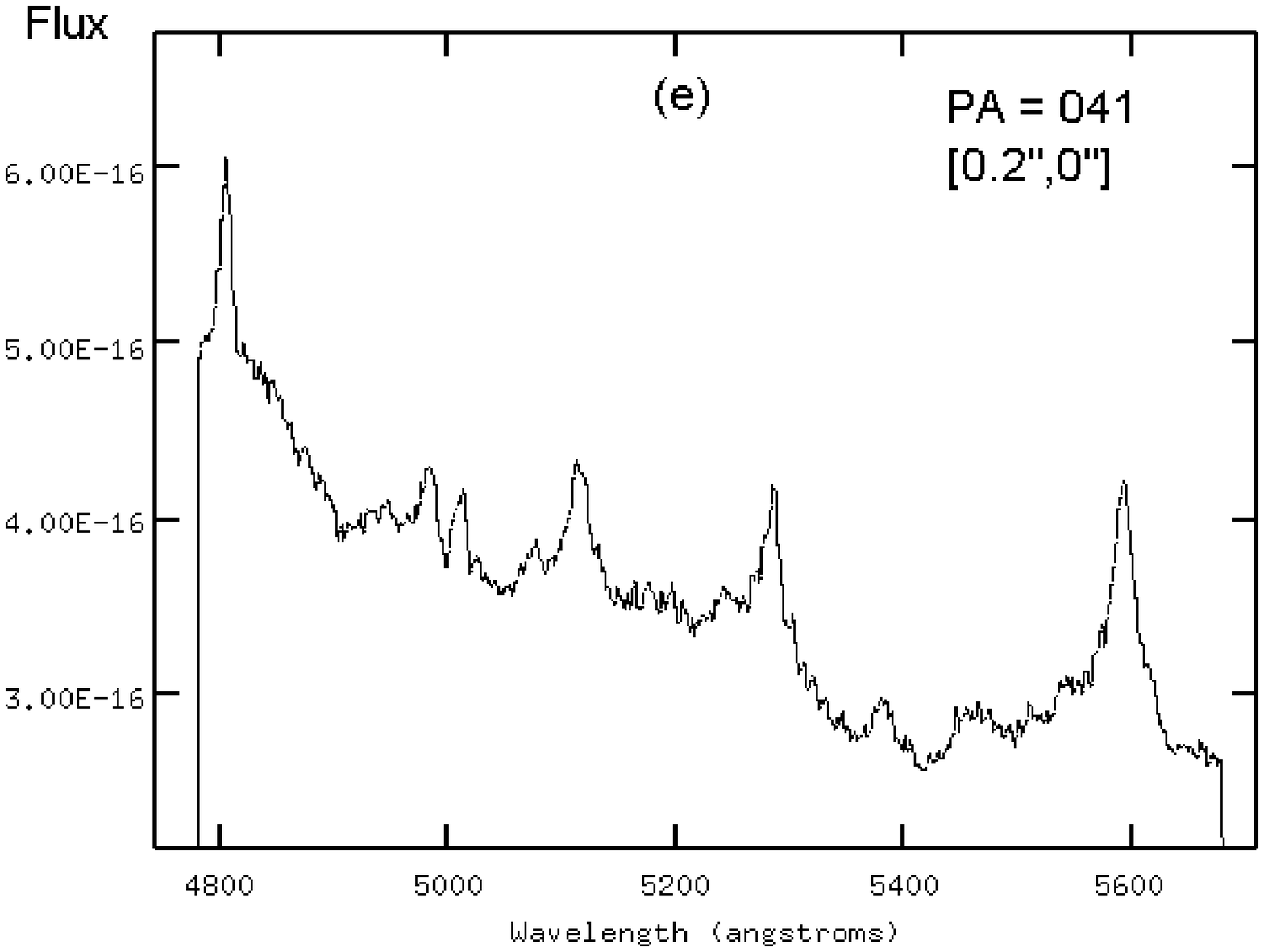}& 
\includegraphics{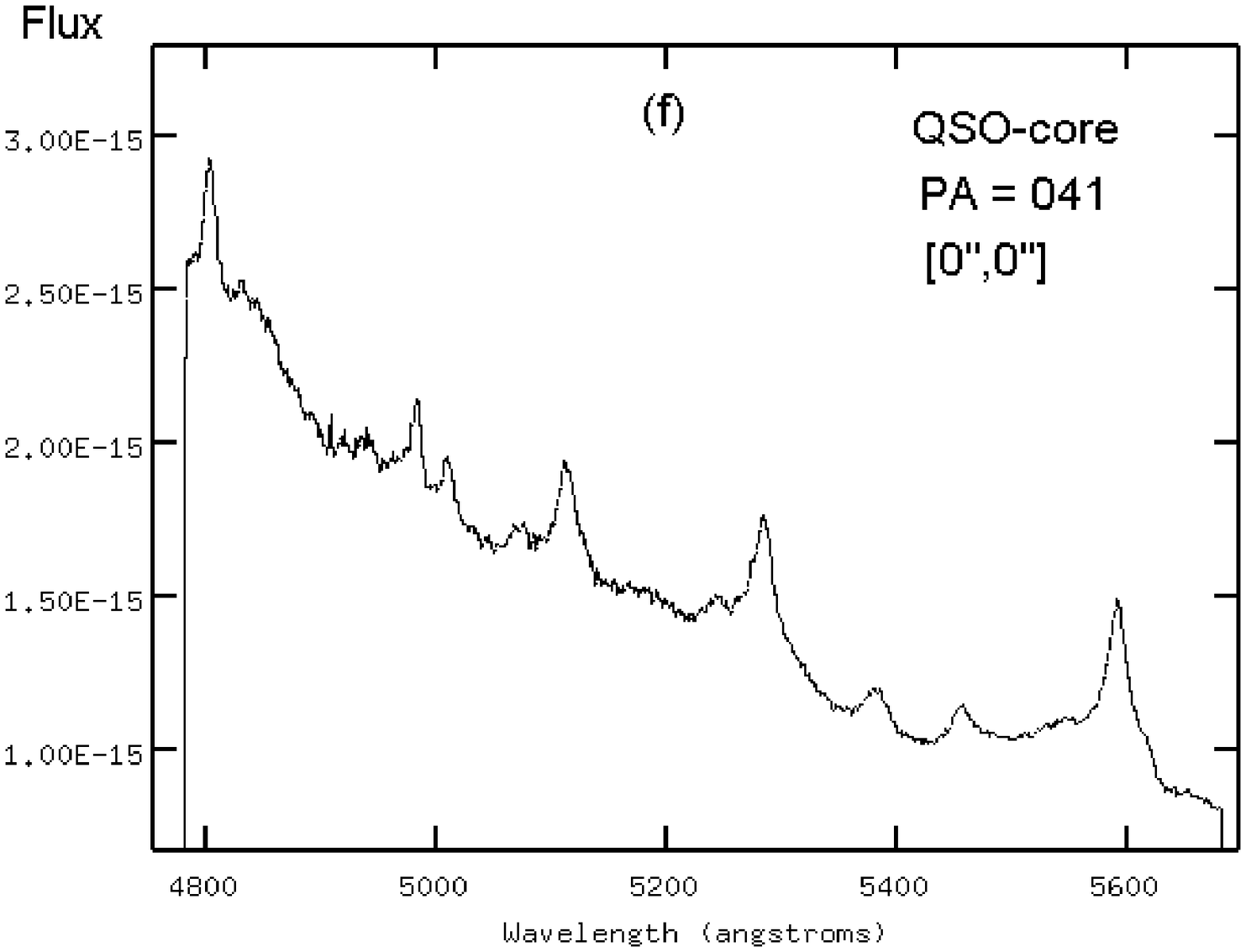} \cr
\end{tabular}
\vspace{8.0 cm}
\caption {
Sequence of individual GMOS-IFU spectra at PA $=$ 041$^{\circ}$ and for the
wavelength range of [O {\sc ii}]$\lambda$3727--H$\gamma$. 
The offset positions are from the QSO-core, and
in the GMOS  X and Y-axis (the Y-axis was located at PA $=$ 131$^{\circ}$). 
 }
\label{fig21}
\end{figure*}

\clearpage

\begin{figure*}
\vspace{12.0 cm}
\begin{tabular}{cc}
\includegraphics{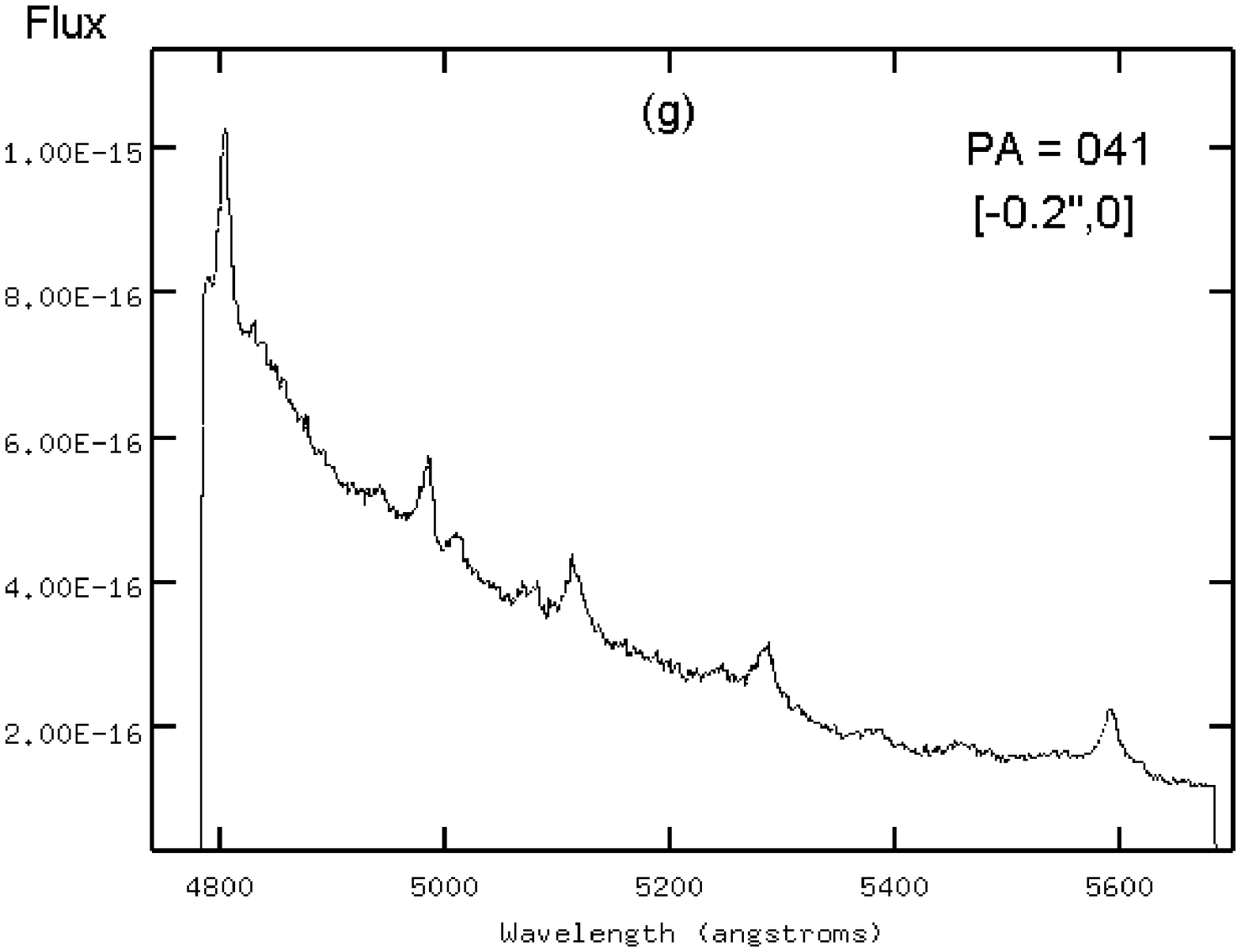}& 
\includegraphics{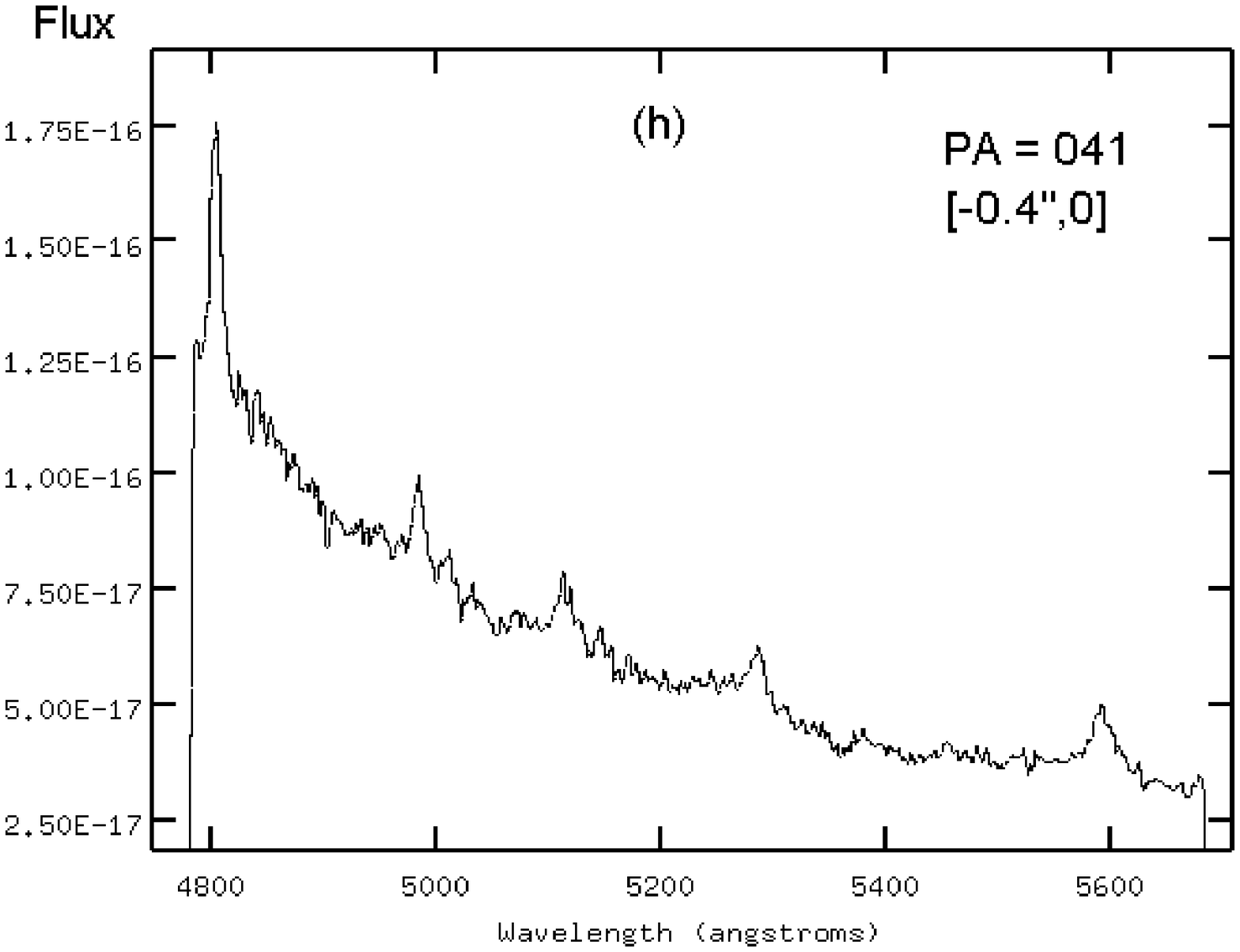} \cr
\includegraphics{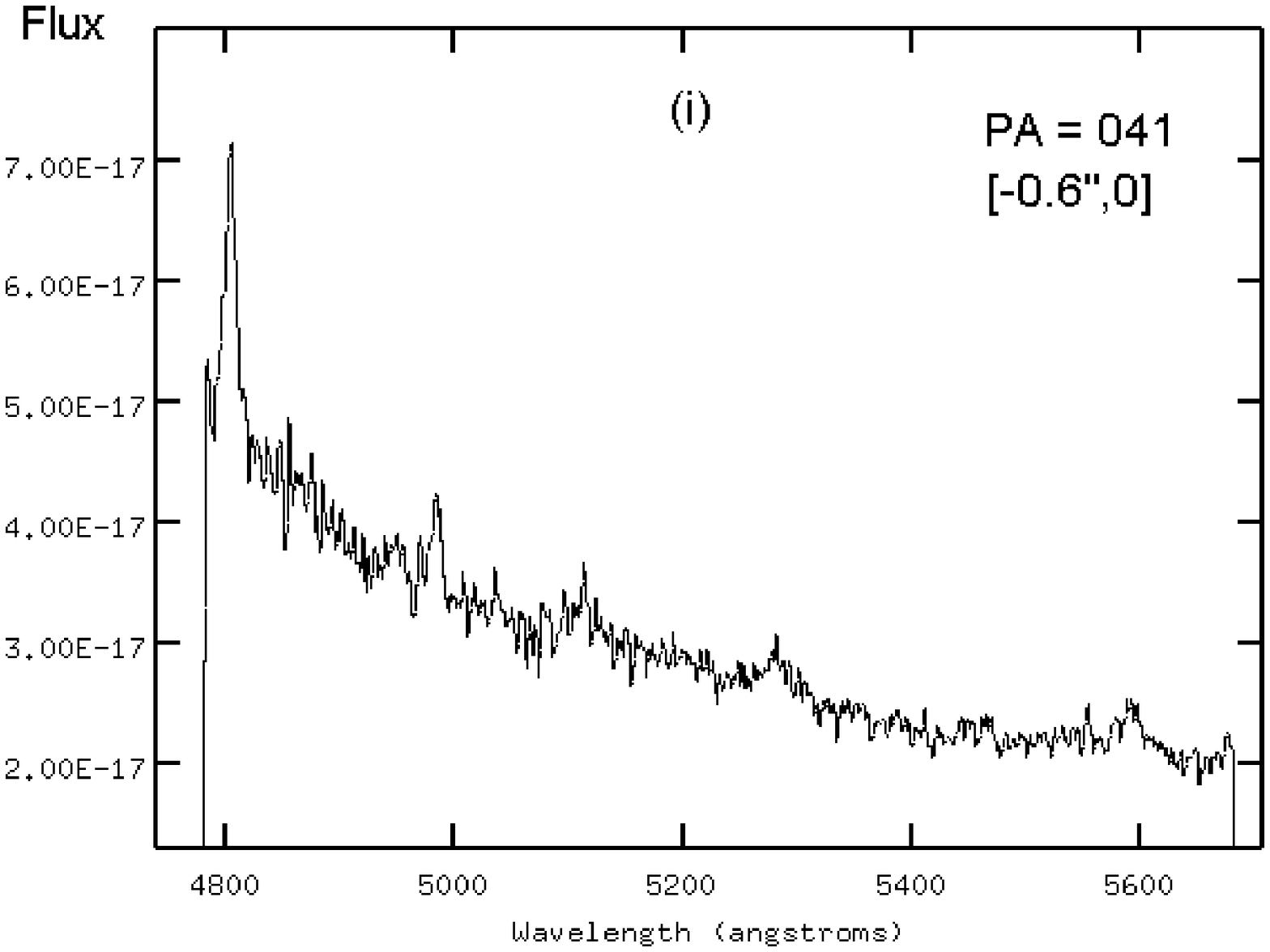}&
\includegraphics{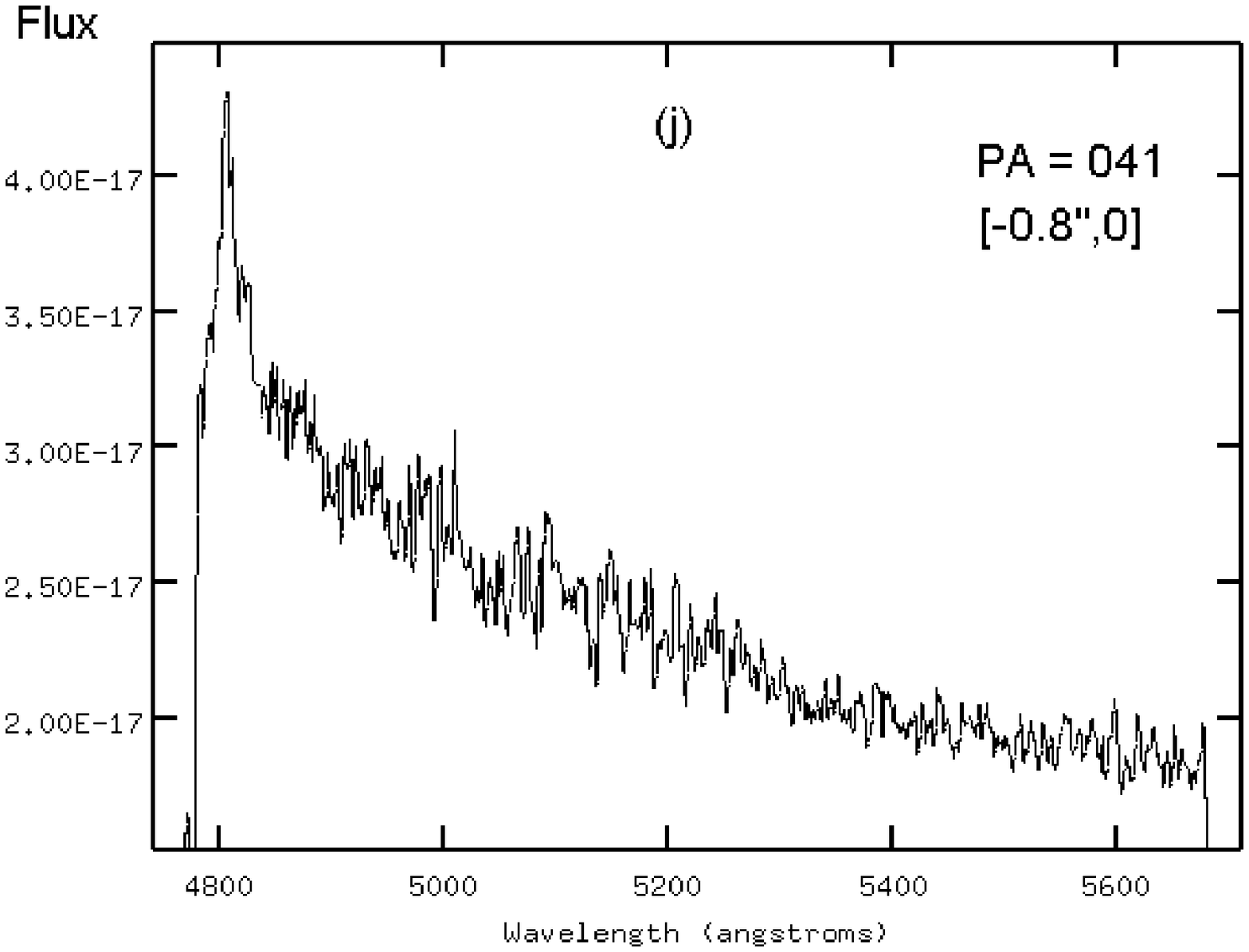} \cr
\end{tabular}
\vspace{8.0 cm}
\addtocounter{figure}{-1}
\caption {Contin.
 }
\label{fig21c}
\end{figure*}


\clearpage

\begin{figure*}
\vspace{12.0 cm}
\begin{tabular}{cc}
\includegraphics{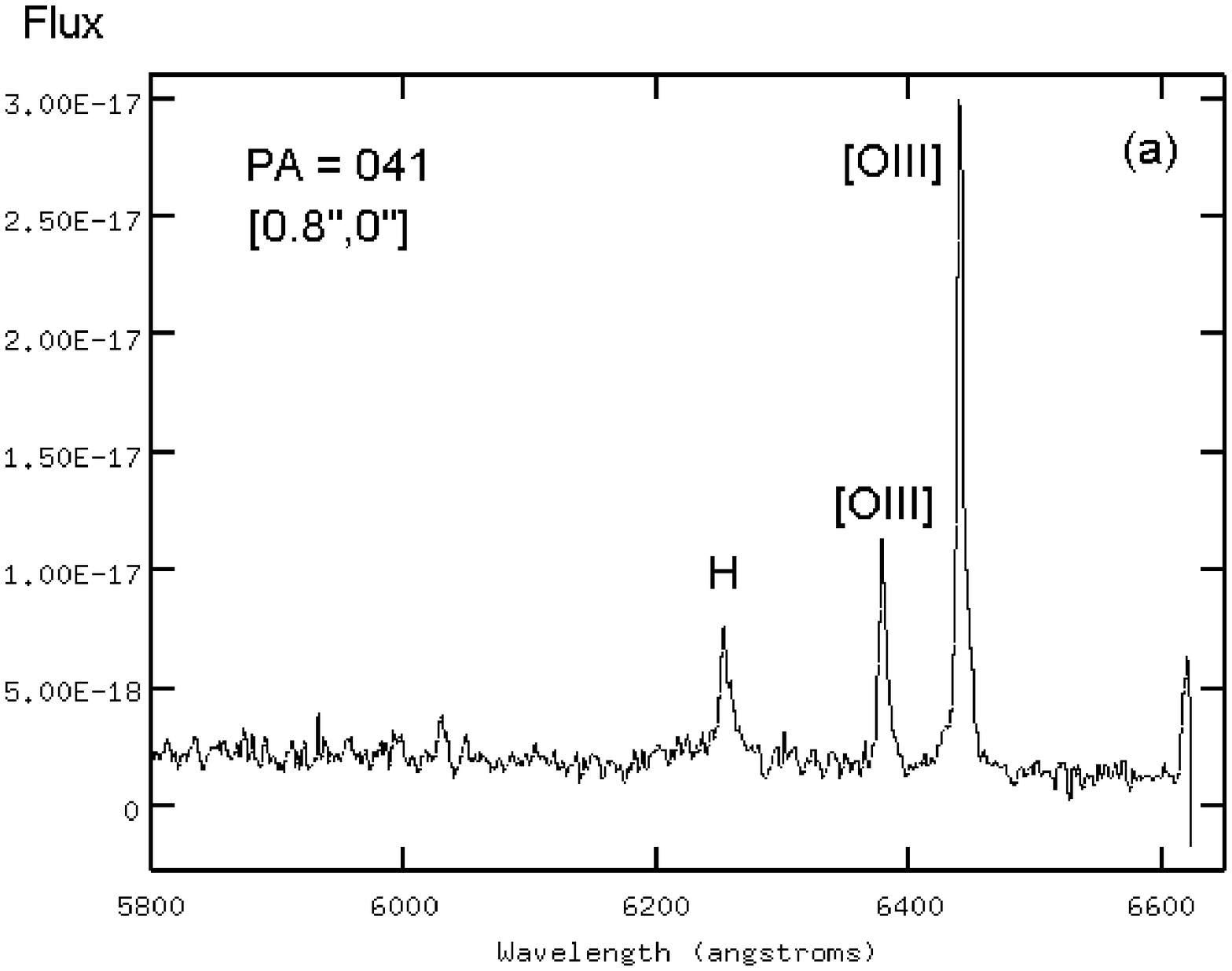}& 
\includegraphics{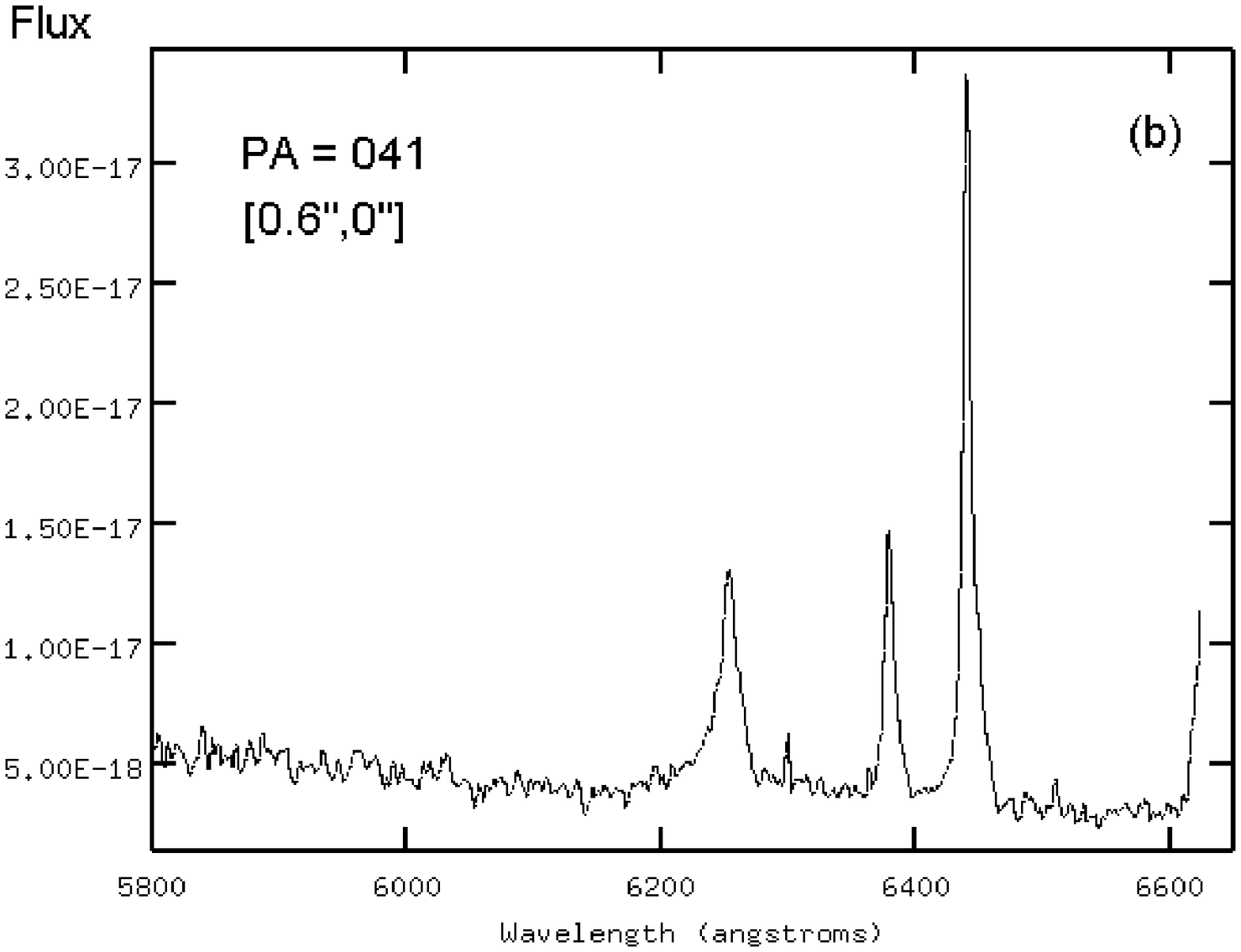} \cr
\includegraphics{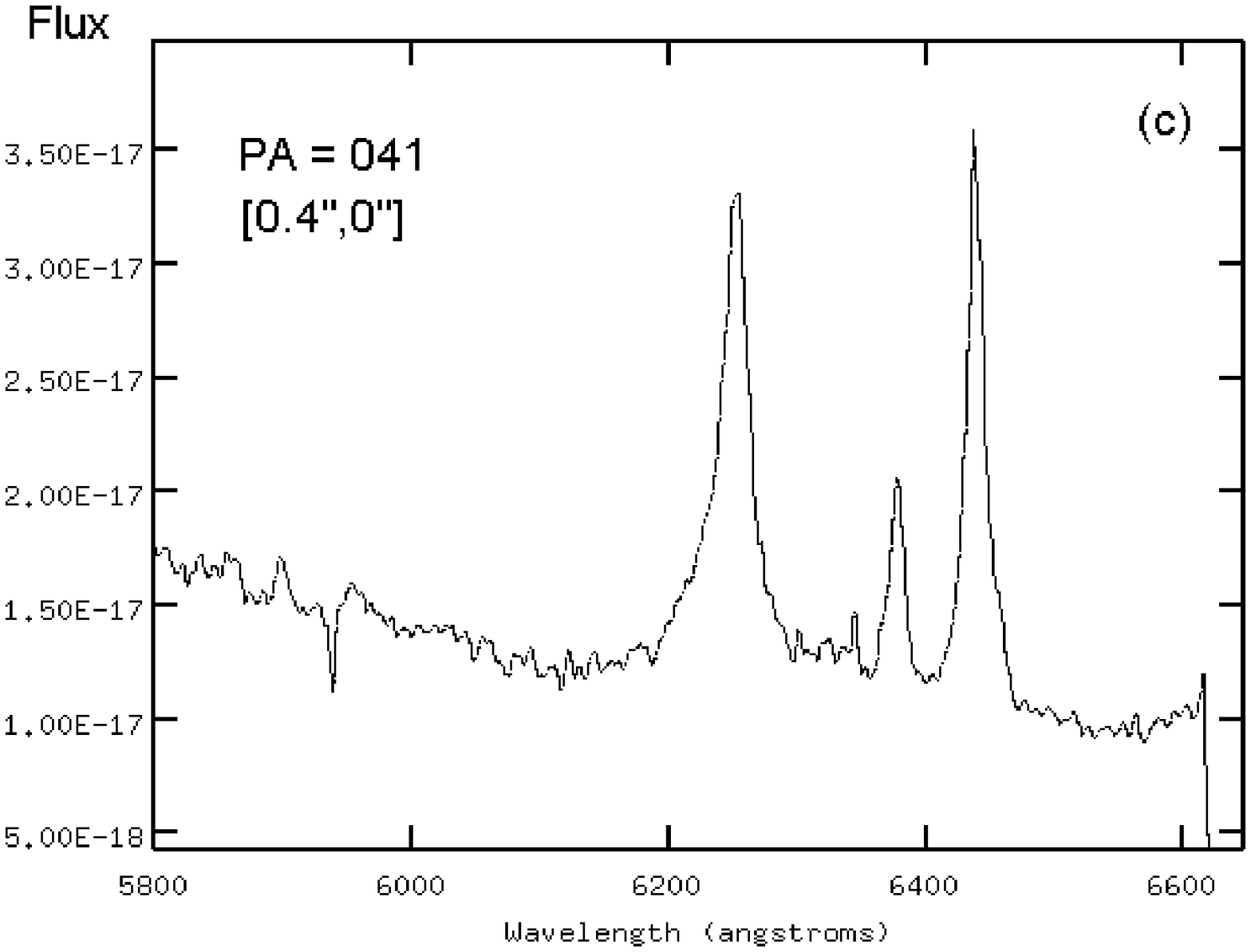}&
\includegraphics{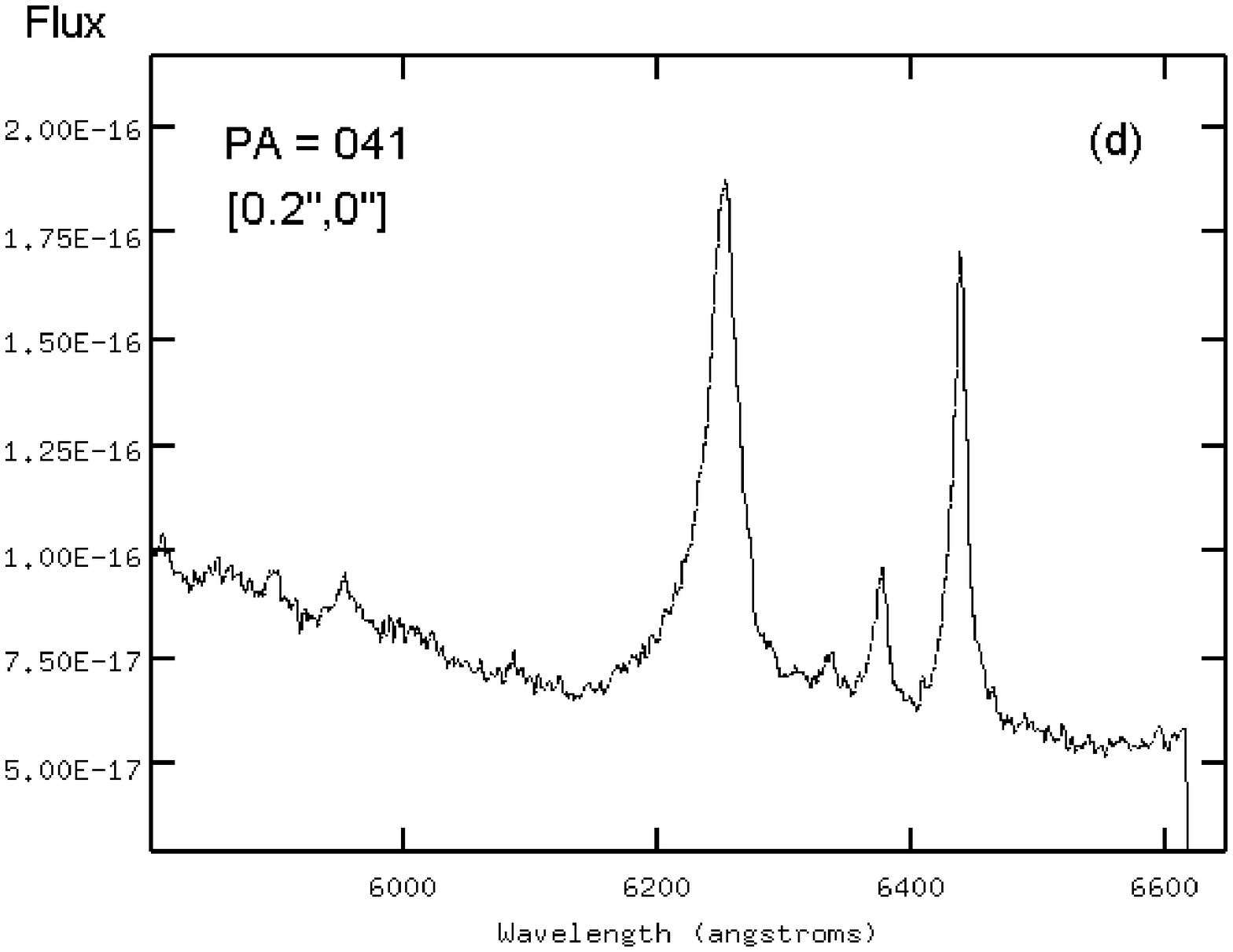} \cr
\includegraphics{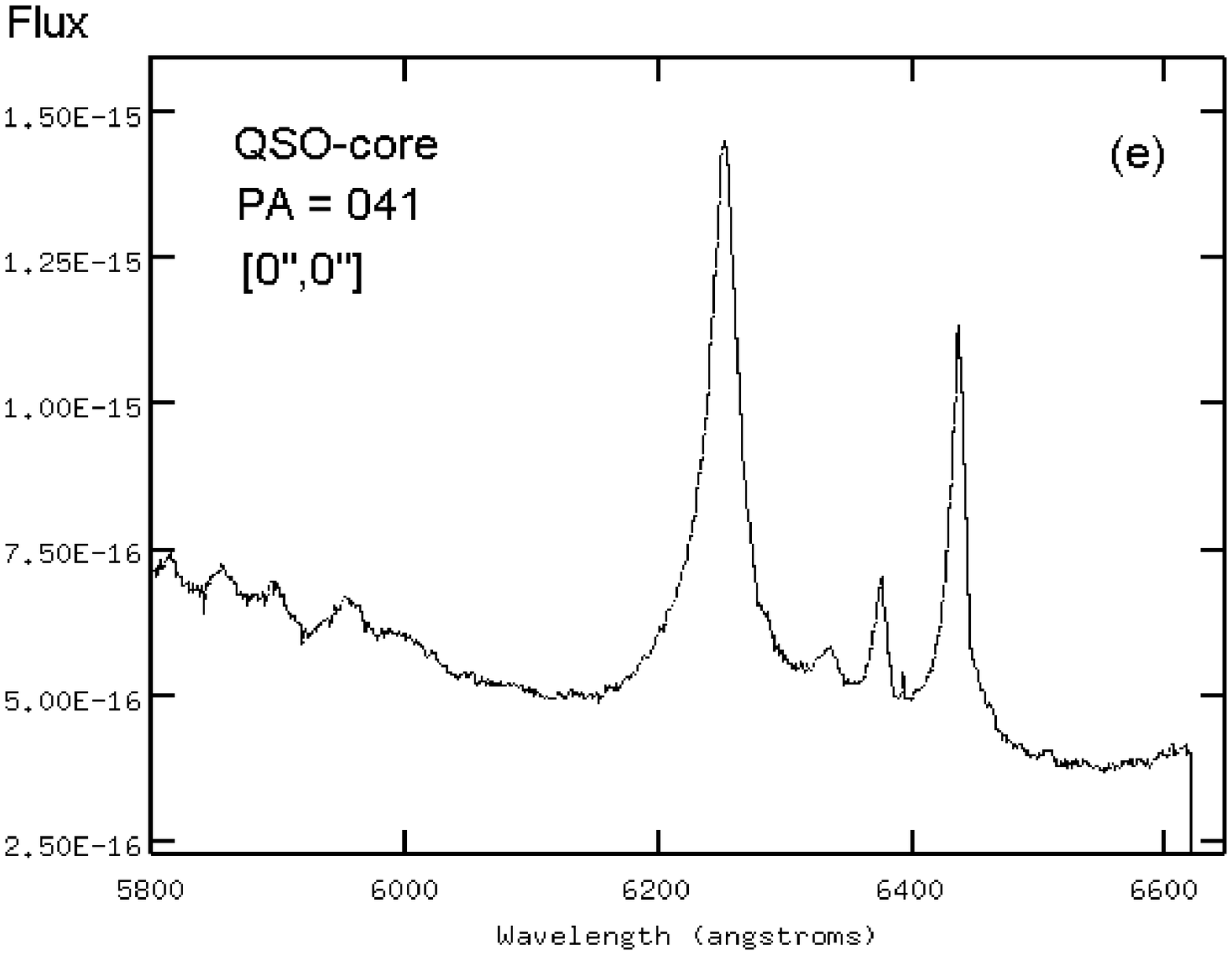}& 
\includegraphics{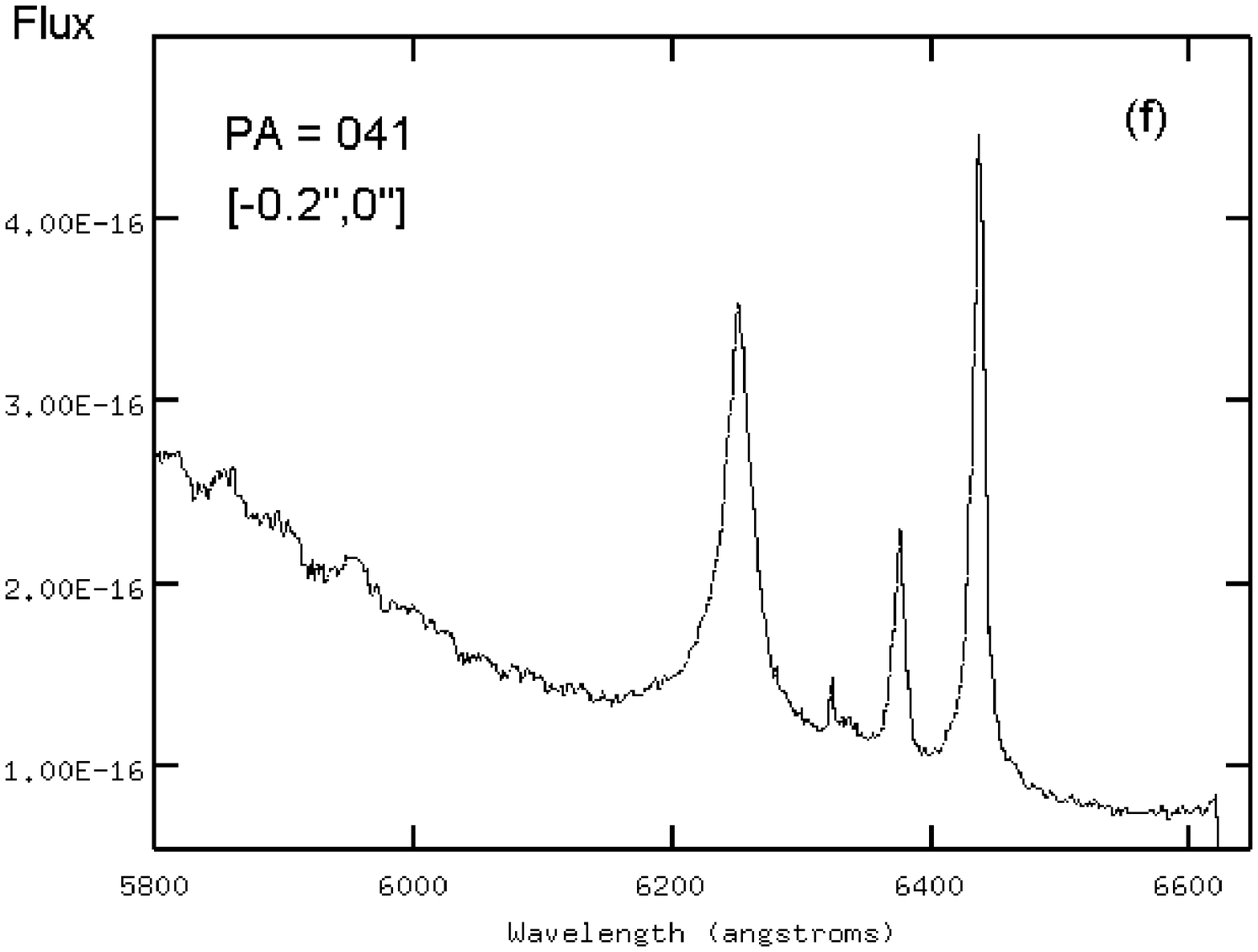} \cr
\end{tabular}
\vspace{8.0 cm}
\caption {
Sequence of individual GMOS spectra at PA $=$ 041$^{\circ}$ and for the
wavelength range of H$\beta$ + [O {\sc iii}]$\lambda$5007 + Fe {\sc ii}. 
The offset positions are from the QSO-core, and
in the GMOS  X and Y-axis (the Y-axis was located at PA $=$ 131$^{\circ}$). 
 }
\label{fig22}
\end{figure*}

\clearpage

\begin{figure*}
\vspace{12.0 cm}
\begin{tabular}{cc}
\includegraphics{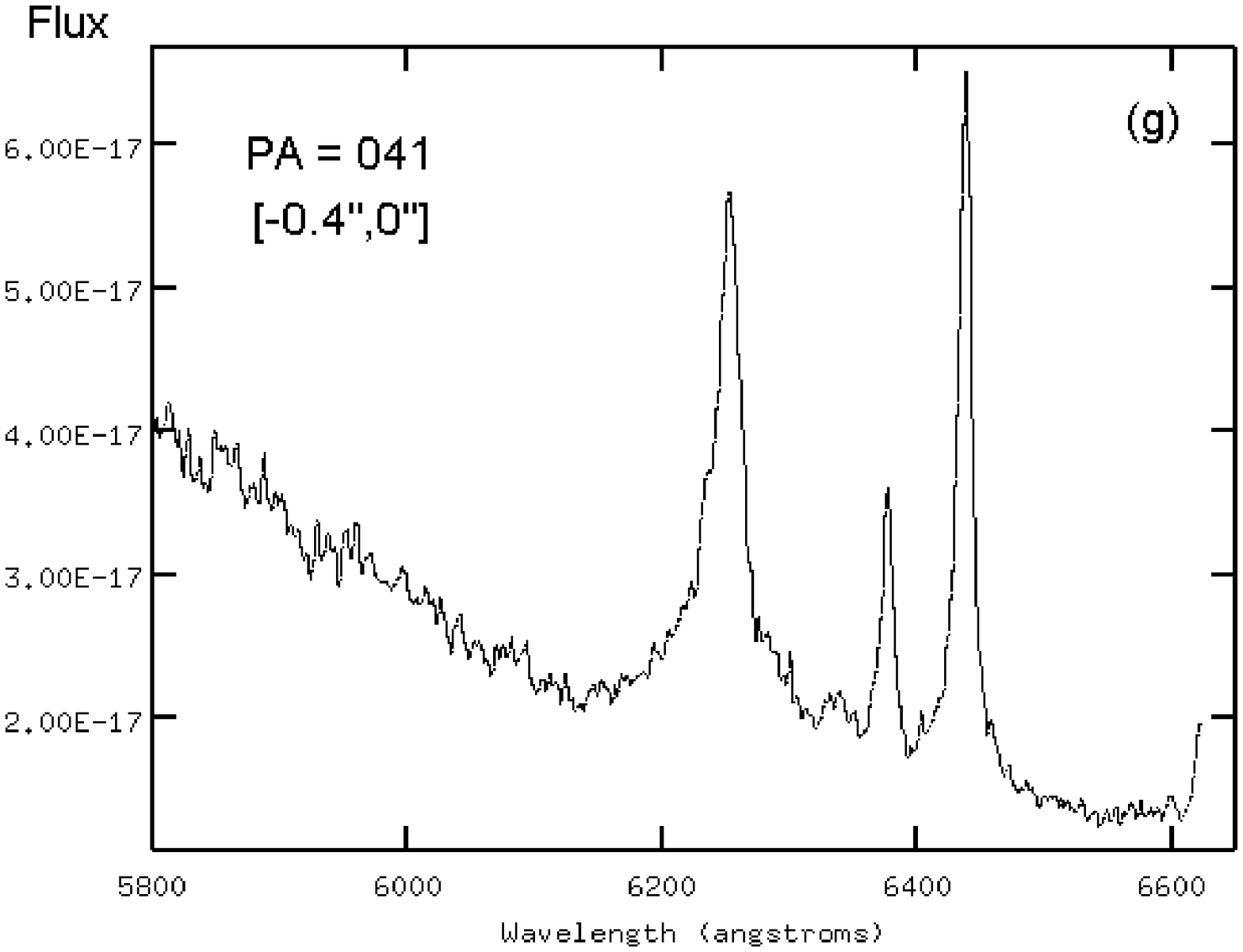}& 
\includegraphics{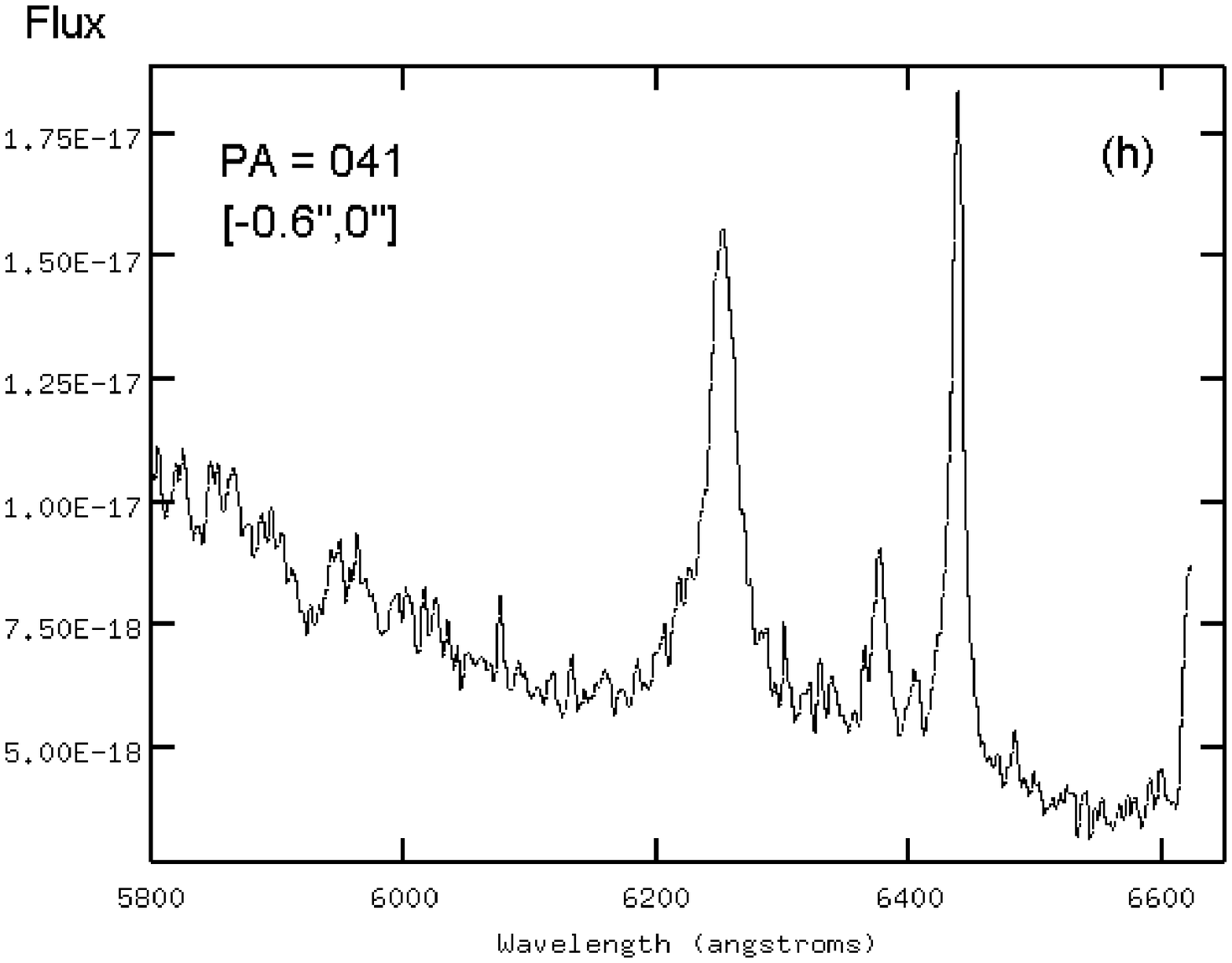} \cr
\includegraphics{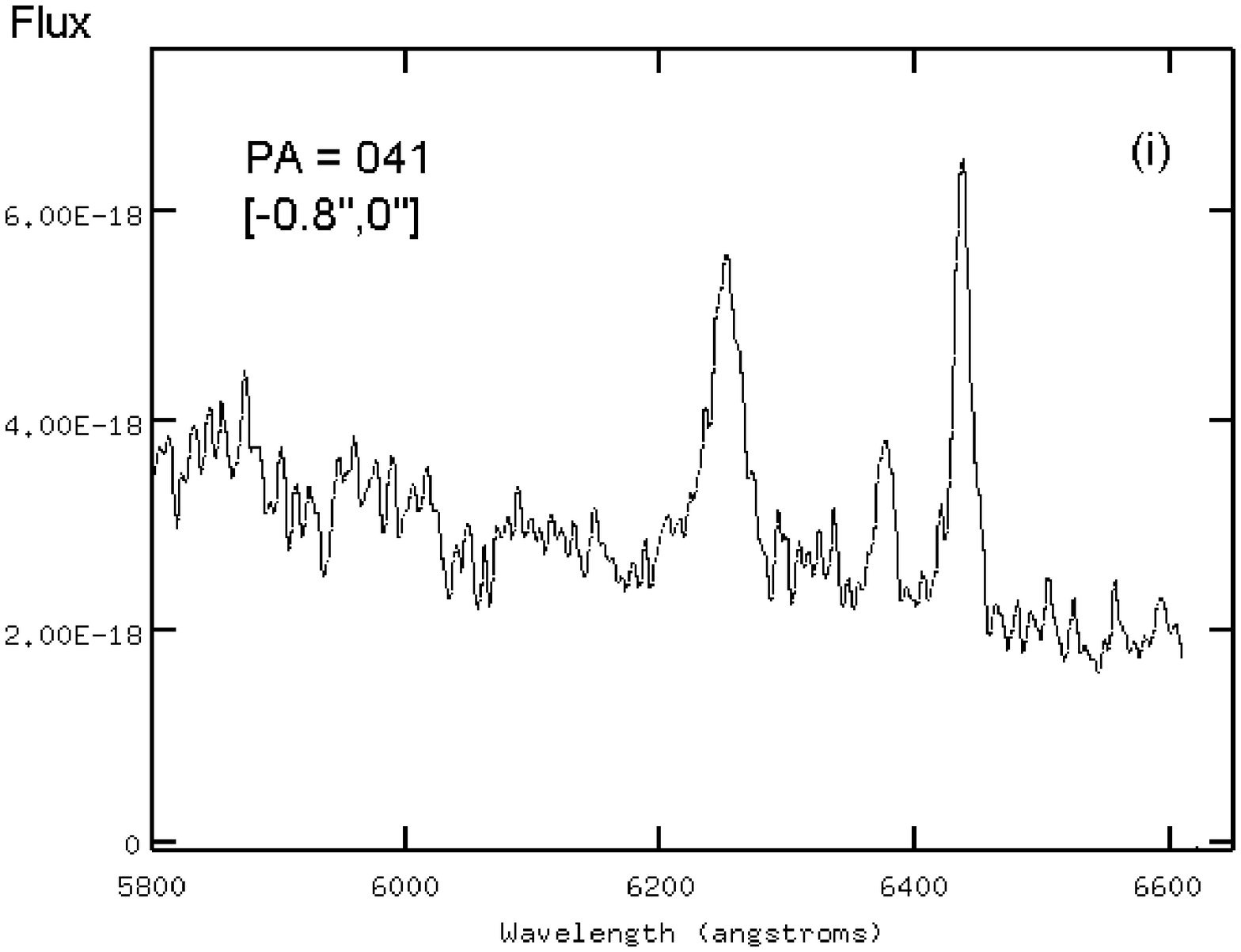}&
\includegraphics{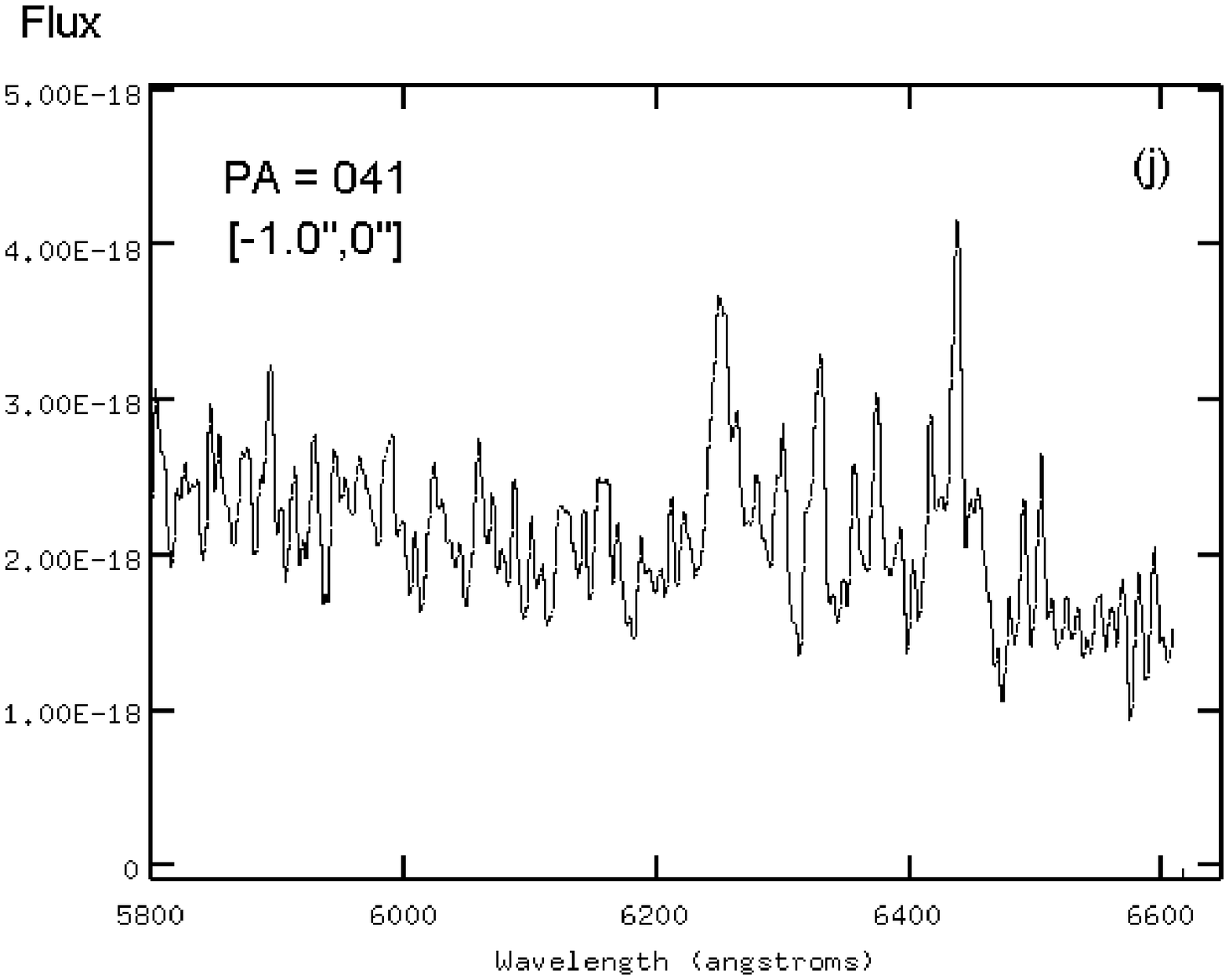} \cr
\end{tabular}
\vspace{8.0 cm}
\addtocounter{figure}{-1}
\caption {Contin.
 }
\label{fig22c}
\end{figure*}


\clearpage

\begin{figure*}
\vspace{12.0 cm}
\begin{tabular}{c}
\includegraphics{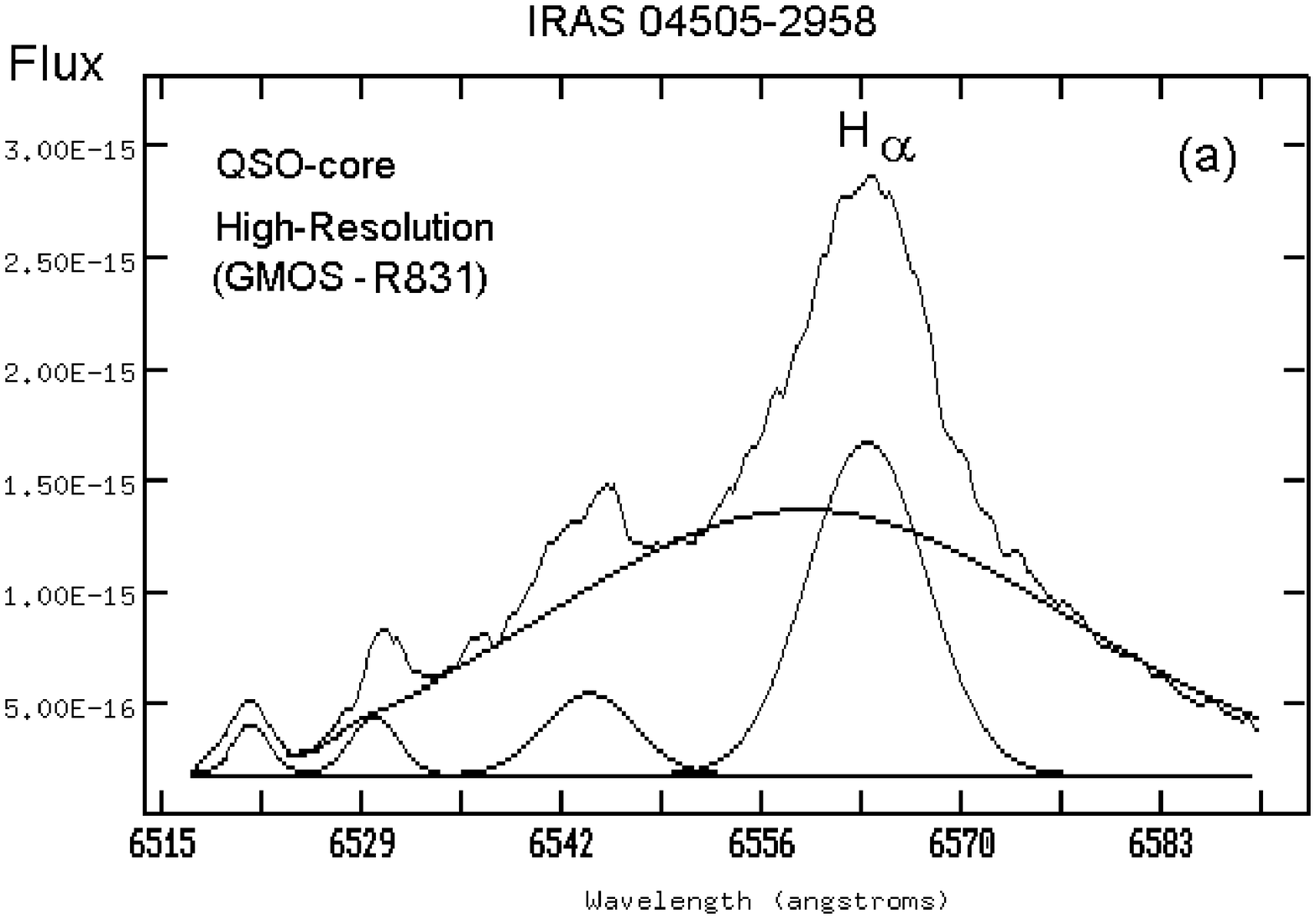}\cr
\end{tabular}
\vspace{8.0 cm}
\caption {
Fitting of the GMOS Spectra of the QSO-core (of IRAS 04505-2958), for
H$\alpha$ (a) and H$\beta$ (b), with pixel of 0.2$''$ and for a
seeing 0.4$''$. The plots show two different fit of the GMOS spectra:
for H$\alpha$, the fit was performed using a broad, an intermediate,
and OF components (for R831 high spectral resolution); and
for H$\beta$ using a broad and an intermediate components (for B600
medium resolution). See for details the text.
 }
\label{fig23}
\end{figure*}

\clearpage

\begin{figure*}
\vspace{12.0 cm}
\begin{tabular}{c}
\includegraphics{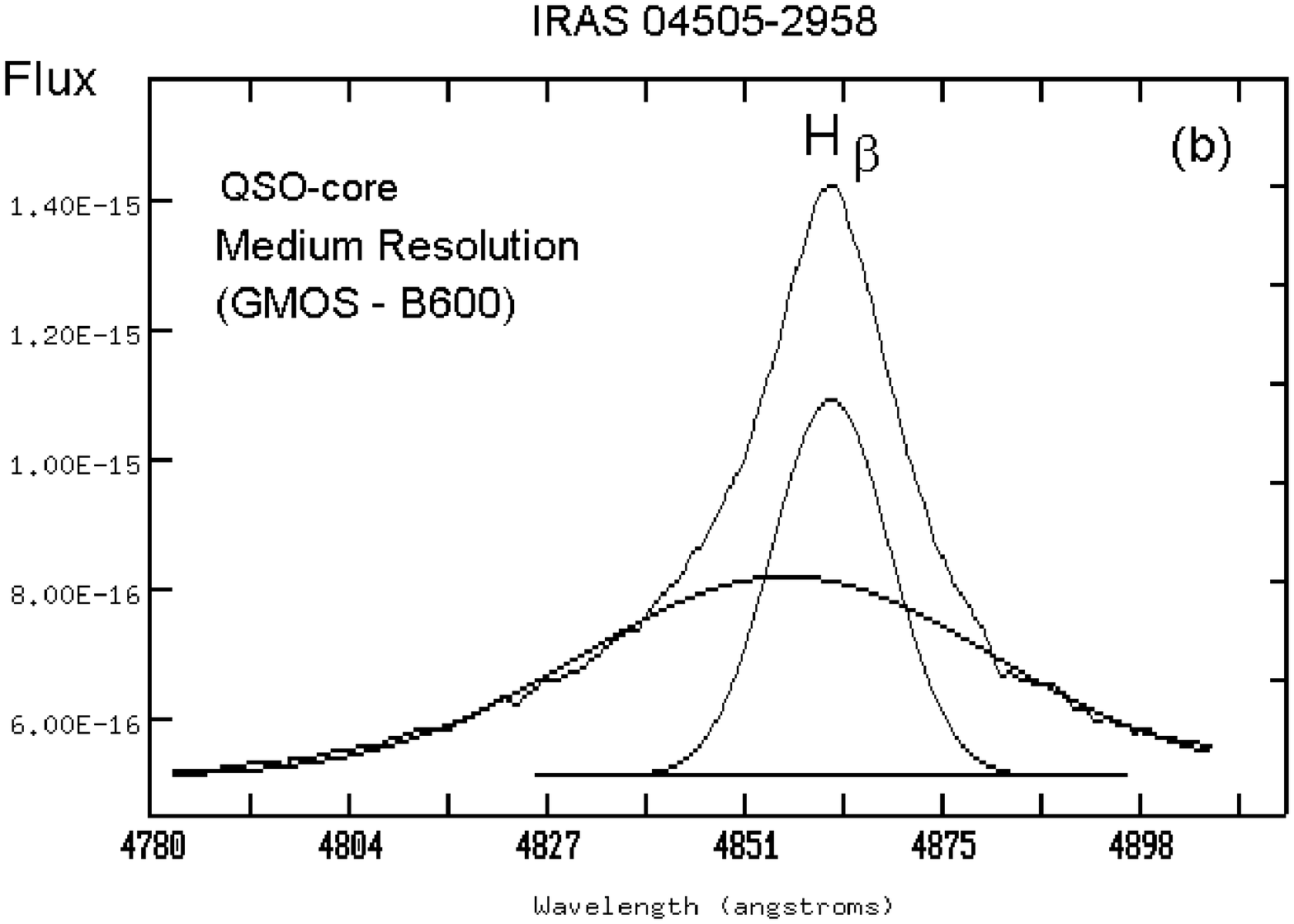}\cr
\end{tabular}
\vspace{8.0 cm}
\addtocounter{figure}{-1}
\caption {Cont.
}
\label{fig23c}
\end{figure*}


\clearpage

\begin{figure*}
\vspace{12.0 cm}
\begin{tabular}{c}
\includegraphics{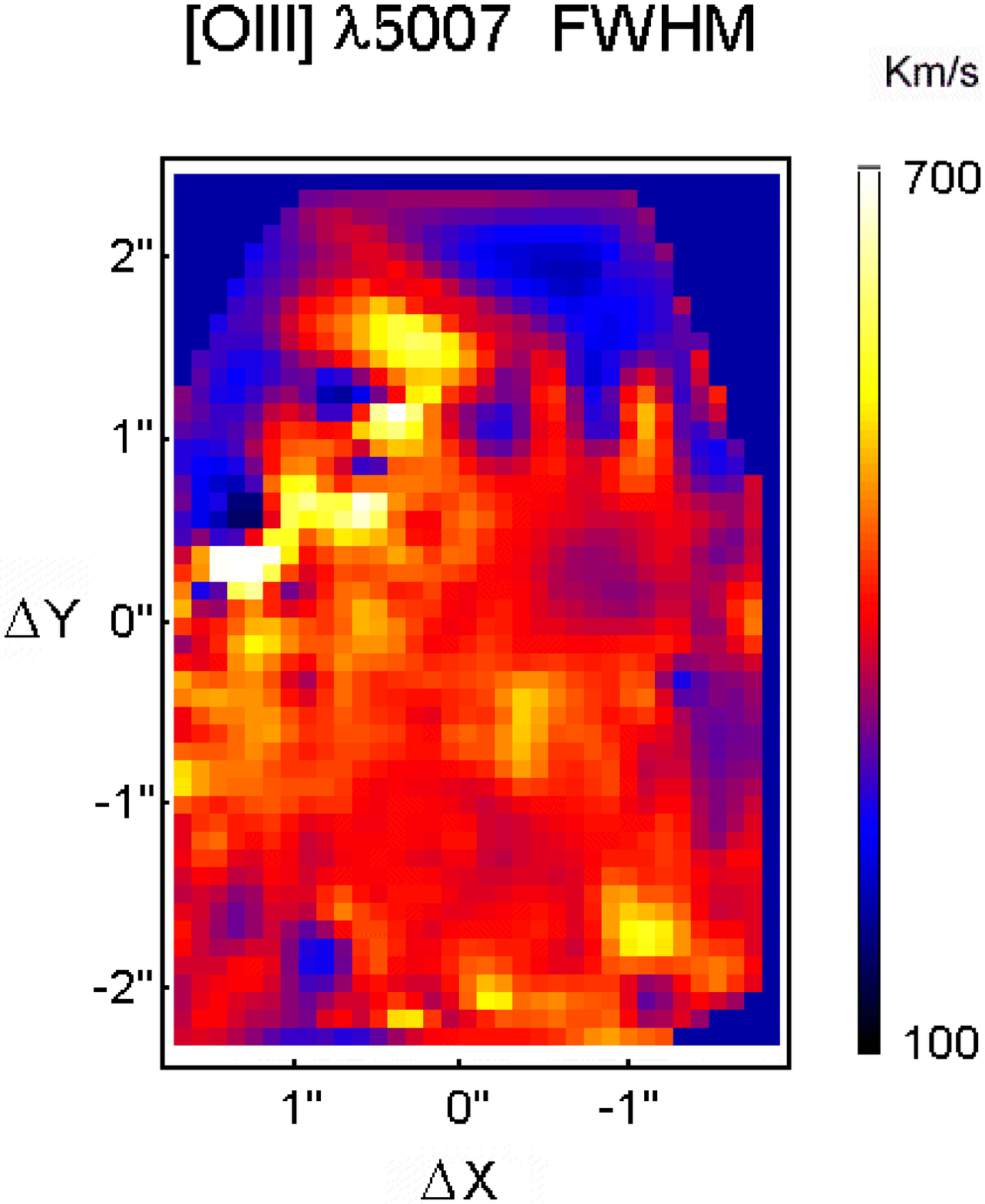}\cr
\end{tabular}
\vspace{8.0 cm}
\caption {
GMOS map of the width/FWHM of the [O {\sc iii}]$\lambda$5007 emission
line, for IRAS 04505-2958.
Showing high values of width/FWHM in the region of the shell S3. For details
see the text.
 }
\label{fig24}
\end{figure*}


\clearpage

\begin{figure*}
\vspace{12.0 cm}
\begin{tabular}{c}
\includegraphics{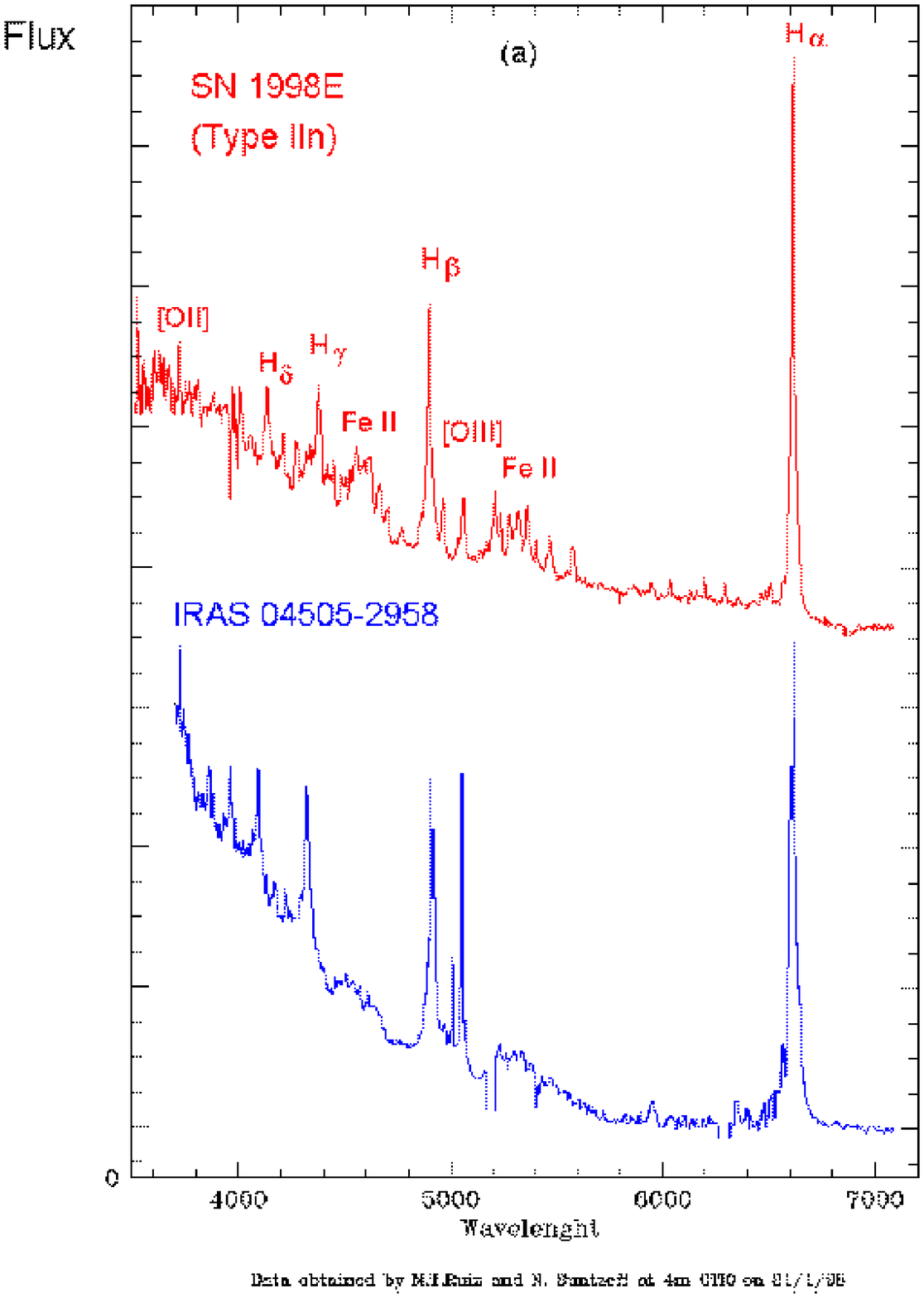}\cr
\end{tabular}
\vspace{8.0 cm}
\caption {
Optical spectra of SN 1998E, and IRAS 04505-2958 (a), plus the prototype
of NLS1 {\sc i} ZW 1 (b).
The SN data are from M. T. Ruiz and Sunzeff (2009, private communication),
obtained at CTIO on 1998 January 31, with 4 mt telescope.
}
\label{fig25}
\end{figure*}

\clearpage

\begin{figure*}
\vspace{12.0 cm}
\begin{tabular}{c}
\includegraphics{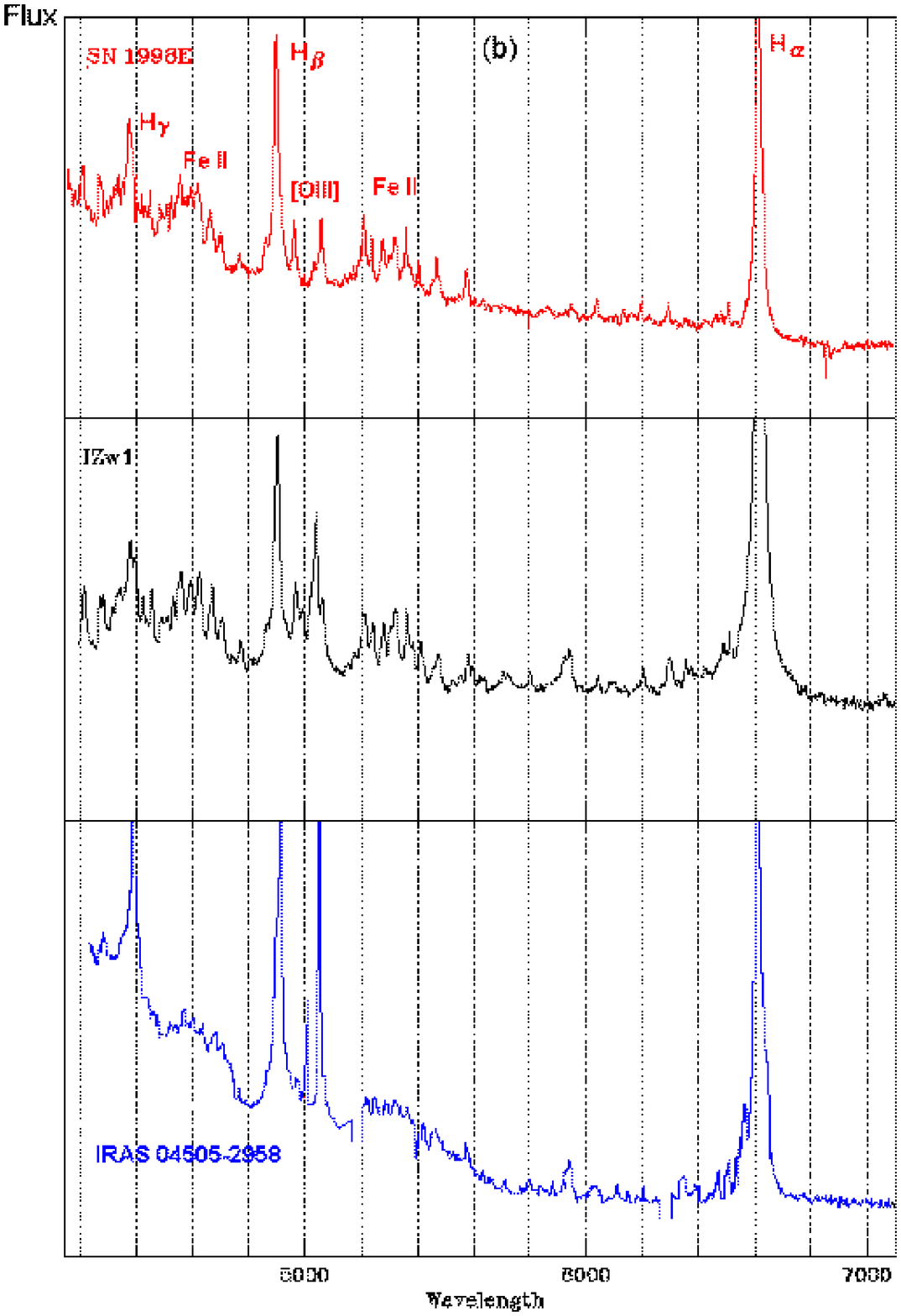}\cr
\end{tabular}
\vspace{8.0 cm}
\addtocounter{figure}{-1}
\caption {Cont.
}
\label{fig25c}
\end{figure*}

\end{document}